\documentclass[12pt]{nmsuth01}

\usepackage{latexsym}
\usepackage{amsmath,amsthm}
\usepackage[all]{xy}
\usepackage{graphicx}
\usepackage[most]{tcolorbox}

\usepackage[utf8]{inputenc}
\usepackage{amsmath,amssymb,amsfonts,mathrsfs,bbold}
\usepackage{yfonts}
\usepackage{amssymb, bm, amsmath, graphicx}
\usepackage{lipsum}
\usepackage{slashed}
\usepackage[normalem]{ulem}
\usepackage{url}
\usepackage{physics}

\usepackage[margin=1in]{geometry}

\newcommand{\pubitem}[1]{\noindent #1 \par~~\par}

\newcommand{\rvec}[1]{\overrightarrow{#1}}
\newcommand{\lvec}[1]{\overleftarrow{#1}}
\newcommand{\dlangle}{\Big\langle\!\!\Big\langle}
\newcommand{\drangle}{\Big\rangle\!\!\Big\rangle}

\theoremstyle{plain}

\theoremstyle{definition}
\theoremstyle{remark}

\newtheorem*{convention*}{Convention}
\renewcommand{\uline}[1]{\underline{#1}}
\newlength{\singlespace}
\newlength{\doublespace}
\begin{document}

	\makeatletter

	\renewcommand*\l@table{\@dottedtocline{1}{1.5em}{2.3em}}
	\renewcommand*\l@subsection{\@dottedtocline{2}{3.8em}{3.2em}}
	\renewcommand*\l@subsubsection{\@dottedtocline{3}{7em}{3.2em}}

	\renewcommand\thesection      {$\boldsymbol{\@arabic\c@section}$}
	\renewcommand\thesubsection   {\thesection $\boldsymbol{.\@arabic\c@subsection}$}
	\newcommand\thesectionub      {\@arabic\c@section}
	\newcommand\thesubsectionub   {\thesectionub .\@arabic\c@subsection}
	\renewcommand\thesubsubsection{\thesubsectionub .\@arabic\c@subsubsection}

	\makeatother
	
	\setlength{\baselineskip}{\doublespace}
	
	
	%
	%
	\pagenumbering{roman}
	\pagestyle{empty}
\renewcommand{\rmdefault}{2}
~\\~\\
\vspace{-48pt}
\begin{center}
	A NEW LOOK AT SMALL-$x$ HELICITY PHENOMENOLOGY: INVESTIGATING UNCERTAINTIES AND THE IMPACT OF POLARIZED PROTON-PROTON DATA     \\~\\
	\vspace{0.075in}
	BY\\
	\vspace{-10pt}
	Nicholas Baldonado, B.S., M.S.
\end{center}

\vspace{.65in}
\begin{center}
	A dissertation submitted to the Graduate School\\
	\vspace{-10pt}
	in partial fulfillment of the requirements\\
	\vspace{-10pt}
	for the degree \\
	\vspace{-10pt}
	Doctor of Philosophy
\end{center}

\vspace{.65in}
\begin{center}
	Major: Physics\\
	\vspace{-10pt}
	Concentration: Nuclear Theory\\
\end{center}

\vspace{.65in}
\begin{center}
	NEW MEXICO STATE UNIVERSITY\\
	\vspace{-10pt}
	LAS CRUCES, NEW MEXICO\\
	\vspace{-10pt}
	April 2025
\end{center}


\pagebreak
\hspace{0pt}
\vfill
\begin{center}
	\copyright~Copyright by \\
	Nicholas Baldonado \\
	2025
\end{center}
\vfill
\hspace{0pt}
\pagebreak

	\pagestyle{plain}
\noindent

\setlength{\baselineskip}{\singlespace}
\vspace{0.1in}
\begin{flushleft}
	
	\small
	Nicholas Baldonado
	
	\vspace{4pt}
	\hrule
	\vspace{4pt}
	
	\large
	\textit{Candidate}
	
	\vspace{0.3in}
	
	\small
	\textit{Physics}
	
	\vspace{4pt}
	\hrule
	\vspace{4pt}
	
	\large
	\textit{Major}
	
	\vspace{0.4in}
	
	\small
	This Dissertation is approved on behalf of the faculty of New Mexico State University, and it is acceptable in quality and form for publication:
	\begin{center} 
	\end{center}
	
	
	\large
	\textit{Approved by the thesis Committee:}
	
	\vspace{0.2in}
	
	\small
	\textit{Dr. Matthew Sievert}
	\vspace{4pt}
	\hrule
	\vspace{4pt}
	\large
	\textit{Chairperson}
	
	\vspace{0.2in}
	
	\small
	\textit{Dr. Igor Vasiliev}
	\vspace{4pt}
	\hrule
	\vspace{4pt}
	\large
	\textit{Committee Member}
	
	\vspace{0.2in}
	
	\small
	\textit{Dr. Michael Paolone}
	\vspace{4pt}
	\hrule
	\vspace{4pt}
	\large
	\textit{Committee Member}
	
	\vspace{0.2in}
	
	\small
	\textit{Dr. Joseph Burchett}
	\vspace{4pt}
	\hrule
	\vspace{4pt}
	\large
	\textit{Committee Member}
	
	
	
\end{flushleft}

\setlength{\baselineskip}{\doublespace}





	%
\begin{center}
	DEDICATION
\end{center}

I dedicate this dissertation to my wife, Elsa. She has been an immeasurable source of inspiration, motivation, support, and love, throughout my academic journey and I could not have achieved this goal without her. I could only begin to pay her the respect she is due if this dedication page were as long as this entire dissertation.


	%
\begin{center}
	ACKNOWLEDGMENTS
\end{center}

This culmination of my work would not be possible without the guidance of my advisor, Dr. Matthew Sievert. He always made himself available for discussion no matter his workload, and his ability to provide positive, constructive feedback on every aspect of my academic career was invaluable. I thank him for his efforts, and I am proud to be among his first set of graduate students. 

I also could not have made it this far without the help of Daniel Adamiak who taught me most of what I know regarding phenomenology. Thank you for responding to my dozens of emails and messages and for your helpful advice from grad student (and doctor) to grad student. The same goes for Daniel Pitonyak and Yuri Kovchegov who deserve recognition for taking the time to discuss with me the many nuanced aspects of this field.

I'd of course like to thank my family; my parents, Ernie and Nadine, and my brothers, Alex and Noah, always believed I'd become a doctor, and I hope to have made them proud in the act of completing this dissertation. 

I am grateful to my group of friends (I'm not naming each of you because I want this to fit on one page) for providing me some much needed comic relief, and our hours of gaming made this academic journey all the more bearable (except those nights when we lost a lot).

Lastly, I reiterate my immense gratitude to my wife, Elsa; I dedicated this dissertation to you and you also deserve a second thanks here. I could not have accomplished a fraction of my work without you by my side, and I cannot possibly thank you enough in so few words. 

I am thankful to you all for helping me get to his point. In return you each get a portion of this acknowledgments page, which you have to admit is pretty neat. 


	%
\begin{center}
	VITA
\end{center}
\begin{flushleft}
	\begin{tabular}{ll}
		April 27, 1998 &  Born in Albuquerque, New Mexico, USA
		\\
		& \\
		2016-2020        &  B.S., New Mexico Institute of Mining and Technology, \\
		& Socorro, New Mexico, USA
		\\
		& \\
		2020-2023        &  M.S., New Mexico State University, Las Cruces, New Mexico, USA
		
		\\
		& \\
		2021-Present       & Graduate Research Assistant, New Mexico State University, \\
		& Las Cruces, New Mexico, USA
		
	\end{tabular}
\end{flushleft}
\vspace{0.1in}
\begin{center}
	PROFESSIONAL  AND HONORARY SOCIETIES
\end{center}
\begin{flushleft}
	American Physical Society
\end{flushleft}
\vspace{0.1in}
\begin{center}PUBLICATIONS
\end{center}
\setlength{\baselineskip}{\singlespace}
\pubitem{Julia I Deitz, Joshua D Sugar, Boris Kiefer, Nicholas S Baldonado, Andrew A Allerman, and Mary H Crawford. \newblock Correlative electron energy-loss spectroscopy bandgap mapping and dft
	modeling in algan diodes. \newblock Microscopy and Microanalysis, 28(S1):2010–2011, 08 2022.}
\pubitem{Daniel Adamiak, Nicholas Baldonado, Yuri V. Kovchegov, W. Melnitchouk, Daniel Pitonyak, Nobuo Sato, Matthew D. Sievert, Andrey Tarasov, Yossathorn Tawabutr \newblock Global analysis of polarized DIS \& SIDIS data with improved small-$x$ helicity evolution. \newblock Phys. Rev. D, 108(11):114007, 2023.}
\pubitem{Matthew Sievert and Nicholas Baldonado. \newblock Global Analyses of Polarized DIS \& SIDIS from Small-x Evolution. \newblock PoS, SPIN2023:185, 2024}
\pubitem{Daniel Adamiak, Nicholas Baldonado, Yuri V. Kovchegov, Ming Li, W. Melnitchouk, Daniel Pitonyak, Nobuo Sato, Matthew D. Sievert, Andrey Tarasov, Yossathorn Tawabutr \newblock First study of polarized proton-proton scattering with small-$x$ helicity evolution. \newblock arXiv: 2503.21006 [hep-ph]}
\setlength{\baselineskip}{\doublespace}
\begin{center}
	FIELD OF STUDY
\end{center}
\begin{flushleft}
	Major Field: Nuclear Theory, High Energy Quantum Chromodynamics
\end{flushleft}
	%
\begin{center}
	ABSTRACT
\end{center}
\vspace{0.3in}
\begin{center}
	A NEW LOOK AT SMALL-$x$ HELICITY PHENOMENOLOGY: INVESTIGATING UNCERTAINTIES AND THE IMPACT OF POLARIZED PROTON-PROTON DATA
	\\
	BY
	\\
	Nicholas Baldonado, B.S., M.S.
\end{center}
\vspace{0.3in}
\begin{center}
	Doctor of Philosophy
	
	New Mexico State University
	
	Las Cruces, New Mexico, 2025
	
	Dr. Matthew D. Sievert , Advisor
	
\end{center}
\vspace{0.3in}
\hspace{\parindent}
This dissertation highlights the contributions I have made to the field of theoretical nuclear physics, specifically in the realm of high-energy Quantum Chromodynamics (QCD). High-energy QCD itself is a robust subject, and my research is more refined to the sub-field of small-$x$ spin physics; small-$x$ physics is characterized by high-energy collisions, high density nuclear targets, and is well-suited for the Color Glass Condensate (CGC) effective field theory. Small-$x$ \textit{spin} physics takes the ultra-relativistic description of high-energy QCD and gives special attention to the spin-dependent interactions that are suppressed by powers of the (large) center-of-mass energy. My expertise, as described by the contents of this manuscript, lies in exploring the theoretical development and phenomenological implications of the KPS-CTT small-$x$ helicity evolution equations, a mathematical rubric that allows one to make predictions of the quarks' and gluons' distributions of spin at energies beyond those capable by current or future particle colliders. These predictions are heavily influenced by the initial conditions provided to the mathematical formulae, and the initial conditions are determined through analyses of world polarized (spin-dependent) data. My contributions focus on Bayesian parameter analysis used in an effort to understand theoretical uncertainties, numerical and analytical calculations to discretize and cross-check the small-$x$ helicity evolution equations, and the incorporation of a new observable into the pool of global data we analyze. The results of such work show that the net amount of spin that quarks and gluons contribute to the proton in the small-$x$ regime is predicted to be negative and/or potentially small; a global analysis of polarized deep-inelastic scattering (DIS) and semi-inclusive deep-inelastic scattering (SIDIS) data resulted in a net small-$x$ spin prediction that can potentially be large and negative, but new results with the inclusion of data for single-inclusive jet production in polarized proton-proton ($pp$) collisions now estimate that the net amount of parton spin at small $x$ is small, with 1-$\sigma$ uncertainty that spans zero.

	\tableofcontents
	\newpage
	\listoftables
	\addcontentsline{toc}{section}{LIST OF TABLES}
	
	\newpage
	\listoffigures
	\addcontentsline{toc}{section}{LIST OF FIGURES}
	\newpage
	\pagenumbering{arabic}

\section{$\boldsymbol{\mathrm{INTRODUCTION}}$}\label{intro}

\subsection{$\boldsymbol{\mathrm{Proton~Spin~Puzzle}}$}

Our comprehensive theoretical understanding of the atom has had a tremendous impact on the world as we know it; through our studies of Quantum Electrodynamics (QED) we have developed modern technologies for communication, medicine, astronomy, and more just from the interactions that occur between a nucleus and its electrons in an atom. It is the analogous wealth of knowledge hidden within the structure of nucleons themselves that nuclear physicists explore via Quantum Chromodynamics (QCD). The field of nuclear physics has only been accelerating since its inception, and it is currently in an era of new growth heralded by the announcement of the Electron-Ion Collider, a large step forward in our attempts to map out and understand the structure, behavior, and properties of hadrons \cite{Accardi:2012qut}. One such property of which we are sorely in need of more information is the proton's spin; more accurately, how the proton obtains its spin from the dynamics of its constituent quarks and gluons. It is well established that the proton is a fermion with a spin of 1/2 (measured in units of $\hbar$), and in the low-energy quark model picture of a proton this can be readily explained as a sum two spin-up quarks and one spin-down quark (quarks as well carrying a spin of 1/2); the issue with this assumption was famously exposed by the European Muon Collaboration (EMC) in the 1980's. While the 3-valence quark model would predict that the proton's spin comes solely from the spins of the constituent quarks, the EMC found that quarks contribute less than 24\% of the proton's spin \cite{EuropeanMuon:1987isl, Ashman:1989ig}. Nearly 40 years on from the EMC's discovery, the search for the proton's missing spin continues, with a sample of reviews found in Refs.~\cite{Aidala:2012mv, Accardi:2012qut, Leader:2013jra, Aschenauer:2013woa, Aschenauer:2015eha, Boer:2011fh, Proceedings:2020eah, Ji:2020ena, AbdulKhalek:2021gbh}

There are multiple ways to separate the channels through which partons can contribute to the proton's spin, but I turn the reader's focus towards the simple and intuitive Jaffe-Manohar spin sum rule \cite{Jaffe:1989jz} below
\begin{equation}\label{spin_sum}
	S_q + L_q + S_G + L_G = \frac{1}{2}.
\end{equation}
Here the distribution of a proton's spin is quite simple: the total spin of the proton is a sum of the total angular momentum of the partons, those being the orbital angular momentum of the quarks and gluons, $L_{q/G}$, and the spin of the quarks and gluons, $S_{q/G}$. In this way I know that I must not only give attention to the quarks' spins, but also the spins of the gluons (which are bosons with an integer spin of 1$\hbar$), as well as the 3D dynamics of the quarks and gluons and how their orbital angular momentum (OAM) is distributed.

\begin{figure}[ht]
	\begin{center}
		\includegraphics[width=0.5\textwidth]{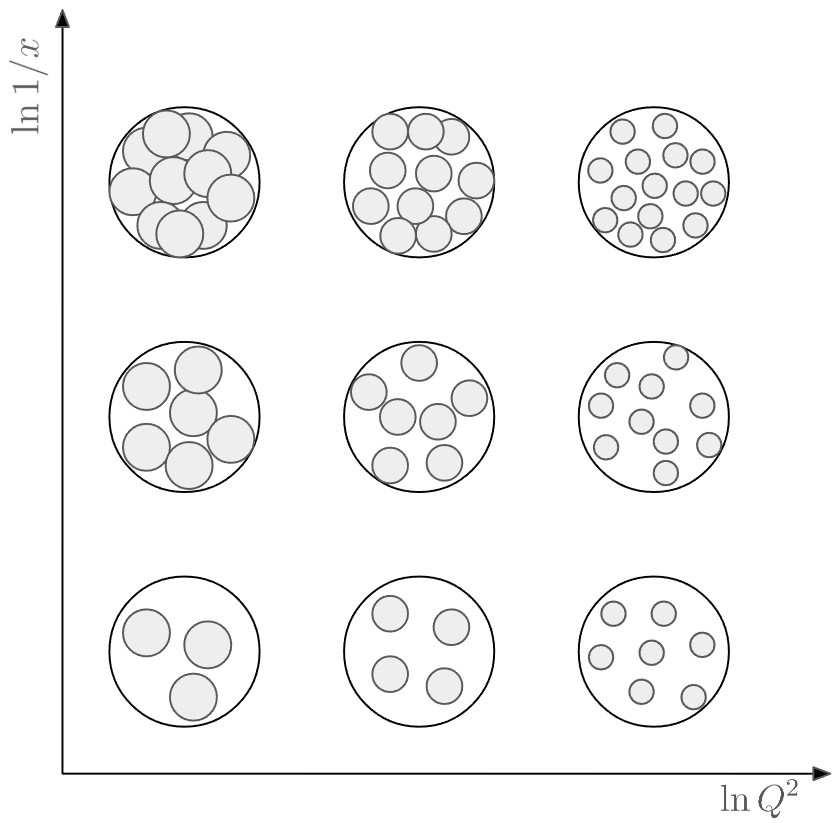} 
		\caption{The distribution of partons in the transverse plane as a function of $\ln 1/x$ and $\ln Q^2$.}
		\label{fig:Q2_x_Diagram}
	\end{center}
\end{figure}

The focus of my research, and thus this dissertation, lies with the spin contributions of the quarks and gluons, $S_{q/G}$, so I will take a short moment to discuss them in more detail. First, it is important to note that the picture of the proton is not static, which is to say that the composition of the proton looks different depending on a number of factors; two factors which are ever-present in this discussion around small-$x$ physics is the momentum scale called $Q^2$, and the titular variable $x$. These variables will be given more attention to detail in the next section, but suffice it to say for now that at large $Q^2$ we are essentially looking at a high-resolution picture of the proton which resolves a larger number of partons in the proton wavefunction than at small $Q^2$, and at small $x$ we are looking at a picture of the proton that appears much more densely populated by quarks and gluons than at large-$x$, as exemplified by Fig.~\ref{fig:Q2_x_Diagram}. With this information in mind, I can discuss the total spin contributions in a more nuanced way; the spin contributions that I aim to explore are those given at a fixed momentum scale of $Q^2$, and are determined by the integrals
\begin{equation}\label{spin_integrals}
	S_q(Q^2) = \frac{1}{2}\sum_q \int\limits_0^1 dx\,\Delta q (x,Q^2) = \frac{1}{2}\int\limits_0^1 dx\,\Delta\Sigma (x,Q^2) \,, \qquad S_G = \int\limits_0^1 dx \, \Delta G (x,Q^2) \,,
\end{equation}
where I have now introduced new objects $\Delta q$, $\Delta\Sigma$, and $\Delta G$, called helicity parton distribution functions (hPDFs). The hPDFs shown here are the main subject of research as they provide the distribution of a parton's helicity as a function of the two variables $x$ and $Q^2$. Helicity is the projection of a particle's spin onto its momentum direction; for the purposes of this work, I consider the proton to be longitudinally polarized and thus treat helicity to be interchangeable with spin. There are methods of extracting spin information about the proton through differently polarized systems, i.e. from the distributions of transversely or longitudinally polarized partons from transversely polarized protons/nucleons, but the relationship is not as intuitive or clear as is the case for helicity PDFs from longitudinally polarized protons/nucleons. The goal is clear then: I aim to determine what contribution the quarks' and gluons' spins provide to the spin of the proton by uncovering the (small) $x$-dependence of the hPDFs and subsequently integrating them. 

With the goal at hand, I can now discuss some of the more technical details surrounding small-$x$ physics (and of course small-$x$ \textbf{spin} physics) to bring the reader up to speed on what makes small-$x$ physics so important from a theory perspective, and how it is even more influential in the polarized (spin-dependent) sector.

\subsection{$\boldsymbol{\mathrm{A~Natural~Starting~Point\!:~Deep~Inelastic~Scattering}}$}

\begin{figure}[ht]
	\begin{center}
		\includegraphics[width=350 pt]{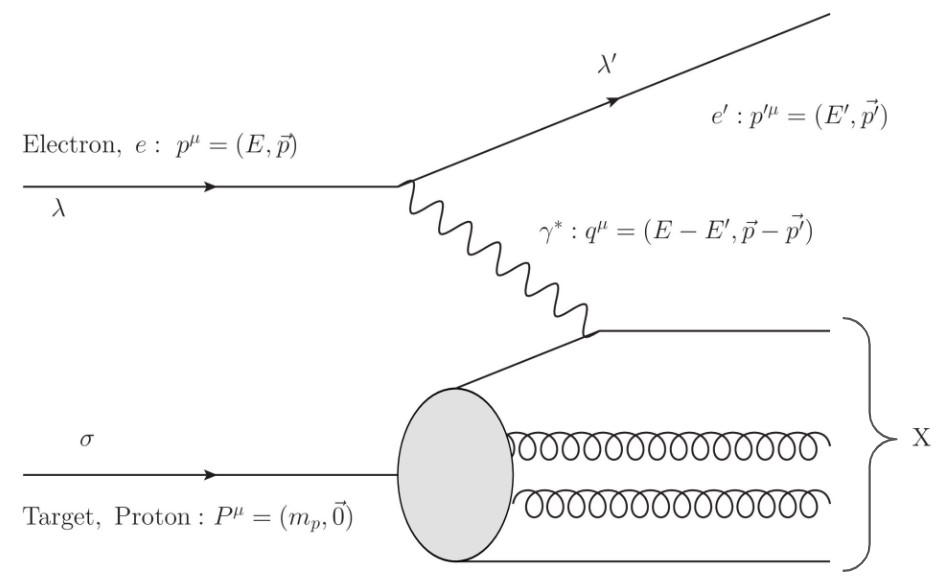} 
		\caption{Deep Inelastic Scattering of an electron scattering off an internal quark/antiquark inside the proton wavefunction.}
		\label{fig:Lowx_DIS}
	\end{center}
\end{figure}

Any discussion on small-$x$ physics, be it polarized or unpolarized, will typically begin the conversation around the process known as Deep Inelastic Scattering or DIS. DIS processes are very useful because it allows for some powerful manipulation of the process that can be used to highlight distinct behaviors of QCD, especially as the collision energy increases. One can learn a lot about a given process by studying the Lorentz invariant quantities, so let us analyze the DIS process in the proton's rest frame as depicted in Fig.~\ref{fig:Lowx_DIS}.

In this diagram I have drawn an electron of momentum $p^{\mu} = (E,\vec{p})$ scattering off of a proton with momentum $P^{\mu} = (m_p,\vec{0})$; this process is mediated by a virtual photon of momentum $q^{\mu} = (E-E', \vec{p}-\vec{p'})$ which does not scatter directly off the proton itself, but of a quark (or antiquark) inside the proton's wavefunction. The final state of the interactions sees a recoiled electron with momentum $p^{\prime \mu} = (e', \vec{p'})$, and an arbitrary hadronic final state denoted by $\mathrm{X}$. With three independent four-momenta I can construct three Lorentz-invariant quantities which are typically defined as: the virtuality of the photon $Q^2 \equiv -q^2$, Bjorken-$x$ (often referred to just as ``$x$") $x_{Bj}\equiv \tfrac{Q^2}{2P\cdot q}$, and the unnamed variable $y = \tfrac{E-E'}{E}$. As you may have determined for yourself, I have re-introduced the important variables $x$ and $Q^2$ that appear in the hPDFs above! While there are a number more invariants to be defined, I will only bring attention to the Mandelstam variable $s$, which provides the center-of-mass energy squared of a given interaction; in this case I will specify the virtual photon-proton interaction such that $s = (P+q)^2 \equiv m_X^2$, where $m_X$ is the final state invariant mass. Expanding the binomial recovers $s = 2P\cdot q + q^2 + m_p^2$. If I take the opportunity now to assume that I am working with a high-virtuality collision such that $Q^2 \gg m_p^2$, I can write the Lorentz invariant quantity
\begin{equation}
	Q^2 = 2P\cdot q + m_p^2 - s \approx 2P\cdot q - s \\
\end{equation}
which signifies the inelastic nature of the process, $m_X > m_p$. I can also define another invariant
\begin{align}\label{Bjorken_x}
	x_{Bj} &= \frac{Q^2}{2P\cdot q} = \frac{Q^2}{Q^2 + s}\notag \\
	&\therefore\; 0\leq x_{Bj} \leq 1.
\end{align}
If $Q^2$ is very large (deeply virtual), then the process must also be deeply inelastic, meaning large-$s$. I can further consider the limit where $s$ is very large, $s \gg Q^2$, known as the Regge limit, in which $x \approx \tfrac{Q^2}{s} \ll 1$. This last inequality is what allows us to also call this the small-$x$ limit.

Collinear factorization \cite{Collins:1989gx} is the process of defining an arbitrary factorization scale $\mu^2$ such that a given cross-section can be broken down into a non-perturbative ``soft" part and a high-energy ``hard" part that can be computed perturbatively. In DIS this is very useful because it allows to choose the momentum scale $Q^2$ as our factorization scale, and thus the perturbative part of our cross-section would be the high-energy scattering of the virtual photon off the struck quark (or antiquark) in the proton's wavefunction, with the non-perturbative part being the probability of that struck quark (or antiquark) being found in the proton's wave function with the given kinematics of $x$ and $Q^2$. This factorization is possible by the assumption that the virtuality $Q^2$ is very large; the DIS cross-section can nominally be expanded in powers of $1/Q^2$, where the leading term is factorized as described above and ``mixing" terms that are not factorized are higher-order and can thus be ignored since $1/Q^2 \gg (1/Q^2)^n$. Given that the leptonic part of the interaction is comparably trivial, the hadron part of the DIS cross-section, at a large fixed $Q^2$, in collinear factorization is written as
\begin{equation}\label{DIS_CS}
	\sigma^{\gamma^*+P} \propto F_2(x,Q^2) = \sum_f\int\limits_0^1d\xi \; q_f(x,Q^2)\;C_f(\frac{x}{\xi})
\end{equation}
where $q_f(x,Q^2)$ is the parton distribution function and $C_f$ is a perturbatively calculable coefficient function. The integral itself defines the unpolarized structure function of the proton, $F_2(x,Q^2)$, which carries information about the internal structure of the proton. The parton distribution functions gives the probability of finding a parton (written above as a quark with flavor $f$) in the proton with longitudinal momentum fraction $x$ (non-trivially equivalent to Bjorken-$x$ when $Q^2$ is large). For polarized DIS the form would be quite similar, with the exceptions that the cross-section would be proportional to the polarized structure function $g_1$, one would no longer sum over all parton polarization states, and the parton distribution function would be replaced with the helicity PDF $\Delta q_f$ (or $\Delta q$ if one absorbs the flavor dependence into the nomenclature such that $\Delta q_u \equiv \Delta u$).

Our discussion is becoming quite refined at this point because I have laid out our subjects for research in a relatively straightforward way. As mentioned at the beginning of the introduction, this research is attempting to understand the helicity parton distribution functions, however in our discussion on collinear factorization I pointed out that the parton distribution functions are non-perturbative. As a theorist, this means that I do not calculate the PDFs (or hPDFs) directly, but must instead investigate them through the other two parts of Eq.~\eqref{DIS_CS}, the cross-section (structure functions) and the perturbatively calculable hard-scattering process. The information one gains on the cross-section comes from experimental data and is a crucial piece that categorizes this research as phenomenology; though not yet defined, phenomenology is a method of testing predictions made by theory against experimental data. Data from particle colliders are typically in the form of structure functions or spin asymmetries; there are two unpolarized structure functions for DIS on a proton, $F_1$ and $F_2$, which are related by the Callan-Gross relation
\begin{equation}
	F_2(x,Q^2) = 2\,x\,F_1(x,Q^2).
\end{equation}
From Eq.~\eqref{DIS_CS} it is shown that we can learn information about the PDF $q_f(x,Q^2)$ in through the experimental measurements of the structure function and the perturbatively calculable coefficient functions. In the same way, we can learn about the helicity PDFs through experimental data of polarized observables like spin asymmetries or polarized structure functions and the perturbatively calculable hard-scattering part of the polarized DIS cross-section. The methodology of the phenomenological work will be expanded up in future sections, but for now I want to provide a statement that should be relatively unsurprising: experimental data is quite variable. By this I mean that the kinematics in a diagram like Fig.~\ref{fig:Lowx_DIS} are set by particle colliders and experimentalists, and as such the non-perturbative PDFs and hPDFs are only strictly measured at specific values of $Q^2$ and $x$ based on the various parameters of the given experiment. As a theorist, and in the effort of making phenomenological predictions, it would be useful to know the $x$ and $Q^2$ dependence of the PDFs and hPDFs, such that if data can provide a measurement at $Q_0^2$ or $x_0$, I could then compute their values at arbitrary $Q^2$ or $x$. This exact dependence is explored through evolution equations, which are the main topic for the following section.

\subsection{$\boldsymbol{\mathrm{Parton~Evolution}}$}\label{DGLAP_evo}

While the parton distribution functions are non-perturbative objects, I can at least treat part of the DIS scattering amplitude as perturbative - specifically, the hard scattering. Taking a closer look at Fig.~\ref{fig:Lowx_DIS}, I want to break the diagram into three main parts: the emission of a virtual photon from the incoming electron (a process well understood through QED), the interaction of the virtual photon with the struck quark (the hard scattering process), and the ``emission" of the struck quark from the proton (determined by the non-perturbative PDF). 

\begin{figure}[ht]
	\begin{center}
		\includegraphics[width=200 pt]{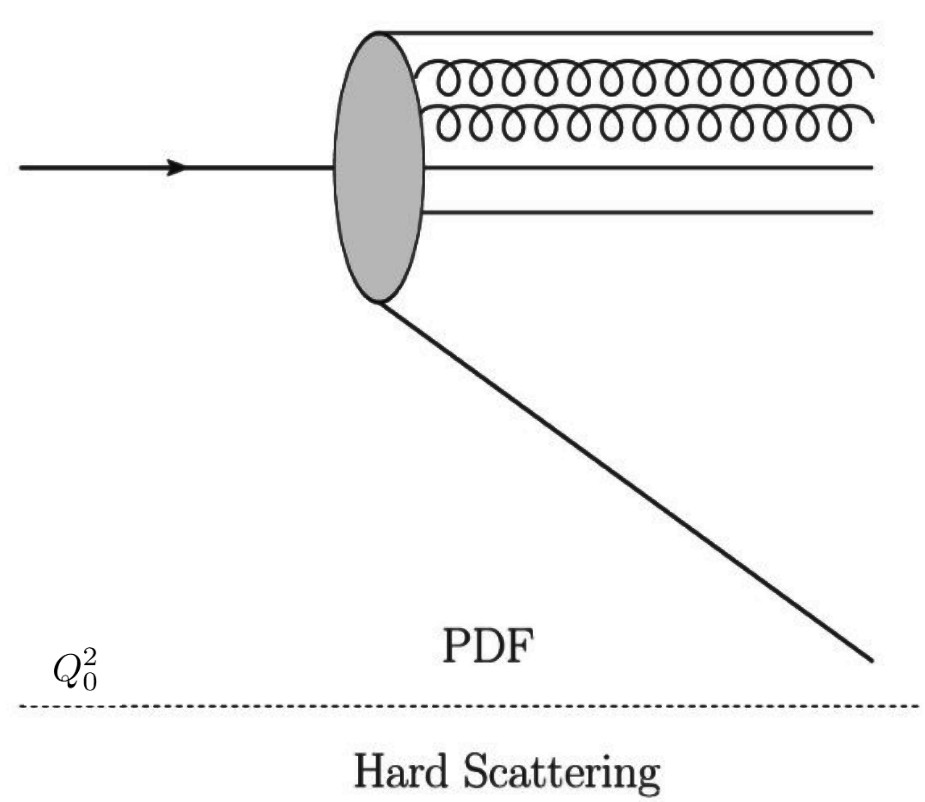} 
		\caption{A diagrammatic representation of the proton's wavefunctions. Above the dashed line is the amplitude that relates to the quark parton distribution function, while below the dashed line is where the quark undergoes the hard-scattering. The separation of scales is defined by $Q_0^2$.}
		\label{fig:PDF_LO}
	\end{center}
\end{figure}

\begin{figure}[ht]
	\begin{center}
		\includegraphics[width=450 pt]{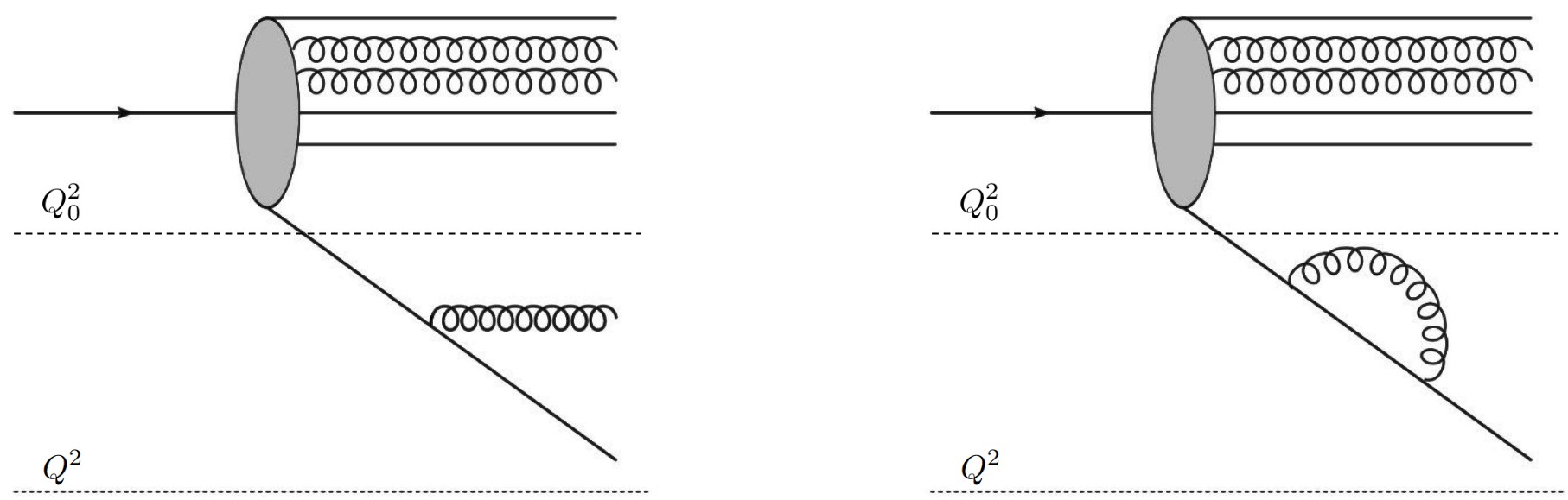} 
		\caption{A radiative correction to the leading-order parton distribution functions, shown as a radiative gluon emission from the proton's wavefunction prior to the hard scattering process. On the left is a real correction, while on the right is a virtual correction whose complex conjugate is not shown for brevity. The two horizontal lines represent the evolution from $Q_0^2$ to $Q^2$ and how that affects the PDF.}
		\label{fig:PDF_LO+gluon}
	\end{center}
\end{figure}

In Figs.~\ref{fig:PDF_LO} and \ref{fig:PDF_LO+gluon} are depictions of the proton's wavefunction. The horizontal dashed line is given to show the separation of the PDF and the hard scattering process; note that the diagram shown explicitly is only the proton wavefunction and that I would need to square it with its complex conjugate to form the PDF. The separation of the hard scattering process and the PDF is determined by the scale $Q^2$. Note that a perturbatively calculated correction at some value of $Q_0^2$ (like gluon emission) can effectively be absorbed into the definition of the PDF at some higher $Q^2 > Q_0^2$ as depicted by the two horizontal lines in Fig.~\ref{fig:PDF_LO+gluon}. It should be clear that the diagrams above show the quark parton distribution function, whereas a diagram with an outgoing gluon separated from the spectator partons would represent the gluon parton distribution function. To capture all radiative emissions of this sort would require four distinct vertices to account for the various emission processes; these are considered the splitting functions, and they describe the splitting of a quark into an outgoing quark with a gluon correction, $P_{qq}(z)$, the splitting of a gluon into quark with a quark emission $P_{qG}(z)$, the splitting of a quark into an outgoing gluon with a quark emission $P_{Gq}(z)$, and the splitting of gluon into an outgoing gluon with a gluon emission $P_{GG}(z)$. With these splitting functions defined, one now sees that Fig.~\ref{fig:PDF_LO+gluon} represents a real and virtual contribution to the $P_{qq}(z)$ splitting function. In this case $z$ is the fraction of the intermediate parton's longitudinal momentum (the parton pre-splitting) carried by the outgoing parton. In computing all diagrammatic contributions and subsequently integrating over the transverse momenta of all partons, it is found that these LO corrections to the PDFs are proportional to $\ln (Q^2/\Lambda_{QCD}^2)$, and carry one factor of $\alpha_s$. The log of $Q^2$ comes from the fact that $Q^2$ is treated as a UV cutoff for the transverse momentum integrals, and thus the larger $Q^2$ gets, the larger the phase-space of the interaction; this is a crucial aspect, considering that a larger $Q^2$ means a higher energy, and through the asymptotic freedom of QCD means a smaller coupling constant. As a result, the correction scales with the resummation parameter $\alpha_s \ln(Q^2/\Lambda_{QCD}^2)\sim1$; the term resummation parameter comes from the notion that diagrams of this type must be summed to all orders in perturbation theory since $(\alpha_s \ln(Q^2/\Lambda_{QCD}^2))^2 \sim \alpha_s \ln(Q^2/\Lambda_{QCD}^2) \sim 1$. The resummation is handled by the Dokshitzer-Gribov-Lipatov-Altarelli-Parisi (DGLAP) \cite{Gribov:1972ri, Altarelli:1977zs,Dokshitzer:1977sg}, and are summarized very neatly in \cite{Kovchegov:2012mbw} as
\begin{equation}\label{DGLAP}
	\frac{\partial}{\partial \ln Q^2}
	\begin{pmatrix}
		\Sigma(x,Q^2) \\
		G(x,Q^2)
	\end{pmatrix}
	= \frac{\alpha_s(Q^2)}{2\pi}\int\limits_x^1 \frac{dz}{z}
	\begin{pmatrix}
		P_{qq}(z) & P_{qG}(z) \\
		P_{Gq}(z) & P_{GG}(z)
	\end{pmatrix}
	\begin{pmatrix}
		\Sigma(x/z,Q^2) \\
		G(x/z,Q^2)
	\end{pmatrix}.
\end{equation}
The solution of the DGLAP evolution equations allows one to perform the exact analysis I set forth to achieve: take a PDF ($\Sigma(x,Q^2)$ for the sum of quarks and antiquarks of all flavors, $G(x,Q^2)$ for gluons) as probed by a photon of virtuality $Q_0^2$, and provide the result of the PDF when probed by a photon of an arbitrary higher virtuality $Q^2$. In effect, the DGLAP evolution equations allow one to travel from the left side of Fig.~\ref{fig:Q2_x_Diagram} to the right, increasing the transverse resolution of the proton through the relation $\Delta x_{\perp} \sim 1/Q$. The DGLAP evolution written above is for the unpolarized PDFs, while the polarized DGLAP evolution is largely similar, with the PDFs being replaced by the hPDFs $\Delta\Sigma(x,Q^2)$ and $\Delta G(x,Q^2)$, and the splitting functions replaced with the polarized splitting functions $\Delta P_{ij}(z)$.

The efficacy of the DGLAP evolution equations (at LO and with NLO corrections \cite{Mertig:1995ny, Vogelsang:1995vh, Vogelsang:1996im}) has been tested over the course of many PDF and hPDF extractions \cite{Gluck:2000dy, Leader:2005ci, deFlorian:2009vb, Leader:2010rb, Jimenez-Delgado:2013boa, Ball:2013lla, Nocera:2014gqa, deFlorian:2014yva, Leader:2014uua, Sato:2016tuz, Ethier:2017zbq, DeFlorian:2019xxt, Borsa:2020lsz, Zhou:2022wzm, Cocuzza:2022jye}, and while it has been extensively proven to predict the $Q^2$ dependence of PDFs and hPDFs, it has also been shown that the extrapolation of the PDFs towards smaller values of $x$ is not a strength of DGLAP evolution. Indeed, DGLAP is not intended to evolve PDFs to small $x$. DGLAP extractions that have been extrapolated to small $x$ see the uncertainty grow rapidly outside the region of $x$ where data was used to constrain the DGLAP initial conditions (see later chapters on the results of small-$x$ helicity global analyses). Just as an investigation of radiative corrections to the proton wavefunction led to the single-logarithmic resummation parameter $\alpha_s\ln (Q^2/\Lambda_{QCD}^2)$ and the DGLAP evolution equations, it is time to explore radiative corrections that bring in logarithmic enhancements when $x$ is small, and in turn find evolution equations that allow small-$x$ extractions of PDFs from large-$x$ initial conditions. 

\subsection{$\boldsymbol{\mathrm{Small}\text{-}x~\mathrm{Physics~and~Color~Dipoles}}$}\label{unpol_evo}

The subject of small-$x$ physics is important enough to warrant its own dedicated section; I will get to the unpolarized small-$x$ evolution equations shortly, but I must first establish some overarching changes to the formalism that occur when $x$ is small.

As mentioned before, the subject of this dissertation revolves around the behavior of hPDFs at small $x$ and fixed $Q^2$, and from that statement alone, one can see by arranging Eq.~\eqref{Bjorken_x} that small-$x$ is synonymous with high-energy:
\begin{equation}
	s \approx \frac{Q^2}{x}.
\end{equation}
This statement is simple, but it has far-reaching effects. With fixed $Q^2$ I am essentially fixing the Mandelstam variable $t$, meaning that the momentum transfer in a collision is fixed while at the same time the center-of-mass energy of the collision is allowed to grow very large. For light-like particles colliding, this essentially means that incoming particles' momenta are dominated by their longitudinal momenta, which is referred to as the eikonal limit. The eikonal limit can also be thought of as the limit where only leading powers of the center-of-mass energy are considered. Thus, in the eikonal limit, any scattering process that scales with powers of $1/s$ are ignored as being sub-eikonal - this is an important property that will be tossed aside in the following chapter.

The discussion of DGLAP evolution began with a focus on the parton distribution functions themselves since the $Q^2$ dependence was readily apparent as the UV cutoff of the transverse momenta integrals. For the case of small-$x$ evolution, it is more easily determined by finding corrections to the LO high-energy scattering diagrams. What I find is that at high energies the dominating QCD interactions are $t$-channel gluon exchanges in strong interactions, and the corrections to these interactions are not enhanced by logarithms of $\ln(Q^2/\Lambda_{QCD}^2)$, but instead come with the resummation parameter $\alpha_s\ln(1/x)\sim 1$. Therefore, to capture the $x$ dependence of these high-energy interactions, one must resum all diagrams that bring in a factor of $\ln(s)$ or $\ln(1/x)$. This is accomplished by the Balitsky-Fadin-Kuraev-Lipatov (BFKL) \cite{Balitsky:1978ic, Kuraev:1977fs} evolution, which evolves in $x$. In DGLAP evolution, I established that the logs of $Q^2$ came from a UV cutoff of the transverse momenta; the logs of $x$ in BFKL evolution instead come from an integral over the longitudinal momenta of the $s$-channel gluon emission. This integral is cut off by the rapidity interval of the interaction, defined by $Y = \ln(1/x)$, which corresponds to the maximum possible case where the longitudinal momentum of the emitted gluon carries all the longitudinal momentum of the proton. In Fig.~\eqref{fig:Q2_x_Diagram} one sees that small-$x$ evolution takes us from the bottom to the top of the diagram to generate more partons as $x$ get small; this shown rather intuitively in a diagrammatic representation of the BFKL ladder (more graciously depicted in \cite{Kovchegov:2012mbw}), which is shown below as being absorbed by the proton's wavefunction in a manner similar to the gluon emissions discussed in the previous section.

\begin{figure}[ht]
	\begin{center}
		\includegraphics[width=225 pt]{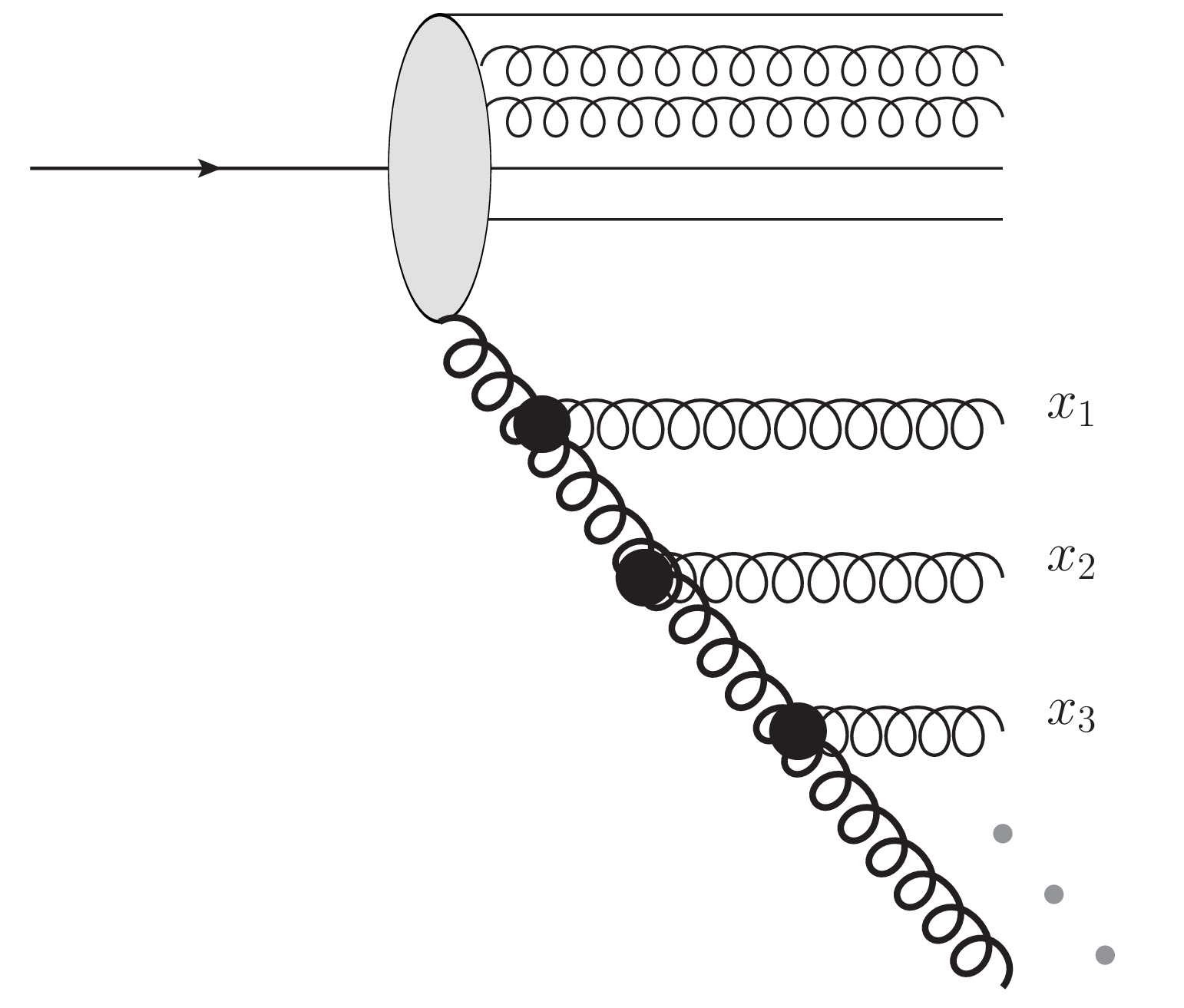} 
		\caption{The BFKL ladder absorbed into the proton wavefunction. It depicts a gluon emission that subsequently emits softer and softer gluons.}
		\label{fig:BFKL_Ladder}
	\end{center}
\end{figure}

This ladder shows subsequent emissions of gluons which are ordered as $x_1 \gg x_2 \gg x_3 ~ \cdot\cdot\,\cdot$; the increase of parton density as $x$ gets small is clear, as one can see that large-$x$ gluons will emit smaller-$x$ gluons in a repeating cascade. This cascade would eventually be halted by non-linear effects captured in the Balitsky-Kovchegov evolution equation (BK) \cite{Balitsky:1995ub, Balitsky:1998ya, Kovchegov:1999yj, Kovchegov:1999ua} that also describes the recombination of gluons. In summary, the picture is clear: at small $x$ the picture of the proton appears very dense and dominated by gluons, a state also referred to as the Color-Glass-Condensate \cite{Iancu:2000hn, Iancu:2001ad}.

\begin{figure}[ht]
	\begin{center}
		\includegraphics[width=300 pt]{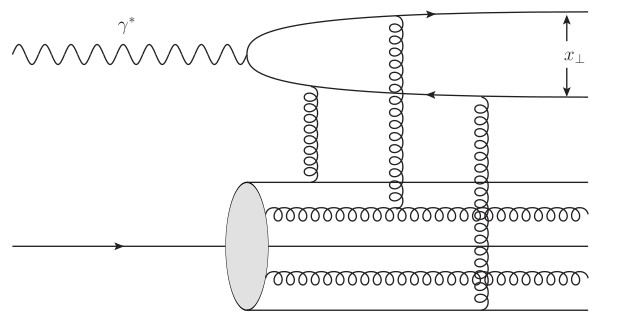} 
		\caption{Small-$x$ ``dipole picture" of DIS where the virtual photon splits into a $q\bar{q}$ pair that proceeds to interact with partons in the nuclear target's wavefunction.}
		\label{fig:smallx_DIS}
	\end{center}
\end{figure}

With the $Q^2$ and $1/x$ evolutions explained away, the mention of the Color-Glass-Condensate (CGC) brings to light the last part of the introduction: a shift in our degrees of freedom in high-energy QCD. Up until now, I have been discussing various corrections to the parton picture of the nucleus and how the DIS process involves virtual photons striking quarks in the proton wavefunction; the remainder of this dissertation will continue with a more recently developed approach using interactions between nuclei and color dipoles. Consider a virtual photon in the DIS process where I can write the virtual photon's momenta using the rules of light-cone perturbation theory (LCPT). In light-cone notation, any four-vector $v^{\mu}$ can be defined such that the light-cone plus and minus directions are $v^{\pm} = (v^0 \pm v^3)/\sqrt{2}$, leaving the transverse components to be a two-vector $\vec{v}_{\perp} = (v^1, v^2)$ with the magnitude fixed via $v^2 = v^+v^- - \vec{v}_{\perp}^2$. In this notation, at high energy, I can write the four-momentum of the virtual photon as
\begin{equation}
	q^{\mu} = \big(q^+, -\frac{Q^2}{2q^+}, \vec{0}\big),
\end{equation}
where it has momentum $q^+$ in the positive light-cone direction, effectively zero transverse momenta, and thus has momentum $-Q^2/2q^+$ in the light-cone minus direction. The longitudinal coherence length is given as $x^+ \approx \tfrac{2}{mx}$, which for a small-$x$ process is much larger than the size of the proton (or nucleus). The long coherence length implies that a quantum fluctuation that sees the virtual photon split into a $q\bar{q}$ pair would also be longer than the nucleus. With this in mind, one can have a substantially different picture of a small-$x$ DIS interaction: Fig.~\ref{fig:Lowx_DIS} looks like a virtual photon probing into a proton or nucleus and striking a valence quark, whereas at small $x$ one has a case as depicted in Fig.~\ref{fig:smallx_DIS} where a $q\bar{q}$ pair, or color dipole, interacts with what is now an essentially frozen, dense proton/nucleus that is dominated by small-$x$ partons (gluons dominate the nucleon wavefunction but can still split into quarks through the usual process).

The change to small-$x$ dipoles as our degrees of freedom is a big one, and the properties of the dipoles are important for the remainder of this dissertation, so let me mention some key components of color dipoles and their implications.

\begin{figure}[ht]
	\begin{center}
		\includegraphics[width=425 pt]{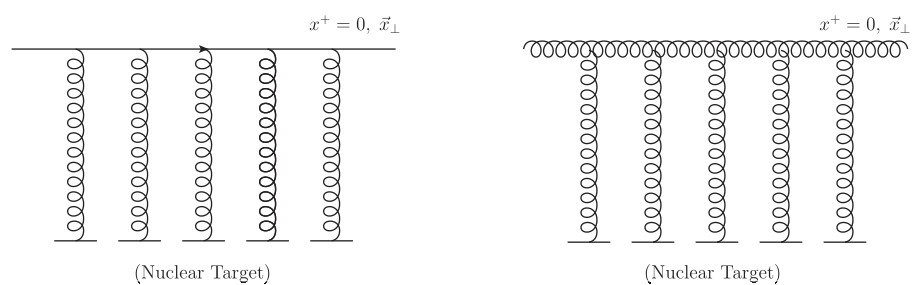} 
		\caption{Diagrams depicting an eikonal quark (left) or eikonal gluon (right) interacting with the gluon field of a nuclear target through the exchange of soft gluons, i.e., quark and gluon Wilson lines. The incoming quark or gluon line on top moves in the light-cone plus direction while the hadron/nucleus it interacts with is traveling along the light-cone minus direction.}
		\label{fig:unpol_Wilson}
	\end{center}
\end{figure}

First, let me formulate a definition for dipoles that is more appropriate for the types of interactions that occur in small-$x$ physics. As seen from Fig.~\ref{fig:BFKL_Ladder}, and then assuming that kind of soft gluon cascade would occur in Fig.~\ref{fig:smallx_DIS}, one can see that this scattering process is dominated by exchanges of soft gluons between the DIS projectile and the target; at small $x$ the coherence length of the projectile is long enough that I can treat the dense target as a frozen ``shock-wave." In resumming these interactions, I use the useful definitions of eikonal Wilson lines. The fundamental Wilson line sums all soft-gluon exchanges from an eikonal quark,
\begin{equation}\label{quark_Wilson}
	V_{\vec{x}_{\perp}}[b^-, a^-] = \mathrm{P}\,\mathrm{exp}\Biggl[i\,g\,\int\limits_{a^-}^{b^-}dx^-\,A^+(x^+=0, x^-, \vec{x}_{\perp})\Biggr]
\end{equation}
and the adjoint Wilson line sums soft-gluon exchanges from an eikonal gluon,
\begin{equation}\label{gluon_Wilson}
	U_{\vec{x}_{\perp}}[b^-, a^-] = \mathrm{P}\,\mathrm{exp}\Biggl[i\,g\,\int\limits_{a^-}^{b^-}dx^-\,\mathcal{A}^+(x^+=0, x^-, \vec{x}_{\perp})\Biggr]
\end{equation}
where the integral runs over the light-cone minus direction and the object being integrated is the gluon field of the target in the fundamental or adjoint representation in the $A^+ = 0$ light-cone gauge of the target, and $g$ is the strong-coupling parameter with $g^2\propto \alpha_s$. With these in hand, I see that in the eikonal limit, where it is assumed that both $Q^2$ and $1/x$ are large and thus any corrections that are suppressed by powers of the energy are neglected, that a color dipole can be constructed as a pair of Wilson lines. Very conveniently, it has been shown that in the 't Hooft large-$N_c$ limit \cite{tHooft:1973alw, tHooft:1974pnl} (a mathematical approximation that the number of colors is large, allowing us to ignore contributions sub-leading in powers of $N_c$) that dipoles defined in this way evolve with the BK small-$x$ evolution equations. Fig.~\ref{fig:unpol_Wilson} gives the diagrams that denote quark and gluon Wilson lines.

And with that, I have completed our short and sweet introduction to small-$x$ spin physics! In this chapter I defined my goal of understanding the helicity parton distributions functions, proposed a method of evolving objects from their small-$Q^2$ or large-$x$ initial conditions to a large-$Q^2$ or small-$x$ final state using evolution equations, and discussed some important distinctions made to a process like DIS when the energy increases towards a small-$x$ and eikonal limit. Notice that I have (maybe not so subtly) emphasized the eikonal limit in defining our new degrees of freedom as dipoles/Wilson lines; this is because at leading order in powers of the energy, Wilson lines do not carry any helicity or spin-dependent information. Spin-dependent interactions only occur at the sub-eikonal level, and thus I must investigate sub-eikonal corrections to the Wilson lines that do transfer spin information between the projectile and target. Only once I have defined a polarized Wilson line will I be able to explore polarized DIS scattering at small $x$, and further investigate how to evolve polarized dipoles towards small $x$. This is the subject of the next chapter.
	\newpage
	\section{$\boldsymbol{\mathrm{Spin~at~Small~}x}$}\label{chapter_2}

The small-$x$ regime brought with it its own shift in perspective. To summarize the important bits, we know that the high-energy of the problem allows the virtual photon in a DIS process to split into a $q\bar{q}$ pair that subsequently interacts with the nuclear target, that the nuclear target is now frozen and dense with the sum of interactions being denoted by a ``shockwave", and that small-$x$ unpolarized evolution resums the parameter $\alpha_s\ln(1/x)$. I will now move forward and discuss the case of polarized DIS (as well as semi-inclusive DIS, called SIDIS), and determine the caveats that come with tagging the polarization-dependent interactions and subsequently evolving the scattering amplitudes. This chapter will summarize the relevant work developed in small-$x$ helicity theory \cite{Kovchegov:2015pbl, Hatta:2016aoc, Kovchegov:2016zex, Kovchegov:2016weo, Kovchegov:2017jxc, Kovchegov:2017lsr, Kovchegov:2018znm, Kovchegov:2019rrz, Cougoulic:2019aja, Kovchegov:2020hgb, Cougoulic:2020tbc, Chirilli:2021lif, Kovchegov:2021lvz, Cougoulic:2022gbk}, with my contributions to the phenomenology covered in the aptly named phenomenology chapters.

\subsection{$\boldsymbol{\mathrm{Sub\text{-}Eikonal~Corrections~and~Polarized~Wilson~Lines}}$}

\begin{figure}[ht]
	\begin{center}
		\includegraphics[width=425 pt]{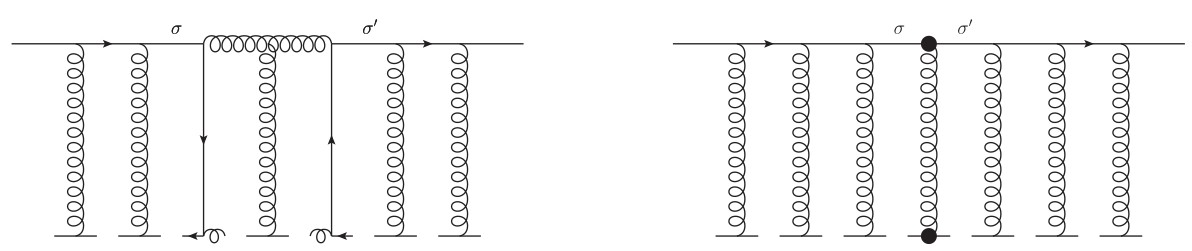} 
		\caption{Diagrams representing polarized quark Wilson lines. The dark vertex denotes a sub-eikonal interaction. $\sigma$ and $\sigma'$ are the polarization states of the quark before and after the sub-eikonal interaction; the incoming quark line on top moves in the light-cone plus direction while the hadron/nucleus it interacts with is traveling along the light-cone minus direction.}
		\label{fig:pol_Wilson}
	\end{center}
\end{figure}

\begin{figure}[ht]
	\begin{center}
		\includegraphics[width=425 pt]{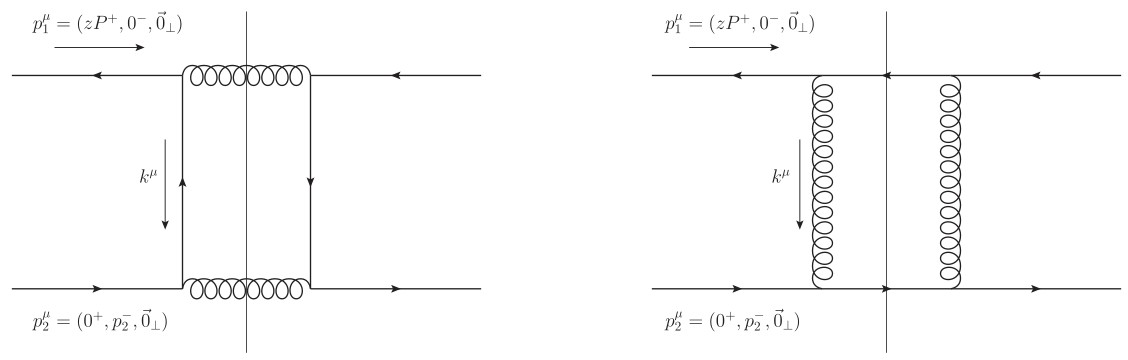} 
		\caption{Tree-level diagrams of the Reggeon exchange (left) and the Pomeron exchange. The incoming antiquark is traveling in the light-cone plus direction while the quark it scatters off of is traveling in the light-cone minus direction. The exchanged quark or gluon is carrying the momentum $k^{\mu}$. The narrow line down the middle represents the final state cut separating the amplitude from its complex conjugate.}
		\label{fig:Reg_Pom}
	\end{center}
\end{figure}

The dipole method goes hand-in-hand with small-$x$ physics, yet the correlators of eikonal Wilson lines do not transfer spin information. I can show this by considering what a spin-dependent interaction inside a Wilson line would look like. If I take the left panel of Fig.~\ref{fig:unpol_Wilson}, I can introduce a quark loop where I do not sum over the quark polarization states and thus I am able to ``follow" the spin; such a diagram would look like the left panel of Fig.~\ref{fig:pol_Wilson}. 
Though not exactly the same, one can see the similarity between this spin-dependent quark exchange and the tree-level spin-dependent Reggeon exchange amplitude as shown in the left panel of Fig.~\ref{fig:Reg_Pom}. 

In fact, I can also consider the case where one of the gluons exchanged in the right panel of Fig.~\ref{fig:unpol_Wilson} is similarly replaced with a tree-level spin-dependent Pomeron exchange as in the right panel of Fig.~\ref{fig:Reg_Pom}. I bring this up because the polarized cross-sections for tree-level Reggeon and Pomeron exchanges, denoted as $\Delta\sigma = \sigma^{++} - \sigma^{+-}$ with the plus and minus referring to the polarization states of the interacting quark and antiquark, are known \cite{Kovchegov:2015pbl}: 
\begin{equation}\label{born_reg_pom}
	\frac{\Delta\sigma_{reggeon}^{Born}}{d^2k} = -2\frac{\alpha_s^2C_F^2}{N_c}\,\frac{1}{zs}\,\frac{1}{k_T^2}\,\delta_{-\sigma,\sigma'}, \qquad \frac{\Delta\sigma_{pomeron}^{Born}}{d^2k} = 2\frac{\alpha_s^2C_F}{N_c}\,\frac{1}{zs}\,\frac{1}{k_T^2}.
\end{equation}
The polarized cross-sections each have the specific scaling factor of $1/(zs)$ compared to their spin-summed counterparts, where $z$ is the longitudinal momentum fraction of the quark/antiquark interacting with the nuclear target (though at tree-level the nuclear target is replaced by a quark line) and $s$ is the typical center-of-mass energy squared again. From this scaling factor, we know that enforcing a spin-dependent interaction into a Wilson line constitutes a sub-eikonal (energy-suppressed) correction. Similar quark loops or sub-eikonal gluon exchanges (denoted by the dark circles in Fig.~\ref{fig:pol_Wilson}) can be imposed onto the gluon Wilson lines as shown below.

\begin{figure}[ht]
	\begin{center}
		\includegraphics[width=425 pt]{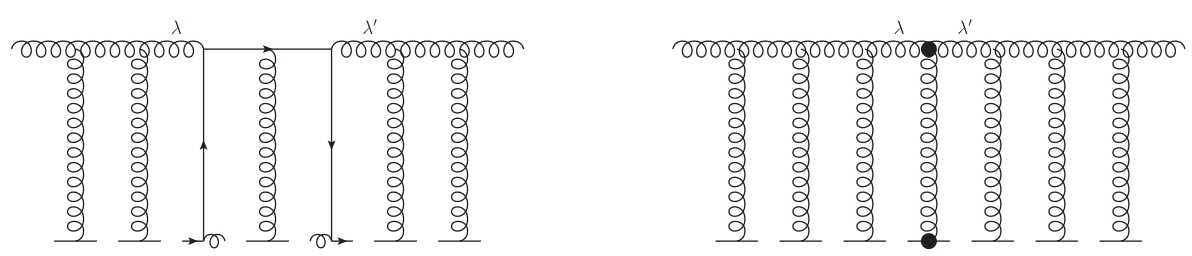} 
		\caption{Diagrams representing polarized gluon Wilson lines. The dark vertex denotes a sub-eikonal interaction. $\lambda$ and $\lambda'$ are the polarization states of the gluon before and after the sub-eikonal interaction; the incoming gluon line moves in the light-cone plus direction while the hadron/nucleus it interacts with is traveling along the light-cone minus direction.}
		\label{fig:pol_Wilson_gluon}
	\end{center}
\end{figure}

Figures~\ref{fig:unpol_Wilson},~\ref{fig:pol_Wilson},~\ref{fig:pol_Wilson_gluon} are adapted from \cite{Kovchegov:2018znm}, which goes into great detail to define the polarized Wilson line operators. In a direct comparison between the unpolarized and polarized cases for something like the SIDIS cross-section, in the dipole formalism I can decompose the unpolarized cross-section as dependent on the S-matrix for unpolarized dipole scattering, written in terms of the unpolarized Wilson lines \cite{Kovchegov:2013cva}
\begin{equation}
	\sigma^{SIDIS} \propto \int dz\,S(\vec{x}_{\perp},\vec{y}_{\perp}) \to \int dz \,\frac{1}{N_c}\Big\langle \mathrm{tr}[V_{\vec{x}_{\perp}}V_{\vec{y}_{\perp}}^{\dagger}]\Big\rangle
\end{equation}
where the angled brackets represent the averaging over the nuclear target. This averaging, also called ``CGC" averaging, simplifies the expectation value notation
\begin{equation}\label{CGC_averaging}
	\big\langle \cdot\, \cdot \,\cdot \big\rangle \equiv \frac{1}{2}\sum_{S_L} \frac{1}{2P^+V^-}\langle P,S_L|\cdot\,\cdot\,\cdot|P,S_L\rangle 
\end{equation}
where $P$ and $S_L$ are the momentum and longitudinal spin of the nuclear target and $V^-$ is an (infinite) volume factor. As is typical, I would integrate over the relevant momenta, one of which is the longitudinal momenta of the eikonal quark or gluon, which are now written in terms of Wilson lines. The polarized cross-section can be decomposed similarly \cite{Kovchegov:2015pbl}, however it is now understood that the polarized Wilson lines are inherently scaled by the sub-eikonal factor $1/(zs)$. If I write explicitly that one of the integrals is over the longitudinal momentum, we can see where one logarithm is generated
\begin{align}\label{double_angle}
	\Delta\sigma^{SIDIS} &\propto \int dz\, \Big\langle \mathrm{tr}[V_{\vec{x}_{\perp}}V_{\vec{w}_{\perp}}^{\dagger}(\sigma')]+\mathrm{tr}[V_{\vec{w}_{\perp}}(\sigma')V_{\vec{y}_{\perp}}^{\dagger}]\Big\rangle (z) \notag \\
	&\propto \frac{1}{s}\int \frac{dz}{z}\, \Big\langle\!\!\Big\langle \mathrm{tr}[V_{\vec{x}_{\perp}}V_{\vec{w}_{\perp}}^{\dagger}(\sigma')]+\mathrm{tr}[V_{\vec{w}_{\perp}}(\sigma')V_{\vec{y}_{\perp}}^{\dagger}]\Big\rangle\!\!\Big\rangle(z),
\end{align}
where the change in angles brackets is simply there to factor out the energy dependence of the polarized dipole operator. The momentum fraction $z$ for the polarized dipole operator (the object contained in the double-angled brackets) always corresponds to the polarized Wilson line, which is written specifically to have dependence on the polarization state $\sigma'$. From the integrated, we see that at least one logarithm is generated by the sub-eikonal polarized Wilson and the longitudinal momentum integral $\int dz/z$. This highlights perfectly how the sub-eikonal nature of polarized evolution will differ from the unpolarized case, especially in terms of the logarithmic enhancements.

The diagrams in Figs.~\ref{fig:pol_Wilson} and \ref{fig:pol_Wilson_gluon} only depict a subset of the possible ways to insert a sub-eikonal operator into an unpolarized Wilson line. All polarized Wilson lines do share a similar structure in the sense that each one is written not as a continuous Wilson line that carries spin information, but instead as an unpolarized Wilson line traveling from $-\infty$ up to the spin-dependent interaction where a sub-eikonal operator is inserted and then continues as an unpolarized Wilson line from the interaction to $+\infty$. Therefore, the different ways of constructing polarized Wilson lines come down to the different possible polarization-dependent sub-eikonal operators that can be inserted in between unpolarized Wilson lines. The operators listed above are those that have explicit spin-dependence, meaning they appear with prefactors composed of the polarization states $\sigma\delta_{\sigma,\sigma'}$ or $\lambda\delta_{\lambda,\lambda'}$, however there are some operators that do not have this explicit spin-dependent factor but still interact with the nuclear target in a spin-dependent way. Additionally, as exemplified by Figs.~\ref{fig:pol_Wilson} and \ref{fig:pol_Wilson_gluon}, the sub-eikonal operators can either transfer spin through quark exchanges (left panels) or gluon exchanges (right panel). Together, this means that I can categorize the quark and gluon Wilson lines into four sub-categories each: polarized Wilson lines with explicit spin-dependence, and polarized Wilson lines without it, and both of these Wilson lines can have a quark contribution and a gluon contribution. 

Summarizing the results of \cite{Kovchegov:2017hjq, Kovchegov:2015pbl, Kovchegov:2016zex} that culminated in \cite{Cougoulic:2022gbk}, there are 8 distinct polarized Wilson lines. Note that the unpolarized Wilson lines \eqref{quark_Wilson} and \eqref{gluon_Wilson} are defined at specific transverse positions $\vec{x}_{\perp}$; this is the case for most of the polarized Wilson lines, however there are two polarized Wilson lines that are non-local in the transverse plane and thus have two transverse positions $\vec{x}_{\perp}$ and $\vec{y}_{\perp}$ referring to the transverse positions of the unpolarized Wilson lines directly before and directly after the insertion of a non-local (in the transverse plane) sub-eikonal operator. To merge my notation with the relevant literature I will denote the transverse position two-vector as $\vec{x}_{\perp} \equiv \underline{x}$, and also define the transverse position difference between two partons (or similarly the two Wilson lines that create a dipole) as $\underline{x}_{ij} = \underline{x}_i - \underline{x}_j$ with magnitude $x_{ij} = |\underline{x}_{ij}|$. The quark Wilson lines are then split based on their spin-dependent prefactor, denoted as \cite{Kovchegov:2018znm, Kovchegov:2021iyc}
\begin{equation}
	V_{\underline{x},\underline{y};\sigma,\sigma'} \Bigg|_{\mathrm{sub-eikonal}} \equiv \sigma\delta_{\sigma.\sigma'}V_{\underline{x}}^{{\mathrm{pol}[1]}}\delta^2(\underline{x}-\underline{y}) + \delta_{\sigma,\sigma'}V_{\underline{x},\underline{y}}^{\mathrm{pol}[2]}
\end{equation}
where I further separate these components into their quark and gluon contributions
\begin{equation}
	V_{\underline{x}}^{\mathrm{pol}[1]} = V_{\underline{x}}^{\mathrm{G}[1]} + V_{\underline{x}}^{\mathrm{q}[1]}, \qquad V_{\underline{x},\underline{y}}^{\mathrm{pol}[2]} = V_{\underline{x},\underline{y}}^{\mathrm{G}[2]}\delta^2(\underline{x}-\underline{y}) + V_{\underline{x},\underline{y}}^{\mathrm{q}[2]}\delta^2(\underline{x}-\underline{y}).
\end{equation}
These polarized quark Wilson lines have the following sub-eikonal operators and definitions:
\begin{subequations}\label{quark_pol_Wilson}
	\begin{align}
		V_{\underline{x}}^{\mathrm{G}[1]} &= \frac{igP^+}{s}\int\limits_{-\infty}^{\infty}dx^-\,V_{\underline{x}}[\infty, x^-]\,F^{12}(x^-,\underline{x})\,V_{\underline{x}}[x^-,-\infty], \\
		V_{\underline{x}}^{\mathrm{q}[1]} &= \frac{g^2P^+}{2s}\int\limits_{-\infty}^{\infty}dx_1^-\int\limits_{x_1^-}^{\infty}dx_2^-\,V_{\underline{x}}[\infty,x_2^-]\,t^b\,\psi_{\beta}(x_2^-,\underline{x})\,U_{\underline{x}}^{ba}[x_2^-,x_1^-]\,[\gamma^+\gamma^5]_{\alpha\beta}\notag \\
		& \qquad\qquad\qquad\qquad\qquad \times \bar{\psi}_{\alpha}\,t^a\,V_{\underline{x}}[x_1^-,-\infty], \\
		V_{\underline{x},\underline{y}}^{\mathrm{G}[2]} &= -\frac{iP^+}{s}\int\limits_{-\infty}^{\infty}dz^-\,d^2z\,V_{\underline{x}}[\infty,z^-]\,\delta^2(\underline{x}-\underline{z})\,\lvec{D}^i(z^-,\underline{z})\,D^i(z^-,\underline{z})\,V_{\underline{y}}[z^-,-\infty]\,\delta^2(\underline{y}-\underline{z}), \label{V_G2_1} \\
		V_{\underline{x}}^{\mathrm{q}[2]} &= -\frac{g^2P^+}{2s}\int\limits_{-\infty}^{\infty}dx_1^-\int\limits_{x_1^-}^{\infty}dx_2^-\,V_{\underline{x}}[\infty,x_2^-]\,t^b\,\psi_{\beta}(x_2^-,\underline{x})\,U_{\underline{x}}^{ba}[x_2^-,x_1^-]\,[\gamma^+]_{\alpha\beta}\notag \\
		& \qquad\qquad\qquad\qquad\qquad \times \bar{\psi}_{\alpha}(x_1^-, \underline{x})\,t^a\,V_{\underline{x}}[x_1^-,-\infty],
	\end{align}
\end{subequations}
where $V_{\underline{v}}$ and $U_{\underline{v}}^{ab}$ are the typical unpolarized quark and gluon (fundamental and adjoint) Wilson lines, $\psi$ and $\bar{\psi}$ are the quark fields, and $t^a$ are the generators of SU($N_c$) in the fundamental representation such that the gluon fields inside fundamental Wilson line \eqref{quark_Wilson} has the definition $A^{\mu} = A^{\mu}_a\,t^a$, and $g$ is the strong coupling constant. The operators $D^i = \partial - igA^i$ and $\lvec{D}^i = \lvec{\partial}^i + igA^i$ are the right-acting and left-acting covariant derivatives, respectively. The gluon field strength operator in the fundamental representation $F^{12}$ is seen in $V_{\underline{x}}^{\mathrm{G}[1]}$ is defined by $F^{12} = \partial^1A^2 - \partial^2A^1 + ig[A^1, A^2]$. 

I can similarly define the gluon Wilson lines by splitting them into spin-dependent components
\begin{equation}
	(U_{\underline{x},\underline{y};\lambda,\lambda'})^{ba}\Bigg|_{\mathrm{sub-eikonal}} \equiv \lambda\delta_{\lambda,\lambda'}\,(U_{\underline{x}}^{\mathrm{pol}[1]})^{ba}\delta^2(\underline{x}-\underline{y}) + \delta_{\lambda,\lambda'}(U_{\underline{x},\underline{y}}^{\mathrm{pol}[2]})^{ba}
\end{equation}
and subsequently decompose them into their quark and gluon contributions
\begin{equation}
	U_{\underline{x}}^{\mathrm{pol}[1]} = U_{\underline{x}}^{\mathrm{G}[1]}+U_{\underline{x}}^{\mathrm{q}[1]}, \qquad U_{\underline{x},\underline{y}}^{\mathrm{G}[2]} + U_{\underline{x},\underline{y}}^{\mathrm{q}[2]}\delta^2(\underline{x}{-\underline{y}})
\end{equation}
These polarized gluon Wilson lines have the following sub-eikonal operators and definitions:
\begin{subequations}\label{gluon_pol_Wilson}
	\begin{align}
		(U_{\underline{x}}^{\mathrm{G}[1]})^{ba} &= \frac{2igP^+}{s}\int\limits_{-\infty}^{\infty}dx^-(U_{\underline{x}}[\infty,x^-])^{bb'}\,
		(\mathcal{F}^{12})^{b'a'}(x^-,\underline{x})\,(U_{\underline{x}}[x^-,-\infty])^{a'a}, \\
		(U_{\underline{x}}^{\mathrm{q}[1]})^{ba} &= \frac{g^2P^+}{2s}\int\limits_{-\infty}^{\infty}dx_1^-\int\limits_{x_1^-}^{\infty}dx_2^-\,(U_{\underline{x}}[\infty,x_2^-])^{bb'}\,\bar{\psi}(x_2^-,\underline{x})\,t^{b'}\,V_{\underline{x}}[x_2^-,x_1^-]\,\gamma^+\gamma^5\,\notag \\
		&\qquad\qquad\qquad\qquad\qquad \times t^{a'}\psi(x_1^-,\underline{x})\,(U_{\underline{x}}[x_1^-,-\infty])^{a'a} + \,\mathrm{c.c.}, \\
		(U_{\underline{x},\underline{y}}^{\mathrm{G}[2]})^{ba} &= -\frac{iP^+}{2}\int\limits_{-\infty}^{\infty}dz^- d^2z\,(U_{\underline{x}}[\infty,z^-])^{bb'}\,\delta^2(\underline{x}-\underline{z})\;\lvec{\mathcal{D}}_i^{b'c}(z^-,\underline{z})\,\mathcal{D}_i^{ca'}(z^-,\underline{z})\notag \\
		& \qquad\qquad\qquad\qquad\qquad \times (U_{\underline{y}}[z^-,\infty])^{a'a}\delta^2(\underline{y}-\underline{z}), \label{U_G2_1} \\
		(U_{\underline{x}}^{\mathrm{q}[2]})^{ba} & = -\frac{g^2P^+}{2s}\int\limits_{-\infty}^{\infty}dx_1^-\int\limits_{x_1^-}^{\infty}dx_2^-\,(U_{\underline{x}}[\infty,x_2^-])^{bb'}\,\bar{\psi}(x_2^-,\underline{x})\,t^{b'}\,V_{\underline{x}}[x_2^-,x_1^-]\,\gamma^+\,\notag \\
		&\qquad\qquad\qquad\qquad\qquad t^{a'}\,\psi(x_1^-,\underline{x})\,(U_{\underline{x}}[x_1^-,-\infty])^{a'a} - \,\mathrm{c.c.},
	\end{align}
\end{subequations}
where the $\mathcal{F}^{12}$ is the adjoint gluon fiend strength tensor defined as $\mathcal{F}^{12} = F^{a12}\,T^a$ where $T^a$ are the generators of SU($N_c$) in the adjoint representation, and likewise $\mathcal{D}_i^{ab}$ are the adjoint covariant derivatives, $\mathcal{D}_i^{ab} = \partial_i - ig(T^c)_{ab}A_i^c$ and $\lvec{\mathcal{D}}_i^{ab} = \lvec{\partial}_i + ig(T^c)_{ab}A_i^c$ with $(T^c)_{ab} = -if^{abc}$ where $f^{abc}$ are the SU($N_c$) structure constants.

Before I jump into the next section, I will predict the future and define two new polarized Wilson lines, dubbed polarized Wilson lines ``of the second kind" when introduced in \cite{Cougoulic:2022gbk}. In Eqs.~\eqref{V_G2_1} and \eqref{U_G2_1} we see the operators $\lvec{D}^iD^i$ and $\lvec{\mathcal{D}}_i^{bc}\mathcal{D}_i^{ca}$ inserted between two Wilson lines; as will be show in \textbf{Chapter}~\ref{KPS-CTT}, this operator is simplified to $\lvec{D}^i - D^i$ (and its adjoint counterpart) through small-$x$ evolution. While these operators do not have explicit spin-dependence, they are related to the Jaffe-Manohar gluon hPDF \cite{Jaffe:1989jz} and are found to mix with the explicit spin-dependent operators $F^{12},\, \mathcal{F}^{12}$ and $\gamma^+\gamma^5$ through evolution. These operators can be interpreted as coupling the gluon probe's OAM to the spin of the proton/nuclear target, and thus contribute to small-$x$ helicity evolution \cite{Adamiak:2023yhz}. The polarized Wilson lines of the second kind are defined as
\begin{subequations}
	\begin{align}
		V_{\underline{z}}^{i\,\mathrm{G}[2]} &\equiv \frac{P^+}{2s}\int\limits_{-\infty}^{\infty}dz^- V_{\underline{z}}[\infty,z^-]\,\Big[D^i(z^-,\underline{z})-\lvec{D}^i(z^-,\underline{z})\Big]\,V_{\underline{z}}[z^-,-\infty] \label{V_G2_2} \\
		(U_{\underline{z}}^{i\,\mathrm{G}[2]})^{ba} &\equiv \frac{P^+}{2s}\int\limits_{-\infty}^{\infty}dz^- (U_{\underline{z}}[\infty,z^-]\,\Big[\mathcal{D}^i(z^-,\underline{z})-\lvec{\mathcal{D}}^i(z^-,\underline{z})\Big]\,U_{\underline{z}}[z^-,\underline{z}])^{ba} \label{U_G2_2},
	\end{align}
\end{subequations}
where they are differentiated from their similar brothers $V_{\underline{x},\underline{y}}^{\mathrm{G}[2]}$ and $(U_{\underline{x},\underline{y}}^{\mathrm{G}[2]})^{ba}$ by an un-summed vector index $i$ and dependence on only one transverse vector instead of two. With all of the polarized Wilson lines laid out, I can continue onto the next section to determine the benefit they provide.

\subsection{$\boldsymbol{\mathrm{Helicity~PDFs~and~the~Polarized~Structure~Function}}$}\label{hPDFs_stf_sec1}

At this point, I have now delved into the energy-suppressed, sub-eikonal corrections to the Wilson lines, but I have not yet shown why or how these are useful; I did, however, tease the subject when discussing the polarized SIDIS cross-section. As will shortly be shown, I can use the polarized Wilson lines to construct polarized dipole amplitudes, which I can then use to define helicity PDFs and even the polarized structure function of protons and neutrons.

Following the template laid out in \cite{Cougoulic:2022gbk}, let me begin with the gluon helicity PDF, $\Delta G(x,Q^2)$. Actually, let me begin first with the gluon helicity transverse momentum distribution function (TMD) in the small-$x$ dipole formalism, $g_{1L}^G(x,k_T^2)$, since it is more general and will provide the gluon hPDF once I integrate it over the transverse momentum. The gluon TMD is defined as \cite{Bomhof:2006dp, Cougoulic:2022gbk}
\begin{align}
	g_{1L}^{G\,(dipole)}(x,k_T^2) &= \frac{-2i}{xP^+}\frac{1}{(2\pi)^3}\frac{1}{2}\sum_{S_L}S_L\int dw^-d^2w_{\perp}\,e^{ixP^+w^-}\,e^{-i\underline{k}\cdot \underline{w}} \\
	& \qquad\qquad \times \big\langle P,S_L\big|\epsilon^{ij}\,\mathrm{tr}\Big[F^{+i}(0)\,\mathcal{U}^{[+]}[0,w_{\perp}]\,F^{+j}(w_{\perp})\,\mathcal{U}^{[-]}[w_{\perp},0]\Big]\big|P,S_L\big\rangle_{w^+=0}
\end{align}
where $w^{\mu}$ is a dummy transverse position that I integrate over, $k_T = k_{\perp} = |\underline{k}|$ is the magnitude of the transverse momentum of the produced gluon, $S_L$ is the longitudinal spin ($\pm 1$), $P$ is the four-momentum of the target proton, and $\mathcal{U}^{[\pm]}$ are future/past pointing gauge links (a modified type of Wilson line that connects two points in the longitudinal axis like a normal Wilson line, but also allows for a transverse jump). Through an intense derivation detailed in \cite{Cougoulic:2022gbk}, the gluon TMD can be rewritten as 
\begin{align}\label{gluon_TMD_Wilson}
	g_{1L}^{G\,(dipole)}(x,k_T^2) &= \frac{-2is}{P^+V^-g^2}\frac{1}{(2\pi)^3}\frac{1}{2}\sum_{S_l}S_L\epsilon^{ij}k^i\int d^2\underline{x}_0d^2\underline{x}_1\,e^{-i\underline{k}\cdot(\underline{x}_1-\underline{x}_0)} \notag \\
	&\qquad\qquad \times \big\langle P,S_L\big|\mathrm{tr}\Big[V_{\underline{x}_0}^{\dagger}V_{\underline{x}_1}^{j\,\mathrm{G}[2]}-\big(V_{\underline{x}_0}^{j\,\mathrm{G}[2]}\big)^{\dagger}V_{\underline{x}_1}\Big]\big|P,S_L\big\rangle
\end{align}
where I have introduced a volume factor in the denominator $V^- = \int dx^-d^2x$ in order to integrate over two spatial variables $x_0$ and $x_1$, which are now used to describe the transverse positions of the polarized and unpolarized Wilson lines. It is not common practice to include vector notation in the differentials $d^2\underline{x}_0$ and $d^2\underline{x}_1$, but I included them here just to emphasize the fact that the differentials are indeed over the transverse positions of the Wilson lines. We see here why I defined the polarized Wilson lines of the second kind \eqref{V_G2_2}, because they appear directly in the definition of the gluon TMD! From this point forward I will again modify the notation of the polarized and unpolarized Wilson lines, where the transverse positions $\underline{x}_0$ and $\underline{x}_1$ are implied to be understood as such, so I denote $V_{\underline{x}_a}$ as $V_{a}$. I can take Eq.~\eqref{gluon_TMD_Wilson} and employ the CGC averaging and sub-eikonal factorization seen in Eqs.~\eqref{CGC_averaging} and \eqref{double_angle}, and re-write the gluon TMD 
\begin{align}\label{gluon_TMD_dipole}
	g_{1L}^{G\,(dipole)}(x,k_T^2) &= \frac{-4i}{g^2(2\pi)^3}\epsilon^{ij}k^i\int d^2\underline{x}_0d^2\underline{x}_1\,e^{-i\underline{k}\cdot(\underline{x}_1-\underline{x}_0)} \notag \\
	&\qquad\qquad \times \dlangle \mathrm{tr}\Big[V_0^{\dagger}V_1^{j\,\mathrm{G}[2]}+\big(V_1^{j\,\mathrm{G}[2]}\big)^{\dagger}V_0\Big]\drangle
\end{align}
where I can at last define a specific polarized dipole amplitude
\begin{equation}\label{G10}
	G_{10}^j(zs) \equiv \frac{1}{2N_c}\dlangle \mathrm{tr}\Big[V_0^{\dagger}V_1^{j\,\mathrm{G}[2]}+\big(V_1^{j\,\mathrm{G}[2]}\big)^{\dagger}V_0\Big]\drangle
\end{equation}
where $z$ is the longitudinal momentum fraction of the virtual photon carried by the polarized Wilson line. Remember that I have defined the system such that the longitudinal momentum of the photon is equivalent to the photon's light-cone minus momentum, and the nuclear target or proton has longitudinal momentum that is effectively its light-cone plus momentum. This polarized dipole amplitude is given in its impact-parameter integrated state, but I can decompose it into its symmetric and antisymmetric components $G_{10}^j \to x_{10}^j G_1(x_{10}^2,zs) + \epsilon^{ij}(x_{10}^i)G_2(x_{10}^2,zs)$ \cite{Kovchegov:2017lsr}, where only $G_2(x_{10}^2,zs)$ survives since the gluon TMD is already antisymmetric. In the end I can combine the transverse coordinates as described above ($\underline{x}_{10} = \underline{x}_1 - \underline{x}_0$), simplify the integrand by differentiating the Fourier factor and integrating by parts, and integrate over the transverse momentum (remembering to cutoff the integral at $Q^2$), and I am left with the dipole definition of the gluon helicity PDF
\begin{align}\label{DeltaG}
	\Delta G(x,Q^2) &= \int\limits^{Q^2} d^2 k_T\,g_{1L}^{G\,(dipole)}(x,k_T^2) \notag \\
	&= \int\limits^{Q^2} d^2 k_T \bigg[\frac{8iN_c}{g^2(2\pi)^3}\int d^2x_{10}\,e^{-i\underline{k}\cdot\underline{x}_{10}}\,\underline{k}\cdot\underline{x}_{10}\,G_2\Big(x_{10}^2,zs = \frac{Q^2}{x}\Big)\bigg] \notag \\
	& = \int\limits^{Q^2} d^2 k_T \bigg[\frac{N_c}{\alpha_s2\pi^4}\int d^2x_{10}\,e^{-i\underline{k}\cdot\underline{x}_{10}}\Big[1+x_{10}^2\frac{\partial}{\partial x_{10}^2}\Big]\,G_2\Big(x_{10}^2,zs = \frac{Q^2}{x}\Big)\bigg] \notag \\
	&= \frac{2N_c}{\alpha_s\pi^2}\Big[\Big(1+x_{10}^2\frac{\partial}{\partial x_{10}^2}\Big)G_2\Big(x_{10}^2, zs  =\frac{Q^2}{s}\Big)\Big]_{x_{10}^2 = \frac{1}{Q^2}}.
\end{align}
This is the beauty of the polarized dipole formalism: I have now successfully written the gluon helicity PDF in terms of the polarized dipole amplitude $G_2(x_{10}^2,zs)$, and I can repeat this exercise to write the quark helicity TMD (and thus the quark helicity PDF) in terms of polarized dipole amplitudes as well. Due to its direct relationship to the gluon helicity PDF and TMD, the polarized dipole amplitude $G_2(x_{10}^2,zs)$ is named the gluon polarized dipole amplitude (in the fundamental representation, given its construction of `$V$'-type Wilson lines).

Let me begin the discussion of the quark helicity TMD, $g_{1L}^q(x,k_T^2)$, with its definition \cite{Mulders:1995dh, Kovchegov:2018znm}
\begin{equation}\label{quark_TMD}
	g_{1L}^q(x,k_T^2) = \frac{1}{(2\pi)^3}\frac{1}{2}\sum_{S_L}S_L\int d^2 r_{\perp}\,dr^-\,e^{i\underline{k}\cdot \underline{r}}\,\big\langle P,S_L\big| \bar\psi(0)\,\mathcal{U}[0,r_{\perp}]\,\frac{\gamma^+\gamma^5}{2}\,\psi(r_{\perp})\big|P,S_L\big\rangle 
\end{equation}
where we can identify helicity projection operator $\tfrac{1}{2}\gamma^+\gamma^5$ and the process-dependent gauge-link $\mathcal{U}[0,r_{\perp}]$. Again, I direct the reader to Ref.~\cite{Cougoulic:2022gbk} for the full derivation; it involves the expansion of the equation above and a thorough exploration of each term and their respective contributions (or lack thereof) depending on their small-$x$ asymptotics, eikonality, or simple cancellation. The resulting expression reads (for the sum of quarks and antiquarks)
\begin{align}\label{quark_TMD_dipole}
	g_{1L}^S(x,k_T^2) &= -\frac{8N_c}{(2\pi)^5}\int\limits_{\Lambda^2/s}^1 \frac{dz}{z}\int d^2 x_0\, d^2 x_1\,e^{i\underline{k}\cdot\underline{x}_{10}}\, \Bigg\{\frac{-i\underline{x}_{10}}{x_{10}^2}\cdot\frac{\underline{k}}{k_T^2}\,Q_{q,10}(zs) \notag \\
	& + \frac{\epsilon^{\ell j}k^{\ell}}{k_T^2} \bigg[k^i\frac{x_{10}^i}{x_{10}^2} + i\frac{\delta^{ij}x_{10}^2 - 2x_{10}^ix_{10}^j}{x_{10}^4}\bigg]G_{10}^i(zs)\Bigg\}
\end{align}
where we see that the gluon polarized dipole amplitude has revealed its contribution to the quark helicity TMD, along with a newly defined (impact parameter integrated) polarized dipole amplitude $Q_{q,10}(zs)$. Appropriately named the quark polarized dipole amplitude, its composition of Wilson lines is given below,
\begin{equation}\label{Q10}
	Q_{q,10}(zs) = \frac{1}{2N_c}\mathrm{Re}\,\dlangle \mathrm{T~tr}\big[V_0V_1^{\mathrm{pol}[1]\dagger}\big]+\mathrm{T~tr}\big[V_1^{\,\mathrm{pol}[1]}V_0^{\dagger}\big]\drangle,
\end{equation}
where the notation $\mathrm{T~tr}[\;]$ implies strict ordering. The subscript $q$ signifies that there is flavor dependence of the quark polarized dipole amplitude, however, this dependence does not come from the Wilson lines themselves, but from the initial condition of the scattering process. I bring in the impact parameter dependence as I did before, and I take note that only the antisymmetric part of $G_{10}^i(zs)$ contributes. After integrating over the transverse momentum, I finally define the quark helicity PDFs in terms of polarized dipole amplitudes \cite{Cougoulic:2022gbk}
\begin{equation}\label{DeltaSigma}
	\Delta \Sigma(x,Q^2) = \int\limits^{Q^2}d^2k_T\, g_{1L}^S (x,k_T^2) = 
	\sum_q\Delta q^+(x,Q^2),
\end{equation}
where I have defined the $C$-even quark hPDFs,
\begin{equation}\label{Deltaqp}
	\Delta q^+(x,Q^2) = \sum_{q,\bar{q}}(\Delta q + \Delta \bar{q})(x,Q^2) =  -\frac{N_c}{2\pi^3}\int\limits_{\Lambda^2/s}^1 \frac{dz}{z}\int\limits_{1/zs}^{\mathrm{min}\{\frac{1}{zQ^2},\frac{1}{\Lambda^2}\}}\frac{dx_{10}^2}{x_{10}^2}\,\big[Q_q(x_{10}^2,zs) + G_2(x_{10}^2,zs)\big].
\end{equation}
It is important to note here some constraints placed on the integration limits that come as a direct result of the small-$x$ dipole formalism. First, note that a polarized dipole has natural physical constraints on its transverse size. I restrict the size of the dipole from above by characterizing the transverse size of the target as $1/\Lambda$, such that a dipole can only interact with the target if it has transverse size $x_{10} < 1/\Lambda$, with $\Lambda \approx 1 \mathrm{GeV}$ for a proton target. Along with this, lifetime ordering (a property that will be touched on more in the following section) is enforced by $x_{10}^2 < \frac{1}{z Q^2}$. The dipoles are also constrained by the smallest possible scale in the problem, $1/s$ where $s\to \infty$ as $x \to 0$; in order to avoid a singularity at $z \to 0$ for the longitudinal momentum fraction of polarized Wilson line, it is cutoff by $\tfrac{(\Lambda^2)}{s} < z$, meanwhile the transverse separation of the polarized dipole also must skirt its singularity as $x_{10}^2 \to 0$, so it too must have a cutoff, imposed by $\tfrac{1}{(z)s} < x_{10}^2$. With these constraints in hand, I have now given definitions of the helicity PDFs for both quarks and gluons in terms of the polarized dipole amplitudes. Both $\Delta\Sigma(x,Q^2)$ and $\Delta G(x, Q^2)$ are hPDFs that can be extracted from inclusive DIS cross-sections, and the last inclusive DIS observable that I will discuss before shifting over to semi-inclusive processes is the $g_1$ structure function.

To study the polarized structure function of the proton,  I start with the DIS cross-section, which can be written (using the kinematics of Fig.~\ref{fig:Lowx_DIS}) as 
\begin{equation}
	\frac{d\sigma}{d^3p'} = \frac{\alpha_{EM}^2}{EE'\,Q^4}L^{\mu\nu}W_{\mu\nu}
\end{equation}
where $L^{\mu\nu}$ is the well-known leptonic tensor describing the electron's process, and $W_{\mu\nu}$ is the hadronic tensor that I wish to investigate more closely. The hadronic tensor itself can be written as \cite{ParticleDataGroup:2020ssz, Lampe:1998eu}
\begin{equation}
	W_{\mu\nu} \equiv \frac{1}{4\pi M_P}\int d^4 x \,e^{iq\cdot x}\, \big\langle P,S_L\big| j_{\mu}(x)j_{\nu}(0)\big|P,S_L \big\rangle
\end{equation}
where $M_P$ is the proton mass, $j_{\mu,\nu}$ is the quark electromagnetic current operator, and $P,\, S_L$ are the proton's momentum and longitudinal spin. The hadronic tensor can be split into symmetric and antisymmetric parts, which correlate to the spin-independent and spin-dependent parts, respectively. Choosing to work in the proton's rest frame, $P^{\mu} = (M_p, \vec{0})$, in which the virtual photon has momentum $q^{\mu} = (-Q^2/2q^-, q^-, \underline{0}_{\perp})$, only transverse photon polarizations contribute to the spin-dependent photon-proton cross-section. Defining the polarization state of the photon as $\lambda$, the spin-dependent cross-section is
\begin{equation}
	\Delta\sigma^{\gamma^*P}(\lambda,S_L) = \frac{4\pi^2\alpha_{EM}}{q^0}W_{\mu\nu}\,\epsilon_{T\lambda}^{*\,\mu}\,\epsilon_{T\lambda}^{\nu} = -\frac{8\pi^2\alpha_{EM}\,x}{Q^2\,M_P}\,\lambda\, S_L\Bigg[g_1(x,Q^2) - \frac{4x^2M_P^2}{Q^2}g_2(x,Q^2)\Bigg]
\end{equation}
where we see that the spin-dependent part of the hadronic tensor has been decomposed into the structure functions $g_1(x,Q^2)$ and $g_2(x,Q^2)$ in the same way that the spin-independent part of the hadronic tensor is decomposed into the unpolarized structure functions $F_1(x,Q^2)$ and $F_2(x,Q^2)$. We also see more confirmation of the sub-eikonal nature of polarized physics by the leading pre-factor of $x$ in front of the structure functions. I take this one step further and surmise that in the small-$x$ limit $x^2M_P^2/Q^2 \ll 1$, and thus the polarized cross-section is effectively a function of $g_1(x,Q^2)$ such that 
\begin{equation}\label{g1_collinear}
	g_1(x,Q^2) = -\frac{Q^2}{16\,\pi^2\,\alpha_{EM}\,x}\Big[\sigma^{\gamma^*P}(+,+) - \sigma^{\gamma^*P}(+,-)\Big]= -\frac{Q^2}{16\,\pi^2\,\alpha_{EM}\,x}\,\Delta\sigma^{\gamma^*P}.
\end{equation}
The remainder of this derivation mirrors the previous two, where the exercise pertains to writing the polarized cross-section $\Delta\sigma^{\gamma^*P}$ in terms of polarized Wilson lines and subsequently polarized dipole amplitudes.  

\begin{figure}[ht]
	\begin{center}
		\includegraphics[width=325 pt]{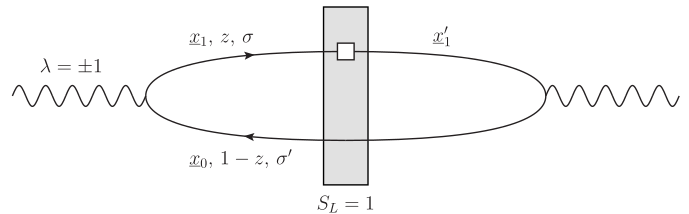} 
		\caption{Diagram representing the polarized DIS cross-section in the dipole-shockwave formalism. The gray shaded box is the proton shock wave, while the white shaded box denotes the sub-eikonal interaction.}
		\label{fig:pDIS_shockwave}
	\end{center}
\end{figure}

Fig.~\ref{fig:pDIS_shockwave} shows a diagrammatic representation of the polarized dipole interacting with the proton target. The softer, polarized Wilson line's interaction with the target is denoted by the shaded white box, similar to the dark vertex in Fig.~\ref{fig:pol_Wilson}. The proton is treated as a ``shockwave" and denoted by the gray shaded box; this is to imply that the proton is effectively frozen, and all dipole-target interactions of this type are summed together inside the gray shaded box. As discussed above in defining Eqs.~\eqref{Q10} and \eqref{G10}, the softer Wilson lines interacts in a spin-dependent way with the target, having transverse position $\underline{x}_1$ before interacting and transverse position $\underline{x}_{1'}$ after the interaction, and the polarized dipole itself has transverse size $x_{10} = |\underline{x}_1 - \underline{x}_0|$. The derivation is detailed again in Ref.~\cite{Cougoulic:2022gbk}, but I present the result below
\begin{equation}\label{g_1}
	g_1(x,Q^2) = -\sum_q\frac{N_cZ_q^2}{4\pi^3}\int\limits_{\Lambda^2/s}^1\frac{dz}{z}\int\limits_{1/zs}^{\mathrm{min}\{\frac{1}{zQ^2},\frac{1}{\Lambda^2}\}}\frac{dx_{10}^2}{x_{10}^2}\,\big[Q_q(x_{10}^2,zs) + 2G_2(x_{10}^2,zs)\big]
\end{equation}
where $Z_q$ is the fractional quark charge, terms that do not generate two ($x$-enhanced) logs have been excluded, and PT-symmetry has been imposed. Eq.~\eqref{g_1} can combine with Eq.\eqref{Deltaqp} and recreate the well-known LO collinear factorization result \cite{Lampe:1998eu}
\begin{equation}\label{g1_LO}
	g_1(x,Q^2) = \frac{1}{2}\sum_q Z_q^2\,\Delta q^+(x,Q^2).
\end{equation}
To recap, I have now been able to express the quark and gluon helicity PDFs $\Delta\Sigma(x,Q^2)$ and $\Delta G(x,Q^2)$ as well as the $g_1(x,Q^2)$ structure functions in terms of the polarized dipole amplitudes $Q_q(x_{10}^2,zs)$ and $G_2(x_{10}^2,zs)$. Considering the goal of this research is to perform phenomenology, which is to use experimental data to extract the hPDFs, I briefly note that the link between this theory and experiment lies in the virtual photon spin asymmetry $A_1 \propto g_1/F_1$ (see \textbf{Chapter}~\ref{Observables} for more details). From this relation, along with Eqs.~\eqref{Deltaqp} and \eqref{g_1}, we see that experimental inclusive DIS data will only provide extractions of the flavor-singlet quark helicity distribution $\Delta q^+(x,Q^2)$, but not individual quark flavor hPDFs. To do so, I shift your focus to semi-inclusive DIS observables, which are sensitive to the flavor nonsinglet quark helicity distribution.

The difference between the flavor nonsinglet case compared to the flavor singlet case is that the flavor nonsinglet case \textit{subtracts} diagrams in which the quark particle number flows in the opposite direction in the quark loop; for example, if the diagram in Fig.~\ref{fig:pDIS_shockwave} were to show a SIDIS process where the one of the quarks became a tagged hadron in the final state, there would be a second diagram depending on if that tagged hadron came from the quark or the antiquark, which would be summed in the flavor singlet case but subtracted in the flavor nonsinglet case. Subtracting these diagrams provides the flavor nonsinglet quark helicity PDF defined very similarly to Eq.~\eqref{Deltaqp},
\begin{equation}\label{Deltaqm}
	\Delta q^-(x,Q^2) = (\Delta q - \Delta \bar{q})(x,Q^2) = \frac{N_c}{2\pi^3}\int\limits_{\Lambda^2/s}^1\frac{dz}{z}\int\limits_{1/zs}^{\mathrm{min}\{\frac{1}{zQ^2},\frac{1}{\Lambda^2}\}}\frac{dx_{10}^2}{x_{10}^2}\,G_q^{\mathrm{NS}}(x_{10}^2,zs),
\end{equation}
where $G_q^{\mathrm{NS}}(x_{10}^2,zs)$ is quite literally the flavor nonsinglet analog to $Q_q(x_{10}^2,zs)$, defined in its impact parameter integrated state as \cite{Kovchegov:2016zex}
\begin{equation}
	G_{q,10}^{\mathrm{NS}}(z) \equiv  \frac{1}{2N_c} \dlangle \mathrm{tr}\Big[V_0\,V_1^{\mathrm{pol}[1]\dagger}\Big] - \mathrm{tr}\Big[V_1^{\mathrm{pol}[1]}\,V_0^{\dagger}\Big] \drangle (z).
\end{equation}
The SIDIS spin asymmetry has a similar relation as in DIS, $A_1^h \propto g_1^h/F_1^h$, where $g_1^h$ and $F_1^h$ are polarized and unpolarized SIDIS structure functions, dependent on the final state tagged hadron $h$. 
The derivation of the SIDIS structure function at small $x$ begins similarly to that of the DIS structure function $g_1(x,Q^2)$ from Eq.~\eqref{g1_collinear}, where $\Delta\sigma^{\gamma^* P}$ is written more explicitly as 
\begin{align}
	\Delta\sigma^{\gamma^*P} \equiv \sum_{\lambda = \pm} \lambda \, \sigma^{\gamma^*
		\,p \to X} (\lambda, +)\,,
\end{align}
where $\sigma^{\gamma^*p\to X}(\lambda,+)$ is the virtual photon-proton part of the DIS cross-section where the proton has helicity $\Sigma = +$ and the virtual photon is transversely polarized with polarization $\lambda$.

The SIDIS process involves the scattering of a virtual photon with a proton (or hadron) target where a specific hadron is tagged in the final state; at small $x$ I have to re-draw this process in the dipole formalism as shown in Fig.~\ref{fig:smallx_SIDIS} below.
%
\begin{figure}[ht]
	\begin{center}
		\includegraphics[width=0.6 \textwidth]{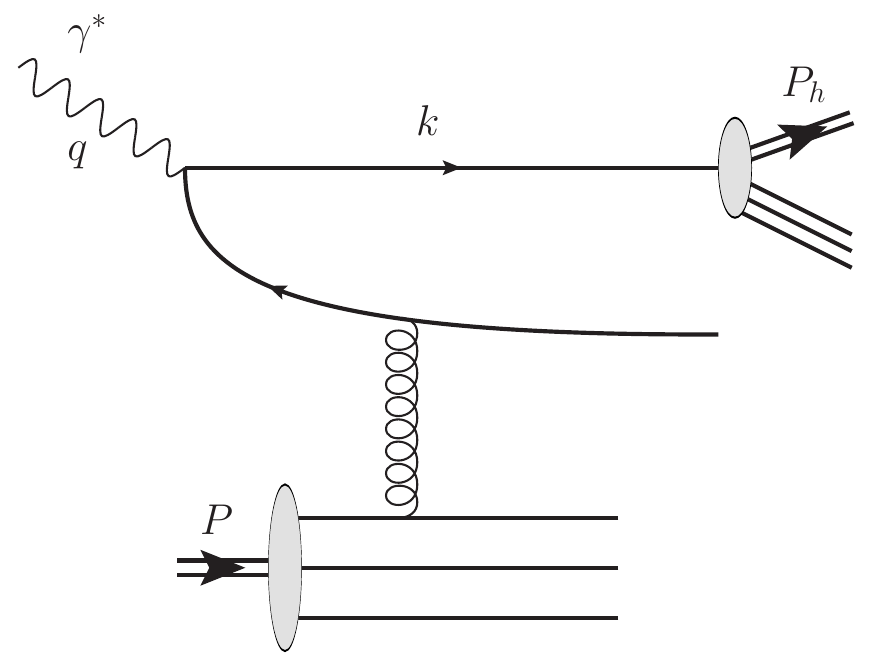} 
		\caption{The SIDIS process at small $x$.  An incoming virtual photon with momentum $q$ decays into a quark--antiquark dipole which interacts with the target proton carrying momentum $P$.  The quark and antiquark then fragment into hadrons, one of which is tagged in the final state with momentum $P_h$.}
		\label{fig:smallx_SIDIS}
	\end{center}
\end{figure}
%
In this figure, the virtual photon has momentum $q$, the dipole interacts with the proton target in such a way that the (anti)quark with momentum $k$ will fragment into a specific tagged hadron that carries momentum $P_h$, and the proton has momentum $P$. The produced hadron will have a fixed value of $z = P\cdot P_h / P \cdot q$; typically, small-$x$ work is done in the frame where the proton's momentum is dominated by its light-cone plus component and the virtual photon's momentum is dominated by its light-cone minus component. In this case, $z\approx P_h^-/q^-$ is effectively the fraction of the virtual photon's (minus) momentum carried by the produced hadron. I can then follow the collinear approximation used to write Eq.~\eqref{g1_collinear} ~\cite{deFlorian:1995fd,Bacchetta:2006tn, Signori:2013mda} and, by analogy, write the SIDIS cross-section as 
\begin{align}\label{g1h_xsect}
	g_1^h (x, z, Q^2) = - \frac{Q^2}{16 \pi^2 \alpha_{EM} \, x} \, \sum_{\lambda = \pm} \lambda \, &\int d^2 k_\perp \, d^2 P_{h\perp} \, \delta^{(2)} \!\left( z \, {\bf k}_\perp - {\bf P}_{h\perp} \right) \, \\
	& \qquad\times\sum_{q , {\bar q}}  \frac{d \sigma^{\gamma^* \,p\to q\,X}}{d^2 k_\perp} (\lambda, +) \,  D_1^{h/q}(z,Q^2)\,, \notag
\end{align}
with the transverse momentum two-vectors of the (anti)quark and the produced hadron, $\mathbf{k}_{\perp}$ and $\mathbf{P}_{h\,\perp}$, respectively. The collinear fragmentation functions $D_1^{h/q}(z,Q^2)$ give the probability that a given parton $q$ will result in a specific hadron with momentum fraction $z$, and the sum goes over both quarks and antiquarks. To arrive at Eq.~\eqref{g1h_xsect} I assumed the DLA-dominant aligned jet configuration, which sees the quark and virtual photon's (minus) momentum as the same $P_h^-/k^- \approx P_h^-/q^- = z$. In the small-$x$ dipole picture the quark and antiquark must be separated by a large rapidity interval, so to ensure that the produced hadron comes from the quark (as depicted in Fig.~\ref{fig:smallx_SIDIS}) and not the antiquark, I must assume that $z$ is not too small such that the produced hadron is produced at forward rapidity (in the direction of the virtual photon). The momentum scale in the argument of the fragmentation functions could be $k_{\perp}^2$, but considering that with the small-$x$ kinematics $k_{\perp}^2\sim Q^2$, $Q^2$ is used as the argument instead. After performing the transverse momentum integrals, I arrive at 
\begin{equation}
	g_1^h(x,z,Q^2) = -\frac{Q^2}{16\,\pi^2\,\alpha_{\mathrm{EM}}\,x}\sum_{\lambda = \pm }\lambda\,\sum_{q,\bar{q}}\sigma^{\gamma^*\,p\to q\,X}(\lambda,+)\,D_q^{h/q}(z,Q^2)
\end{equation}
within which one can find Eq.~\eqref{g1_collinear}, substitute it with Eq.~\eqref{g1_LO}, and ultimately derive
\begin{equation}\label{g1h}
	g_1^h(x,z,Q^2) = \frac{1}{2}\sum_{q,\bar{q}}Z_q^2\,\Delta q(x,Q^2)\,D_1^{h/q}(z,Q^2).
\end{equation}
From Eq.~\eqref{g1h} we see that SIDIS data is sensitive to individual quark flavor hPDFs, rather than the inclusive flavor singlet sum $\Delta q^+(x,Q^2)$, because the fragmentation functions are dependent on the quark flavor. One needs both the flavor nonsinglet distribution and the flavor singlet distribution in order to describe SIDIS data since $\Delta q(x,Q^2) = \tfrac{1}{2}(\Delta q^+(x,Q^2) + \Delta q^-(x,Q^2))$.

To summarize this section, I have intensively studied polarized DIS and SIDIS processes in the small-$x$ regime and have found a direct correlation between helicity observables like hPDFs and polarized structure functions and polarized dipole amplitudes. This completes a very necessary first step of my process, but only a first step; describing the hPDFs via polarized dipoles is useful, but extracting the small-$x$ behavior of the hPDFs necessitates the knowledge of the polarized dipole amplitudes themselves at small $x$. \textbf{Chapter}~\ref{unpol_evo} covered the idea of small-$x$ unpolarized evolution, and how the resummation parameter was $x$-enhanced as $\alpha_s\ln(\tfrac{1}{x})\sim 1$. This resummation parameter has one logarithm of energy ($\tfrac{1}{x} \propto s$), and thus is considered a single logarithmic approximation, or SLA. I discussed how this logarithm of energy is generated from the BFKL ladder, and how the integral over the longitudinal momentum is cutoff by the rapidity interval $Y = \ln(\tfrac{1}{x})$. I will show in the next section that \textit{polarized} small-$x$ evolution resums contributions with the parameter $\alpha_s\ln^2(\tfrac{1}{x})\sim 1$, and thus is a double logarithmic approximation, or DLA, and is more sensitive to the small-$x$ regime given by the square of the logarithm of energy. In fact, the physical constraints employed above are a result of this DLA, where the integrals over the longitudinal momentum and transverse dipole size in Eq.~\eqref{Deltaqp} each generate a logarithm of energy. Small-$x$ helicity evolution is therefore governed by the resummation of all corrections to the polarized dipole amplitudes that contribute two logarithms of energy for every instance of the strong coupling constant $\alpha_s$. As an end-cap to this section, I will combine all the hPDFs and structure functions below in the DLA (which I specify only to note that the definition of $\Delta G(x,Q^2)$ will now ignore the derivative term $\partial/\partial x_{10}^2$, as it will remove one logarithm of energy):
\begin{subequations}\label{all_hPDFs}
	\begin{align}
		\Delta q^+(x,Q^2) &= -\frac{N_c}{2\pi^3}\int\limits_{\Lambda^2/s}^1 \frac{dz}{z}\int\limits_{1/zs}^{\mathrm{min}\{\frac{1}{zQ^2},\frac{1}{\Lambda^2}\}}\frac{dx_{10}^2}{x_{10}^2}\,\big[Q_q(x_{10}^2,zs) + 2G_2(x_{10}^2,zs)\big], \\
		\Delta G(x,Q^2) &= \frac{2N_c}{\alpha_s\pi^2}\,G_2\Big(x_{10}^2 = \frac{1}{Q^2}, zs = \frac{Q^2}{x}\Big), \\
		g_1(x,Q^2) &= -\sum_q\frac{N_c\,Z_q^2}{4\pi^3}\int\limits_{\Lambda^2/s}^1\,\frac{d z}{z}\int\limits_{1/zs}^{\mathrm{min}\{1/zQ^2,1/\Lambda^2\}}\,\frac{d x_{10}^2}{x_{10}^2}\,\bigl[Q_q(x_{10}^2,zs) + 2G_2(x_{10}^2,zs)\bigr], \\
		\Delta q^-(x,Q^2) &= \frac{N_c}{2\pi^3}\int\limits_{\Lambda^2/s}^1 \frac{dz}{z}\int\limits_{1/zs}^{\mathrm{min}\{\frac{1}{zQ^2},\frac{1}{\Lambda^2}\}}\frac{dx_{10}^2}{x_{10}^2}\,G_q^{\mathrm{NS}}(x_{10}^2,zs), \\
		g_1^h(x,z,Q^2) &= \frac{1}{s}\sum_{q,\bar{q}}Z_q^2\,\Delta q(x,Q^2)\,D_1^{h/q}(z,Q^2),
	\end{align}
\end{subequations}
as well as the polarized dipole amplitudes for completeness
\begin{subequations}\label{dipole_defs}
	\begin{align}
		Q_{q,10}(zs) &\equiv \frac{1}{2N_c}\,\mathrm{Re}\,\dlangle \mathrm{T~tr}\big[V_0V_1^{\mathrm{pol}[1]\dagger}\big] + \mathrm{T~tr}\big[V_1^{\mathrm{pol}[1]}V_0^{\dagger}\big] \drangle \\
		G_{10}^i(zs) & \equiv \frac{1}{2N_c}\,\dlangle \mathrm{tr}\big[V_0^{\dagger}V_1^{i\,\mathrm{G}[2]} + \big(V_1^{i\,\mathrm{G}[2]}\big)^{\dagger}V_0\big] \drangle \\
		G_{q,10}^{\mathrm{NS}}(zs) &\equiv \frac{1}{2N_c}\,\mathrm{Re}\,\dlangle \mathrm{tr}\big[V_0V_1^{\mathrm{pol}[1]\dagger}\big] - \mathrm{tr}\big[V_1^{\mathrm{pol}[1]}V_0^{\dagger}\big] \drangle
	\end{align}
\end{subequations}
where $G_{10}^i$ will be replaced by $G_2$ via the antisymmetry decomposition, and all dipoles above are given in their impact parameter integrated state. 

Before moving on to the small-$x$ evolution I note here that the result for $\Delta q^{\pm}(x,Q^2)$ above are specific to the use of the operator language in light cone time-ordered Feynman diagrams, called the light-cone operator treatment (LCOT) formalism \cite{Kovchegov:2018znm, Kovchegov:2017lsr, Mueller:2012bn, Cougoulic:2022gbk}, and as such constitutes its own scheme coined as the ``polarized DIS scheme." The small-$x$ helicity phenomenology in the polarized DIS scheme is explored in \textbf{Chapter}~\ref{pheno_1}. Recent works have modified the lower integration limit of the transverse dipole size $x_{10}^2$ in the definitions of the flavor singlet and flavor nonsinglet quark hPDFs, bringing them more in line with results derived in the $\overline{\mathrm{MS}}$ scheme. The $\overline{\mathrm{MS}}$-modified hPDFs are given below,
\begin{equation}
	\Delta q^+(x,Q^2) = -\frac{N_c}{2\pi^3}\int\limits_{\Lambda^2/s}^1 \frac{dz}{z}\int\limits_{\mathrm{max}\{1/Q^2,1/zs\}}^{\mathrm{min}\{\frac{1}{zQ^2},\frac{1}{\Lambda^2}\}}\frac{dx_{10}^2}{x_{10}^2}\,\big[Q_q(x_{10}^2,zs) + G_2(x_{10}^2,zs)\big],
\end{equation}
\begin{equation}
	\Delta q^-(x,Q^2) = \frac{N_c}{2\pi^3}\int\limits_{\Lambda^2/s}^1 \frac{dz}{z}\int\limits_{\mathrm{max}\{1/Q^2,1/zs\}}^{\mathrm{min}\{\frac{1}{zQ^2},\frac{1}{\Lambda^2}\}}\frac{dx_{10}^2}{x_{10}^2}\,G_q^{\mathrm{NS}}(x_{10}^2,zs),
\end{equation}
and the small-$x$ helicity phenomenology in this scheme is explored in \textbf{Chapter}~\ref{pheno_2}. Notice that this new lower bound on $x_{10}^2$ is only applied to $\Delta q^{\pm}(x,Q^2)$ and not to $g_1(x,Q^2)$, meaning in this scheme Eq.~\eqref{g1_LO} is not valid.

\subsection{$\boldsymbol{\mathrm{KPS\text{-}CTT\!:~Small\text{-}}x~\mathrm{Helicity~Evolution~Equations}}$}\label{KPS-CTT}

The entire discussion up until now has laid out a powerful foundation for the remainder of my theoretical work; I have established the importance of small-$x$ physics, touched on parton evolution and resummation, and cast the helicity PDFs in terms of the new degrees of freedom, the polarized dipole amplitudes. This section will combine all of these features in order to determine the small-$x$ evolution of helicity PDFs.

\begin{figure}[ht]
	\begin{center}
		\includegraphics[width=450 pt]{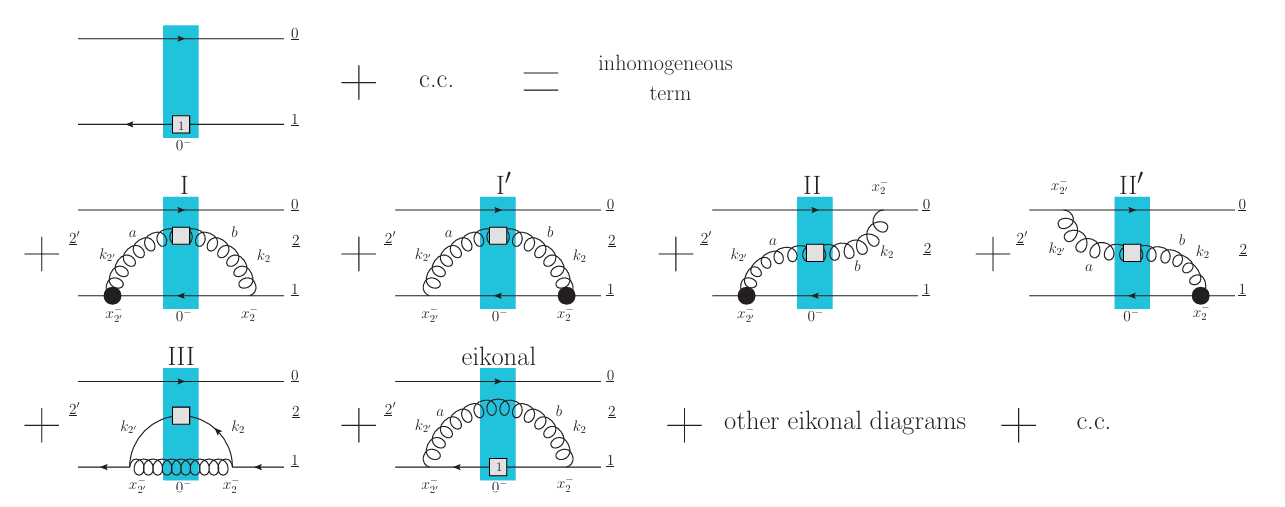} 
		\caption{Diagram representing one step of evolution of the $Q_{q,10}$ polarized dipole amplitude. The blue band represents the proton shockwave (compare to the gray band in Fig.~\ref{fig:pDIS_shockwave}) and the white box represents a polarized interaction with the target. The dark vertex is the sub-eikonal vertex as discussed in Fig.~\ref{fig:pol_Wilson}, and the label `1' inside the shaded box indicates the specific sub-eikonal interaction described by $V_{\underline{x}}^{\mathrm{pol}[1]}$.}
		\label{fig:Q10_evo_diagram}
	\end{center}
\end{figure}

Helicity PDF evolution is really an indirect process, where I do not evolve the hPDFs themselves, but instead evolve the polarized dipole amplitudes; of course, due to Eqs.~\eqref{all_hPDFs}, evolving the polarized dipoles will inherently evolve the hPDFs, but the distinction should still be understood. When discussing small-$x$ unpolarized evolution I mentioned that the BFKL evolution equation (and their cousins which include non-linear contributions such as the BFKL  and JIMWLK \cite{JalilianMarian:1997dw, JalilianMarian:1997gr, Weigert:2000gi, Ferreiro:2001qy, Iancu:2000hn, Iancu:2001ad} evolution equations) applies to unpolarized dipoles by exploring radiative corrections to their scattering amplitudes, and I will do the same here but for the polarized dipole scattering amplitudes. I did not discuss in detail how small-$x$ (or rapidity) evolution worked, but the principle is similar to the DGLAP evolution in the sense that I want to consider an infinitesimal rate of change of the operators. In the case of small-$x$ polarized dipole evolution, I essentially want to understand $\tfrac{\partial}{\partial Y}G(Y)$, where $Y$ is the rapidity, previously defined as $Y = \ln(\tfrac{1}{x})$. The leading contribution of small-$x$ radiative corrections in the unpolarized sector was already found to be the emission and subsequent reabsorption of gluons, and this remains true in the polarized case, with the quarks also contributing at sub-eikonal order though suppressed by powers of $N_c$. The sub-eikonal order is a crucial part of small-$x$ polarized evolution. As I discussed when defining the polarized Wilson lines, only energy-suppressed interactions are capable of transferring spin information. The existing small-$x$ helicity literature has provided many diagrammatic representations showing the quantum corrections to polarized dipole amplitudes of this type, and I will borrow a diagram from  Ref.~\cite{Cougoulic:2022gbk}, labeled here as Fig.~\ref{fig:Q10_evo_diagram}, that contains a mountain of information. 

In Fig.~\ref{fig:Q10_evo_diagram} I show one step of evolution of the $Q_{10}$ polarized dipole amplitude, where each correction on the right-hand side depicts a possible quark or gluon emission and reabsorption. The labels $\underline{0},\underline{1}, \underline{2}$ on the left indicate the transverse coordinate of the Wilson lines, and the blue shaded region represents the proton shockwave and the white box denotes a polarized interaction as in Fig.~\ref{fig:pDIS_shockwave}. Note here that, though not shown explicitly, a step in evolution corresponds to the generation of more dipoles; using the first term on the right-hand-side of Fig.~\ref{fig:Q10_evo_diagram}, we see that one step of evolution now consists of two polarized dipole amplitudes: a polarized dipole made of a quark Wilson line at position $\underline{0}$ and polarized gluon Wilson line at position $\underline{2}$, and then a dipole constructed out of the polarized gluon Wilson line at position $\underline{2}$ and the unpolarized quark Wilson line at position $\underline{1}$\footnote{The operator related to the gluon going through the shockwave more complex than the polarized dipole operator, but in the dilute limit the operator is decomposed into a sum of discrete, typical polarized dipole operators.}. The main property that appears when investigating emissions like this is that the sub-eikonal corrections involved in each diagram now mean that the leading contributions are those that generate two logarithms of energy for every power of $\alpha_s$, an important distinction from the single-logarithmic enhancement of unpolarized evolution.

As discussed in the previous section, we can identify where both of these logarithms are generated. The first logarithm is generated when integrating over the longitudinal momentum of the softest Wilson line in the polarized dipole; imposing lifetime ordering of each subsequent soft emission (see Fig.~\ref{fig:BFKL_Ladder}) results in a UV divergence when $z\to0$, so I must impose a cutoff via a non-perturbative momentum scale $\Lambda > \Lambda_{\mathrm{QCD}}$ and the center-of-mass energy squared $s$. This is the same source of logarithmic enhancements that appear in unpolarized small-$x$ evolution. The salient difference between unpolarized evolution and polarized evolution is that the integral over the transverse dipole size (or perhaps the transverse momentum) is regulated naturally in the unpolarized case. Following an example from Ref.~\cite{Kovchegov:2012mbw}, the $S$-matrix operator for the unpolarized dipole-nucleus scattering amplitude is defined as 
\begin{equation}\label{unpol_dipole_op}
	\hat{S}_{10} = \frac{1}{N_c}\,\mathrm{tr}\big[V_1V_0^{\dagger}]
\end{equation}
and then the unpolarized evolution of this operator (in the 't Hooft large-$N_c$ limit) is written as 
\begin{equation}
	\frac{\partial}{\partial Y}\langle \hat{S}_{10}\rangle_Y \propto \int d^2 x_2 \frac{x_{10}^2}{x_{20}^2x_{21}^2}\Big[\langle\hat{S}_{12}\rangle_Y\langle\hat{S}_{20}\rangle_Y - \langle\hat{S}_{10}\rangle_Y\Big]
\end{equation}
where I have employed the CGC averaging angled brackets. See that we approach a singularity in the case that the dipole $\hat{S}_{21}$ has a transverse size $x_{21} \to 0$, however, in the same limit we get $\underline{2} \to \underline{1}$ and thus $V_1V_2^{\dagger}\to V_1V_1^{\dagger}\to 1$ since Wilson lines are unitary matrices. The remaining two terms then cancel each other out and remove the singularity altogether since the integrand itself vanishes. This exact property of the Wilson lines does not translate to polarized Wilson lines, and thus the integral over transverse dipole size has an unregulated divergence unless I implement a cutoff at the smallest transverse scale in the problem, determined again by the energy $s$. This is the reason there are two specific integration limits imposed in Eqs.~\eqref{all_hPDFs}, and why the resummation parameter for small-$x$ polarized evolution is $\alpha_s\ln^2(x)$, which is known as the double logarithmic approximation or DLA. The DLA nature of polarized evolution is especially important for my goal of performing phenomenology, because the double enhancement through $x$ means that $x$ is ``small" for much larger values when analyzing polarized data; consider that if the SLA resummation parameter $\alpha_s\ln(1/x^{\mathrm{unpol.}})\sim 1$, then for the same coupling constant $\alpha_s$, the DLA resummation parameter can achieve $\alpha_s\ln^2(1/x^{\mathrm{pol.}})\sim 1$ for $x^{\mathrm{pol.}}>x^{\mathrm{unpol.}}$. This means that there is a larger selection of ``small-$x$" data for polarized asymmetries than small-$x$ unpolarized asymmetries, and that small-$x$ effects may be identified through polarized data more easily than unpolarized data.

Now the meaning of Fig.~\ref{fig:Q10_evo_diagram} has been made clearer: I want to establish what a step of small-$x$ evolution looks like for polarized dipole amplitudes by counting processes that emit and reabsorb partons and contribute two logarithms of energy for one power of $\alpha_s$. This is made more complicated when we realize that 1) the evolution of a polarized dipole amplitude itself relies on more, various combinations of polarized and unpolarized Wilson lines as labeled by $\mathrm{I}^{(')}, \mathrm{II}^{(')}, \mathrm{III}$, and 2) I cannot stop at an arbitrary order because for a resummation parameter $\alpha_s\ln^2(1/x)\sim 1$, a higher order diagram (in the resummation parameter) is equally important, $\alpha_s^2\ln^4(1/x) = (\alpha_s\ln^2(1/x))^2 \sim 1^2 = 1$. The second reason above is handled by the fact that an evolution equation is built to inherently resum all orders of the resummation parameter, just like how the DGLAP evolution equations resums all orders of $\alpha_s\ln(Q^2/\Lambda_{QCD}^2)$ and BFKL resums all orders of $\alpha_s\ln(1/x)$. The first point is harder to contend to, as we see that, for example, diagram $\mathrm{I}$ in Fig.~\ref{fig:Q10_evo_diagram} has its own ``daughter" polarized dipole amplitudes (constructed out of the quark line at position $\underline{0}$ and the polarized gluon line at position $\underline{2}$) that in turn would need to be evolved, and so on. To solve this issue, I can implement some well-known approximations that get rid of the infinite cascade of dipole amplitudes: the 't Hooft large-$N_c$ limit \cite{tHooft:1973alw} and Veneziano's large-$N_c\&N_f$ limit \cite{Veneziano:1976wm}. $N_c$ refers to the number of quark colors and $N_f$ refers to the number of quark flavors; to be clear, I am not assuming that there are an infinite number of colors or flavors, but instead I take the limit that $N_c\to\infty$ (with $\alpha_sN_c$ is small) or the limit $N_c, N_f\to\infty$ (with $\alpha_sN_c$ and $\alpha_sN_f$ small and $N_c/N_f$ is fixed), to neglect terms that are sub-leading in powers $1/N_c$ and/or not $N_f$ enhanced. The number of distinct polarized dipole amplitudes becomes finite in either of these limits, and thus the evolution equations are closed and solvable.

The large-$N_c$ limit is also known as the ``pure-glue" limit because all contributions with soft quark emissions are $N_c$-suppressed and thus neglected, and there is no external flavor dependence. Additionally, the large-$N_c$ allows gluon lines to be approximated as a color nonsinglet dipole of quarks. These soft quark emissions are, however, enhanced by the prefactor $N_f$, and are thus brought back in when the large-$N_c\&N_f$ limit is taken. This is a more accurate limit, and the one that I will proceed with, since the flavor dependence is necessary for resolving individual quark flavor hPDFs.

The last thing I need to establish before quoting the results of \cite{Cougoulic:2022gbk} and providing the small-$x$ helicity evolution equations is the adjoint analogue of the quark polarized dipole amplitude, which would look quite similar to Fig.~\ref{fig:Q10_evo_diagram}, but with the quark Wilson lines replaced with gluon Wilson lines (and of course each of the sub-eikonal interactions would be different owing to the differences between Eqs.~\eqref{quark_pol_Wilson} and \eqref{gluon_pol_Wilson}). This would nominally have a straightforward definition of 
\begin{equation}\label{G_adj}
	G_{10}^{adj}(zs) = \frac{1}{2(N_c^2-1)}\,\mathrm{Re}\,\dlangle \mathrm{T~Tr}\Big[U_0\,U_1^{\mathrm{pol}[1]\dagger}\Big] + \mathrm{T~Tr}\Big[U_1^{\mathrm{pol}[1]}\,U_0^{\dagger}\Big]\drangle (zs)
\end{equation}
however I can define a new polarized Wilson line $W_{\underline{x}}^{\mathrm{pol}[1]}$ as
\begin{equation}\label{W_Wilson}
	W_{\underline{x}}^{\mathrm{pol}[1]} = V_{\underline{x}}^{\mathrm{G}[1]} + \frac{g^2p_1^+}{4s}\int\limits_{-\infty}^{\infty}dx_1^-\int\limits_{x_1^-}^{\infty}dx_2^-\,V_{\underline{x}}[\infty,x_2^-]\,\psi_{\alpha}(x_s^-,\underline{x})\,\Big(\frac{1}{2}\gamma^+\gamma_5\Big)_{\beta\alpha}\,\bar\psi_{\beta}(x_1^-,\underline{x})\,V_{\underline{x}}[x_1^-,-\infty]
\end{equation}
such that in the large-$N_c\&N_f$ limit I have the relation
\begin{equation}\label{U_lNcNf}
	\Big(U_{\underline{x}}^{\mathrm{pol}[1]}\Big)^{ba}\, \Bigg|_{\mathrm{large}\textrm{-}N_c\&N_f}  = 4\,\mathrm{tr}\Big[W_{\underline{x}}^{\mathrm{pol}[1]\dagger}\,t^b\,V_{\underline{x}}\,t^a\Big] + 4\,\mathrm{tr}\Big[V_{\underline{x}}^{\dagger}\,t^b\,W_{\underline{x}}^{\mathrm{pol}[1]}\,t^a\Big]
\end{equation}
and can rewrite Eq.~\eqref{G_adj} as 
\begin{equation}
	G_{10}^{adj}(zs) = 4\hat{S}_{10}(zs)\,\widetilde{G}_{10}(zs).
\end{equation}
In the above equation, I have made use of Eq.~\eqref{unpol_dipole_op} and defined a new (impact parameter integrated) polarized dipole amplitude $\widetilde{G}_{10}(zs)$, aptly called the adjoint polarized dipole amplitude. It has the official definition
\begin{equation}\label{Gtilde}
	\widetilde{G}_{10}(zs) = \frac{1}{2N_c}\,\mathrm{Re}\,\dlangle\mathrm{T~tr}\,\Big[V_0\,W_1^{\mathrm{pol}[1]\dagger}\Big] + \mathrm{T~tr}\Big[W_1^{\mathrm{pol}[1]}\,V_0^{\dagger}\Big]\drangle(zs).
\end{equation}
I took us through this slightly convoluted construction simply to make the evolution equations easier to solve; the diagrams in Fig.~\ref{fig:Q10_evo_diagram}, as well as those that would appear when $G_2(zs)$, can use as much simplification as possible, and $\widetilde{G}_{10}(zs)$ is more useful in deriving them than $G_{10}^{adj}(zs)$.

In the end, all DLA contributions to the large-$N_c\&N_f$ evolution of $Q_q(x_{10}^2,zs)$, $\widetilde{G}(x_{10}^2,zs)$, and $G_2(x_{10}^2,zs)$ have been found and calculated through the combined effort authors Kovchegov, Pitonyak, and Sievert, and updated by authors Cougoulic, Tarasov, and Tawabutr. The culmination of their work results in the KPS-CTT small-$x$ helicity (flavor singlet and flavor nonsinglet) evolution equations \cite{Kovchegov:2015pbl, Hatta:2016aoc, Kovchegov:2016zex, Kovchegov:2016weo, Kovchegov:2017jxc, Kovchegov:2017lsr, Kovchegov:2018znm, Kovchegov:2019rrz, Cougoulic:2019aja, Kovchegov:2020hgb, Cougoulic:2020tbc, Chirilli:2021lif, Kovchegov:2021lvz, Cougoulic:2022gbk}: 

\setlength{\baselineskip}{\singlespace}
{\allowdisplaybreaks
	\begin{subequations}\label{eq_LargeNcNf}
		\begin{align}
			& Q_q (x^2_{10},zs) = Q_q^{(0)}(x^2_{10},zs) + \frac{N_c}{2\pi} \int\limits^{z}_{1/x^2_{10}s} \frac{d z'}{z'}   \int\limits_{1/z's}^{x^2_{10}}  \frac{d x^2_{21}}{x_{21}^2}  \ \alpha_s \!\!\left( \frac{1}{x_{21}^2} \right) \label{Q_q-xspace} \\
			&\hspace*{3cm} \times \, \bigg[ 2 \, {\widetilde G}(x^2_{21},z's)+ 2 \, {\widetilde \Gamma}(x^2_{10},x^2_{21},z's)+ Q_q (x^2_{21},z's) \notag \\
			&\hspace*{5cm}-  \overline{\Gamma}_q (x^2_{10},x^2_{21},z's) + 2 \, \Gamma_2(x^2_{10},x^2_{21},z's) + 2 \, G_2(x^2_{21},z's)   \bigg] \notag  \\
			&\hspace*{2cm}+ \frac{N_c}{4\pi} \int\limits_{\Lambda^2/s}^{z} \frac{d z'}{z'}   \int\limits_{1/z's}^{\min \left[ x^2_{10}z/z', 1/\Lambda^2 \right]}  \frac{d x^2_{21}}{x_{21}^2} \ \alpha_s\!\! \left( \frac{1}{x_{21}^2} \right) \, \bigg[Q_q (x^2_{21},z's) + 2 \, G_2(x^2_{21},z's) \bigg] ,  \notag  \\[0.5cm]
			&\overline{\Gamma}_q (x^2_{10},x^2_{21},z's) = Q^{(0)}_q (x^2_{10},z's) + \frac{N_c}{2\pi} \int\limits^{z'}_{1/x^2_{10}s} \frac{d z''}{z''}   \int\limits_{1/z''s}^{\min[x^2_{10}, x^2_{21}z'/z'']}  \frac{d x^2_{32}}{x_{32}^2}   \ \alpha_s\!\! \left( \frac{1}{x_{32}^2} \right)   \\
			&\hspace*{3cm} \times \; \, \bigg[ 2\, {\widetilde G} (x^2_{32},z''s)+ \; 2\, {\widetilde \Gamma} (x^2_{10},x^2_{32},z''s) +  Q_q (x^2_{32},z''s) \notag \\
			&\hspace*{5cm}-  \overline{\Gamma}_q (x^2_{10},x^2_{32},z''s) + 2 \, \Gamma_2(x^2_{10},x^2_{32},z''s) + 2 \, G_2(x^2_{32},z''s) \bigg] \notag \\
			&\hspace*{2cm}+ \frac{N_c}{4\pi} \int\limits_{\Lambda^2/s}^{z'} \frac{d z''}{z''}   \int\limits_{1/z''s}^{\min \left[ x^2_{21}z'/z'', 1/\Lambda^2 \right] }  \frac{d x^2_{32}}{x_{32}^2} \ \alpha_s\!\! \left( \frac{1}{x_{32}^2} \right) \, \bigg[Q_q (x^2_{32},z''s) + 2 \, G_2(x^2_{32},z''s) \bigg] , \notag \\[0.5cm]
			& {\widetilde G}(x^2_{10},zs) = {\widetilde G}^{(0)}(x^2_{10},zs) + \frac{N_c}{2\pi}\int\limits^{z}_{1/x^2_{10}s}\frac{d z'}{z'}\int\limits_{1/z's}^{x^2_{10}} \frac{d x^2_{21}}{x^2_{21}} \ \alpha_s\!\! \left( \frac{1}{x_{21}^2} \right)  \\
			&\hspace*{3cm}  \times \; \bigg[3 \, {\widetilde G}(x^2_{21},z's)+ \; {\widetilde \Gamma}(x^2_{10},x^2_{21},z's) +  2\,G_2(x^2_{21},z's) \notag \\
			&\hspace*{5cm}+  \Big(2 - \frac{N_f}{2N_c}\Big) \Gamma_2(x^2_{10},x^2_{21},z's) - \frac{1}{4N_c}\, \sum_q \overline{\Gamma}_q (x^2_{10},x^2_{21},z's)\bigg] \notag \\
			&\hspace*{1.5cm}- \frac{1}{8\pi}  \int\limits_{\Lambda^2/s}^z \frac{d z'}{z'}\int\limits_{\max[x^2_{10},\,1/z's]}^{\min \left[ x^2_{10}z/z' , 1/\Lambda^2 \right]} \frac{d x^2_{21}}{x^2_{21}} \ \alpha_s\!\! \left( \frac{1}{x_{21}^2} \right) \, \bigg[ \sum_q  Q_q (x^2_{21},z's) +     2 \, N_f \, G_2(x^2_{21},z's)  \bigg] , \notag \\[0.5cm]
			& {\widetilde \Gamma} (x^2_{10},x^2_{21},z's) = {\widetilde G}^{(0)}(x^2_{10},z's) + \frac{N_c}{2\pi}\int\limits^{z'}_{1/x^2_{10}s}\frac{d z''}{z''}\int\limits_{1/z''s}^{\min[x^2_{10},x^2_{21}z'/z'']} \frac{d x^2_{32}}{x^2_{32}} \ \alpha_s\!\! \left( \frac{1}{x_{32}^2} \right) \,   \\
			&\hspace*{3cm} \times \;\bigg[3 \, {\widetilde G} (x^2_{32},z''s) + {\widetilde \Gamma}(x^2_{10},x^2_{32},z''s) + 2 \, G_2(x^2_{32},z''s)  \notag \\
			&\hspace*{5cm}+  \Big(2 - \frac{N_f}{2N_c}\Big) \Gamma_2(x^2_{10},x^2_{32},z''s) - \frac{1}{4N_c} \sum_q \overline{\Gamma}_q (x^2_{10},x^2_{32},z''s)  \bigg] \notag \\
			&\hspace*{0.6cm}- \frac{1}{8\pi}  \int\limits_{\Lambda^2/s}^{z'x^2_{21}/x^2_{10}} \frac{d z''}{z''}\int\limits_{\max[x^2_{10},\,1/z''s]}^{\min \left[ x^2_{21}z'/z'', 1/\Lambda^2 \right]} \frac{d x^2_{32}}{x^2_{32}} \ \alpha_s\!\! \left( \frac{1}{x_{32}^2} \right) \bigg[ \sum_q \, Q_q (x^2_{32},z''s) +  2  \,  N_f \, G_2(x^2_{32},z''s)  \bigg] , \notag \\[0.5cm]
			& G_2(x_{10}^2, z s)  =  G_2^{(0)} (x_{10}^2, z s) + \frac{N_c}{\pi} \, \int\limits_{\Lambda^2/s}^z \frac{d z'}{z'} \, \int\limits_{\max \left[ x_{10}^2 , 1/z' s \right]}^{\min \left[ x_{10}^2 z/z', 1/\Lambda^2 \right] } \frac{d x^2_{21}}{x_{21}^2} \ \alpha_s\!\! \left( \frac{1}{x_{21}^2} \right) \\
			&\hspace*{9cm} \times \, \bigg[ {\widetilde G} (x^2_{21} , z' s) + 2 \, G_2 (x_{21}^2, z' s)  \bigg] , \notag \\[0.5cm]
			& \Gamma_2 (x_{10}^2, x_{21}^2, z' s)  =  G_2^{(0)} (x_{10}^2, z' s) + \frac{N_c}{\pi} \!\! \int\limits_{\Lambda^2/s}^{z' x_{21}^2/x_{10}^2} \frac{d z''}{z''} \!\!\!\!\!\!\!\!\! \int\limits_{\max \left[ x_{10}^2 , 1/z''s \right]}^{\min \left[  z'x_{21}^2/z'', 1/\Lambda^2 \right] } \!\! \frac{d x^2_{32}}{x_{32}^2} \ \alpha_s\!\! \left( \frac{1}{x_{32}^2} \right) \\
			&\hspace*{9cm} \times \; \bigg[ {\widetilde G} (x^2_{32} , z'' s) + 2 \, G_2(x_{32}^2, z'' s)  \bigg] \notag \\
			& G_q^{\mathrm{NS}}(x_{10}^2, z )  =  G_q^{\mathrm{NS}\,(0)} (x_{10}^2, z ) + \frac{N_c}{4\pi} \, \int\limits_{\Lambda^2/s}^z \frac{d z'}{z'} \, \int\limits_{1/z's}^{\min \left[ z x_{10}^2/z', 1/\Lambda^2 \right] } \frac{d x^2_{21}}{x_{21}^2} \ \alpha_s\!\! \left( \frac{1}{x_{21}^2} \right) \; G_q^{\mathrm{NS}}(x_{21}^2,z') \label{nonsinglet_evo}. 
		\end{align}
	\end{subequations}
}
\setlength{\baselineskip}{\doublespace}

Each polarized dipole amplitude integrates over its respective ``daughter" polarized dipole amplitudes as described above; the daughter dipoles are integrated over their respective dipole size and longitudinal momentum fraction, $x_{21}^2$ and $z'$. What has been newly introduced, however, are the ``neighbor" polarized dipole amplitudes $\overline{\Gamma}_q$, $\widetilde{\Gamma}$, and $\Gamma_2$. As can be inferred by their chosen variables, there is a close relationship between the neighbor polarized dipole amplitudes and their fundamental counterparts $Q_q$, $\widetilde{G}$, and $G_2$. The neighbor dipoles are born out of the necessity of life-time ordering; the life-time of any daughter dipole, $\sim x_{21}z'$, must always be shorter than that of the parent dipole, $\sim x_{10}z$, and this is enforced by the neighbor dipoles who, by construction, only exist in the region $x_{10}^2\,z \gg x_{21}^2\,z'$ \cite{Kovchegov:2015pbl, Kovchegov:2016zex, Kovchegov:2018znm, Cougoulic:2022gbk}. Outside of this region, any evolution would be beyond DLA accuracy and ignored. The neighbor dipoles only exist in the context of being a daughter dipole themselves, and as such, they integrate over a subsequent daughter dipole with transverse size $x_{32}^2$ and longitudinal momentum fraction $z''$. Note that the initial conditions, or inhomogeneous terms, of the neighbor dipoles are the same as the fundamental polarized dipole amplitudes, meaning that the evolution of polarized dipole amplitudes only depends on five distinct initial conditions.

One may note that the strong coupling constant $\alpha_s$ is not actually fixed; the running of the coupling has been absorbed into the evolution kernel, and it runs with the transverse size of the daughter dipole (which, one remembers, means it still scales with the energy as expected). The running of the coupling is determined by the one-loop expression 
\begin{equation}\label{running}
	\alpha_s(Q^2) = \frac{12\pi}{11N_c - 2N_f}\,\frac{1}{\ln(Q^2/\Lambda_{\mathrm{QCD}}^2)}\Bigg|_{Q^2\sim1/x_{ij^2}}
\end{equation}
where the QCD confinement scale $\Lambda_{\mathrm{QCD}}$ is taken to be $241~\mathrm{MeV}$ such that $\alpha_s(M_Z^2) = 0.1176$ for $N_f=3$, with $M_Z$ the mass of the $Z$ boson~\cite{Albacete:2009fh} (note that the IR cutoff is such that $\Lambda > \Lambda_{\mathrm{QCD}}$). For the $Q^2$ range I consider in the analyses that follow, I compute the running coupling with $N_c = 3$ and $N_f = 3$.

Additionally, it is a notable feature that the small-$x$ evolution of polarized dipole amplitudes is linear; there are no products or powers of polarized dipole amplitudes. This in principle means the evolution of polarized dipoles generated from an arbitrary set of initial condition is equal to a linear combination of evolutions generated from each initial condition separately; for example, the evolved dipole amplitudes that result from an initial condition state of $[Q_u^{(0)} = a,\;G_2^{(0)}=b]$ are equal to the sum of evolved dipoles from an evolution with the initial condition $[Q_u^{(0)} = a]$ and evolved dipole amplitudes from an evolution with the initial condition $[G_2^{(0)} = b]$. This is an especially useful feature when it comes to phenomenology, as I will be discussing shortly.

It is only now that I have reached the theoretical goal set out before us; I have the power to determine the small-$x$ behavior of the polarized dipole amplitudes $Q_q(x_{10}^2,zs)$ and $G_2(x_{10}^2,zs)$, and thus can determine the small-$x$ behavior of the helicity PDFs $\Delta q(x, Q^2)$ and $\Delta G(x,Q^2)$. However, that is not to say that the solution to small-$x$ spin is ready and available; the evolution equations above are all dependent on the initial conditions. Just as I discussed with the DGLAP evolution equations, I must know what the polarized dipole amplitudes look like at some starting value of $x_0$, and only then can I input those initial conditions in Eqs.~\eqref{eq_LargeNcNf} and then compute their values at some smaller value $x < x_0$. I present two ways that I can determine the initial conditions, and in fact, I will show how they can be used in tandem. The first option is to use a Born-level cross-section as the initial condition. This was conducted in \cite{Kovchegov:2015pbl, Kovchegov:2016zex, Cougoulic:2022gbk}, where a lowest-order toy-model was computed that treated the target simply as a single quark, with the resultant initial conditions 
\begin{subequations}\label{Born_IC}
	\begin{align}
		Q_q^{(0)}(x_{10}^2,zs) &= \widetilde{G}^{(0)}(x_{10}^2,zs) = \frac{\alpha_s^2C_F}{2N_c}\pi\Big[C_F\,\ln\frac{zs}{\Lambda^2} - 2\,\ln(zsx_{10}^2)\Big], \\
		G_2^{(0)}(x_{10}^2,zs) &= \frac{\alpha_s^2C_F}{2N_c}\pi\,\ln\frac{1}{x_{10}^2\Lambda^2}
	\end{align}
\end{subequations}
where $C_F = (N_c^2-1)/2N_c$, and it is assumed that the projectile is much smaller than the target such that $x_{10} \ll 1/\Lambda$. Of course, these initial conditions are not expected to accurately represent a proton or nuclear target, but it is a good starting point. The second option is to use data to determine the initial conditions, which has been used widely used in DGLAP phenomenology \cite{Gluck:2000dy, Leader:2005ci, deFlorian:2009vb, Leader:2010rb, Jimenez-Delgado:2013boa, Ball:2013lla, Nocera:2014gqa, deFlorian:2014yva, Leader:2014uua, Sato:2016tuz, Ethier:2017zbq, DeFlorian:2019xxt, Borsa:2020lsz, Zhou:2022wzm, Cocuzza:2022jye, Borsa:2023tqr, Hunt-Smith:2024khs}. I will show in the following chapter my methodology for conducting phenomenological analyses of the KPS-CTT small-$x$ helicity evolution equations, using global data to determine appropriate initial conditions, and subsequently using those initial conditions to extract small-$x$ hPDFs.

	\newpage
	\section{$\boldsymbol{\mathrm{Global~Analysis~and~Methodology}}$}\label{chapter_3}

\subsection{$\boldsymbol{\mathrm{Numerics,~Initial~Conditions,~and~Basis~Functions}}$}\label{numerics_etc}

Eqs.~\eqref{eq_LargeNcNf} are very powerful, and if I am able to determine the initial conditions for the polarized dipole amplitudes, I will be able to predict their $x$-dependence. The issue, however, is that the evolution equations are very difficult to solve analytically; the large-$N_c$ evolution equations have been solved analytically \cite{Borden:2024jpd}, and the flavor nonsinglet evolution can be solved analytically when the strong coupling is fixed \cite{Adamiak:2023yhz}, but there is currently no analytic solution for the large-$N_c\&N_f$ evolution with the running coupling. Thus, the best course of action is to discretize the evolution equations and solve them numerically. This has been done in Refs.~\cite{Adamiak:2021ppq, Adamiak:2023yhz, Adamiak:2025dpw}, and I will summarize the procedure here.

First, note that the evolution equations can be cast into a simpler form if I define new logarithmic variables
\begin{equation}\label{log_vars}
	\eta^{(n)} = \sqrt{\frac{N_c}{2\pi}}\ln\Big(\frac{z^{(n)}s}{\Lambda^2}\Big), \qquad s_{ij} = \sqrt{\frac{N_c}{2\pi}}\ln\Big(\frac{1}{x_{ij}^2\Lambda^2}\Big)
\end{equation}
where $\eta^{(n)} = \eta, \eta', \eta''$ in correspondence with the various momentum fractions $z^{(n)} = z, z', z''$. This will alter the definition of the running of the coupling
\begin{equation}
	\alpha_s(s_{ij}) = \sqrt{\frac{N_c}{2\pi}}\frac{12N_c}{(11N_c - 2N_f)}\,\frac{1}{s_{ij} + s_0}, \qquad s_0 = \sqrt{\frac{N_c}{2\pi}}\ln\Big(\frac{\Lambda^2}{\Lambda_{\mathrm{QCD}}^2}\Big),
\end{equation}
which conveniently avoids the Landau pole at $s_{ij} = -s_0$ since I have established that $\Lambda > \Lambda_{\mathrm{QCD}}$ and that $x_{ij} < 1/\Lambda$, resulting in both $s_0,~s_{ij} > 0$. 

Now, as written, Eqs.~\eqref{eq_LargeNcNf} do not \textit{enforce} the constraint that $x$ must be small, it was simply derived under the assumption that terms enhanced in the small-$x$ regime dominate. I can use the variables defined above to enforce a small-$x$ constraint; treating these variables in the most conservative case where $z$ extends all the way up to its maximum of 1, and treating the parent dipole's size as $x_{10}^2 \sim 1/Q^2$, I can define a ``rapidity" variable $y \equiv \eta - s_{10} = \sqrt{\tfrac{N_c}{2\pi}}\ln\tfrac{1}{x}$, remembering that $x \approx Q^2/s$. Doing so allows me to define a starting condition $y_0 = \sqrt{\tfrac{N_c}{2\pi}}\ln\tfrac{1}{x_0}$, where  I can ensure that $x$ is small by choosing a cutoff $x_0$ such that $x < x_0$; I enforce this with my logarithmic variables by using the relation $\eta - s_{10} > y_0$, as well as their daughter dipole analogs $\eta'-s_{21} > y_0$ and $\eta'' - s_{32} > y_0$.  I will discuss later how I make the decision regarding what value of $x_0$ can be considered small. I mentioned above when defining the hPDFs that the transverse dipole size must also be constrained; there is a constraint from above by the IR cutoff $x_{ij} < 1/\Lambda$, but the dipoles are also not physical for transverse sizes that are smaller than the smallest length scale in the process, so I cut it off from below by $x_{10} > 1/zs$. After the change of variables, these constraints are written as $0 < s_{10} < \eta$. Though not explicitly stated before, it is also reasonable to enforce constraints on the neighbor dipoles, $s_{10} < \eta$, $s_{21} < \eta'$, $s_{32} < \eta''$ and that they keep their ordering, $s_{32} > s_{21} > s_{10} > 0$. To highlight an example, I can rewrite Eq.~\eqref{Q_q-xspace} as 
\begin{align}\label{Q_q-etaspace}
	Q_q(s_{10},\eta) &= Q_q^{(0)}(s_{10},\eta) + \frac{1}{2}\int\limits_{y_0}^{\eta}d\eta' \!\!\!\!\! \int\limits_{\max\{0, s_{10} + \eta'-\eta\}}^{\eta'-y_0}\!\!\!\!\!\!\!ds_{21}\, \alpha_s(s_{21})\big[Q_q(s_{21},\eta) + 2\,G_2(s_{21},\eta)\big] \\
	&  + \int\limits_{s_{10}+y_0}^{\eta} d\eta' \int\limits_{s_{10}}^{\eta'-y_0}ds_{21} \, \alpha_s(s_{21})\big[2\,\widetilde{G}(s_{21},\eta') + 2\,\widetilde{\Gamma}(s_{10},s_{21},\eta') + Q_q(s_{21},\eta') \notag \\
	& \qquad\qquad\qquad - \overline{\Gamma}_q(s_{10},s_{21},\eta') + 2\,\Gamma_2(s_{10},s_{21},\eta') + 2\,G_2(s_{10},s_{21},\eta')\big] , \notag
\end{align}
where the lifetime ordering of $s_{10} < s_{21} < \eta'$ is imposed upon the neighbor dipoles. The other polarized dipole amplitudes are similarly re-cast, but I have not included them for brevity; the full set of evolution equations in terms of $s_{10}$ and $\eta$ and with the above constraints enforced can be found in Eqs. A1 of Ref.~\cite{Adamiak:2023yhz}. I ultimately wish to discretize these equations, and in that spirit I make the decision to format the polarized dipole amplitudes as an evenly spaced $s_{10}$-$\eta$ grid with a step-size of $\Delta\eta^{(n)} = \Delta s_{ij} \equiv \Delta$, with end-points of $\eta^{\max} = s_{10}^{\max} = \eta(z=1)\propto \ln(Q^2/x)$ where I can determine the value of $x$ I wish to extend the grid down to. I then denote the polarized dipole amplitudes as $G(s_{10},\eta) \to G(i\Delta, j\Delta) \to G[i,j]$, and the neighbor dipoles as $\Gamma(s_{10},s_{21},\eta) \to \Gamma(i\Delta, k\Delta, j\Delta) \to \Gamma[i,k,j]$; I make note that when $i=j$ the neighbor dipoles reduce to their fundamental counterparts, $\Gamma[i,k=i,j] = G[i,j]$. I could discretize both of these integrals into a numerical approximation like a Riemann sum, however, I can actually remove one integral using a recursion relation. This starts with a Taylor expansion in $\eta$, 
\begin{equation}
	Q_q(s_{10},\eta + \Delta) = Q_q(s_{10},\eta) + \Delta \frac{\partial}{\partial{\eta}}Q_q(s_{10},\eta) + \mathcal{O}(\Delta^2)
\end{equation}
where I ignore higher order $\Delta^2$ terms (since the step-size is small $\Delta \ll 1$). Using this approach I can rewrite Eq.~\eqref{Q_q-etaspace} as 
\begin{align}
	&Q_q(s_{10},\eta+\Delta) = Q_q(s_{10},\eta) + Q_q^{(0)}(s_{10},\eta+\Delta) - Q_q^{(0)}(s_{10},\eta) \\
	& \qquad\;\; +\Delta\int\limits_{s_{10}}^{\eta-y_0}ds_{21}\,\alpha_s(s_{21})\Big[\frac{3}{2}\,Q_q(s_{21},\eta) + 2\,\widetilde{G}(s_{21},\eta) + 2\,\widetilde{\Gamma}(s_{10},s_{21},\eta)  \notag\\
	& \qquad\qquad\qquad\qquad\qquad\qquad - \overline{\Gamma}_q(s_{10},s_{21},\eta) + 2\,\Gamma_2(s_{10}s_{21},\eta) + 3\,G_2(s_{21},\eta) \Big] \notag \\
	&\qquad\;\; + \frac{1}{2}\Delta\int\limits_{\eta-s_{10}}^{\eta}d\eta'\;\alpha_s(s_{10}+\eta'-\eta)\,\Big[Q_q(s_{10}+\eta'-\eta,\eta') + 2\,G_2(s_{10}+\eta'-\eta,\eta')\Big]  , \notag
\end{align}
and have successfully removed one integral at the expense of one step-size of the grid, having $Q_q(s_{10}, \eta+\Delta)$ dependent on $Q_q(s_{10},\eta)$. This same process can be repeated for the other polarized dipole amplitudes, except for the caveat that the neighbor dipoles are doubly recursive in $s_{21}$ and $\eta'$, such that
\begin{equation}
	\Gamma(s_{10},s_{21}+\Delta, \eta' + \Delta) = \Gamma(s_{10},s_{21},\eta') + \Delta \frac{\partial}{\partial{\eta}}\Gamma(s_{10},s_{21},\eta')+ \Delta \frac{\partial}{\partial{s_{21}}}\Gamma(s_{10},s_{21},\eta') + \mathcal{O}(\Delta^2).
\end{equation}
The remaining integrals are then discretized using a left-hand Riemann sum \footnote{Technically any numerical approximation for an integral would work and potentially be more accurate, however in practice a left-hand sum was more accurate than the right-hand and midpoint sums, both of which are significantly quicker to compute than the trapezoidal approach.} and a step-size of $\Delta$. After both steps, I can obtain the discretized form of the large-$N_c\&N_f$ evolution equations with the small-$x$ and dipole size constraints imposed: 
{\allowdisplaybreaks
	\begin{subequations}\label{disc_evo}
		\begin{align}
			& Q_q[i,j] = Q_q[i,j-1] + Q_q^{(0)}[i,j] - Q_q^{(0)}[i,j-1]  \\
			&\qquad\ +\Delta^2\sum\limits_{i' = i}^{j-2-y_0}\alpha_s[i']\Big[\frac{3}{2}\,Q_q[i',j-1] + 2\,\widetilde{G}[i',j-1] + 2\,\widetilde{\Gamma}[i,i',j-1] \notag \\
			&\qquad\qquad\qquad\;-\overline{\Gamma}_q[i,i',j-1] + 3\,G_2[i',j-1] + 2\,\Gamma_2[i,i',j-1]\Big] \notag \\
			&\qquad + \frac{1}{2}\Delta^2\sum\limits_{j' = j-1-i}^{j-2}\alpha_s[i+j'-j+1]\Big[Q_q[i+j'-j+1,j'] + 2\,G_2[i+j'-j+1,j']\Big] \notag \\[0.5cm]
			& \overline{\Gamma}_q[i,k,j] = \overline{\Gamma}_q[i,k-1,j-1] + Q_q^{(0)}[i,j] - Q_q^{(0)}[i,j-1]  \\
			&\qquad\ +\Delta^2\sum\limits_{i' = i}^{j-2-y_0}\alpha_s[i']\Big[\frac{3}{2}Q_q[i',j-1] + 2\,\widetilde{G}[i',j-1] + 2\,\widetilde{\Gamma}[i,i',j-1] \notag \\
			&\qquad\qquad\qquad\;-\overline{\Gamma}_q[i,i',j-1] + 3\,G_2[i',j-1] + 2\Gamma_2[i,i',j-1]\Big] \notag \\[0.5cm]
			&\widetilde{G}[i,j] = \widetilde{G}[i,j-1] + \widetilde{G}^{(0)}[i,j] - \widetilde{G}^{(0)}[i,j-1] \\
			&\qquad +\Delta^2\sum\limits_{i'=i}^{j-2-y_0}\alpha_s[i']\Big[2\,\widetilde{G}[i',j-1] + \widetilde{\Gamma}[i,i',j-1] +2\,G_2[i',j-1] \notag \\
			&\qquad\qquad\qquad\; + \big(2-\frac{N_f}{2N_c}\big)\,\Gamma_2[i,i',j-1] - \frac{1}{4N_c}\sum\limits_q\overline{\Gamma}_q[i,i',j-1]\Big] \notag \\
			&\qquad  - \Delta^2\frac{1}{4N_c}\sum\limits_{j' = j-1-i}^{j-2}\alpha_s[i+j'-j+1] \notag \\
			&\qquad\qquad\qquad\qquad\qquad\; \times \Big[\sum\limits_qQ_q[i+j'-j+1,j'] + 2N_f\,G_2[i+j'-j+1,j']\Big] \notag \\[0.5cm]
			&\widetilde{\Gamma}[i,k,j] = \widetilde{\Gamma}[i,k-1,j-1] + \widetilde{G}^{(0)}[i,j] - \widetilde{G}^{(0)}[i,j-1] \\
			&\qquad +\Delta^2\sum\limits_{i'=i}^{j-2-y_0}\alpha_s[i']\Big[2\,\widetilde{G}[i',j-1] + \widetilde{\Gamma}[i,i',j-1] +2\,G_2[i',j-1] \notag \\
			&\qquad\qquad\qquad\; + \big(2-\frac{N_f}{2N_c}\big)\,\Gamma_2[i,i',j-1] - \frac{1}{4N_c}\sum\limits_q\overline{\Gamma}_q[i,i',j-1]\Big] \notag \\[0.5cm] 
			&G_2[i,j] = G_2[i,j-1] + G_2^{(0)}[i,j] - G_2^{(0)}[i,j-1]  \\
			&\qquad + 2\Delta^2\sum\limits_{j' = j-1-i}^{j-2}\alpha_s[i+j'-j+1]\Big[\widetilde{G}[i+j'-j+1,j'] +2\,G_2[i+j'-j+1,j']\Big] \notag \\[0.5cm]
			&\Gamma_2[i,k,j] = \Gamma_2[i,k-1,j-1] + G_2^{(0)}[i,j] - G_2^{(0)}[i,j-1] \\[0.5cm]
			&G_q^{\mathrm{NS}}[i,j] = G_q^{\mathrm{NS}}[i,j-1] + G_q^{\mathrm{NS}\,(0)}[i,j] - G_q^{\mathrm{NS}\,(0)}[i,j-1] \\
			&\qquad + \frac{1}{2}\Delta^2\Bigg[\sum\limits_{i'=i}^{j-2-y_0}\alpha_s[i']G_q^{\mathrm{NS}}[i',j-1] + \sum\limits_{j' = j-1-i}^{j-2}\alpha_s[i+j'-j+1]\,G_q^{\mathrm{NS}}[i+j'-j+1,j']\Bigg], \notag \label{GNS_disc}
		\end{align}
	\end{subequations}
}
which can be computed numerically using any kind of programming language, assuming that I can provide an initial condition state. 

I discussed above that the Born-level quark target model provided a baseline initial condition. I can rewrite Eqs.~\eqref{Born_IC} in terms of the logarithmic variables:
\begin{subequations}
	\begin{align}
		Q_q^{(0)}(s_{10},\eta) &= \widetilde{G}^{(0)}(s_{10},\eta) = \frac{\alpha_s^2\,C_F\,\pi}{2\,N_c}\sqrt{\frac{2\pi}{N_c}}\Big[\big(C_F-2\big)\,\eta + 2\,s_{10}\Big] \to A\,\eta + B\,s_{10} \\
		G_2^{(0)}(s_{10},\eta) & = \frac{\alpha_s^2\,C_f\,\pi}{2\,N_c}\sqrt{\frac{2\pi}{N_c}}\,s_{10} \to A'\,s_{10} \\ 
		G_q^{\mathrm{NS}\,(0)}(s_{10},\eta) & = \frac{\alpha_s^2\,C_F^2\,\pi}{N_c}\,\sqrt{\frac{2\pi}{N_c}}\,\eta \to A''\,\eta
	\end{align}
\end{subequations}
where it is clear that each of the initial conditions for the polarized dipole amplitudes is linear in $s_{10}$ and $\eta$. I then define a linear combination \textit{ansatz} \cite{Kovchegov:2015pbl, Kovchegov:2016zex, Adamiak:2021ppq, Adamiak:2023yhz, Adamiak:2025dpw}, in which I take the toy model initial conditions and generalize them to allow for a generalized set of initial conditions:
\begin{subequations}\label{IC_params}
	\begin{align}
		Q_q^{(0)}(s_{10},\eta) &= a_q\,\eta + b_q\,s_{10} + c_q \\
		\widetilde{G}^{(0)}(s_{10},\eta) &= \widetilde{a}\,\eta + \widetilde{b}\,s_{10} + \widetilde{c} \\
		G_2^{(0)}(s_{10}, \eta) &= a_2\,\eta + b_2\,s_{10} + c_2 \\
		G_q^{\mathrm{NS}\,(0)}(s_{10},\eta) &= a_q^{\mathrm{NS}}\,\eta + b_q^{\mathrm{NS}}\,s_{10} + c_q^{\mathrm{NS}}.
	\end{align}
\end{subequations}
This way I have a theoretical justification for making the initial conditions dependent on the logarithmic variables $s_{10}$ and $\eta$ without limiting ourselves to the Born level calculation. This does mean, however, that there are 24 free parameters that control the initial conditions of the small-$x$ helicity evolution equations. 

\begin{figure}[h!]
	\begin{center}
		\includegraphics[width=450 pt]{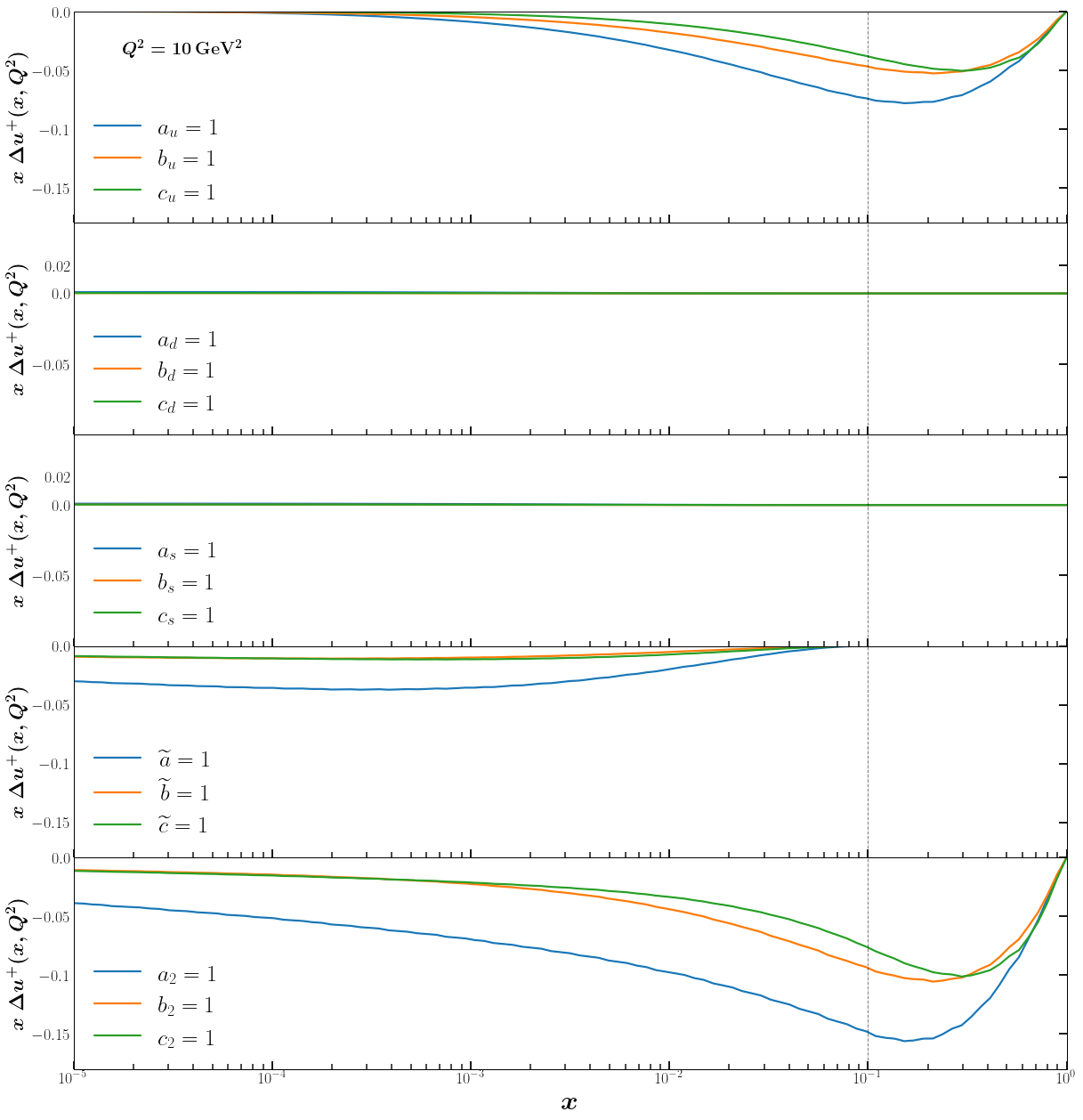} 
		\caption{Basis functions of the $C$-even hPDF $\Delta u^{+}(x,Q^2)$ Vs. $x$ for a fixed $Q^2 = 10~\mathrm{GeV}^2$. There are 5 plots for each of the 5 initial condition dipoles $Q_{u,d,s}^{(0)}$, $\widetilde{G}^{(0)}$, and $G_2^{(0)}$, each with 3 lines corresponding to each of the parameters from the linear combination \textit{ansatz} Eq.~\eqref{IC_params}.}
		\label{fig:Deltaqp_basis}
	\end{center}
\end{figure}

To help inform me of the influence of each individual parameter, I construct what are referred to as basis functions. The idea behind a basis function is to assume that the initial condition state, which can be thought of as a 24-parameter vector, is a unit vector in the direction of only one parameter. That is to say, I investigate the final output of the evolution equations given an initial condition state that looks like $a_u = 1$ and $b_u = c_u = a_d = b_d =\, ... = c_2 = 0$. As can be seen from Eqs.~\eqref{eq_LargeNcNf}, because the large-$N_c\&N_f$ evolution equations are closed, linear, and all coupled to each other, any single initial condition parameter will result in all evolved polarized dipole amplitudes being non-zero. Because of this fact, hPDFs can be constructed entirely from any single initial condition parameter; an $a_u = 1$ initial condition will result in evolved $Q_u(x,Q^2)$, $Q_d(x,Q^2)$, $Q_s(x,Q^2)$ and $G_2(x,Q^2)$ dipole amplitudes which can be combined via Eq.~\eqref{Deltaqp}, for example, to generate any flavor of $\Delta q^+(x,Q^2)$ or the $g_1(x,Q^2)$ structure function. This process can be repeated for each of the initial condition parameters, although I note here that the flavor singlet and flavor nonsinglet evolution equations are decoupled, so an $a_u^{\mathrm{NS}} = 1$ initial condition will not generate $\Delta q^+(x,Q^2)$ - it can however generate $\Delta q^-(x,Q^2)$ as seen from Eq.~\eqref{Deltaqm}. With this information in hand, I can present to you the $\Delta u^+(x,Q^2)$ basis functions for a fixed $Q^2 = 10~\mathrm{GeV}^2$ in Fig.~\ref{fig:Deltaqp_basis} below as an example.
I make the note that while the contributions from $Q_{d,s}^{(0)}$ appear to be non-existent, this is simply because the peak contributions of those initial conditions are roughly 1\% of the size of the $Q_u^{(0)}$ initial conditions. 

The basis functions are the ultimate link that brings us to the topic of phenomenology. Particle colliders can provide data describing what the hPDFs and structure functions look like in nature at and around the small-$x$ cutoff of $x_0$, so the goal of phenomenology is to find a linear combination of the basis function curves such that the resultant curve agrees with the data\footnote{To be more accurate, the data itself is proportional to $g_1^{(h)}(x,Q^2,(z))$, so the goal is really to find the correct combination of $g_1^{(h)}(x,Q^2,(z))$ basis functions, rather than $\Delta q^+(x,Q^2)$ basis functions. Since the hPDFs and structure functions are constructed using the same polarized dipole amplitudes \eqref{all_hPDFs}, the difference in these statements is trivial.}. From the initial condition state that agrees with larger-$x$ data, I can use the evolution equations to compute and effectively predict the values of the polarized dipole amplitudes (and in turn the hPDFs) at an arbitrary value of $x < x_0$.

\subsection{$\boldsymbol{\mathrm{JAM\!:~Bayesian\text{-}Monte~Carlo~Analysis}}$}

To conduct my phenomenological work and perform an analysis of world polarized data using small-$x$ helicity evolution, I use the Jefferson Angular Momentum (JAM) framework. Given any arbitrary initial condition state (the aforementioned 24-element vector) one can use the KPS-CTT evolution equations to generate theoretical asymmetries; the JAM framework allows one to compare the theoretical asymmetries generated from the arbitrary initial condition state to real world collider data. From a true random selection of the 24 initial condition parameters the theoretical asymmetries are not statistically likely to agree with collider data, but through JAM I use a Bayesian Monte-Carlo procedure to refine the initial condition parameters until the resultant theoretical asymmetry does agree with collider data \cite{Sato:2016wqj, Sato:2016tuz, Zhou:2022wzm, Hunt-Smith:2024khs}. The JAM framework is robust, but it can be effectively broken into four parts: the input, the data, the theory, and the output, with one unifying goal of minimizing the reduced $\chi^2$ of a given fit. The end goal is quite straightforward to discuss, and that is where I will start.

When comparing a set of theoretical asymmetries and true data, one can compute the reduced $\chi^2$, otherwise called $\chi^2_{\mathrm{red}}$, which is a measure of how good or poorly the theoretical asymmetries agree with the true data. The $\chi^2_{\mathrm{red}}$ is calculated by the equation
\begin{equation}\label{red_chi2}
	\chi^2_{\mathrm{red}} = \frac{1}{N_{\mathrm{pts}}}\sum_{e,i}\bigg(\frac{d_{e,i} - \mathrm{E}_{e,i}^{\mathrm{thy}}}{\alpha_{e,i}}\bigg)^2
\end{equation}
where $d_{e,i}$ refers to a specific data value from dataset $e$ and data point $i$, and $\mathrm{E}_{e,i}^{\mathrm{thy}}$ in the context of these analyses is the theoretical asymmetry, shifted and normalized by the correlated uncertainties of the relevant dataset/data point, and $\alpha_{e,i}$ is the uncorrelated uncertainty for the data point. The uncertainties of the data are outside my control, but the theoretical asymmetry $E_{e,i}^{\mathrm{thy}}$ is controlled largely by the initial condition parameters. A reduced $\chi^2$ is considered ``good" if it is near 1, and thus if an arbitrary initial condition state generates theoretical asymmetries with $\chi_{\mathrm{red}}^2 \gg 1$, then that specific initial condition state is not valid to describe world data. The method, then, is to find initial condition states that result in a $\chi^2_{\mathrm{red}}\approx 1$, using a Bayesian Monte-Carlo process to achieve this. Bayes theorem states that the posterior probability, $\mathcal{P}(A|B)$, which is the probability that event $A$ will occur given that $B$ is true, is dependent on the conditional probability $\mathcal{P}(B|A)$, and the prior probability $\mathcal{P}(A)$, written neatly as 
\begin{equation}
	\mathcal{P}(\textbf{A}|B) \propto \mathcal{P}(B|A)\,\mathcal{P}(\textbf{A}).
\end{equation}
In short, the posterior probability that I am after is the distribution of parameters ($\textbf{A}$) as determined by the data ($B$). Since that amounts to having large probability of determining good parameters that fit to data, the goal is to maximize the conditional probability and prior probability; the conditional probability, that is the probability of the data $B$ being represented well by a given initial condition vector $\textbf{A}$, would clearly be inversely related to the $\chi_{\mathrm{red}}^2$, and the prior probability is optimized if I know the general range in which each of the initial condition parameters can be found. Together that means that if I want to determine the best initial conditional state, I need to investigate an initial guess for the initial condition state and then iterate through the parameters of that state until the $\chi^2_{\mathrm{red}}$ has been minimized. The Monte-Carlo part of this analysis occurs here. In order to avoid possible bias in my initial condition state priors, the initial condition guess is randomly sampled (Monte-Carlo sampling). This process is inherently recursive; given some initial condition guess, the parameters are iterated until the $\chi^2_{\mathrm{red}}$ has been minimized, at which point the newly optimized distribution of parameters can be used as a new initial guess. After enough iterations, the posterior probability should be largely similar to that of the prior probability, indicating that the $\chi^2_{\mathrm{red}}$ has been minimized optimally. 

The method through which I minimize the $\chi^2_{\mathrm{red}}$ from an initial guess is to use a least-squares algorithm with a Trust-Region-Reflective (`trf') cost function. In effect it takes one parameter of the initial condition vector at a time, shifts it by a small amount, and only keeps the shift if the resultant $\chi^2_{\mathrm{red}}$ is smaller than it was before the shift; this process is then repeated for each parameter in the initial condition guess, completing one iteration of the least-squares algorithm. This least-squares procedure occurs recursively, using the output from each iteration as the input for the next, until the overall change in $\chi^2_{\mathrm{red}}$ is less than $0.1\%$. Only now would the fitting procedure have provided a final state posterior initial condition vector as determined by the original initial guess; the linear combination determined by the posterior initial condition state is called a replica. From here, the entire process is repeated from the beginning with a new initial guess that subsequently generates a new (ideally similar) replica. In my experience of performing a small-$x$ helicity evolution analysis, I consider a statistically stable number of replicas to be above 300, but as with most Monte-Carlo style analyses, more is better. An important thing to note is that the initial condition guess cannot be completely randomly sampled (there are an infinite number of scalars of course), but instead must be a pseudo-random guess that is constrained by the prior distribution of parameters and the parameter configurations; simply put, the pseudo-random guess is influence by the input for the JAM framework.

\begin{figure}[h!]
	\begin{center}
		\includegraphics[width=475 pt]{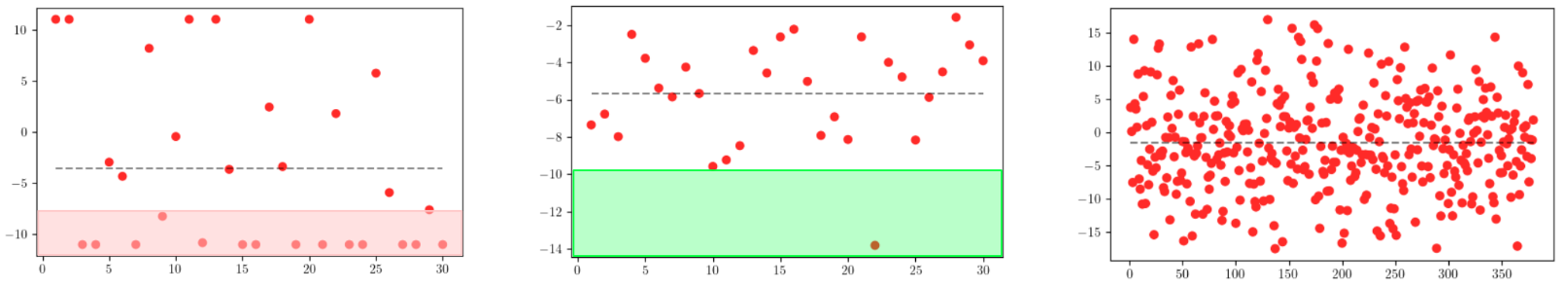} 
		\caption{Examples of posterior parameter distributions inside the allowable parameter ranges; each plot shows one single parameter ($a_u$, for example) for each replica of a given fit, with the $x$-axis showing the sequential number of the replica and the $y$-axis showing the value of the parameter. The left-most figure shows a poorly chosen parameter range since the posterior parameters are bunching at the lower boundary (highlighted in red), suggesting that the data would prefer a more negative lower boundary. The center plot shows an improved parameter range without bunching, but the empty space in green implies that the data is constraining the parameter values more than the chosen range. The right-most plot shows a good choice for the parameter range, seeing as the posterior distribution of parameters fills the space evenly without bunching.}
		\label{fig:param_dist}
	\end{center}
\end{figure}

The input for JAM is determined in part by the datasets in use, the 24-variable parameter space as determined by Eq.~\eqref{IC_params}, and any other decision made with regard to the model attempting to be fit. First and foremost, I have 24 parameters that make up an initial condition state, and each one \textit{a priori} is undetermined. As such, I must start with some allowable parameter range that needs to be large enough to contain the ``true" initial condition for that parameter, but small enough that I can make a meaningful prediction of small-$x$ extractions. These parameter ranges essentially constitute $\mathcal{P}(\textbf{A})$, and I can widen, narrow, or shift the parameter ranges such that the parameter values from all posterior replicas are uniformly distributed within their respective ranges. Early, large parameter-range tests of the global analysis process found that the spin asymmetries from data can be constructed from linear combinations of $|a|,\,|b|,\,|c|,\, < 50$; I refine the parameter ranges by iteratively shrinking the allowable parameter range, re-fitting to data, and ensuring that the distribution of posterior parameters are uniformly distributed within the allowable range. I know that the parameter ranges have been constrained improperly if there is bunching along either boundary, which implies that the data would prefer a larger/smaller parameter than is currently allowed by the parameter range. An example of a bad parameter range with bunching and a good parameter range with uniformly distributed posterior parameters is given in Fig.~\ref{fig:param_dist}.
In a perfect world the theory would be precise enough and the data would be constraining enough for there to be a definitive ``answer" for what each parameter is ``supposed" to be; in reality, the parameterization of the initial conditions is an approximation, and the data's distribution in $x$ has a finite amount of information to constrain all 24 parameters. As a result, each of the replicas is considered optimized if its linear combination of basis functions agrees with data, but the possible states that provide a good fit to data are largely degenerate, meaning that the distribution of each parameter can be constrained to a finite but not infinitesimal range. Therefore, it is my goal to find the \textit{narrowest} range of parameters in which the posterior distributions are uniformly distributed. JAM's input is also influenced by the models attempting to be fit. By this I mean that there may be a theoretical or phenomenological justification that, for example, the strange hPDF $\Delta s^{+}$ should be negligible compared to the up and down hPDFs, so I can choose to run a fit to global data with a model in which set the strange-related parameters are zero, $a_s^{(\mathrm{NS})} = b_s^{(\mathrm{NS})} = c_s^{(\mathrm{NS})} = 0$, and investigate the effect on the remaining fitted parameters. Similar model decisions, like whether the running of the coupling is implemented, are considered a part of the JAM framework's input. The last, but not least, important part of JAM's input is the choice of ``priors". Consider that the $g_1^h(x,Q^2)$ structure function measured in SIDIS collisions is dependent on the hadron fragmentation functions $D_1^{h/q}(z,Q^2)$ as seen in Eqs.~\eqref{all_hPDFs}. These fragmentation functions, similar to the hPDFs, are objects that need to be extracted themselves from other phenomenological studies (like DGLAP global analyses). Since those fragmentation functions are not the subject of study for small-$x$ helicity evolution analyses, I simply take the extracted fragmentation functions from JAM-DGLAP analyses as ``priors" and use those in my analysis. 

The data part of the JAM framework is easier to cover; polarized DIS and polarized SIDIS collisions provide longitudinal spin asymmetries. A longitudinal spin asymmetry is effectively the difference in DIS or SIDIS cross-sections when the lepton probe's spin is parallel to the nucleon beam axis and when the lepton probe's spin is anti-parallel to the nucleon beam axis. Subtracting one from the other provides information that is dependent on the helicity of the nucleon, and because of this, those asymmetries are related to the polarized structure functions $g_1(x,Q^2)$ and $g_1^h(x,z,Q^2)$. The two arguments of the structure functions are critical to the data chosen, and the cuts implemented on the data's $x$ and $Q^2$ (and $z$ for SIDIS) will affect the global analysis and the resultant replicas' parameters. In the next section I will discuss in more detail how the structure functions and asymmetries are related to each other. I will also save the explanation on what cuts I impose on the $x$ and $Q^2$ of data, but suffice it to say that the small-$x$ helicity evolution equations are only valid to describe small-$x$ data, so only DIS and SIDIS data with $x < x_0$ are considered, where $x_0$ is treated (in early small-$x$ helicity phenomenology) as another free parameter to be fitted from data.

The theory part of the JAM framework is determined by the theorists and the phenomenology being attempted. By this, I mean that the theorists define the theoretical asymmetries that are compared to global data. As discussed briefly above, polarized DIS and SIDIS data are related to the polarized structure functions $g_1(x,Q^2)$ and $g_1^h(x,Q^2)$, and as was determined by Eqs.~\eqref{all_hPDFs} and \eqref{eq_LargeNcNf}, I can define those polarized structure functions using the polarized dipole amplitudes and their respective evolution equations. In the same way that small-$x$ helicity theory is used to fit the initial conditions of polarized dipoles' evolution and predict the small-$x$ behavior, other JAM collaborations have used, for example, a DGLAP based approach to fit to data make extractions/predictions for large $Q^2$ observables (remembering that DGLAP evolves in $Q^2$ instead of $x$). This means that the small-$x$ theory must be ready to accept the data's kinematics, which is the case in Eq.~\eqref{all_hPDFs} since the only free variables remaining on the right-hand-side are the $x$ and $Q^2$ (and $z$ for SIDIS-related objects).

The last part of the JAM framework is the output, and how I determine a fit to be good or bad, and interpret the respective results. The process of randomly sampling, iterating, and optimizing the initial condition parameters is not fool-proof; sometimes the converged "optimal" combination of parameters has a minimum $\chi^2_{\mathrm{red}} > 1$, which could indicate that the allowable parameter ranges are poorly chosen. Perhaps instead this means that the theory, as implemented, is incapable of fitting to the selected data and it would then be up to me and my human intelligence to determine if the fault lies in the data chosen, i.e. if the $x_0$ cutoff chosen is too high, or if there is some fundamental incompatibility between the theory and data, such as if the theory is meant to describe mid-rapidity observables but the data ranges from mid to forward rapidity. I must analyze the fit's results, make the proper modifications to theory, datasets, kinematics, etc., and repeat the fitting process again. I consider a fit to be good if the overall average $\chi^2_{\mathrm{red}}$ of all replicas is $\chi^2_{\mathrm{red}} \approx 1$, but I take into account any replicas who's individual $\chi^2_{\mathrm{red}} < 10$; since the least-squares algorithm is set to minimize the $\chi^2_{\mathrm{red}}$, there should be far fewer large-$\chi^2_{\mathrm{red}}$ replicas than small one. I perform a cross-check on the results of a given fit by making sure the distribution of $\chi^2_{\mathrm{red}}$ is approximately gaussian and centered near 1. If there is a gaussian distribution of good-$\chi^2_{\mathrm{red}}$ replicas centered near 1, and the parameters are uniformly distributed within the prior allowable ranges, then I can be convinced that the fit to data is good and begin analyzing the implications of those results. The overall goodness-of-fit is then determined from the average of all replicas, which to say for each given data point I compute the theoretical asymmetry from the average of all replicas, and use that average to compute the overall $\chi^2_{\mathrm{red}}$ of the entire fit using Eq.~\eqref{red_chi2}. If the overall $\chi^2_{\mathrm{red}}$ is close to one, then I continue on to make predictions based on those results. In a sense, the global analysis is a two-step process: 1) randomly sample the parameters to construct theoretical values for the structure functions and hPDFs to compare to data, and 2) use the optimized parameters to construct hPDFs and structure functions at values of $x$ lower than those provided from data. This is what makes the JAM framework so powerful, because once it provides the resultant good-fitting combinations of parameters, I can use those parameters in the small-$x$ helicity evolutions and predict/extract the small-$x$ behaviors of hPDFs and structure functions beyond the region of $x$ where data exists.

I mentioned above that in my experience, a set of 300 replicas is sufficiently informative and stable; in that context, I am referring to converged, good-$\chi^2_{\mathrm{red}}$ replicas. Since some initial condition guesses are too poor to optimize within the $\chi^2_{\mathrm{red}} < 10$ cutoff, it is likely to lose some replicas throughout the fitting process. I generally submit 500 or more replicas to be computed, and once the fitting procedure is complete and I am left with at least 300 replicas (hopefully more), with which I can then plot the hPDFs and structure functions and determine their standard deviation, thus quantifying the small-$x$ predictions. These results are ultimately the final output of a small-$x$ helicity evolution phenomenological study.

\subsection{$\boldsymbol{\mathrm{Observables~and~Data}}$}\label{Observables}

In preparation for the presentation of the small-$x$ helicity phenomenology for which I contributed, I will lay out the observables and how I approximate the theoretical asymmetries. The phenomenology chapters are two-fold, in accordance with the two manuscripts I co-authored; the first study was a global analysis that simultaneously fitted polarized DIS and SIDIS data, whereas the second study was the first analysis of polarized DIS, SIDIS, and polarized proton-proton ($pp$) scattering data using small-$x$ helicity evolution. Let me first begin with the DIS and SIDIS observables.

Starting with the longitudinal DIS asymmetry $A_{\parallel}$ (see, {\it e.g.}, Refs.~\cite{Sato:2016tuz,COMPASS:2007qxf})
\begin{align}\label{eq: DSA}
	A_{\parallel} &=\frac{\sigma^{\downarrow \Uparrow}-\sigma^{\uparrow \Uparrow}}{\sigma^{\downarrow \Uparrow}+\sigma^{\uparrow \Uparrow}}
	=  D(A_1 + \eta A_2)\,, 
\end{align}
the asymmetry is defined explicitly as a ratio of the polarized cross-section to the unpolarized cross-section. The polarized cross-section is the difference between the cross-section when the lepton is anti-aligned with the beam axis ($\downarrow\Uparrow$) and the cross-section where the lepton is aligned to the beam axis ($\uparrow\Uparrow$), the double-arrow $\Uparrow$ representing the polarization along the beam axis. See that this asymmetry is decomposed into two further asymmetries, $A_1$ and $A_2$ from the virtual photon-target interaction, as well as the kinematic variables $D$ and $\eta$. The kinematic variables are functions $x$, $Q^2$, the nucleon mass $M$, the fractional energy transfer of the lepton in the nucleon target's rest frame $y$, and the ratio of longitudinal virtual photoproduction cross-section to transverse virtual photoproduction cross-section $R$. They read as
\begin{align}
	D &= \frac{y(2-y)(2+\gamma^2 y)}{2(1+\gamma^2)y^2+(4(1-y)-\gamma^2y^2)(1+R)},
	\ \ \ 
	\eta = \gamma \frac{4(1-y)-\gamma^2 y^2}{(2-y)(2+\gamma^2y)}\,.
\end{align}
In the small-$x$ limit $\gamma \ll 1$, and the virtual photon-target asymmetries are
\begin{align}\label{A1_fin}
	A_1 &= \frac{g_1 -\gamma^2 g_2}{F_1} \approx \frac{g_1(x,Q^2)}{F_1(x,Q^2)}, \ \ \ A_2 = \gamma \, \frac{g_1 + g_2}{F_1} \ll 1\,,
\end{align}
ultimately producing the small-$x$ limit DIS asymmetry
\begin{equation}
	A_{\parallel}\approx D\,\frac{g_1(x,Q^2)}{F_1(x,Q^2)}.
\end{equation}
Following a similar process for the polarized SIDIS asymmetry $A_1^h$ with tagged final-state hadron $h$, I acquire the expression \cite{Ethier:2017zbq, Leader:2010rb}
\begin{align}\label{A1h}
	A_1^h = \frac{g_1^h-\gamma^2 g_{2}^h}{F_1^h} \approx \frac{g_1^h(x,z,Q^2)}{F_1^h(x,z,Q^2)}\,.
\end{align}
Between the two asymmetries above and the repository of data accessible through JAM, I have fit to ten distinct observables: two DIS observables related to either a proton/deuteron target or ${}^3\mathrm{He}$ target, and eight SIDIS observables related to either a proton/deuteron or ${}^3\mathrm{He}$ target with four possible tagged hadrons $K^{\pm}$ or $\pi^{\pm}$. This is a crucial victory from the perspective of phenomenology because I have 10 observables but only eight "unknowns" (the five flavor-singlet and three flavor-nonsinglet polarized dipole amplitudes)\footnote{Note that while there are 24 initial condition parameters, there are three parameters for one polarized dipole amplitude. The data should only be strictly sensitive to the combined initial condition of each polarized dipole amplitude.}. As discussed thoroughly above, both $g_1(x,Q^2)$ and $g_1^h(x,z,Q^2)$ are computed via the polarized dipole amplitudes, and through the basis functions and linear-combination \textit{ansatz} I can use data at `large' $x$ to fit the appropriate initial conditions and subsequently use the same initial condition parameter vector to predict the hPDFs and structure functions at any smaller value of $x$.

Quite noticeably the asymmetries above do not only depend on the polarized structure functions $g_1(x,Q^2)$ and $g_1^h(x,z,Q^2)$, but are also dependent on the \textit{unpolarized} DIS and SIDIS structure functions $F_1(x,Q^2)$ and $F_1^h(x,z,Q^2)$. In the same way that the polarized structure functions are heavily related to the helicity PDFs, the unpolarized structure functions too are heavily related to the unpolarized parton distributions functions for quarks, $q(x,Q^2)$, and gluons, $g(x,Q^2)$, and in the case of the SIDIS structure functions, the hadronic fragmentation functions $D_1^{h/q}(z,Q^2)$. These are not small-$x$ specific objects and are subject to their own extractions via analogous, unpolarized, phenomenological work. Fitting the unpolarized PDFs and structure functions is outside the scope of this current work, so this work uses extractions computed from NLO collinear factorization and DGLAP evolution from the JAM analyses Refs.~\cite{Cocuzza:2022jye} and \cite{Anderson:2024evk}. The former JAM-DGLAP extraction was used in the phenomenological work of \textbf{Chapter}~\ref{pheno_1}, whereas the latter JAM-DGLAP extraction was used in the more recent phenomenology covered in \textbf{Chapter}~\ref{pheno_2}; the reason for the two separate JAM-extractions is that the work of Ref.~\cite{Anderson:2024evk} found an error in Ref.~\cite{Cocuzza:2022jye} where one source of data used in extracting the fragmentation functions incorrectly labeled a systematic uncertainty as a normalization uncertainty. This error was corrected in Ref.~\cite{Anderson:2024evk}, and the corrected extractions were used in the work covered in \textbf{Chapter}~\ref{pheno_2}. Using the JAM-DGLAP extractions for unpolarized functions is justified by the fact that unpolarized small-$x$ evolution is singly logarithmic, whereas polarized small-$x$ evolution is doubly logarithmic. This means that small-$x$ evolution will affect the polarized numerators of the spin asymmetries more so than the unpolarized denominators where small-$x$ evolution is sub-leading. A simultaneous treatment of small-$x$ evolution for the polarized and unpolarized parts of the spin asymmetries is left for future work.

The scope of this dissertation up to this point has been largely centered around inclusive and semi-inclusive deep inelastic scattering, which is supported by the journey taken in \textbf{Chapters}~\ref{intro} and \ref{chapter_2}. A salient feature of my research, however, concerns the first analysis of the impact of polarized proton-proton data on small-$x$ helicity evolution phenomenology. While the details of this analysis will be covered in depth in \textbf{Chapter} \ref{pheno_2}, I include the discussion of its relevant observable here. The JAM framework does not have access to all collider data, but the polarized proton-proton data that is available to the JAMsmallx collaboration is single-inclusive jet production data from polarized proton-proton collisions. The data is in the form of the double-longitudinal spin asymmetry $A_{LL}^{\mathrm{jet}}$, which reads
\begin{equation}\label{ALL_ratio}
	A_{LL}^{\mathrm{jet}} = \frac{\sigma^{\Downarrow\Uparrow} - \sigma^{\Uparrow\Uparrow}}{\sigma^{\Downarrow\Uparrow} + \sigma^{\Uparrow\Uparrow}} \equiv \frac{(d\Delta\sigma/d^2p_T\, d y)^{pp\to \mathrm{jet} X}}{(d\sigma/d^2p_T\, d y)^{pp\to \mathrm{jet} X}}\,
\end{equation}
where the notation is similar to Eq.~\eqref{eq: DSA} in which the superscript $\Downarrow\Uparrow$ refers to the jet production cross-section where the two protons' spins are anti-aligned, and the superscript $\Uparrow\Uparrow$ implies the cross-section comes from a collision between two protons whose spins are aligned. I denote the polarized jet production cross-section in the numerator as $\Delta\sigma^{pp\to \mathrm{jet}\,X}$, and the unpolarized cross-section in the denominator as $\sigma^{pp\to \mathrm{jet}\,X}$. The polarized jet production cross-section and its importance to small-$x$ helicity phenomenology is chronicled in \textbf{Chapter}~\ref{pheno_2}, but the unpolarized jet production cross-section can be explained away in a manner similar to the unpolarized structure functions. Small-$x$ unpolarized evolution is still sub-leading, and a NLO collinear factorization JAM-DGLAP analysis of the single-inclusive jet production cross-section in $pp$ collisions has already been completed; I thus use the JAM-DGLAP extractions from Ref.~\cite{Zhou:2022wzm} as a faithful proxy for the unpolarized denominator of Eq.~\eqref{ALL_ratio}. 
	\newpage
	\setlength{\arrayrulewidth}{0.4pt}
\setlength{\arraycolsep}{2pt}
\setlength{\tabcolsep}{1.5pt}
\setlength{\doublerulesep}{2pt}
\renewcommand{\arraystretch}{1.2}
\section{$\boldsymbol{\mathrm{Phenomenology~I\!:~DIS~and~SIDIS~Data}}$}\label{pheno_1}

As a prologue to the phenomenological work covered in this chapter I make an observation about how the accuracy of the small-$x$ dipole evolution formalism goes beyond the typical collinear factorization approach. We can see from Eq.~\eqref{g1_LO} that I have successfully recreated the LO expression for $g_1$ in the collinear factorization approach, however I must make clear that the small-$x$ version contains more information than just the LO contribution since the polarized dipole amplitudes resum all orders of perturbation theory with the parameter $\alpha_s\ln^2(1/x)$. I can prove this by comparing the definitions of $\Delta G(x,Q^2)$ and $g_1(x,Q^2)$ in Eqs.~\eqref{all_hPDFs}: there is a direct correlation between the gluon hPDF $\Delta G$ and the polarized dipole amplitude $G_2$, and also the structure function $g_1$ has a dependency on $G_2$. By association that means the $g_1$ structure function has implicit contributions from $\Delta G$ in addition to the explicit contribution from $\Delta q^+$ in Eq.~\eqref{g1_LO}. The gluon hPDF only contributes to the $g_1$ structure function at next-to-leading order in typical collinear factorization, written as \cite{Altarelli:1977zs,Dokshitzer:1977sg,Zijlstra:1993sh,Mertig:1995ny,Moch:1999eb,vanNeerven:2000uj,Vermaseren:2005qc,Moch:2014sna,Blumlein:2021ryt,Blumlein:2021lmf,Davies:2022ofz,Blumlein:2022gpp} 
\begin{align}\label{g1_coef_ftns}
	g_1 (x, Q^2)  = \frac{1}{2} \sum_q \, Z^2_q \, \left\{ \Delta q^+ (x, Q^2) + \int\limits_x^1 \frac{d{z}}{z} \, \left[ \Delta c_q (z) \,  \Delta q^+\! \left( \frac{x}{z} , Q^2\right) +  \Delta c_G (z) \,  \Delta G\! \left( \frac{x}{z} , Q^2\right) \right] \right\},
\end{align}
with $\Delta c_q(z)$ and $\Delta c_G (z)$ being coefficient functions calculated order-by-order in perturbation theory. Thus the $\Delta G$ contributions to the NLO calculation of $g_1$ are already incorporated via evolution in the small-$x$ formalism, and beyond that, all orders of perturbation theory are included so long as they are accompanied by the two logarithms of energy. If I write out the $\overline{\mathrm{MS}}$ scheme coefficient functions \cite{Zijlstra:1993sh},
\begin{subequations}\label{coef_functions}
	\begin{align}
		& \Delta c_q (z) = \frac{\alpha_s N_c}{4 \pi} \, \ln \frac{1}{z} + \frac{5}{12} \, \left( \frac{\alpha_s N_c}{4 \pi} \right)^2 \left[ 1 - 4 \, \frac{N_f}{N_c} \right] \, \ln^3 \frac{1}{z} + {\cal O} (\alpha_s^3)\,, \label{Cq_full}  \\
		& \Delta c_G (z) = - \frac{\alpha_s}{2 \pi} \, \ln \frac{1}{z} - \frac{11}{2} \, \left( \frac{\alpha_s }{4 \pi} \right)^2 \, N_c \, \ln^3 \frac{1}{z} + {\cal O} (\alpha_s^3\,), \label{dCg}
	\end{align}
\end{subequations}
it becomes clear that the order-$\alpha_s$ terms are accompanied by log terms, which, when integrated as per Eq.~\eqref{g1_coef_ftns}, become terms that scale with $\alpha_s\ln^2(1/x)$ as allowed by the DLA power counting. I have now shown that the small-$x$ evolution formalism does indeed contain information far beyond LO as may have been incorrectly assumed by Eq.~\eqref{g1_LO} alone.

\subsection{$\boldsymbol{\mathrm{Previous~phenomenology~and~Subject~of~this~work}}$}

My mission statement to understand the proton spin puzzle effectively amounts to extracting the hPDFs $\Delta\Sigma(x,Q^2)$ and $\Delta G(x,Q^2)$. Up until recently the standard way to do this was by using a collinear factorization approach along with the spin-dependent DGLAP evolution equations that were discussed in \textbf{Chapter}~\ref{intro} to extract the $Q^2$ dependent hPDFs. There have been numerous successful phenomenological extractions using DGLAP evolution \cite{Gluck:2000dy, Leader:2005ci, deFlorian:2009vb, Leader:2010rb, Jimenez-Delgado:2013boa, Ball:2013lla, Nocera:2014gqa, deFlorian:2014yva, Leader:2014uua, Sato:2016tuz, Ethier:2017zbq, DeFlorian:2019xxt, Borsa:2020lsz, Zhou:2022wzm, Cocuzza:2022jye}, however they of course do not give a prediction of the $x$ dependence of the hPDFs. These analyses fit to data to constrain their $Q^2_0$ initial conditions which means their extractions nominally describe data well, but without $x$ dependent evolution their uncertainties grow rapidly when extrapolating to values of $x$ smaller than that provided by data. This is the unique benefit of using small-$x$ helicity evolution equations, and is the subject of this phenomenological work.

The first small-$x$ evolution for hPDFs was pioneered by the combined work of Bartels, Ermolaev, and Ryskin (BER) \cite{Bartels:1995iu,Bartels:1996wc} who used the infrared evolution equation formalism (IREE) from Refs.~\cite{Gorshkov:1966ht,Kirschner:1983di,Kirschner:1994rq,Kirschner:1994vc,Griffiths:1999dj}. These evolution equations also resummed double logarithms of energy with the resummation parameter $\alpha_s\ln^2(1/x)$, and they found that the leading small-$x$ asymptotics for the hPDFs are given by the expression
\begin{equation}\label{asymptotics}
	\lim_{x\to0}g_1(x,Q^2)\sim \lim_{x\to0} \Delta \Sigma (x,Q^2) \sim \lim_{x\to 0}\Delta G(x,Q^2)  \sim \Bigl(\frac{1}{x}\Bigr)^{\alpha_h}
\end{equation}
where $\alpha_h$ is what is colloquially known as the helicity intercept. They computed their helicity intercept to be $\alpha_h = 3.66\sqrt{\tfrac{\alpha_s N_c}{2\pi}}$ in the pure-gluon limit (which would be the large-$N_c$ limit), and $\alpha_h = 3.45\sqrt{\tfrac{\alpha_s N_c}{2\pi}}$ when quark contributions are considered and $N_f = 4$ (and $N_c$ = 3). When using the typical fixed strong coupling of $\alpha_s \in [0.2, 0.3]$, this intercept would be large, $\alpha_h >1$, which makes the subsequent integral over $x$ divergent as $x \to 0$. This issue could be dampened through an implementation of the running of coupling, if not by parton saturation effects (see Refs.~\cite{Gribov:1984tu, Iancu:2003xm, Weigert:2005us, JalilianMarian:2005jf, Gelis:2010nm, Albacete:2014fwa, Kovchegov:2012mbw, Morreale:2021pnn,Itakura:2003jp} at very small $x$ which should slow this growth. More recently BER has also contributed to work regarding OAM distributions \cite{Boussarie:2019icw}.  Up until circa 2016 this was the only effort to describe the small-$x$ dependence of hPDFs, after which the world was introduced the works of Kovchegov, Pitonyak, and Sievert and their novel KPS evolution equations.

Over the past decade a new approach to small-$x$ helicity evolution has been developed \cite{Kovchegov:2015pbl, Hatta:2016aoc, Kovchegov:2016zex, Kovchegov:2016weo, Kovchegov:2017jxc, Kovchegov:2017lsr, Kovchegov:2018znm, Kovchegov:2019rrz, Cougoulic:2019aja, Kovchegov:2020hgb, Cougoulic:2020tbc, Chirilli:2021lif, Kovchegov:2021lvz, Cougoulic:2022gbk} which employs the shockwave or $s$-channel evolution formalism that was constructed in Refs.~\cite{Mueller:1994rr,Mueller:1994jq,Mueller:1995gb,Balitsky:1995ub,Balitsky:1998ya,Kovchegov:1999yj,Kovchegov:1999ua,JalilianMarian:1997dw,JalilianMarian:1997gr,Weigert:2000gi,Iancu:2001ad,Iancu:2000hn,Ferreiro:2001qy} for the case of unpolarized eikonal scattering. The main idea behind the works \cite{Kovchegov:2015pbl, Hatta:2016aoc, Kovchegov:2016zex, Kovchegov:2016weo, Kovchegov:2017jxc, Kovchegov:2017lsr, Kovchegov:2018znm, Kovchegov:2019rrz, Cougoulic:2019aja, Kovchegov:2020hgb, Cougoulic:2020tbc, Chirilli:2021lif, Kovchegov:2021lvz, Cougoulic:2022gbk} is that sub-eikonal quantities obey small-$x$ evolution equations similar to the eikonal ones; this was the subject discussed more at length in \textbf{Chapter}~\ref{chapter_2}. The original KPS evolution equations also only closed in the large-$N_c$ (pure-gluon) limit or the large-$N_c\&N_f$ limit (which brings back in quark contributions), and when solved in the large-$N_c$ limit they indeed also predicted the same asymptotics from Eq.~\eqref{asymptotics}, but with an intercept of $\alpha_h \approx 2.31\sqrt{\frac{\alpha_s N_c}{2\pi}}$, which of course is significantly smaller than the pure-gluon intercept of BER. After further investigation of this disagreement it was found that the original KPS evolution equations were missing contributions from what are now called the polarized Wilson lines ``of the second kind" that were discussed in \textbf{Chapter}~\ref{chapter_2} and are interpreted as coupling the gluon probe's OAM to the spin of the proton; authors Cougoulic, Tarasov, and Tawabutr augmented the KPS evolution equations with these new contribution and provided the KPS-CTT evolution equations that are given by Eqs.~\eqref{eq_LargeNcNf} (in the large-$N_c\&N_f$ limit) ~\cite{Kovchegov:2015pbl, Kovchegov:2018znm,Cougoulic:2022gbk}. The KPS-CTT evolution equations were solved numerically and analytically ~\cite{Cougoulic:2022gbk,Borden:2023ugd} in the large-$N_c$ limit and found agreement with the intercept from BER of $\alpha_h = 3.66\sqrt{\tfrac{\alpha_s N_c}{2\pi}}$ up to the third decimal point; the analytic solution revealed that the KPS-CTT evolution equations deviate from BER at four loops. Additionally, there was a similar $2\%-3\%$ disagreement with BER when solving the KPS-CTT evolution equations numerically in the large-$N_c\&N_f$ limit which are dependent on the number of quark flavors $N_f$. These deviances with BER will be discussed further in \textbf{Chapter}~\ref{pheno_2} which covers more recent analyses. The quantitative differences between BER and KPS(-CTT) evolution were modest enough to justify the phenomenological works that proceeded.

The first global analysis of the KPS evolution equations \cite{Adamiak:2021ppq} served as a ``proof-of-principle" analysis to show that world polarized DIS data can be fit by the KPS evolution equations, provided that the small-$x$ cutoff $x_0$ was sufficiently small. In that work, the cutoff $x_0$ was treated as a free parameter, and an analysis of world polarized data was repeated for varying values of $x_0$ from as large as $x_0 = 0.3$ to as small as $x_0 = 0.05$. Since small $x$ is synonymous with high-energy, it is beneficial to have a larger $x_0$ cutoff when possible to allow for more data in the analysis since data becomes quite sparse at small values of $x$. It was determined that the KPS evolution equations were not capable of describing DIS data above $x = 0.25$, and that the $chi^2_{\mathrm{red}}$ was much improved for $x_0 \leq 0.2$; after some consideration, this analysis (and indeed the analyses the followed) considered that small $x$ for polarized evolution was sufficient described as $x < x_0 = 0.1$. This is consistent with the typical unpolarized small-$x$ cutoff of $x_0^{\mathrm{unpol.}} = 0.01$ when one considers that $\alpha_s\ln(1/x_0^{\mathrm{unpol.}}) \approx \alpha_s\ln^2(1/x_0^{\mathrm{pol.}})$ for $\alpha_s \approx 0.25$, $x_0^{\mathrm{unpol.}} \approx 0.01$, and $x_0^{\mathrm{pol.}}\approx0.1$~\cite{Kuraev:1977fs,Balitsky:1978ic,Balitsky:1995ub,Balitsky:1998ya,Kovchegov:1999yj,Kovchegov:1999ua, JalilianMarian:1997dw,JalilianMarian:1997gr,Weigert:2000gi,Iancu:2001ad,Iancu:2000hn,Ferreiro:2001qy}. This first analysis also established my choice to cutoff the virtuality such that $Q^2 > m_c^2\approx 1.69~\mathrm{GeV}^2$, where $m_c$ is the mass of the charm quark which I do not consider in these analyses. This is also the same $Q^2$ cut applied to the JAM-DGLAP fragmentation functions \cite{Cocuzza:2022jye}. Lastly, this analysis completed an EIC impact study; since this analysis was fitting solely to DIS data it was only capable of extracting the $g_1$ structure function, but a fit onto EIC pseudodata allowed an extraction of the $C$-even quark hPDF $\Delta q^+ = \Delta q + \Delta \bar{q}$ (the $C$-odd quark hPDF being defined as $\Delta q^- = \Delta q - \Delta \bar{q}$).

With the foundations laid, this next installment of small-$x$ helicity phenomenology conducted a global analysis of DIS and SIDIS data using the revised and updated large-$N_c\&N_f$ KPS-CTT evolution equations with the running of the coupling \eqref{eq_LargeNcNf}. Semi-inclusive DIS data is sensitive to individual quark flavor hPDFs as seen from the dependence of $\Delta q(x,Q^2)$ in the definition of $g_1^h(x,z,Q^2)$ from Eq~\ref{all_hPDFs}, requiring the need of the flavor nonsinglet evolution equations from \cite{Kovchegov:2016zex}, since individual quark hPDFs can only be defined from the combination of \textit{both} $\Delta q^+(x,Q^2) \propto [Q_f + 2G_2]$ and $\Delta q^-(x,Q^2) \propto G_q^{\mathrm{NS}}$. The running of the coupling was implemented to improve the accuracy of the small-$x$ evolution and to avoid a divergent integral that may have arisen from a large intercept $\alpha_h$; the coupling runs with the transverse size of the daughter dipole which has already been established to scale with the CM energy squared, $s$. I also discussed in \textbf{Chapter}~\ref{chapter_3} that the SIDIS observable is dependent on the hadronic fragmentation functions $D^{h/q}$ since $A_1^h \propto g_1^h \propto D^{h/q}$, and how this first analysis of SIDIS data established the method of adopting JAM-DGLAP extractions of the fragmentation functions (and unpolarized structure functions) instead of attempting a simultaneous small-$x$ extraction of these functions in addition to the hPDFs.

In preparation for the presentation of the results of this analysis, I now present a brief discussion regarding my ability to constrain specifically the gluon polarized dipole amplitude initial conditions $\widetilde{G}^{(0)}$ and $G_2^{(0)}$. I direct the reader to take another look at Fig.~\ref{fig:Deltaqp_basis}, from which I note three important features: 1) each of the basis functions have distinct pre-asymptotic forms (indeed they are linearly independent), 2) $\widetilde{G}^{(0)}$ and $G_2^{(0)}$ initial conditions have, by a large margin, the largest influence over the small-$x$ asymptotics, and 3) $\widetilde{G}^{(0)}$ initial conditions only contribute after evolution ``kicks in". Tackling these points in reverse order, I make the point that feature number three could have been predicted by observing Eqs.~\eqref{all_hPDFs} and noting that the structure functions and hPDFs are ``independent" of $\widetilde{G}$. By the coupled nature of the evolution equations a $\widetilde{G}^{(0)}$ initial condition will eventually result in non-zero $Q_q$ and $G_2$ and thus non-zero hPDFs and a non-zero structure function, but before evolution commences the $\widetilde{G}$ does not directly contribute to the hPDFs. In effect this means that large-$x$ DIS and SIDIS asymmetries will have a difficult time constraining the $\widetilde{G}^{(0)}$. Feature number two is especially important for the goal of this research; if one wants to know the small-$x$ behavior of the hPDFs, then there must be good constraints on both the $\widetilde{G}$ and $G_2$ initial conditions since they control the small-$x$ asymptotics. I have established from feature number three that $\widetilde{G}$ initial conditions will be difficult to constrain, however, I can be slightly more optimistic about constraining $G_2$ initial conditions since $g_1(x,Q^2)$ is directly sensitive to it. Having said that, one should keep in mind that the data is only directly sensitive to the specific combination ($Q_q + 2G_2$), as seen from Eq.~\eqref{g_1}, and not to $Q_q$ or $G_2$ alone. On the positive side, feature number one may allow for data to separate contributions from each of the initial conditions. Since all basis functions are linearly independent with unique (even if similar) $x$ dependence, the initial conditions can be independently fitted, provided there are enough data points distributed over a wide range of $x$ values. It is then fortunate that each of the basis functions' dominant contributions and unique curvatures is the most dominant and dynamic for $x \lesssim 0.1$ (except those from $\widetilde{G}$), the same range in $x$ where the data is found. With challenges laid out, I now detail the selection of DIS and SIDIS data and the results of this analysis.

\subsection{$\boldsymbol{\mathrm{Results~I\!:~Data~vs~Theory}}$}\label{pheno_1_data}

The DIS and SIDIS data analyzed in this work came from a plethora of colliders and collaborations, but the cuts on those data are consistently followed. From all data sets, I only fit to data that simultaneously satisfies the limits $x < 0.1$ and $Q^2 > 1.69~\mathrm{GeV}^2$; because of the relationship $x\approx Q^2/s$, the upper/lower limits of $Q^2$ and $x$ are correlated for a given center-of-mass energy squared $s$, and in practice I find that that all DIS and SIDIS data can be found in the Bjorken-$x$ range of $5\times10^{-3} < x < 0.1$ and the virtuality range $1.69~\mathrm{GeV}^2 < Q^2 < 10.4~\mathrm{GeV}^2$. To be consistent with the first small-$x$ helicity global analysis, $\Lambda$ is set to $1~\mathrm{GeV}$ which roughly characterizes the size of the proton; I make the note that this a choice, and other choices for $\Lambda$ are possible so long as $c$ in the initial condition parameterization is appropriately redefined. For their explicit uses in the running coupling expression \eqref{running}, I use $N_f = N_c = 3$. The range of $z$ for outgoing hadron momentum fractions in polarized SIDIS data is $0.2 < z < 1.0$, and I do not make any further cuts to this range; after applying the $x$ and $Q^2$ cuts the remaining SIDIS data covers the range $0.4 < z < 0.6$, if not integrating over the range $z\in[0.2,1.0]$ or $z\in[0.2,0.85]$.

\begin{figure}[h!]
	\begin{centering}
		\includegraphics[width=\textwidth]{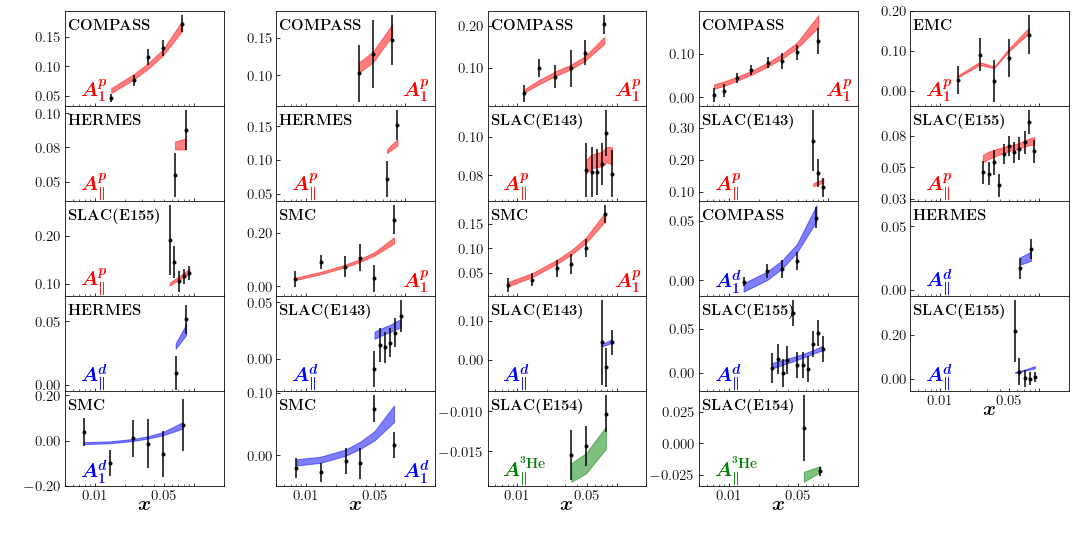}
		\vspace{-0.35cm}
		\caption{Experimental data (black) vs small-$x$ theory for the double-spin asymmetries $A_1$ and $A_\parallel$ in polarized DIS on a proton (red), deuteron (blue) and ${}^3 \mathrm{He}$ (green) target.
			\label{Plot_pidis_data}
			\vspace{-0.1cm}
		}
	\end{centering}
\end{figure}

\begin{figure}[h!]
	\begin{centering}
		\includegraphics[width=\textwidth]{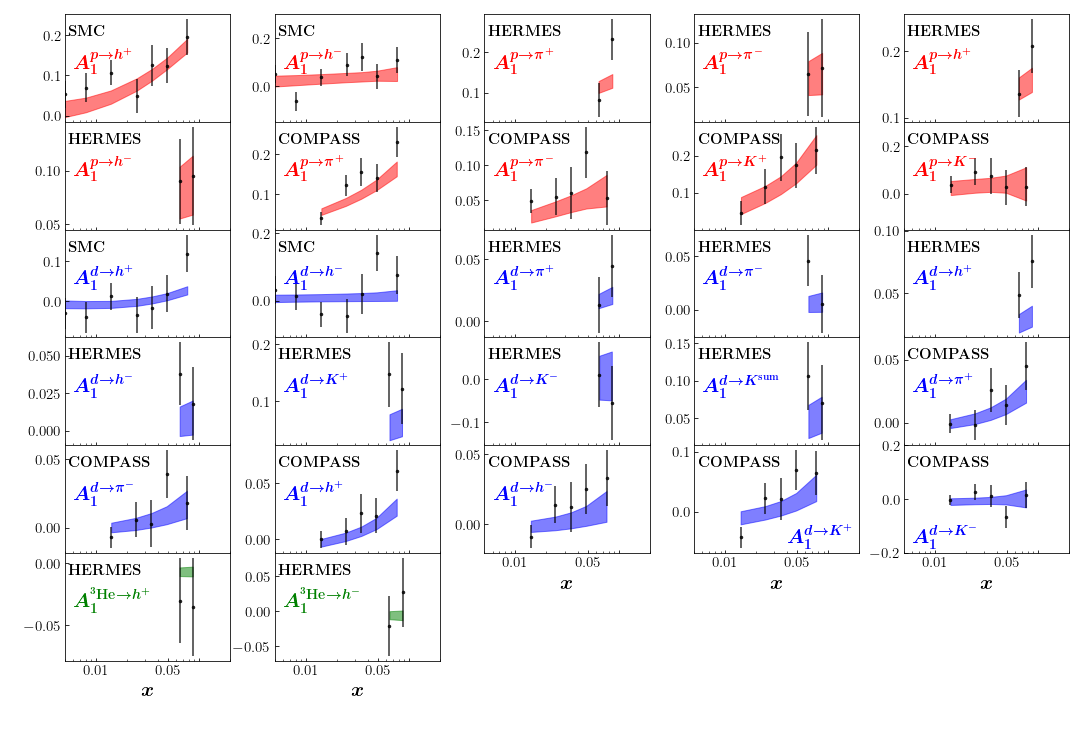}
		\vspace{-0.35cm}
		\caption{Experimental data (black) vs small-$x$ theory for the double-spin asymmetry $A_1^h$ in polarized SIDIS on a proton (red), deuteron (blue) and ${}^3 \mathrm{He}$ (green) target for charged pion, kaon and unidentified hadron final states.
			\label{Plot_psidis_data}
			\vspace{-0.1cm}
		}
	\end{centering}
\end{figure}

The world polarized data for which I apply these cuts comes from SLAC~\cite{Anthony:1996mw, Abe:1997cx, Abe:1998wq, Anthony:1999rm, Anthony:2000fn}, EMC~\cite{Ashman:1989ig}, SMC~\cite{Adeva:1998vv, SpinMuon:1998eqa, Adeva:1999pa}, COMPASS~\cite{Alekseev:2010hc, Adolph:2015saz,Adolph:2016myg}, and HERMES~\cite{Ackerstaff:1997ws, Airapetian:2007mh} for DIS, and SMC~\cite{SpinMuon:1997yns}, COMPASS~\cite{COMPASS:2010hwr, COMPASS:2009kiy}, and HERMES~\cite{HERMES:2004zsh, HERMES:1999uyx} for SIDIS. In the end that leaves 122 polarized DIS data points and 104 polarized SIDIS data points for a total of $N_{\mathrm{pts}} = 226$. I can examine how the 500 individual fits to data and their resultant theoretical asymmetries compare to experimental data via the following tables and figures. In summary, I achieve a reduced $\chi^2$ for the average of all replicas of $\chi^2_{\mathrm{red}} \approx 1.03$, which no significant changes in $\chi^2_{\mathrm{red}}$ for analysis where $x_0 = 0.08$ and $x_0 = 0.05$; Table~\ref{t:Chi2_DIS_Apa} provides the reduced $\chi^2_{\mathrm{red}}$ on a per-dataset basis for the DIS data while Table~\ref{t:Chi2_SIDIS} does the same for the SIDIS data. Shown visually, Figures~\ref{Plot_pidis_data} and \ref{Plot_psidis_data} depicts the experimental data in black against the $1\sigma$ (standard deviation) of replica asymmetries. Both quantitatively and qualitatively it is clear that small-$x$ helicity evolution can describe world polarized data through the method established here, and I can continue onto the next section to learn the implications and results of this analysis.
\begin{table}[t!]
	\caption{Summary of my fit to polarized DIS data, separated into $A_1$  (left) and $A_{\parallel}$ (right), along with the  $\chi^2_{\mathrm{red}}$ for each data set.}
	\vspace{0.4cm}
	\begin{tabular}{l|c|c|c}
		\hline
		{$\boldsymbol{\mathrm{Dataset}}$ ($\boldsymbol{A_1}$)} & 
		~{$\boldsymbol{\mathrm{Target}}$}~ &
		~$\boldsymbol{N_\mathrm{pts}}$~ & 
		~$\boldsymbol{\chi^2_{\mathrm{red}}}$~  
		\\ \hline     
		SLAC (E142) \cite{Anthony:1996mw} &
		${}^3 \mathrm{He}$ &
		$1$       & 
		$0.60$    \\ \hline
		EMC  \cite{Ashman:1989ig} &
		$p$ &
		$5$ &
		$0.20$   \\ \hline
		SMC  \cite{Adeva:1998vv, Adeva:1999pa} &
		$p$ &
		$6$ &
		$1.29$  \\ 
		&
		$p$ &
		$6$ &
		$0.53$  \\
		&
		$d$ &
		$6$ &
		$0.67$  \\
		&
		$d$ &
		$6$ &
		$2.26$  	\\ \hline
		COMPASS   \cite{Alekseev:2010hc} &
		$p$ &
		$5$ &
		$1.02$  \\ \hline
		COMPASS   \cite{Adolph:2015saz} &
		$p$ &
		$17$ &
		$0.74$   \\ \hline
		COMPASS  \cite{Adolph:2016myg} &
		$d$ &
		$5$ &
		$0.88$  \\ \hline
		HERMES   \cite{Ackerstaff:1997ws} &
		$n$ &
		$2$ &
		$0.73$ \\ \hline\hline
		{$\boldsymbol{\mathrm{Total}}$} &    & 59 & 0.91 \\ \hline
	\end{tabular}
	\qquad\;\;
	\begin{tabular}{l|c|c|c} 
		\hline
		{$\boldsymbol{\mathrm{Dataset}}$ ($\boldsymbol{A_{\parallel}}$)}& 
		~{$\boldsymbol{\mathrm{Target}}$}~ &
		~$\boldsymbol{N_\mathrm{pts}}$~ & 
		~$\boldsymbol{\chi^2_{\mathrm{red}}}$~  
		\\ \hline
		SLAC(E155) \cite{Anthony:1999rm} &
		$p$ &
		$16$ &
		$1.28$ 	\\ 
		&
		$d$ &
		$16$ &
		$1.62$      \\ \hline
		SLAC (E143) \cite{Abe:1998wq} &
		$p$ &
		$9$ &
		$0.56$    \\
		&
		$d$ &
		$9$ &
		$0.92$  	\\ \hline
		SLAC (E154) \cite{Abe:1997cx} &
		${}^3 \mathrm{He}$ &
		$5$       &
		$1.09$     \\ \hline
		HERMES    \cite{Airapetian:2007mh} &
		$p$ &
		$4$ &
		$1.54$  \\ 
		&
		$d$ &
		$4$ &
		$0.98$  \\ \hline\hline
		{$\boldsymbol{\mathrm{Total}}$} &   & 63 & 1.19 \\ \hline
	\end{tabular}
	\label{t:Chi2_DIS_Apa}
\end{table}

\begin{table}[h!]
	\begin{center}
		\caption{Summary of the polarized  SIDIS data on $A_1^h$ included in the fit, along with the $\chi^2/N_{\rm pts}$ for each data set.}
		\label{t:Chi2_SIDIS}
		\vspace{0.4cm}
		\begin{tabular}{l|c|c|c|c} 
			\hline
			{$\boldsymbol{\mathrm{Dataset}}$ ($\boldsymbol{A_1^h}$})& 
			~{$\boldsymbol{\mathrm{Target}}$}~ &
			~{$\boldsymbol{\mathrm{Tagged~Hadron}}$}~ &
			~$\boldsymbol{N_\mathrm{pts}}$~ & 
			~$\boldsymbol{\chi^2_{\mathrm{red}}}$~  
			\\ \hline
			SMC  \cite{SpinMuon:1998eqa} &
			$p$ &
			$h^+$ &
			$7$ &
			$1.03$  \\ 
			&
			$p$ &
			$h^-$ &
			$7$ &
			$1.45$  		\\ 
			&
			$d$ &
			$h^+$ &
			$7$ &
			$0.82$  \\ 
			&
			$d$ &
			$h^-$ &
			$7$ &
			$1.49$   \\ \hline
			HERMES  \cite{HERMES:2004zsh} &
			$p$ &
			$\pi^+$ &
			$2$ &
			$2.39$ \\ 
			&
			$p$ &
			$\pi^-$ &
			$2$  &
			$0.01$    \\ 
			&
			$p$ &
			$h^+$ &
			$2$ &
			$0.79$  \\
			&
			$p$ &
			$h^-$ &
			$2$  &
			$0.05$  \\ 
			&
			$d$  &
			$\pi^+$ &
			$2$ &
			$0.47$ 	 \\
			&
			$d$ &
			$\pi^-$ &
			$2$ &
			$1.40$ 	 \\
			&
			$d$ &
			$h^+$  &
			$2$  &
			$2.84$   \\ 
			&
			$d$  &
			$h^-$ &
			$2$  &
			$1.22$ 	 \\ 
			&
			$d$ &
			$K^+$ &
			$2$  &
			$1.81$ 	 \\ 
			&
			$d$  &
			$K^-$ &
			$2$  &
			$0.27$ 	\\ 
			&
			$d$  &
			$K^+ + K^-$ &
			$2$  &
			$0.97$ 	\\ \hline	
			HERMES \cite{HERMES:1999uyx} &
			${}^3 \mathrm{He}$  &
			$h^+$  &
			$2$ &
			$0.49$ 	\\ 
			&
			${}^3 \mathrm{He}$ &
			$h^-$  &
			$2$   &
			$0.29$ 	\\ \hline
			COMPASS \cite{COMPASS:2010hwr} &
			$p$ &
			$\pi^+$ &
			$5$ &
			$1.88$ 		\\ 
			&
			$p$   &
			$\pi^-$ &
			$5$ &
			$1.10$ \\ 
			&
			$p$  &
			$K^+$ &
			$5$   &
			$0.42$ 	\\
			&
			$p$ &
			$K^-$ &
			$5$   &
			$0.31$ 		\\ \hline
			COMPASS \cite{COMPASS:2009kiy}  
			&
			$d$ &
			$\pi^+$ &
			$5$   &
			$0.50$ 	\\ 
			&
			$d$  &
			$\pi^-$ &
			$5$   &
			$0.78$ \\
			&
			$d$   &
			$h^+$ &
			$5$ &
			$0.90$ \\ 
			&
			$d$ &
			$h^-$  &
			$5$  &
			$0.86$ 		\\ 
			&
			$d$     &
			$K^+$ &
			$5$  &
			$1.50$ 		\\
			&
			$d$ &
			$K^-$ &
			$5$ &
			$0.78$ 	\\ \hline\hline
			{$\boldsymbol{\mathrm{Total}}$} &    &    & 104 & 1.01 \\ \hline
		\end{tabular}
	\end{center}
\end{table}


\subsection{$\boldsymbol{\mathrm{Results~II\!:~}g_1^p~\mathrm{and~Asymptotic~Behavior}}$}

\begin{figure}[t!]
	\begin{centering}
		\includegraphics[width=345pt]
		{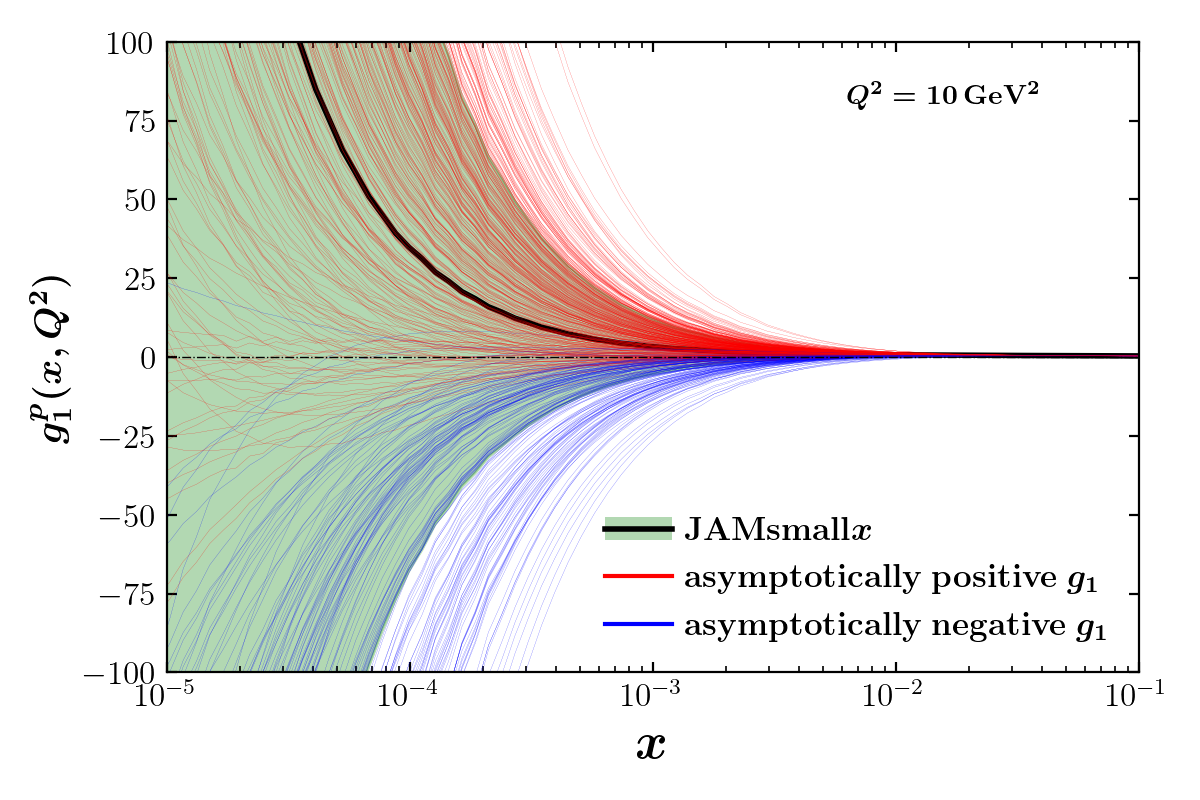}
		\caption{The small-$x$ calculation of the $g_1$ structure function of the proton. The black curve is the mean of all the replicas with the green band giving the 1$\sigma$ uncertainty band. Red and blue curves are solutions that grow asymptotically positive and negative, respectively.
			\label{Plot_g1}
		}
	\end{centering}
\end{figure}

The $g_1$ structure function has a close relationship to the spin asymmetries and to the hPDFs, so it is a good place to start. The results below are all in relation specifically to the structure function of the proton, labeled $g_1^p$ and/or $g_1^p(x,Q^2)$. Shown in Fig.~\ref{Plot_g1} is the result of 500 replicas for the fit to data (each with their own combination of parameters as discussed thoroughly in \textbf{Chapter}~\ref{chapter_3}), whose parameters were each used to compute the $g_1^p$ structure function of the proton (as well as the hPDFs) as a function of $x$ down to $x = 10^{-5}$; plotted along with the replicas is the midpoint of those replicas (black) and their $1\sigma$ uncertainty band (green). These extractions, and the results of the analysis below, were taken at $Q^2 = 10~\mathrm{GeV}^2$. Replicas that grow positive as $x$ approaches zero are shown in red, while replicas that grow negative as $x$ goes to zero are shown in blue; the asymptotics of the replicas are non-trivial and will be discussed in more detail towards the end of this section. 
The most notable feature of Fig.~\ref{Plot_g1} is that, while the uncertainty of the structure function is very well constrained in the region of $x$ where the experimental data lies ($5\times 10^{-3} < x < 0.1$), the uncertainty becomes unconstrained at smaller values of $x$. I have determined the cause of this unconstrained uncertainty to be the insensitivity of DIS and SIDIS data to the polarized dipole amplitudes $\widetilde{G}$ and $G_2$. The small-$x$ uncertainty is not indicative of an issue with the small-$x$ helicity evolution equations, but due to the distribution of replicas growing both asymptotically positive and asymptotically negative; these asymptotics are a non-trivial feature of this analysis, and I will study it extensively.

\subsubsection{\uline{Identifying Asymptotic Bimodality}}

\begin{figure}[t] 
	\begin{centering}
		\includegraphics[width=355pt]{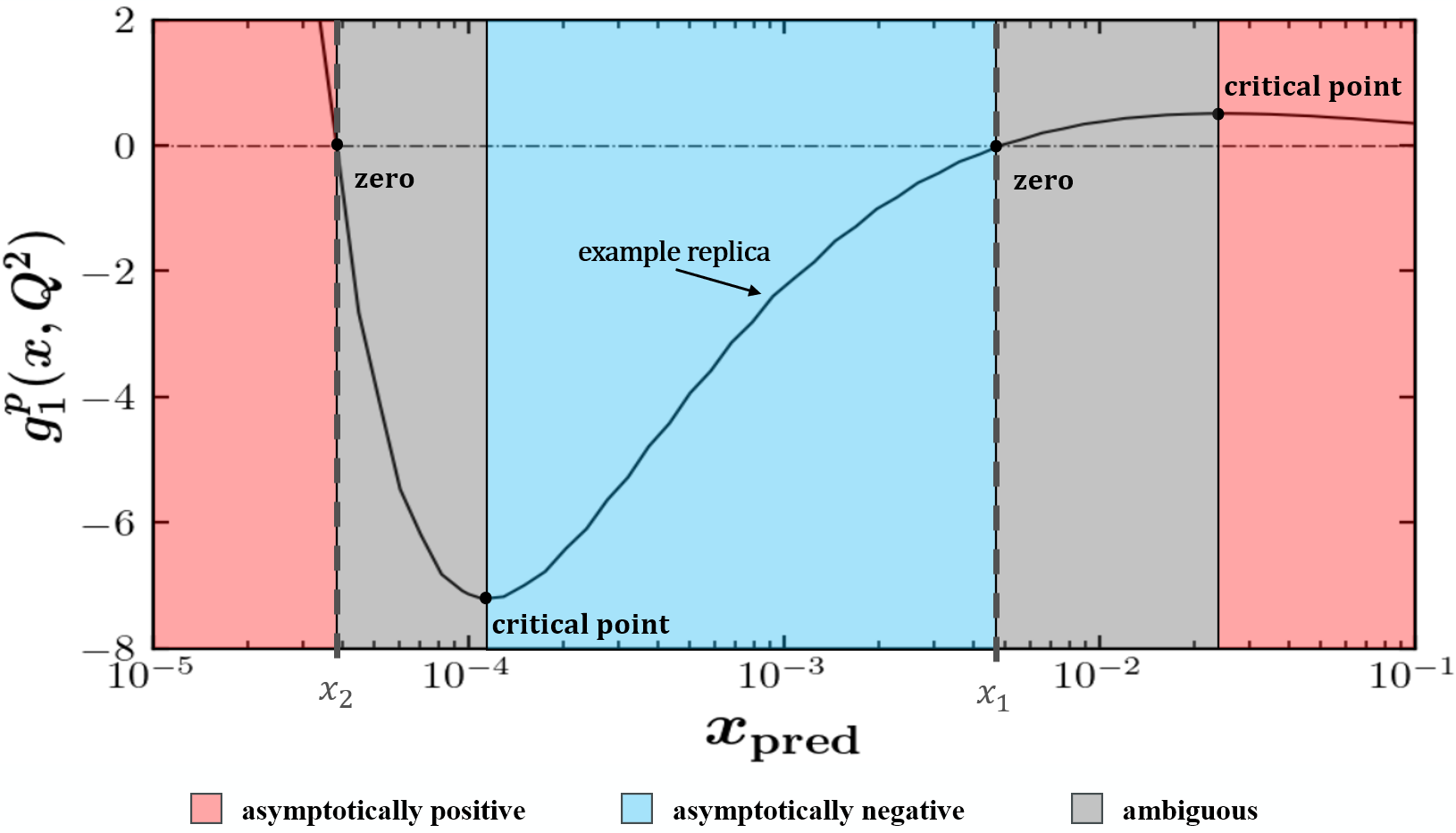}
		\caption{An example replica of $g_1^p(x, Q^2=10\, {\rm GeV^2})$ that demonstrates how the asymptotic sign is dependent on $x_{\mathrm{pred}}$. If $x_{\mathrm{pred}}$ resides in the red (blue) region then the replica will be considered asymptotically positive (negative) according to the sign of its first derivative (as $x$ decreases), and its agreement with its magnitude's sign. If $x_{\mathrm{pred}}$ resides in either gray region then the asymptotic sign is ambiguous due to a disagreement between the sign of the slope and the sign of the magnitude at $x_{\mathrm{pred}}$.
			\label{Ambiguity_Ex}
		}
	\end{centering}
\end{figure}

\begin{figure}[t!]
	\begin{centering}
		\includegraphics[width=\textwidth]{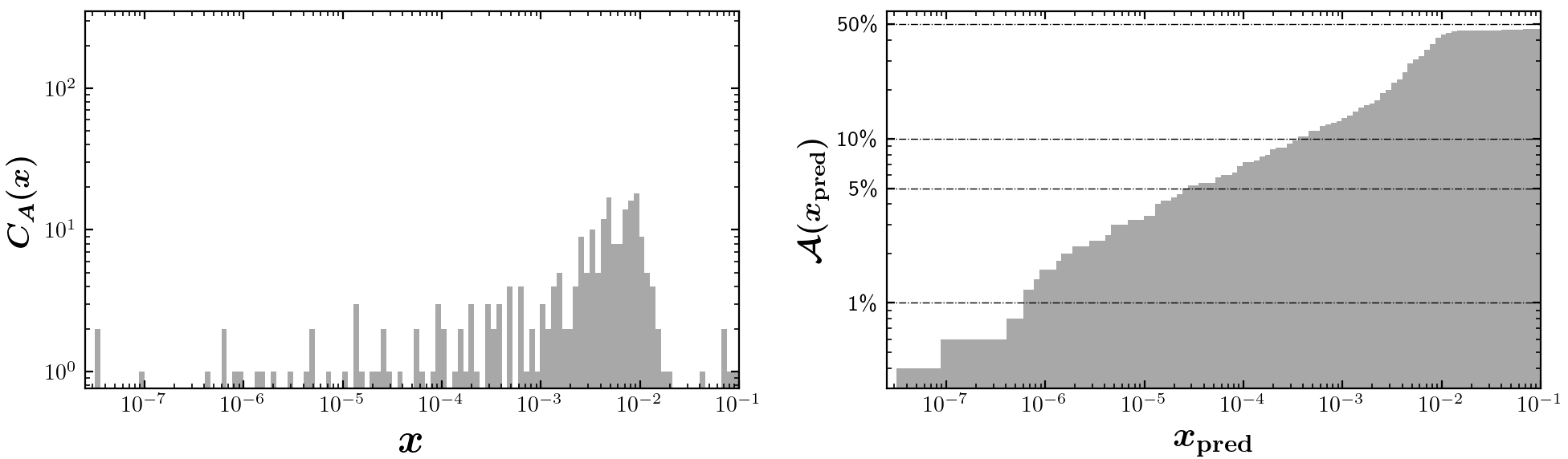}
		\caption{(Left) Histogram that counts the number of replicas with a smallest-$x$ ambiguity in a narrow bin near the given value of $x$. (Right) The running sum of the ambiguity histogram, providing the percentage of replicas that have an ambiguity below a given value of~$x$.
			\label{Plot_g1_ambiguities}
		}
	\end{centering}
\end{figure}

The asymptotic growth is a feature of the small-$x$ helicity evolution equations as seen from Eq.~\eqref{asymptotics}, however the sign of this growth must be determined by the initial conditions, and is thus something that the data must be able to constrain. Investigations into the behavior of each replica reveals that with 90\% accuracy I can predict a given replica's asymptotics by its behavior at $x = 3.5\times 10^{-4}$, and with 95\% accuracy can predict the asymptotics by $x = 2.5\times 10^{-5}$; my method of determining asymptotics is supported by Fig.~\ref{Ambiguity_Ex}, and information on the accuracy of those predictions is given in Fig.~\ref{Plot_g1_ambiguities}. DIS and SIDIS data is not constraining enough for all solutions to share a single asymptotic sign, but it prefers positive solutions which make up approximately $70\%$. This preference for positive $g_1^p$ solutions at small $x$ is consistent with recent analyses of (polarized and unpolarized) DIS and SIDIS structure functions using the anti-de Sitter space/Conformal Field Theory (AdS/CFT) correspondence~\cite{Kovensky:2018xxa,Jorrin:2022lua,Borsa:2023tqr} that make an even stronger statement about the positivity of $g_1^p$ as $x$ approaches zero. Determining the sign of $g_1^p$ and quantifying its ambiguity requires more dedicated examination.

Between Fig.~\ref{Plot_g1} and Eq.~\eqref{asymptotics}, it is clear that the power-law growth as $x \to 0$ is expected, but considering that all basis functions have this same growth but with different combinations (one also must consider that parameters can be negative), determining the sign can be a difficult task. It is not possible to measure or compute the sign of $g_1^p$ directly at $x = 0$, so the color coding of Fig.~\ref{Plot_g1} is based on the slope of the replica as measured at $x_{\mathrm{asymp}} = 10^{-7.5}$: if the slope (as $x$ decreases) at $x_{\mathrm{asymp}}$ is increasing then the replica is colored red and considered ``asymptotically" positive, but if the slope (as $x$ decreases) is negative at $x_{\mathrm{asymp}}$ then it is colored blue and considered ``asymptotically" negative\footnote{The value of $x_{\mathrm{asymp}}$ was chosen simply because it was small enough to be computationally expensive but not inhibiting.}. This, however, brings up the potential issue that replicas can have different ``asymptotic" signs depending on how small the choice for $x_{\mathrm{asymp}}$. Since each replica is a different combination of parameters there can be competition between positive and negative contributions of similarly shaped basis functions which can result in nodes. Shown in Fig.~\ref{Ambiguity_Ex} is an example of a $g_1^p$ replica with two nodes, i.e. points where the slope changes sign; depending on the initial condition state, the number of nodes and the location in $x$ of those nodes can change on a per-replica basis. Also shown in Fig.~\ref{Ambiguity_Ex} is another feature of these $g_1^p$ replicas that is associated with the asymptotic uncertainty, areas of \textit{ambiguity}. A replica's predicted asymptotics are considered ``ambiguous" when it has a positive (negative) slope at a given value of $x_{\mathrm{pred}}$ while simultaneously having a magnitude of the opposite sign. This is the reason that a handful of replicas in Fig.~\ref{Plot_g1} seem to be incorrectly colored as red/blue, because they appear to be growing positive/negative as of $x = 10^{-5}$, but in truth have a critical point ($\tfrac{d \,g_1(x,Q^2)}{dx} = 0$) at some point $x < 10^{-5}$.

Since the goal is to have predictive power at small $x$, quantifying this ambiguity is a logical goal; I do this by calculating its probability density as a function of $x$. I measure this probability density by examining each replica over its computed range of $x_{\mathrm{asymp}} < x < 0.1$ and identifying the smallest-$x$ instance of ambiguity (as determined from a minimum value of $x_{\mathrm{asymp}}$). In Fig.~\ref{Ambiguity_Ex} we see an example of a replica begins positive (which is true for all replicas), after which evolution drives it more positive, then more negative, and then finally more positive again; note again that this behavior is based on the initial conditions and their resultant linear combination of basis functions. For any given replica I can attempt predict its small-$x$ asymptotics by measuring its slope and magnitude at some point $x_{\mathrm{pred}}$. If, for this particular replica, I predict its asymptotics at $x_{\mathrm{pred}} \approx 0.1$, I would consider it ``asymptotically" positive, although it should be obvious that $x \approx 0.1$ is not approaching any asymptotic limit. If instead I examine the replica at a value $x_{\mathrm{pred}}$ between its largest-$x$ critical point and its largest-$x$ zero (the gray region), I would see that it is actively positive but growing negative and therefor it is considered ambiguous. If the replica is probed at decreasing values of $x_{\mathrm{pred}}$ then I see that if $x_{\mathrm{pred}}$ is in the blue region then it would be considered ``asymptotically" negative since its slope is negative (for decreasing $x$) while its magnitude is negative, until $x_{\mathrm{pred}}$ is in the second gray region where it the magnitude is actively positive but the slope is increasing (for decreasing $x$) and is then considered ambiguous again. The smallest-$x$ instance of ambiguity occurs at the second zero-crossing, labeled by the position $x_2$, after which $x_{\mathrm{pred}}$ crosses into the red region and is considered ``asymptotically" positive since its slope is growing positive (for decreasing $x$) and its magnitude is also actively positive. For every replica I then bin its smallest-$x$ instance of ambiguity at $x_2$, thus counting every replica exactly once, with multi-nodal replicas having their most delayed ambiguity counted. 

I quantify this ambiguity through the number of smallest-$x$ ambiguities counted in a particular bin of $x$, called $C_A(x)$ (count of ambiguities), and make a histogram. This histogram, along with the running sum of the histogram, is shown in Fig.~\ref{Plot_g1_ambiguities}. The ambiguity count $C_A(x)$ is normalized such that it sums to the total number of replicas $N_{\mathrm{ambig}}$ containing at least one ambiguity:
\begin{align}   \label{e:ambigsum1}
	\sum_{x = x_{\mathrm{asymp}}}^{x_0} C_A (x) = N_{\mathrm{ambig}} \leq N_{\rm tot}  \: .
\end{align}
Since some replicas are completely unambiguous over the entire range $x_{\mathrm{asymp}} < x < 0.1$, the ambiguity count is less than or equal to the total number of replicas $N_{\mathrm{tot}}$. To estimate the accuracy of my predictions of the small-$x$ asymptotics, I can use $C_A(x)$ to estimate the probability that an unobserved ambiguity can be found in a given replica at $x_{\mathrm{asymp}} < x < x_{\mathrm{pred}}$. This is achieved by the truncated sum 
\begin{align}   \label{e:ambigsum2}
	\mathcal{A}(x_{\mathrm{\mathrm{pred}}}) = \frac{1}{N_{\mathrm{rep}}}\sum\limits_{x=x_{\mathrm{\mathrm{asymp}}}}^{x_{\mathrm{\mathrm{pred}}}} C_A(x) \: .
\end{align}
From Eq.~\eqref{e:ambigsum1} it follows that Eq.~\eqref{e:ambigsum2} is normalized such that a measurement at $x_{\mathrm{pred}} = x_0$ provides the total fraction of replicas containing at least one ambiguity, $\mathcal{A}(x_0) = N_{\mathrm{ambig}}/N_{\mathrm{tot}}$. In the left panel of Fig.~\ref{Plot_g1_ambiguities} the bulk of replicas have chosen their asymptotics over the range near or around the data-range $5\times 10^{-3} < x < 0.1$, and in the right panel $90\%$, $95\%$, and $99\%$ of replicas have chosen their final asymptotic state by $x = 3.5\times 10^{-4},\; 2.5\times 10^{-5},$ and $6 \times 10^{-7}$ respectively. This is a strong justification that $x_{\mathrm{asymp}} = 10^{-7.5}$ is reasonably low enough to capture the true asymptotic sign of the replicas. Replicas that have zero instances of ambiguity can be captured at $x_0 = 0.1$, and according to the right panel of Fig.~\ref{Plot_g1_ambiguities} approximately $50\%$ of replicas choose their asymptotic signs immediately as evolution begins; note that the positive starting point for all replicas implies that only asymptotically positive replicas are those that can have zero instance of ambiguity.

Though not shown in Fig.~\ref{Plot_g1_ambiguities}, I repeated this ambiguity quantification twice more: once on the set of asymptotically negative replicas and once more on the set of asymptotically positive replicas. I found that among the replicas that have at least one instance of ambiguity, asymptotically negative replicas choose their asymptotics earlier than their positive counter parts; $95\%$ of asymptotically negative replicas choose their asymptotic sign by $x\approx 4.3\times 10^{-4}$, while the $25\%$ of asymptotically positive replicas that have at least one ambiguity choose their asymptotics by $x\approx 2\times 10^{-5}$ with $5\%$ accuracy. This suggests that a lower $x_{\mathrm{pred}}$ is more likely to capture asymptotically positive replicas that were incorrectly predicted to be negative at some higher $x_{\mathrm{pred}}$

Lastly (as far as ambiguity is concerned), I performed a similar analysis of the smallest-$x$ critical points (as opposed to their ambiguities), and found that on average the smallest-$x$ critical point occurs $4\%$ earlier in $\ln(1/x)$ than the smallest-$x$ zero. This is precisely the span in $x$ of the ambiguity region. Having such a small percentage indicates that any unresolved ambiguities at small $x$ will be resolved quickly. I thus conclude that, from the perspective of Fig.~\ref{Plot_g1_ambiguities}, having data as low as $x\approx 10^{-5}$ will allow me to predict the small-$x$ asymptotics of $g_1^p$ with a certainty greater than $95\%$. This information has all been helpful in understanding the certainty of the small-$x$ asymptotics, but I still need to better understand the nature of having two distinct asymptotic solutions in the first place.

Following from the analysis above, there are many more $g_1^p$ replicas that adopt their asymptotic signs early than there are replicas that have a delayed sign change at small $x$. This results in the clustering behavior around $x = 5\times 10^{-3}$ in the left panel of Fig.~\ref{Plot_g1_ambiguities} and implies that the replicas share similar rates of growth, i.e. they have similar curvature. Since the majority of replicas have no instances of ambiguity, which effectively means they have critical points at $x = x_0 \equiv 0.1$, this supports the idea that early adoption of asymptotic growth is preferred over delayed asymptotic growth since small-$x$ critical points are less common. Because of this, I expect that the existence of both rapidly growing positive solutions and rapidly growing negative solutions is indicative of a bimodality. This bimodality is a novel result and something that demands further examination. Fig.~\ref{Plot_g1} alone does not provide enough information to conclude that this bimodality of solutions exists; reformatting this plot into a histogram of $g_1^p$ replicas measured at small $x$ ($x = 10^{-3}$) and very small $x$ ($x = 10^{-7.45}$), as done in Fig.~\ref{g1_normdist} below, shows that all solutions are normally distributed and centered near zero.
\begin{figure}[b!]
	\begin{centering}
		\includegraphics[width=\textwidth]{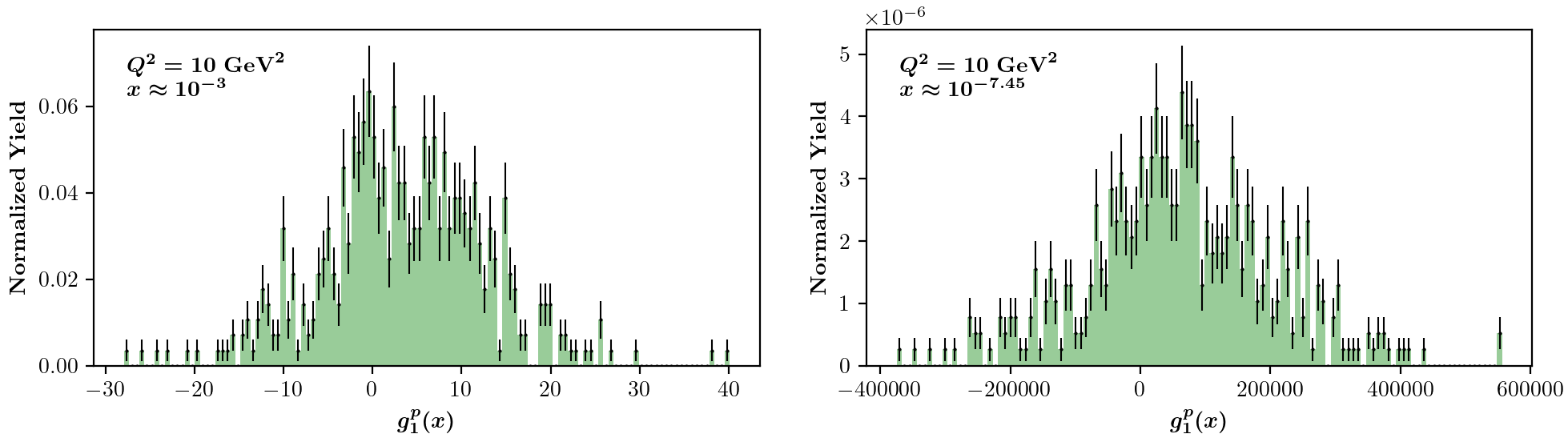}
		\caption{Histograms counting all replicas of $g_1^p$ at $x=10^{-3}$ (left) and $10^{-7.45}$ (right), displaying normal distributions centered slightly above zero.
			\label{g1_normdist}
		}
	\end{centering}
\end{figure}
As hinted above, however, it is not the values of $g_1^p$ that are bimodal, but their curvatures. The curvature of a given replica is sensitive to when in $x$ its asymptotic growth starts to dominate, and I have already established that the asymptotics are given by Eq.~\eqref{asymptotics}. Using this information I can define the generalized $x$-dependent intercept $\alpha_h(x)$ through the logarithmic derivative of $g_1^p$:
\begin{align}   \label{e:logderv1}
	\lim_{x\to0}\,g_1^p(x) \equiv g_1^{p\,(0)} \, x^{-\alpha_h(x)}
	\qquad \therefore \qquad
	\alpha_h(x) \equiv \frac{1}{g_1^p(x)}\frac{d \,g_1^p (x)}{d \ln (1/x)}, 
\end{align}
where $g_1^{p\,(0)}$ is a constant. Since $\alpha_h(x)$ provides information on $g_1^p(x,Q^2)$ itself, I can then determine that, regardless of the magnitude of $g_1^p$ at a given values of $x$, if replicas have the same intercept $\alpha_h(x)$ then they must have the same curvature. To specifically identify the sign of that curvature I generalize the logarithmic derivative as
\begin{align}   \label{e:logderv2}
	\alpha_h(x) = \frac{1}{g_1^p(x)} \frac{d \,g_1^p (x)}{d \ln (1/x)}
	\qquad \Rightarrow \qquad
	\mathrm{Sign}\bigl[g_1^p (x)\bigr]\,\alpha_h (x) = \frac{1}{\big| g_1^p(x) \big|} \frac{d \,g_1^p (x)}{d \ln (1/x)}    \: .
\end{align}
I visualize the behavior of these $x$-dependent intercepts by plotting them too as histograms (counting each replica\footnote{Replicas that have a delayed critical point, and subsequently a delayed zero, will have an artificially large intercept ratio if $g_1^p\,'(x,Q^2) \gg g_1^p(x,Q^2)$. To avoid these outliers, any replicas in Fig.~\ref{Bimodal_Curvature} that have an intercept value outside of 5 standard deviations from the average are excluded.}) at $x\approx10^{-2},\;x\approx 10^{-3},\;x\approx 10^{-5},$ and $x\approx 10^{-7}$ as in the left/right panels of Fig.~\ref{Bimodal_Curvature}.
\begin{figure}[t]
	\begin{centering}
		\includegraphics[width=\textwidth]{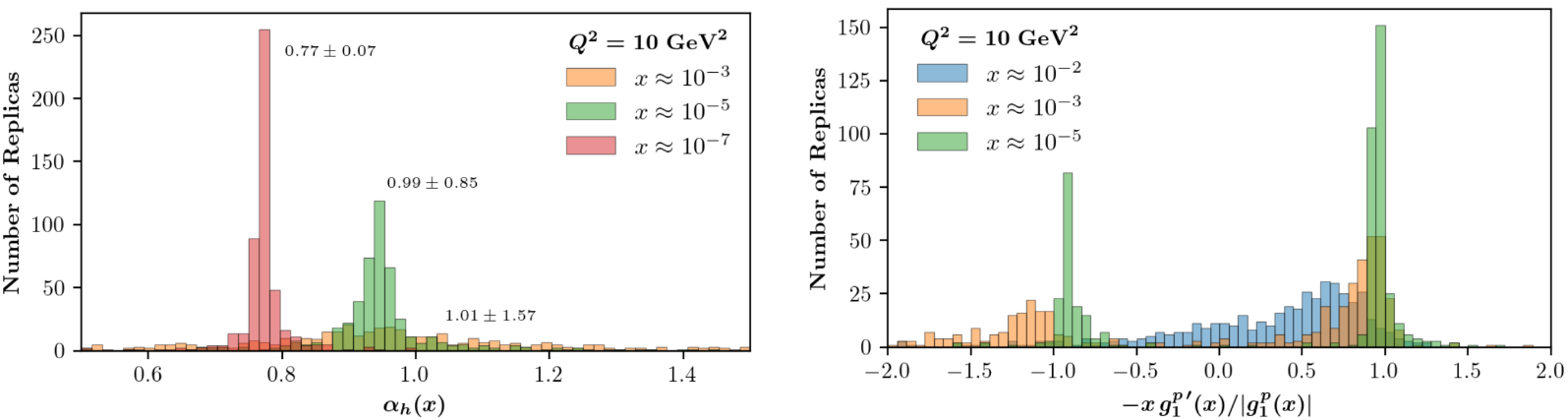}
		\caption{(Left) Histograms utilizing Eq.~\eqref{e:logderv1} showing that the intercept $\alpha_h(x)$ converges as $x$ decreases; this is a consequence of the small-$x$ evolution equations. (Right) Keeping the sign dependence by using Eq.~\eqref{e:logderv2} shows bimodal peaks of $\pm\alpha_h(x)$. At large $x$ there is no asymptotic behavior, and two refined peaks emerge for smaller values of $x$.
			\label{Bimodal_Curvature}
		}
	\end{centering}
\end{figure}
I direct the reader's attention to the right panel of Fig.~\ref{Bimodal_Curvature}. At $x\approx 10^{-2}$ we are quite far away from the asymptotic regime, and because of this there is a skew-normal distribution of the intercept (blue histogram). Probing at smaller values of $x$, however, shows the emergence of two distinct peaks of the $x$-dependent intercept that are positive and negative (yellow histogram), indicating the bimodality of the replicas' curvatures towards the small-$x$ asymptotics. As $x$ decreases further (green histogram), the two peaks become more refined as more replicas adopt their expected asymptotic behaviors of $\alpha_h$. The fact that the intercepts become distinct at $x\approx 10^{-3}$ implies that data sensitive to this curvature at $x$ as large as $x\approx 10^{-3}$, moderately below the current data range, would be enough to identify which bimodal peak $g_1^p$ belongs to. This is a much more optimistic picture than that painted by Fig.~\ref{Plot_g1_ambiguities}, which suggested that the bimodality could only be broken by data at $x\approx 10^{-5}$. Indeed this topic of breaking the bimodality is not only beneficial for identifying the asymptotic sign of $g_1^p(x,Q^2)$, but also the (flavor singlet and $C$-even) hPDFs. To understand why this is, I now need to determine the root cause of asymptotic bimodality, and thus the large small-$x$ uncertainties.

\subsubsection{\uline{Origins of Asymptotic Behavior}}\label{origin_asymp}

A replica's shape and asymptotics are subject to the initial conditions that created it; Fig.~\ref{fig:Deltaqp_basis} shows the $\Delta u^+(x,Q^2)$ hPDF as constructed from singular initial condition parameters, and the same can be done for the $g_1$ structure functions. As a proxy, I can still imagine that a given initial condition state (again, imagine a 24-element vector with one element for each parameter) simply takes each parameter value, multiplies its relevant basis hPDF, and then sums the contributions from each of those lines to produce the final state replica. I can then analyze the distribution of parameters for each replica and search for a pattern that may reveal itself. In the interest of studying the asymptotics, I separate the parameter distributions between replicas that are asymptotically positive and replicas that are asymptotically negative and search for a statistically significant difference between the two.

\begin{figure}[b!]
	\begin{centering}
		\includegraphics[width=0.9\textwidth]{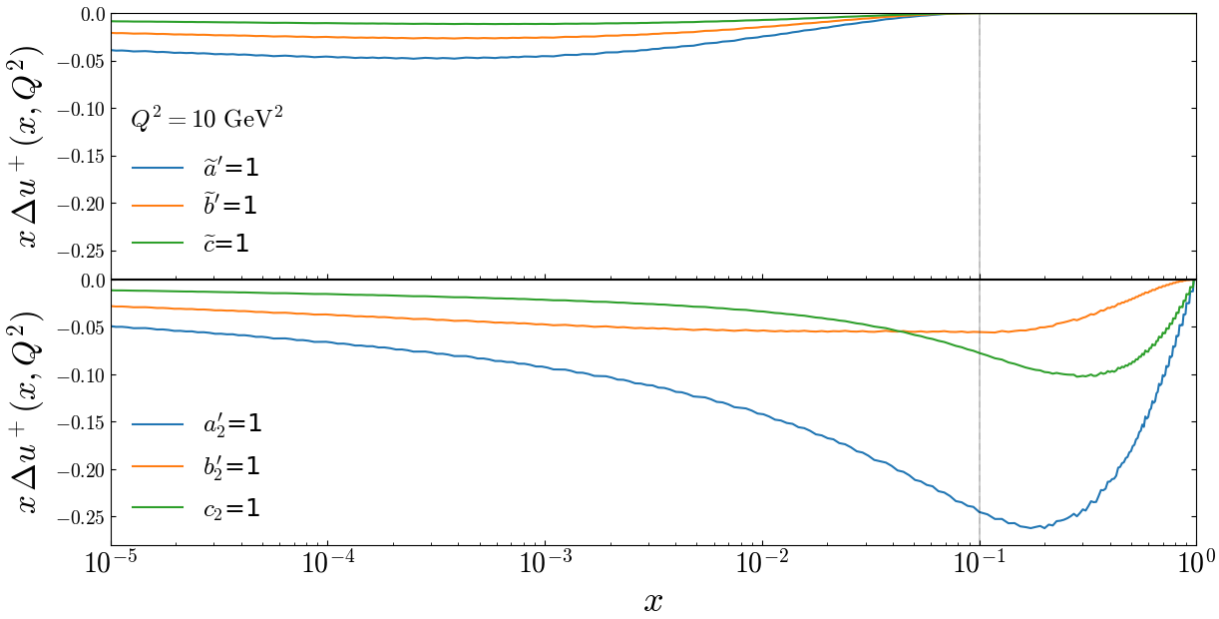}
		\caption{Basis functions analogous to those in Fig.~\ref{fig:Deltaqp_basis}, where instead of plotting the $\eta$, $s_{10}$, and $1$ contributions, I show the contributions of $\eta+s_{10}$, $\eta-s_{10}$, and $1$ displayed as the curves labeled $a'=1$, $b'=1$, and $c'=c=1$. Only the $\widetilde{G}$ and $G_2$ dipole amplitudes are shown here.
			\label{Primed_basisfunctions}
		}
	\end{centering}
\end{figure}

Before taking a look at the unaltered parameter distributions, remember that the $g_1$ structure function (and by extension the spin asymmetry) is only sensitive to the entire polarized dipole amplitude as opposed to individual parameters. See from Eq.~\eqref{all_hPDFs} that $g_1$ is strictly constructed from the evolved polarized dipole amplitudes $Q_q$ and $G_2$; rather than looking at individual parameters as given by Eq.~\eqref{IC_params}, I can reorganize the parameters in such a way that I get a more holistic view of the initial conditions and the overall sign that results from them. There is enhanced sensitivity the sign of $g_1^p$ through the linear combinations defined by $a' \equiv (a + b)/2$ and $b' \equiv (a-b)/2$, which reformats the initial conditions from the state $G^{(0)} = a\,\eta + b\,s_{10} + c$, to the state
\begin{align}
	G^{(0)} &= a' \, (\eta+s_{10}) + b' \, (\eta-s_{10}) + c \,.\label{e:param_transform}
\end{align}
In this way the $a'$ parameter is more sensitive to the asymptotic sign of the overall initial condition since the $\eta$ and $s_{10}$ components of the initial conditions always have the same sign, are the typically the two largest components, and are the two components grow as $x$ decreases. I can also use what we learned earlier about the small-$x$ behavior of the basis functions, namely that the most influential initial conditions are those that arise from $\widetilde{G}^{(0)}$ and $G_2^{(0)}$. I plot the $\Delta u^+(x,Q^2)$ basis functions that arise from the $\widetilde{G}$ and $G_2$ initial conditions as defined in Eq.~\eqref{e:param_transform} in Fig.~\ref{Primed_basisfunctions}.
As can be seen from Fig.~\ref{Primed_basisfunctions}, the $a'$ parameter has been more substantially separated from $b'$ and $c'$, meaning the entire initial condition is largely controlled by this parameter and that this parameter is more constrained by the data. After investigating the distribution of $a'$ parameters for all replicas I find that largest difference between the asymptotically positive/negative replicas is the constrained value of the $\widetilde{a}'$ parameter; asymptotically positive solution have an average value $\widetilde{a}' = -1.56\pm2.32$, while asymptotically negative parameters on average have a value of $\widetilde{a}' = 1.42 \pm 2.34$. This is the only parameter whose average values were more than one standard deviation apart from the other.

I also gain some insight on the sign-correlations from Fig.~\ref{Primed_basisfunctions}; $\widetilde{G}$ and $G_2$ basis functions are negative definite (for positive $a$, $b$, and $c$ parameters) due to the explicit minus sign in Eq.~\eqref{Deltaqp}. Additionally, as was noticed in Fig.~\ref{fig:Deltaqp_basis} and emphasized in Fig.~\ref{Primed_basisfunctions}, $\widetilde{G}$ and $G_2$ basis functions have the largest influence over the small-$x$ behavior; having said that, $\widetilde{G}$ only contributes at small $x$, meanwhile the large contributions of $G_2$ basis functions in the data-range of $x$ means that $G_2^{(0)}$ parameters are more constrained by the data than $\widetilde{G}^{(0)}$ parameters. The data prefers $G_2^{(0)}$ parameters that are negative, $a_2' = -0.98\pm1.00$, which can explain the overall preference for positive $g_1^p$ replicas, and the lack of constraining power the data has over $\widetilde{G}^{(0)}$ parameters explains why both asymptotically positive and negative solutions are equally valid.

\begin{figure}[t!]
	\begin{centering}
		\includegraphics[width=0.9\textwidth]{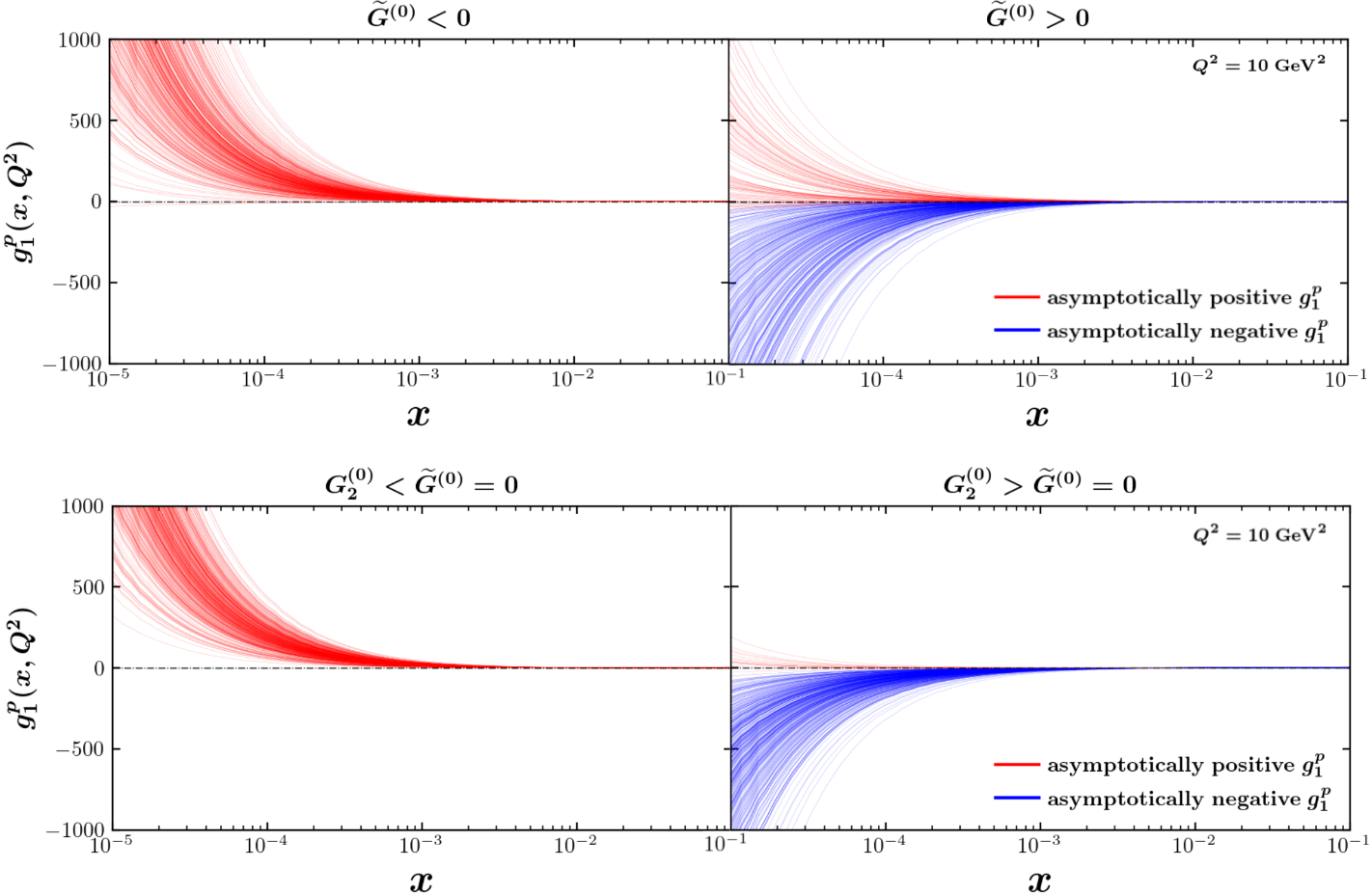}
		\caption{Comparing the effect that $\widetilde{G}$ and $G_2$ have on the overall sign of $g_1^p(x)$ at small $x$. Top row:~the priors are restricted so that (left) $\widetilde{G} \leq 0$  and (right) $\widetilde{G} \geq 0$. Bottom row:~the priors are restricted so that (left) $G_2 < 0$ and also that $\widetilde{G} = 0$, and (right) $G_2 > 0$ while $\widetilde{G} = 0$. All other parameters are fitted identically to the unaltered fit of Fig.~\ref{Plot_g1}. Controlling the sign of $\widetilde{G}^{(0)}$ strongly influences the sign of $g_1^p$, and that the sign of $G_2^{(0)}$ will also influence the sign of $g_1^p$.
			\label{GT_G2_Influence}
		}
	\end{centering}
\end{figure}

I test this hypothesis by altering the input of JAM framework to force $\widetilde{G}^{(0)}$ parameters to randomly sample in sign-specific bins and then fit to the same DIS and SIDIS data. When doing so I find that forcing the $\widetilde{G}^{(0)}$ parameters to be negative-definite resulted in $100\%$ of replicas growing asymptotically positive, while forcing $\widetilde{G}^{(0)}$ parameters to be positive-definite resulted in a $73\%$ majority of replicas to grow asymptotically negative -- note that this is the opposite of the unaltered fit displayed in Fig.~\ref{Plot_g1} which had $\sim70\%$ of replicas growing asymptotically positive. To further verify that $G_2^{(0)}$ parameters also have an influential effect on the asymptotics I repeated the test above but for sign-definite $G_2^{(0)}$ parameters; to control for the influence of $\widetilde{G}^{(0)}$ parameters I set them to zero. The results of these fits are all shown in Fig.~\ref{GT_G2_Influence}.

In truth, the control that $\widetilde{G}$ has on the small-$x$ uncertainty could have been predicted from Eqs.~\eqref{all_hPDFs} and \eqref{asymptotics}, along with the basis function seen in Fig.~\ref{fig:Deltaqp_basis}. Since there is no direct dependence of $\widetilde{G}$ in any of the structure functions (and by extension the spin asymmetries), it only contributes to the basis functions through evolution, and as result has very little contributions in the range of $x$ where data has constraining power. This inherently means that the $\widetilde{G}^{(0)}$ parameters would be largely unconstrained, and thus be a source of theoretical uncertainty. I only know \textit{a posteriori} through solving the evolution equations and plotting the basis functions in Fig.~\ref{fig:Deltaqp_basis} that these unconstrained parameters would also be those that have a very large influence as $x$ approaches zero. Because the data is sensitive to all other initial conditions, this means that the small-$x$ asymptotics are dominated by the contributions and overall sign of the $\widetilde{G}$ initial condition; through evolution even $\widetilde{G}$ initial conditions will result in non-zero values for all other evolved dipole amplitudes. In effect this means that I can compare Eqs.~\eqref{g_1}, \eqref{Deltaqm}, and \eqref{DeltaG} and draw the asymptotic correlation 
\begin{align}\label{e:hpdf_correlation}
	\lim_{x\to 0} g_1^p(x,Q^2) \propto \lim_{x\to 0}\Delta q^+(x,Q^2) &\sim -(Q_q + 2G_2) \to -\widetilde{G} \\
	\lim_{x\to0}\Delta G(x,Q^2) &\sim G_2 \to \widetilde{G} \notag
\end{align}
since through evolution the dipoles will be dominated by $\widetilde{G}$. 
\begin{figure}[t!]
	\begin{centering}
		\includegraphics[width=\textwidth]{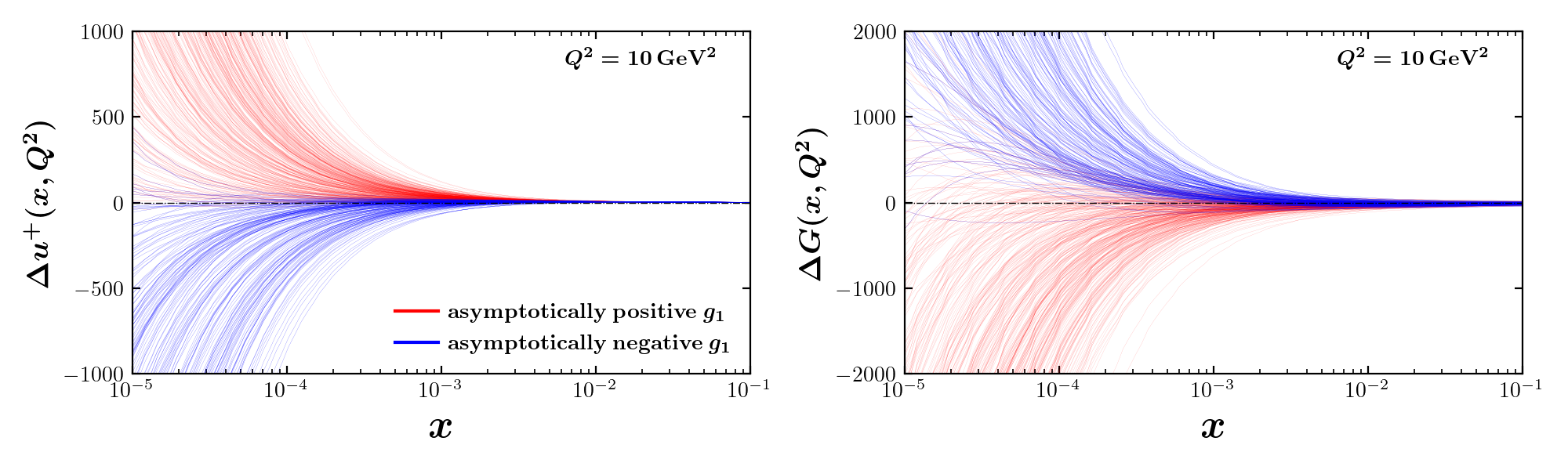}
		\caption{Color coding the hPDF replicas according to the asymptotic sign of $g_1^p$ shows that there is a novel correlation \eqref{e:hpdf_correlation}:~at small $x$, quark hPDFs (left) have the same sign as $g_1^p$  (only $\Delta u^+$ is shown) while the gluon hPDF (right) has the opposite sign as $g_1^p$.
			\label{hPDF_Correlation}
		}
	\end{centering}
\end{figure}
At small $x$ a $\widetilde{G}$-dominated $\Delta q^+$ and a $\widetilde{G}$-dominated $g_1^p$ will have the same sign as each other and an opposite sign as a $\widetilde{G}$-dominated $\Delta G$ (again, see the relative minus signs between Eqs.~\eqref{g_1}, \eqref{Deltaqm}, and \eqref{DeltaG}). Using the same replicas from Fig.~\ref{Plot_g1}, and keeping the same color-coding as determined by the asymptotic sign of $g_1^p$, I plot the replicas for $\Delta u^+$ (though this can be any choice of $\Delta q^+$) and $\Delta G$ in Fig.~\ref{hPDF_Correlation}. In agreement with Eq.~\eqref{e:hpdf_correlation}, all the red/blue replicas grow positive/negative for $\Delta u^+$ while the same red/blue replicas grow asymptotically negative/positive for $\Delta G$. The fact that the small-$x$ asymptotics of $g_1^p$, $\Delta q^+$, and $\Delta G$ are correlated is an important, novel prediction of the small-$x$ helicity evolution framework.\footnote{I note that no such relationship is exhibited by the nonsinglet hPDFs.  When attempting the same strategy of color coding the nonsinglet hPDFs (not shown) according to the asymptotic sign of the proton SIDIS structure function $g_1^{p\to h}$, no correlations could be identified.}$^,\!$
\footnote{In Ref.~\cite{Borsa:2020lsz} a connection was found at small $x$ between $\Delta G(x,Q^2)$ and the $\log Q^2$ derivative of $g_1 (x,Q^2)$: $\Delta G (x, Q^2) \approx - \partial g_1 (x,Q^2) /\partial \ln Q^2$. My result, however, demonstrates anti-correlation of the signs of $\Delta G(x,Q^2)$ and $g_1^p(x,Q^2)$ (and not of the $\log Q^2$ derivative of $g_1^p(x,Q^2)$). In addition, the calculation in Ref.~\cite{Borsa:2020lsz} was in a DGLAP-based NLO perturbative QCD framework, while my calculation involves the all-order DLA-resummed coefficient functions (see the discussion around Eq.~\eqref{g1_coef_ftns}).} This leads to a clear mission forward: find a way to better constrain the polarized dipole amplitude $\widetilde{G}$. One way to do this is through data from the future EIC, with several more options outlined in \textbf{Chapter}~\ref{pheno1_more_testing}; one of which is to use polarized proton-proton collisions, which is the subject of \textbf{Chapter}~\ref{pheno_2}.

\subsection{$\boldsymbol{\mathrm{Results~III\!:~Everything~hPDFs}}$}
\begin{figure}[t!]
	\begin{centering}
		\includegraphics[width=\textwidth]{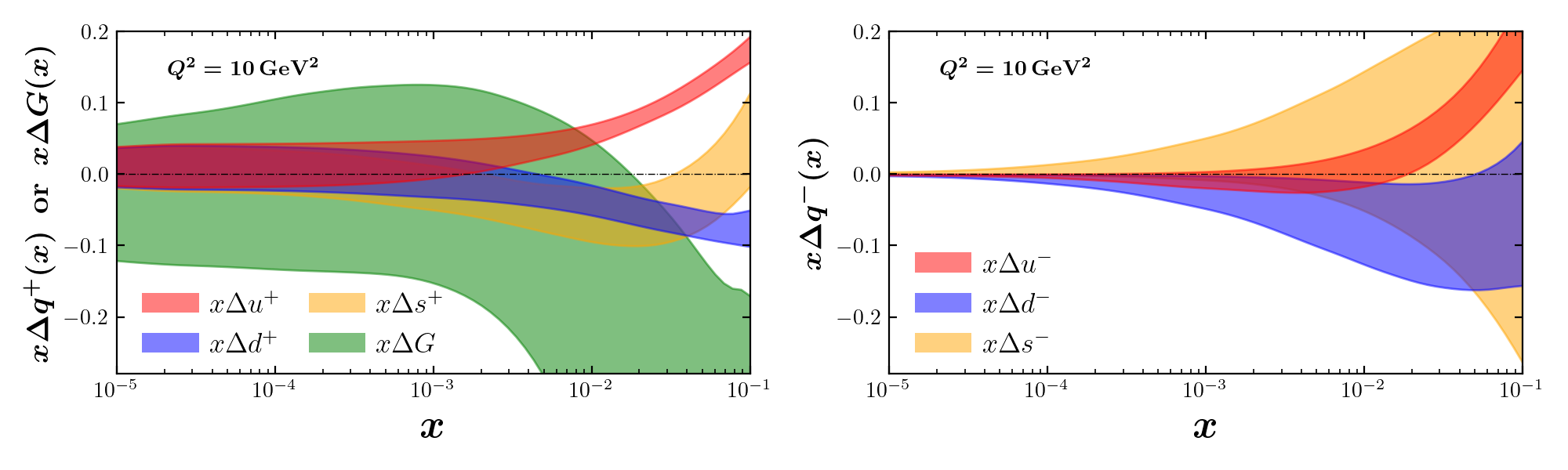}
		\caption{(Left) $C$-even hPDFs $x\Delta u^+$ (red), $x\Delta d^+$ (blue), $x\Delta s^+$ (orange) and $x\Delta G$ (green) extracted from low-$x$ data.
			(Right) Same as left panel but for the flavor nonsinglet $C$-odd hPDFs $x\Delta u^-$ (red), $x\Delta d^-$ (blue), $x\Delta s^-$~(orange). All hPDFs are scaled by $x$ to better observe their small-$x$ behavior.
			\label{hPDF_bands}
		}
	\end{centering}
\end{figure}

\begin{figure}[t!]
	\begin{centering}
		\includegraphics[width=0.7 \textwidth]
		{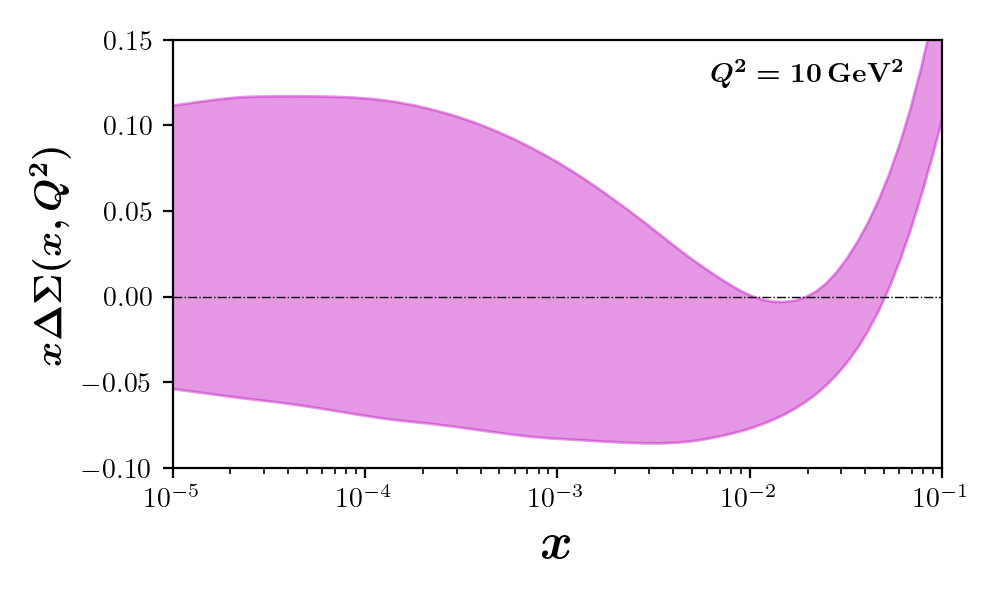}
		\vspace{-0.3cm}
		\caption{Quark flavor singlet helicity distribution $x\Delta \Sigma(x,Q^2 = 10~\mathrm{GeV}^2)$ calculated from hPDFs extracted from low-$x$ data. 
			\label{xDelta_Sigma_plot}
		}
	\end{centering}
\end{figure}
From the same replicas that provided the $g_1$ structure function of the proton in Fig.~\ref{Plot_g1}, I can also reconstruct the hPDFs. Show in the left panel of Fig.~\ref{hPDF_bands} are the $c$-even quark hPDFs, $\Delta u^+$, $\Delta d^+$, and $\Delta s^+$, along with the gluon hPDF $\Delta G$; the right panel of this same figure shows the extractions of the flavor nonsinglet $C$-odd hPDFs $\Delta u^-$, $\Delta d^-$, and $\Delta s^-$. Additionally, I plot the flavor sum of quark hPDFs $\Delta\Sigma(x,Q^2)$ in Fig.~\ref{xDelta_Sigma_plot}. As with the structure function above, the bands represent a $1\sigma$ spread of the replicas, and to better observe the small-$x$ behavior the hPDFs are scaled by $x$; the latter point is the reason why the small-$x$ uncertainties appear much narrower than that in Fig.~\ref{Plot_g1} which is not scaled by $x$. Since evolution begins at $x_0 = 0.1$, the analysis is restricted to below that region in $x$; all extractions were similarly performed at $Q^2 = 10~\mathrm{GeV}^2$. Just as observed in $g_1^p$, the uncertainty bands of the hPDFs become quite large at small $x$ ($x\lesssim10^{-3}$) and span zero, which is to be expected from the correlation established by Eq.~\eqref{e:hpdf_correlation}. Also supported by the discussion in \textbf{Chapter}~\ref{origin_asymp} is the observation that the gluon hPDF $\Delta G$ has the largest uncertainty at small $x$; while $G_2$ can be constrained by the data, it is only ever constrained with $Q_q$ in the specific combination ($Q_q + 2G_2$). $\Delta q^+$ extractions are more explicit connected to the structure function $g_1^p$, and are thus better constrained by spin asymmetry data, through Eq.~\eqref{g1_LO}. This is supported by the fact that the smaller-$x$ region ($x \lesssim 10^{-4}$) sees $\Delta u^+$, $\Delta d^+$, and $\Delta s^+$ exhibit the same approximate error band, while in the larger-$x$ region where data is more constraining and evolution is dominated by the $Q_q$ dipoles, there is flavor separation of the three quark flavors. $\Delta s^+$ has slightly larger uncertainties at large $x$ likely due to the limited SIDIS kaon data; The similar small-$x$ error bands for $\Delta q^+$ distributions are in contrast to the more distinct error bands of $\Delta u^-$, $\Delta d^-$, and $\Delta s^-$ in the right panel of Fig.~\ref{hPDF_bands}. Remember that the flavor nonsinglet hPDFs are driven by the separate flavor nonsinglet evolution \eqref{nonsinglet_evo} which is sensitive to flavor separation through SIDIS data. The different evolution equations see a faster convergence of $x\,\Delta q^-$ towards zero since it has a smaller intercept $\alpha_h$ (explored more in \textbf{Chapter}~\ref{pheno1_apdx}). Since $G_2$ has a larger small-$x$ influence than $Q_q$, the similarity between small-$x$ error bands of $\Delta q^+$ can be reflective of the uncertainty of $G_2$ (and of course $\widetilde{G}$ emphasizes this effect due to its influence over the bimodality).

One important feature of the hPDFs in Fig.~\ref{hPDF_bands} is the rather large magnitudes of $\Delta s^+$ and $\Delta G$ compared to recent JAM-DGLAP analyses~\cite{Ethier:2017zbq, Zhou:2022wzm, Cocuzza:2022jye}. Fig. 6 of Ref.~\cite{Zhou:2022wzm} exhibits a $\Delta s^+$ that is consistent with zero across the entire range $5\times 10^{-3} \leq x \leq 0.9$, whereas my analysis predicts the one standard deviation spread of $\Delta s^+$ to be definitively negative at $x\approx 10^{-2}$. In truth the JAM-DGLAP approach has some fundamental differences, including the use of DGLAP evolution within collinear factorization, a larger set of data across the full range of $x$, and even some fits with imposed SU(2) or SU(3) flavor symmetries, however it is still a valuable cross-check to make sure the JAMsmallx analyses can be consistent with JAM-DGLAP. In this spirit I repeated a fit to the same DIS and SIDIS, while strictly imposing $\Delta s^+ = \Delta s^- = 0$, and still achieved a good fit to data with $\chi^2_{\mathrm{red}} = 1.04$, compared with the statistically similar $\chi^2_{\mathrm{red}} = 1.03$ of the original fit; this slight reduction in the goodness-of-fit comes from a worse fit to the 26 $A_1^h$ data points from tagged kaon SIDIS (out of the total 226 data points), whose data-specific $\chi^2$ reduced from $\chi^2_{\mathrm{red}} = 0.81$ to $\chi^2_{\mathrm{red}} = 1.05$ in the zero-strangeness fit. I can conclude that, since the overall goodness-of-fit was largely unaffected, that the JAM-DGLAP and JAMsmallx analysis are consistent with each other, and that there is a real (but weak) preference for nonzero $\Delta s^+$ at $x\sim0.001$ within the small-$x$ framework.

With the hPDFs all examined, I can finally address the contribution to the proton spin and the axial-vector charges from the small-$x$ regime. The net parton spin contribution amounts to taking the $x$-integral of $\Delta\Sigma(x,Q^2)$ from Fig.~\ref{xDelta_Sigma_plot} and the $x$-integral of $\Delta G(x,Q^2)$ from Fig.~\ref{hPDF_bands}, while the triplet $g_A$ and octet $a_8$ axial vector charges are integrals over specific combinations of quark hPDFs (detailed below). Focusing on the small-$x$ region of interest, $10^{-5} \leq x \leq 0.1$, I can compute the following truncated integrals:
\begin{subequations}\label{SG_xmax}
	\begin{align}
		\left(\frac{1}{2}\Delta \Sigma+\Delta G\right)_{\![x_{\rm max(min)}]} \! & \equiv \int\limits_{x_1}^{x_2} d x \, \left(\frac{1}{2}\Delta \Sigma+\Delta G\right) \!(x, Q^2)\,,\\
		g_{A[x_{\rm max(min)}]}\! \equiv \int\limits_{x_1}^{x_2} d x \,  g_A(x,Q^2)&\equiv\int\limits_{x_1}^{x_2} d x \, \left[ \Delta u^+ (x, Q^2) - \Delta d^+ (x, Q^2) \right], \\
		a_{8[x_{\rm max(min)}]}\! \equiv \int\limits_{x_1}^{x_2} d x \, a_8(x,Q^2)\equiv\int\limits_{x_1}^{x_2} d x \,& \left[ \Delta u^+ (x, Q^2) + \Delta d^+ (x, Q^2)  - 2 \, \Delta s^+ (x, Q^2) \right]    \: .
	\end{align}
\end{subequations}
The integration limits are left arbitrary so I can compute the truncated integrals in two distinct ways: as a function of $x_{\mathrm{min}}$ such that the integral can be computed dynamically over the range $x_{\mathrm{min}} \leq x \leq 0.1$, and as a function of $x_{\mathrm{max}}$ such that the integral is computed over the range $10^{-5} \leq x \leq x_{\mathrm{min}}$. The $Q^2$ arguments on the left-hand side of Eq.~\eqref{SG_xmax} have been dropped for brevity, but as established above $Q^2 = 10~\mathrm{GeV}^2$. 

\begin{figure}[t!]
	\begin{centering}
		\includegraphics[width=0.9 \textwidth]{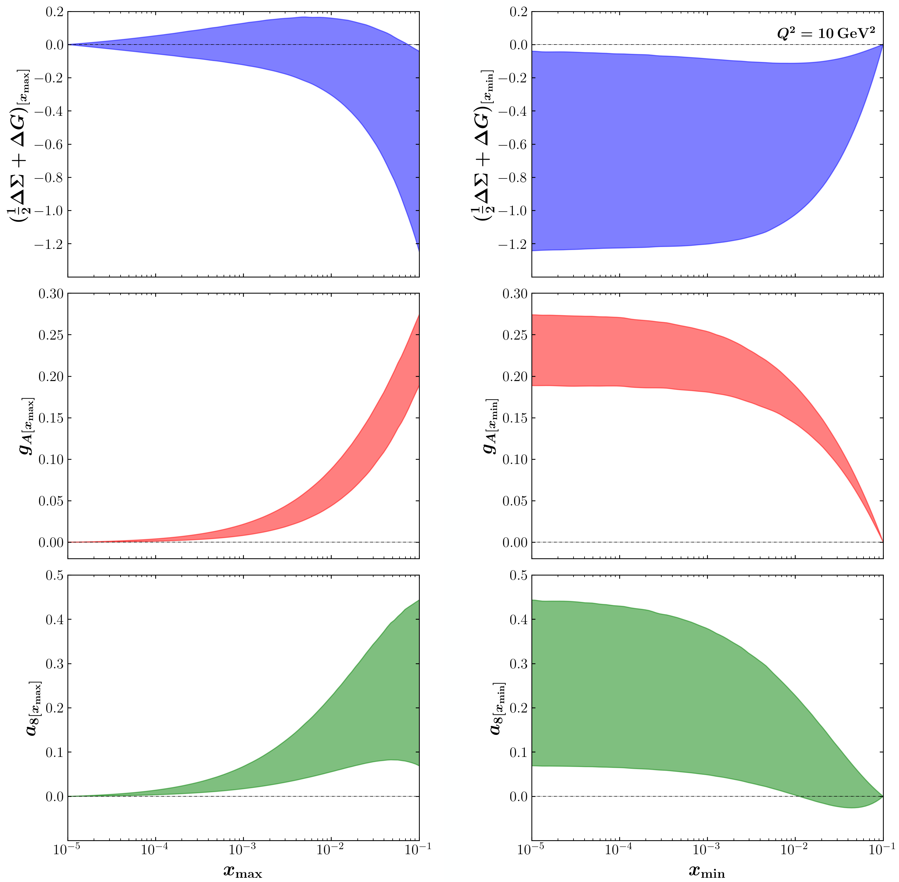}
		\caption{
			Truncated moments of $(\tfrac{1}{2}\Delta\Sigma+\Delta G)(x,Q^2)$, $g_A(x,Q^2)$ and $a_8(x,Q^2)$, defined in Eqs.~(\ref{SG_xmax}), versus $x_{\rm max}$ (left) and $x_{\rm min}$ (right) at $Q^2=10~\mathrm{GeV}^2$. 
			\label{integrated_plots}
		}
	\end{centering}
\end{figure}

The truncated moments, computed either as a function of $x_{\mathrm{min}}$ or as a function of $x_{\mathrm{max}}$ are given in Fig.~\ref{integrated_plots}. From these plots the net parton spin contribution $(\tfrac{1}{2}\Delta\Sigma + \Delta G)$ can be quite large, despite the sizable uncertainties. Since any value within the bands is probable within $1\sigma$ accuracy, this implies that it is possible for small-$x$ partons to contribute more spin than their large-$x$ counterparts. From close examination of the error bands of $\Delta\Sigma(x,Q^2)$ and $\Delta G(x,Q^2)$, the error bands of the truncated integrals are not simply the sum of the two error bands, implying that the uncertainties are correlated -- this is something known from Eq.~\eqref{e:hpdf_correlation}. This correlation also results in a good deal of cancellation between the replicas, which skews the truncated moment $(\tfrac{1}{2}\Delta\Sigma + \Delta G)$ towards more negative values. Additionally one can see that the slopes of the outer boundaries of the truncated helicity band are not completely flat, but are actually growing and have not yet fully saturated at that point in $x$; in contrast the bands for the truncated moments of $g_A$ and $a_8$ have saturated near $x = 10^{-4}$, giving finite and non-negligible contributions from small-$x$ partons. These results are quite striking: small-$x$ partons seem to contribute \textbf{negatively} to the spin of the proton, even when accounting for the full standard deviation of replicas. This would imply that large amounts of parton spin must come from the orbital angular momentum contributions to satisfy the Jaffe-Manohar sum rule \eqref{spin_sum}. This conclusion is supported by similar observations of $g_1^p$ made using AdS/CFT~\cite{Hatta:2009ra,Kovensky:2018xxa,Jorrin:2022lua,Borsa:2023tqr}. I also predict approximately $15\text{-}21\%$ of the known value of $g_A$ and $12\text{-}77\%$ of the known $a_8$ values are generated from small-$x$ partons; these are percentages based on the full range moments $x\in[0,1]$ of the known neutron and hyperon $\beta$-decay values of $g_A = 1.269(3)$ and $a_8 = 0.586(31)$ \cite{Jimenez-Delgado:2013boa}.

The reader is cautioned, however, because this small-$x$ analysis is at the will of the large-$x$ initial conditions of the evolution equations and the error bands shown are strictly statistical in nature. They are an accurate representation of the uncertainty coming from experimental data and from the Monte Carlo sampling procedure, but they do not reflect the systematic bias that results from omitting large-$x$ data that cannot be captured by the small-$x$ formalism. A combined, thorough analysis the merges the small-$x$ formalism with a large-$x$ compatible formalism like JAM-DGLAP ~\cite{Ethier:2017zbq, Zhou:2022wzm, Cocuzza:2022jye} that brings in external input from large-$x$ can therefore, potentially, have a large and systematic effect on the small-$x$ extracted hPDFs beyond their current $1\sigma$ uncertainties; this indeed is crucial for determining the total proton spin budget across the entire range of $x$. Detailed in \textbf{Chapter}~\ref{pheno1_more_testing} is a preliminary matching procedure that provides a glimpse of the effect that such large-$x$ matching can have on small-$x$ extractions. A comprehensive implementation of true large-$x$ matching will be an important (and inevitable) aspect of future analyses. With that vital caveat emphasized, I can quantify the results of my small-$x$ truncated moments for the total parton helicity and the axial-vector charges, integrated over the window $x\in[10^{-5},0.1]$ for $Q^2 = 10~\mathrm{GeV}^2$:
\begin{subequations}    
	\begin{align}
		\label{e:moment_sum}
		&\int\limits_{10^{-5}}^{0.1} d x \, \left(\frac{1}{2}\Delta \Sigma+\Delta G\right) \!(x) = -0.64 \pm 0.60\,, \\[0.5cm]
		&\qquad\qquad \int\limits_{10^{-5}}^{0.1} d x \,g_A(x) = 0.23 \pm 0.04\,, \\[0.5cm]
		&\qquad\qquad \int\limits_{10^{-5}}^{0.1} d x \,a_8(x) = 0.26 \pm 0.19\,. 
	\end{align}
\end{subequations}  
%


\subsection{$\boldsymbol{\mathrm{Further~Testing\!:~EIC~Impact~and~Other~Constraints}}$}\label{pheno1_more_testing}

To study the impact that lower $x$ measurements will have on my ability to predict the behavior of the hPDFs and $g_1^p$ at even smaller values of $x$, I turn towards the future EIC. Though not yet constructed, it is known that the EIC will be able to provide polarized scattering data over the kinematic range $10^{-4} < x < 0.1$ and $1.69 ~\mathrm{GeV}^2 < Q^2 < 50~\mathrm{GeV}^2$. In truth, it will be able to go lower in $x$, but this comes at the expense of going higher in $Q^2$ and this formalism is not expected to be applicable for arbitrarily large-$Q^2$ (DGLAP evolution, however, can describe the $Q^2$ dependence). The EIC yellow report \cite{AbdulKhalek:2021gbh} reveals that will be able to provide data for DIS on the proton with center-of-mass energies $\sqrt{s} = \{29,~45,~63,~141\}~\mathrm{GeV}$ with integrated luminosity $100~\mathrm{fb}^{-1}$, data for DIS on deuteron or ${}^3$He beams with $\sqrt{s} = \{29,~66,~89\}~\mathrm{GeV}$ with integrated luminosity $10~\mathrm{fb}^{-1}$, and data for SIDIS on a proton with $\sqrt{s} = 141~\mathrm{GeV}$. It can do all this with $2\%$ point-to-point uncorrelated systematic uncertainties. From this information I can generate EIC pseudodata by taking the theoretical asymmetries and applying an EIC ``smearing" such that I can create new data points with EIC-like error bars in the kinematic range where the EIC will truly provide data.

\begin{figure}[b!]
	\begin{centering}
		\includegraphics[width=0.7\textwidth]
		{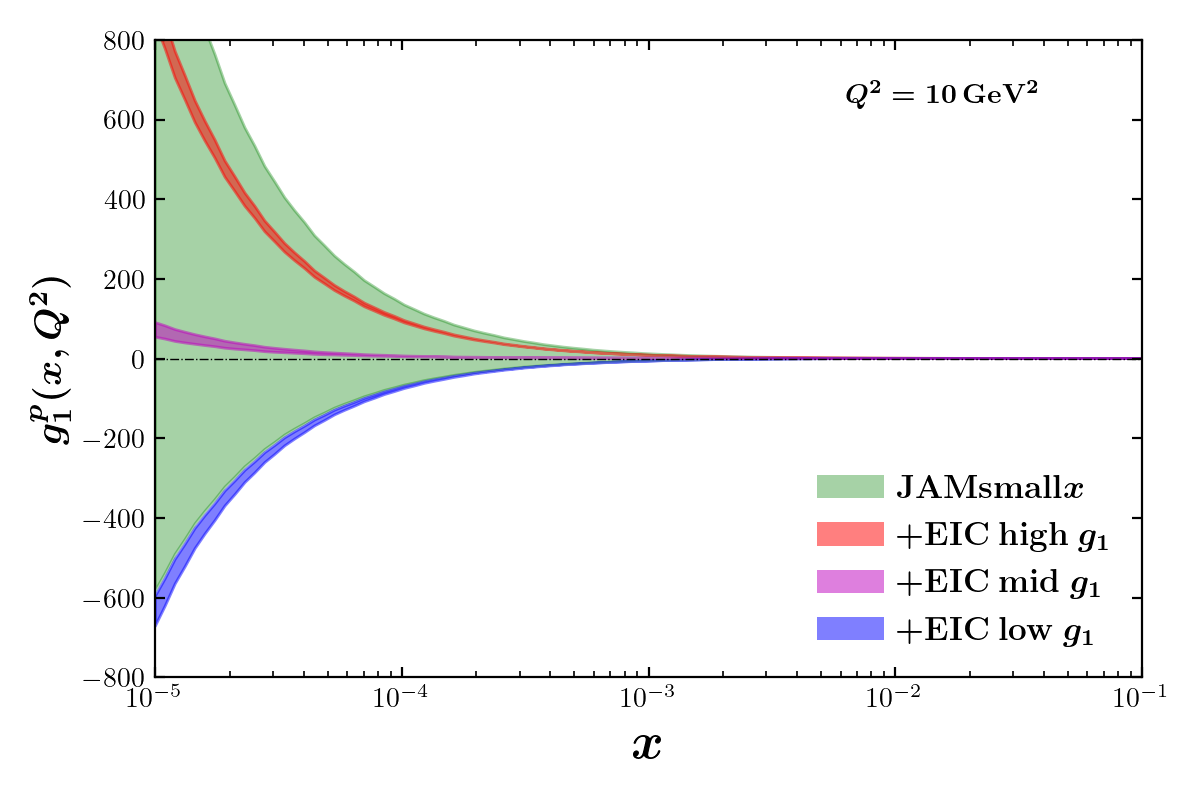}
		\vspace{-0.3cm}
		\caption{Extraction of $g_1^p$ from the current low-$x$ experimental DIS and SIDIS data (green) and with EIC pseudodata generated from the mean of the asymptotically positive $g_1^p$ replicas (red), mean of the asymptotically negative $g_1^p$ replicas (blue), and mean of low-magnitude $|g_1^p(x,Q^2)| < 100 $ at $x=10^{-4}$ replicas (magenta).
			\label{Plot_g1_EIC_impact}
		}
	\end{centering}
\end{figure}

\begin{figure}[b!]
	\begin{centering}
		\includegraphics[width=0.65\textwidth]
		{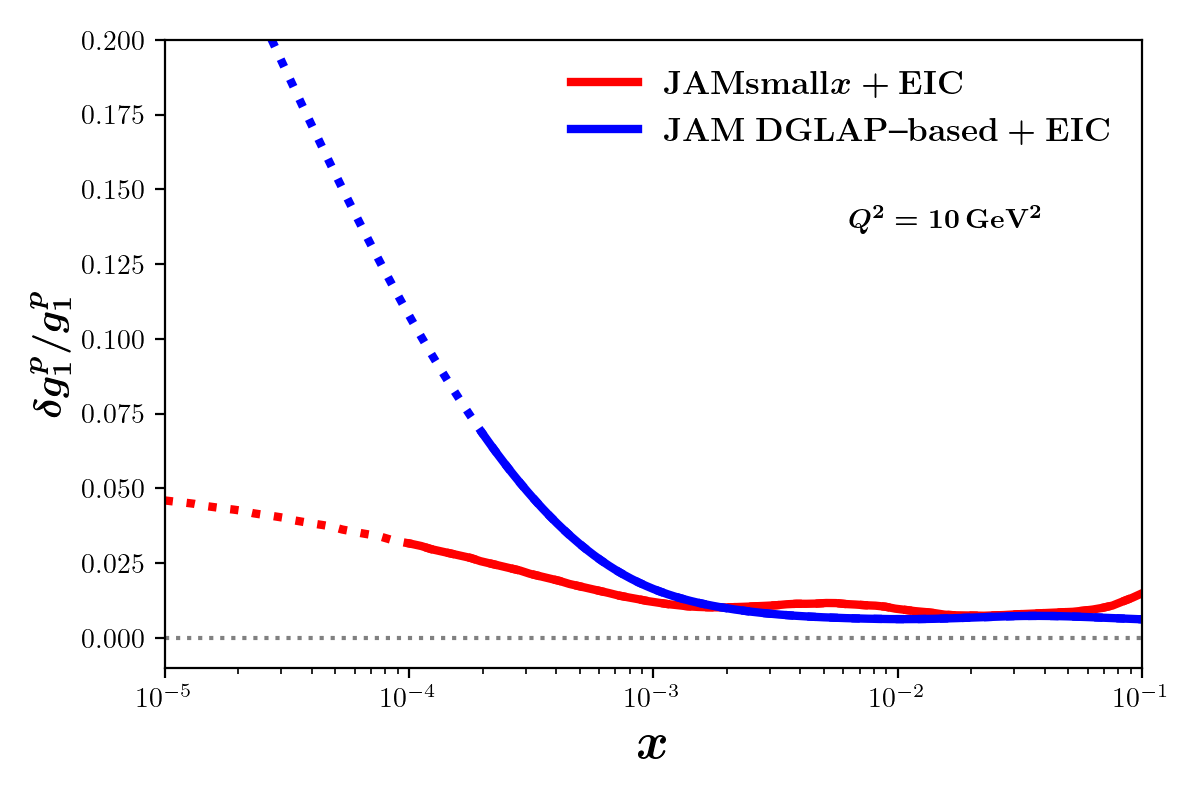}
		\vspace{-0.3cm}
		\caption{Relative uncertainty for my JAMsmallx analysis (red) and a JAM DGLAP-based extraction~\cite{Zhou:2021llj} (blue) with EIC pseudodata generated in the high $g_1^p$ scenario. Dotted lines denote extrapolating beyond lowest $x$ for which pseudodata was generated. For this JAMsmallx pseudodata was generated down to $x=10^{-4}$. For the JAM DGLAP-based fit pseudodata was generated down to $x=2\times 10^{-4}$~\cite{Zhou:2021llj}. 
			\label{Plot_g1_uncertainty}
		}
	\end{centering}
\end{figure}

My analysis of $g_1^p$ makes three logical choices to generate EIC pseudodata from: 1) the mean of all replicas of $g_1^p$ that have a relatively subdued magnitude of $|g_1^p| < 100$ at $x = 10^{-4}$ (``mid $g_1$"), 2) the mean of asymptotically positive replicas (``high" $g_1$), and 3) the mean of asymptotically negative replicas (``low" $g_1$). The second and third options have clear advantages, but I clarify that the first option is restricted to low-magnitude $g_1^p$ replicas in the spirit of taking a conservative approach in the hypothesis that EIC data may be able to distinguish between the two asymptotic solutions. From these three choices I generate three sets of EIC pseudodata and subsequently perform three new analyses: a simultaneous fit to DIS, SIDIS, and EIC mid, high, or low pseudodata. The results of these new fits, and the original JAMsmallx fit, are shown in Fig.~\ref{Plot_g1_EIC_impact}. The inclusion of smaller-$x$ EIC data will evidently have a dramatic effect on the small-$x$ uncertainties, reducing them greatly in the extended data range $10^{-4} < x < 0.1$, and keeping these uncertainties small even far into the asymptotic range of $x$. Additionally, as predicted from Fig.~\ref{Bimodal_Curvature}, these smaller-$x$ EIC pseudodata points are also informative enough to break the asymptotic bimodality and predict explicitly either positive or negative $g_1^p$, even when the pseudodata was generated from theoretical asymmetries that are near zero (``mid $g_1$" scenario). I can compare the relative uncertainty (the ratio of uncertainty to magnitude) to the relative uncertainty of a JAM-DGLAP analysis that also used EIC pseudodata \cite{Zhou:2021llj} and find that in the extrapolation region beyond where EIC pseudodata was generated ($x=10^{-4}$ for JAMsmallx, $x = 2\times 10^{-4}$ for JAM-DGLAP) that JAM-DGLAP uncertainties are uncontrolled, whereas the small-$x$ helicity evolution equations show their predictive power by keeping the uncertainties small even as $x$ decreases well below the data range. Indeed, these DGLAP-based extractions are not expected to perform well outside the range of $x$ where data is constraining, but it is a worthwhile comparison to confirm the genuine predictability of small-$x$ helicity evolution.

The future EIC is not the only avenue to help constrain the parameters and small-$x$ uncertainties; in fact some options are more readily available. The hPDF that is least constrained is the gluon hPDF $\Delta G(x,Q^2)$, so as part of this analysis I explored some options to help constrain it. The first of which is the positivity constraint, which is the statement that the number densities for positive and negative helicity partons must be positive; this is enforced by the constrained $|\Delta G(x,Q^2) < G(x,Q^2)$ where $G(x,Q^2)$ is the unpolarized gluon PDF\footnote{It is not clear whether the positivity constraint must strictly be satisfied under $\overline{\mathrm{MS}}$ renormalization, but I set that argument aside for this analysis ~\cite{Candido:2020yat,Collins:2021vke,Candido:2023ujx}.}. I can impose this constraint as part of the JAM ``input" discussed in \textbf{Chapter}~\ref{chapter_3}; this is done by refitting to the same DIS and SIDIS data as usual, but this time imposing a $\chi^2$ punishment on replicas that generate a gluon hPDF who's magnitude is greater than the unpolarized PDF at all values of $x < x_0$. This effectively gives a skewed preference to the least-squares $\chi^2$ minimization procedure, ultimately resulting in more final state replicas that are in agreement with the positivity constraint. In practice I find that enforcing this constraint has negligible effect on the fits since the JAMsmallx $\Delta G(x,Q^2)$ extractions only break the positive constraints at the small $x$ interface $x\approx 0.1$ and are already of comparable size to the JAM-DGLAP $G(x,Q^2)$ extractions~\cite{Cocuzza:2022jye,Zhou:2022wzm} by the time evolution begins. After evolution the unpolarized PDF grows faster than the polarized PDF anyway, which is expected since the former is eikonal and the latter is sub-eikonal (suppressed by a power of $x$).

\begin{figure}[b!]
	\begin{centering}
		\includegraphics[width=0.9\textwidth]
		{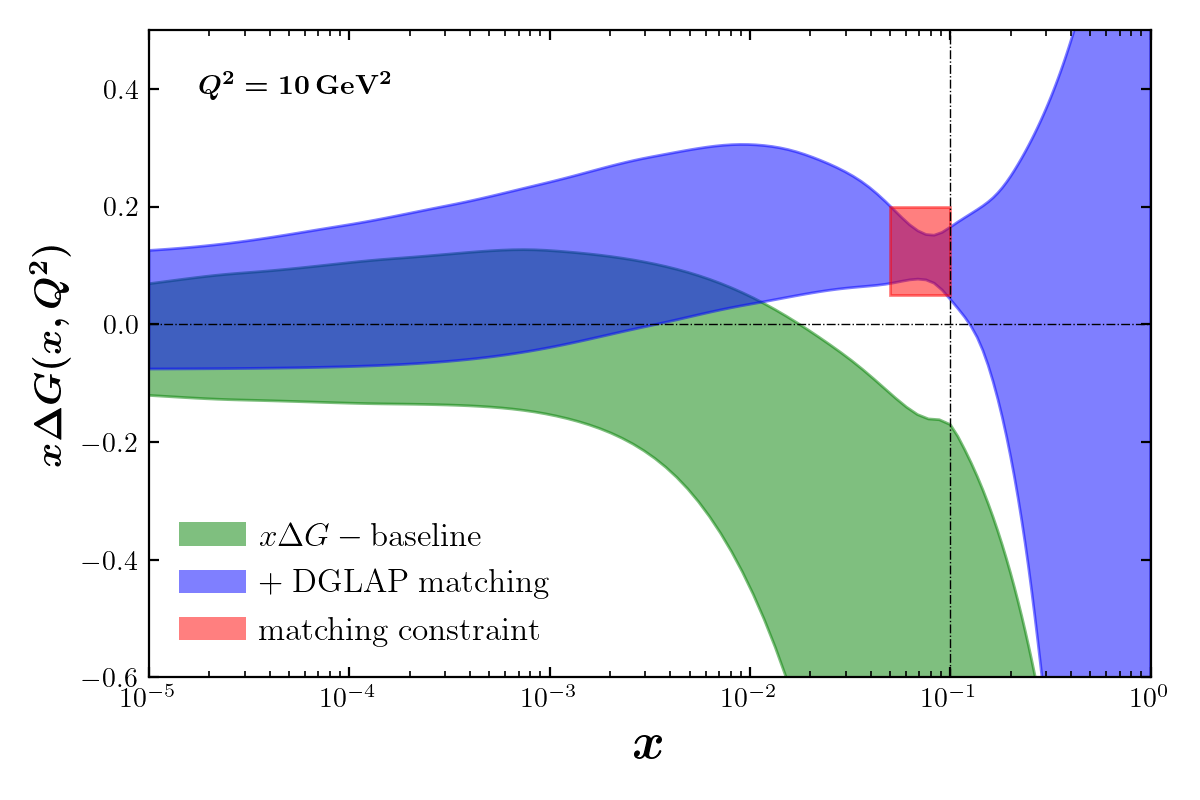}
		\caption{The result of matching onto the $\Delta G(x,Q^2)$ extraction from DGLAP \cite{Zhou:2022wzm,Cocuzza:2022jye} at $x\lesssim0.1$. The green band is my baseline fit. The blue band is the result of matching. The light red square is the region where I enforce matching. 
			\label{Plot: Delta G matching}
		}
	\end{centering}
\end{figure}

A more interesting approach to better constrain $\Delta G(x,Q^2)$ and the initial condition parameters is to perform large-$x$ matching. In truth, a full large-to-small $x$ analysis would require a rigorous formalism, but in the meantime I can attempt a preliminary matching procedure that uses information from JAM-DGLAP fits at the small $x$ boundary of $x\lesssim0.1$. I achieve this matching by loosely quantifying the JAM-DGLAP extractions of $\Delta G(x,Q^2)$ from Refs.~\cite{Zhou:2022wzm,Cocuzza:2022jye}, specifically in their ``SU(3) + positivity scenario". I employ a similar $\chi^2$ punishment as described above, but this time replicas that probe outside the general magnitude of JAM-DGLAP's $\Delta G(x,Q^2)$ are punished; the matching region is defined by a box with boundaries $10^{-1.3} < x < 0.1$ and $0.05 < \Delta G(x,Q^2) < 0.17$. In Fig.~\ref{Plot: Delta G matching}
I show the JAMsmallx baseline fit (green) along with the resultant fit with JAM-DGLAP matching (blue); the matching constraint is depicted by the red box. The motivation is that with proper matching the small-$x$ extractions should be consistent both with the JAM-DGLAP analyses in this region. What I find is that the $\Delta G(x,Q^2)$ extraction are, unexpectedly, more positive at $x \lesssim 0.1$, but also that the extractions are now shifted to be more positive even for smaller values of $x < 10^{-3}$. Note that while the baseline small-$x$ fit was largely negative for $x \to 0.1$, this is only the $1\sigma$ uncertainty band; there still exists a large number of (good fitting) replicas that were positive in this region, so the matching constraint effectively gives a selection preference to those replicas. Remembering the correlation \eqref{e:hpdf_correlation}, it should be no surprise that the now positive $\Delta G(x,Q^2)$ preference has resulted in the $g_1^p(x,Q^2)$ replicas having a new bimodal split of only $40\%$ asymptotically positive and $40\%$ asymptotically negative. These large changes to the small-$x$ extraction support my previous statement that any large-$x$ input has a significant effect even in the small-$x$ region.

Furthermore, I established that the bimodality and small-$x$ uncertainties can be alleviated by better constraints on $\widetilde{G}$. The current issue with constraining $\widetilde{G}$ is that DIS and SIDIS data are not directly sensitive to $\widetilde{G}$, meaning it only contributes through evolution effects which makes its contributions in the data-range of $5\times 10^{-3} < x < 0.1$ quite small when evolution only begins at $x = x_0 \equiv 0.1$. The truth, however, is that I make the active choice to start evolution at $x_0 = 0.1$, when in reality polarized evolution in $x$ begins at $x=1$ but is sub-leading at large-$x$ compared to the dominant DGLAP-driven dynamics. Using the method of matched asymptotic expansions~\cite{MuellerDingle1962, o2014method} suggests that it is possible to start evolution at $x = 1$ and include the DGLAP contributions, but subtract the double counting of logarithms that are present in both KPS-CTT and DGLAP resummations. Starting the evolution earlier may allow for $\widetilde{G}$ contributions to be sizable in the region $5\times 10^{-3} < x <0.1$, and thus more sensitive to the small-$x$ data. This extra sensitivity will result in better constraints on $\widetilde{G}$ and, in turn, better constraints on the predicted small-$x$ behaviors.

Perhaps the best way to directly constrain $\Delta G$ is to include in the analysis an observable that is directly sensitive to it (in such a way that is not $\alpha_s$ suppressed as in Eq.~\eqref{g1_coef_ftns}). The first option is that which is explored in \textbf{Chapter}~\ref{pheno_2}, incorporating polarized proton-proton scattering data into the small-$x$ helicity evolution global analysis. In principle there are two processes that have been used in JAM-DGLAP extractions~\cite{deFlorian:2014yva,Leader:2010rb,Nocera:2014gqa,Zhou:2022wzm,Cocuzza:2022jye}, jet and hadron production in polarized proton-proton collisions. The resultant double-longitudinal spin asymmetry $A_{LL}$ was teased in Eq.~\eqref{ALL_ratio} for the case of single-inclusive jet production, but at the time of this analysis was only formulated as
\begin{align}\label{pp_coll}
	\sigma^{\downarrow \Uparrow}-\sigma^{\uparrow \Uparrow} = \sum\limits_{a,b}\Delta f_{a/A}\otimes f_{b/B}\otimes \sigma_{ab}\,,
\end{align}
for an unknown partonic cross-section $\sigma_{ab}$ (of interacting partons $a$ and $b$ coming from protons $A$ and $B$) convolved with the (h)PDFs $(\Delta)f_{a,b/A,B}$.

Lastly, a future option to constrain $\Delta G$ (especially its large-$x$ initial conditions) is to match onto nonperturbative calculations from lattice QCD. The unique advantage of the DLA polarized evolution allows for $x \sim 0.1$ calculations on the lattice to be used to constrain small-$x$ evolution initial conditions. A relatively recent approach to determine (unpolarized) small-$x$ evolution initial conditions at the level of the proton wave function has been developed in Ref.~\cite{Dumitru:2020gla}, and the possibility that it may be extended to the polarized case is an enticing one.

\subsection{$\boldsymbol{\mathrm{Conclusions\!:~ Phenomenology~I}}$}

In this phenomenological study I presented the first simultaneous analysis of polarized DIS and SIDIS data using the KPS-CTT theoretical framework~\cite{Kovchegov:2015pbl, Kovchegov:2018znm,Cougoulic:2022gbk} for the evolution of hPDFs. This was a significant improvement over the first global analysis \cite{Adamiak:2021ppq} by adding in polarized SIDIS data, including the running of the coupling, and employing the updated KPS-CTT evolution equations over the original KPS equations, and taking the large-$N_c\&N_f$ limit over the less realistic large-$N_c$ limit. The inclusion of SIDIS data, as well as the use of the flavor nonsinglet KPS evolution, allowed an analysis of information sensitive to flavor separation, permitting me to resolve both the $C$-even and $C$-odd quark hPDFs $\Delta q^+$ and $\Delta q^-$, as well as individual quark flavor hPDFs $\Delta q$ and the gluon hPDF $\Delta G$.

Due to my use of the JAM Bayesian Monte Carlo framework I have successfully used small-$x$ helicity evolution to describe world polarized DIS and SIDIS data at $x < x_0 \equiv 0.1$ very well, with an overall $\chi^2_{\mathrm{red}} = 1.03$. This successful analysis is, however, slightly muddied by the substantial uncertainties of my small-$x$ predictions. I identified the source of these uncertainties and found that the inherent sensitivity of polarized DIS and SIDIS data to the polarized dipole amplitudes $\widetilde{G}$ and $G_2$ is the culprit. I conclude that the best course of action to address this fault in the analysis is to incorporate polarized proton-proton scattering data, and I will do so in the analysis that follows. 

Another issue that makes itself apparent is the impact of the axial anomaly on the $g_1$ structure functions and the hPDFs at small $x$. The role of the axial anomaly in the polarized structure functions was originally pointed out in Refs.~\cite{Altarelli:1988nr, Jaffe:1989jz, Shore:1991dv} and revisited more recently in Refs.~\cite{Tarasov:2020cwl, Tarasov:2021yll, Bhattacharya:2022xxw, Bhattacharya:2023wvy}. The effect is markedly different from that of the DLA BER and KPS-CTT evolution and development of the corresponding phenomenology is left for future work.

Ultimately, I find from this analysis that the contribution to total parton helicity from small-$x$ partons is negative-definite and potentially large; this extends to the small-$x$ contributions to the triplet and octet axial-vector charges $g_A$ and $a_8$. The true impact of small-$x$ partons is subject to the large uncertainties as well as the unaccounted-for systematic effect that would arise from the unavoidable large-$x$ matching onto JAM-DGLAP fits that will be implemented in the future. With the present estimates, the Jaffe-Manohar sum rule may need significant contributions from small-$x$ parton orbital angular momentum to be satisfied. In the long term, I expect that data from the future EIC will greatly enhance our understanding of small-$x$ hPDF, as shown from my impact study; the extra data at smaller values of $x$ will enable more precise statements about the distribution of angular momentum within the proton.
\newpage
\subsection{$\boldsymbol{\mathrm{Appendices\!:~ Analytic~ Cross\text{-}checks~ and~ Convergence~ Testing}}$}\label{pheno1_apdx}

\begin{figure}[t!]
	\begin{centering}
		\includegraphics[width=\textwidth]{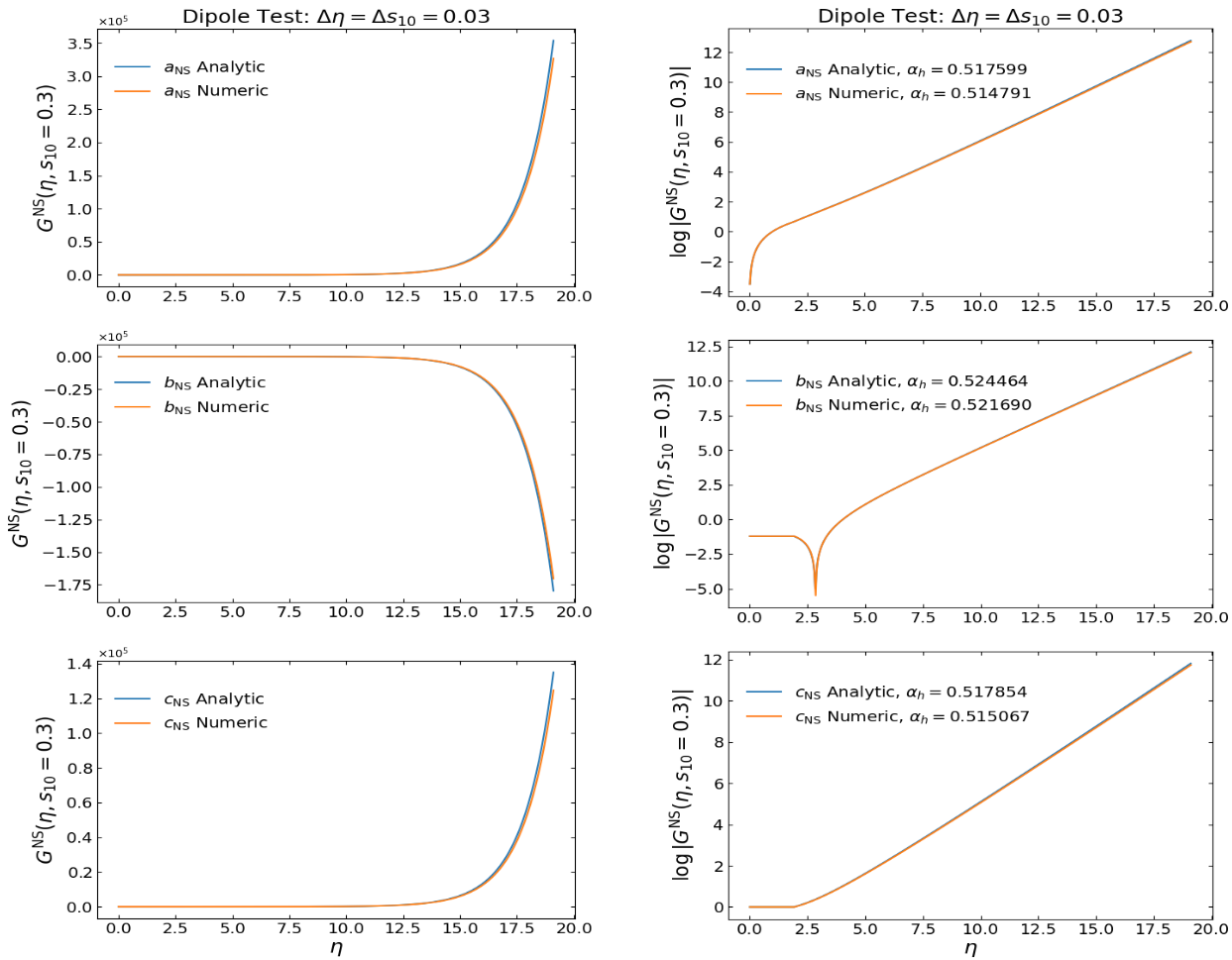}
		\caption{$G^{\textrm{NS}}\,(s_{10}, \eta)$ (left) and $\log{\abs{G^\mathrm{NS}\,(s_{10}, \eta)}}$ (right) plotted as functions of $\eta$ for a fixed value of $s_{10} = 0.3$. The large-$\eta$ behavior corresponds to the small-$x$ behavior, and this allows one to see how and when the numerical solution deviates from the analytic. The absolute value of the logarithm allows us to investigate the sign change (the cusp), and the slope of the logarithmic plot in the large-$\eta$ region will give a measure of the dipole amplitude-level intercept.
			\label{NS_Crosscheck}
		}
	\end{centering}
\end{figure}

The analytical solution for the large-$N_c$ flavor nonsinglet evolution equation that includes all of my physical assumptions and the running of the coupling has not been developed and is outside the scope of this work. However, there is an analytic solution for the large-$N_c$ flavor nonsinglet evolution with fixed coupling \cite{Kovchegov:2016zex} that does not enforce the $1/\Lambda$ IR cutoff (the same cutoff that results in $s_{ij} > 0$). I can perform a limited cross-check by modifying the numerical solution for the case of fixed coupling and allow evolution in the region $s_{ij} \leq 0$; in reference to this latter point I refer to the analytic cross-check regime as the all-$s_{10}$ ($\pm s$) regime. The following use of the subscript $\pm s$ is not to be confused with the center-of-mass energy squared $s$.

The evolution equation for fixed coupling and all $s_{10}$ is given by 
\begin{equation}
	G_{\pm s}^\textrm{NS}(s_{10},\eta) = G_{\pm s}^{\rm NS\,(0)}(s_{10},\eta) + \frac{\alpha_s}{2}\int\limits_0^{\eta}d\eta'\int\limits_{s_{10}-\eta+\eta'}^{\eta'-y_0}d s_{21} \, G_{\pm s}^\textrm{NS}(s_{21},\eta')\,.
\end{equation}
I can then discretize this equation using the same $s_{10}$-$\eta$ grid as described in \textbf{Chapter}~~\ref{numerics_etc}. The new equation becomes
\begin{align}\label{NS_Num_CC}
	G_{\pm s}^\textrm{NS}[i,j] & = G_{\pm s}^\textrm{NS}[i,j-1]+G_{\pm s}^{\rm NS\,(0)}[i,j] - G_{\pm s}^{\rm NS\,(0)}[i,j-1]  \\ 
	&+\frac{\alpha_s}{2}\Delta^2\Biggl[\sum\limits_{i'=i}^{j-2-j_0}G_{\pm s}^\textrm{NS}[i',j-1]+\sum\limits_{j'=0}^{j-2}G_{\pm s}^\textrm{NS}[i-j+1+j',j']\Biggr], \notag
\end{align}
which is especially different from Eq.~\eqref{GNS_disc} by the factored-out strong coupling and the different starting point for the $j'$ sum.

The all-$s_{10}$ solution is solvable using Laplace-Mellin transforms (c.f. Ref.~\cite{Kovchegov:2016zex}; to enforce the small-$x$ assumptions onto my conjugate variables I define the forward and inverse transforms
\begin{subequations}\begin{align}
		G_{\pm s}^\textrm{NS}(s_{10},\eta) & = \int\frac{d\omega}{2\pi i}e^{\omega\eta}\int\frac{d\lambda}{2\pi i}e^{(\eta - s_{10}-y_0)\lambda}G_{\pm s}^\textrm{NS}(\omega,\lambda)\,, \\
		G_{\pm s}^\textrm{NS}(\omega,\lambda)\,, & = \int_0^{\infty}d(\eta-s_{10}-y_0)e^{-\lambda(\eta-s_{10}-y_0)}\int_0^{\infty}d\eta\, e^{-\omega\eta}\, G_{\pm s}^\textrm{NS}(s_{10},\eta)\,. \label{GNS_conj}
\end{align}\end{subequations}
In Mellin-space the solution is given by~\cite{Kovchegov:2016zex}
\begin{equation}
	G_{\pm s}^\textrm{NS}(\omega,\lambda) = \frac{\omega\lambda}{\lambda-\frac{\alpha_s}{2\omega}}\frac{1}{\omega}G_{\pm s}^{\rm NS\,(0)}(\omega,\lambda)\,,
\end{equation}
where, fortunately, the simple initial conditions are already known $G_{\pm s}^{\mathrm{NS}\,(0)}(s_{10},\eta) = \eta,\; s_{10},\; 1$. I tackle the evaluation of each initial condition separately, beginning with $G_{\pm s}^{\mathrm{NS}\,(0)}(s_{10},\eta) = 1$. Plugging this initial condition into Eq.~\eqref{GNS_conj} gives
\begin{align}
	G_{\pm s}^{\rm NS\,(0)}(\omega,\lambda) & = \int_0^{\infty}d(\eta-s_{10}-y_0)\,e^{-\lambda(\eta-s_{10}-y_0)}\int_0^{\infty}d\eta\, e^{-\omega\eta} 
	= \frac{1}{\omega\lambda} \,,
\end{align}
which when plugged into the evolution equation leads to a contour integral with a pole at $\lambda = \alpha/(2\omega)$, which is evaluated via the residue theorem and provides
\begin{align}
	G_{\pm s}^\textrm{NS}(\eta,s_{10}) & = \int\frac{d\omega}{2\pi i}\,e^{\omega\eta}\int\frac{d\lambda}{2\pi i}\,e^{(\eta-s_{10}-y_0)\lambda}\frac{1}{\lambda-\frac{\alpha_s}{2\omega}}\frac{1}{\omega}  
	= \int\frac{d\omega}{2\pi i}e^{\omega\eta+\frac{\alpha_s}{2\omega}(\eta-s_{10}-y_0)}\frac{1}{\omega}\,. 
\end{align}
I can Taylor-expand the singular ($\sim1/\omega$) part of the integral, use the residue theorem again, and get the simplified result
\begin{align}
	G_{\pm s}^\textrm{NS}(\eta,s_{10}) & = \int\frac{d\omega}{2\pi i}e^{\omega\eta}\sum\limits_{n=0}^{\infty}\frac{1}{n!}\left(\frac{\alpha_s (\eta-s_{10}-y_0)}{2\omega}\right)^n\frac{1}{\omega}  
	= \sum\limits_{n=0}^{\infty}\frac{1}{(n!)^2}\left(\frac{\alpha_s}{2}\eta(\eta-s_{10}-y_0)\right)^n \,,
\end{align}
within which I can identify the modified Bessel function of the first kind $I_m(z)$ for $m=0$, and ultimately arrive at 
\begin{equation}
	G_{\pm s}^{NS,\,1}(\eta,s_{10}) = I_0\bigl(\sqrt{\alpha_s}\sqrt{2\eta(\eta-s_{10}-y_0)}\bigr).
\end{equation}

I repeat this process from the beginning but with the $\eta$ initial condition:
\begin{align}
	G_{\pm s}^{\rm NS\,(0)}(\omega,\lambda) & = \int_0^{\infty}d(\eta-s_{10}-y_0)\,e^{-\lambda(\eta-s_{10}-y_0)}\int_0^{\infty}d\eta\, e^{-\omega\eta}\eta  = \frac{1}{\omega^2\lambda} \,
\end{align}
is the initial condition in Mellin space, which results in the same $\lambda  = \alpha/(2\omega)$ pole in the evolution equation, and now gives a slightly different integral 
\begin{equation}
	G_{\pm s}^\textrm{NS}(\eta,s_{10})  = \int\frac{d\omega}{2\pi i}e^{\omega\eta+\frac{\alpha}{2\omega}(\eta-s_{10}-y_0)}\frac{1}{\omega^2}\,.
\end{equation}
I can perform the same Taylor expansion on this equation, identify this time a modified Bessel function of the first kind for $m = 1$, and write the solution as
\begin{equation}
	G_{\pm s}^{NS,\,\eta}(\eta,s_{10}) = \frac{1}{\sqrt{\alpha_s}}\sqrt{\frac{2 \, \eta}{\eta-s_{10}-y_0}} \, I_1\bigl(\sqrt{\alpha_s}\sqrt{2\eta(\eta-s_{10}-y_0)}\bigr).
\end{equation}

Lastly I start with the initial condition $G_{\pm s}^\mathrm{NS}(s_{10},\eta) = s_{10}$, where it is noted that $s_{10} = \eta-(\eta-s_{10}-y_0)-y_0$, and find the Mellin space initial condition to be
\begin{align}
	G_{\pm s}^{\rm NS\,(0)}(\omega,\lambda)  &= \int_0^{\infty}d(\eta-s_{10}-y_0)\,e^{-\lambda(\eta-s_{10}-y_0)}\int_0^{\infty}d\eta\, e^{-\omega\eta}(\eta-(\eta-s_{10}-y_0)-y_0) \\
	&= \frac{\lambda-\omega-y_0\omega\lambda}{(\omega\lambda)^2}\, \notag.
\end{align}
This actually results in two contributing poles $\lambda = 0,\;\alpha_s/(2\omega)$, but there are conveniently no poles in $\omega$ at $\lambda = 0$, so I only need to worry about the other one. The resultant evolution equation is then written as 
\begin{align}
	G_{\pm s}^\textrm{NS}(\eta,s_{10}) & = \int\frac{d\omega}{2\pi i}\,e^{\omega\eta}\int\frac{d\lambda}{2\pi i}\,e^{(\eta-s_{10}-y_0)\lambda}\frac{\omega\lambda}{\lambda-\frac{\alpha_s}{2\omega}}\frac{1}{\omega}\frac{\lambda-\omega-y_0\omega\lambda}{(\omega\lambda)^2}  \\ 
	& = \int\frac{d\omega}{2\pi i} \, e^{\omega\eta+\frac{\alpha_s}{2\omega}(\eta-s_{10}-y_0)}\left(\frac{1}{\omega^2} - \frac{2}{\alpha_s} - \frac{y_0}{\omega}\right), \notag
\end{align}
which is actually just a linear combination of the other two solution plus a new term. After solving for the new term the overall solution for the $s_{10}$ initial condition becomes
\begin{align}
	G_{\pm s}^{NS,\,s_{10}}(\eta,s_{10}) & = \frac{1}{\sqrt{\alpha_s}}\sqrt{\frac{2 \, \eta}{\eta-s_{10}-y_0}}I_1\bigl(\sqrt{\alpha_s}\sqrt{2\eta(\eta-s_{10}-y_0)}\bigr) \\
	&\;\;\; - \frac{1}{\sqrt{\alpha_s}}\sqrt{\frac{2(\eta-s_{10}-y_0)}{\eta}}I_1\bigl(\sqrt{\alpha_s}\sqrt{2\eta(\eta-s_{10}-y_0)}\bigr) \notag \\
	& \;\;\; -y_0\,I_0\bigl(\sqrt{\alpha_s}\sqrt{2\eta(\eta-s_{10}-y_0)}\bigr). \notag
\end{align}
Combining all initial conditions, I arrive at an analytic solution for the flavor nonsinglet evolution equation in the all-$s_{10}$ regime,
\begin{align}\label{e:NS_analytic}
	G_{\pm s}^\textrm{NS}(\eta,s_{10}) & = a^{\mathrm{NS}}\,G_{\pm s}^{NS,\,\eta} + b^{\mathrm{NS}}\,G_{\pm s}^{NS,\,s_{10}} + c^{\mathrm{NS}}\,G_{\pm s}^{NS,\,1} \\
	&=\; a^{\mathrm{NS}} \,  \frac{1}{\sqrt{\alpha_s}}\sqrt{\frac{2 \, \eta}{\eta-s_{10}-y_0}} \, I_1\bigl(\sqrt{\alpha_s}\sqrt{2\eta(\eta-s_{10}-y_0)}\bigr) \notag \\
	&\qquad+ \frac{b^{\mathrm{NS}}}{\sqrt{\alpha_s}}\sqrt{\frac{2 \, \eta}{\eta-s_{10}-y_0}} \, I_1\bigl(\sqrt{\alpha_s}\sqrt{2 \, \eta(\eta-s_{10}-y_0)}\bigr) \notag \\
	&\qquad - \frac{b^{\mathrm{NS}}}{\sqrt{\alpha_s}}\sqrt{\frac{2(\eta-s_{10}-y_0)}{\eta}} \, I_1\bigl(\sqrt{\alpha_s}\sqrt{2\eta(\eta-s_{10}-y_0)}\bigr)\notag \\
	&\qquad - b^{\mathrm{NS}}y_0\, I_0\bigl(\sqrt{\alpha_s}\sqrt{2\eta(\eta-s_{10}-y_0)}\bigr) \notag \\
	&\qquad + \;c^{\mathrm{NS}}\, I_0\bigl(\sqrt{\alpha_s}\sqrt{2\eta(\eta-s_{10}-y_0)}\bigr). \notag 
\end{align}
%

\begin{figure}
	\begin{centering}
		\includegraphics[width=\textwidth]{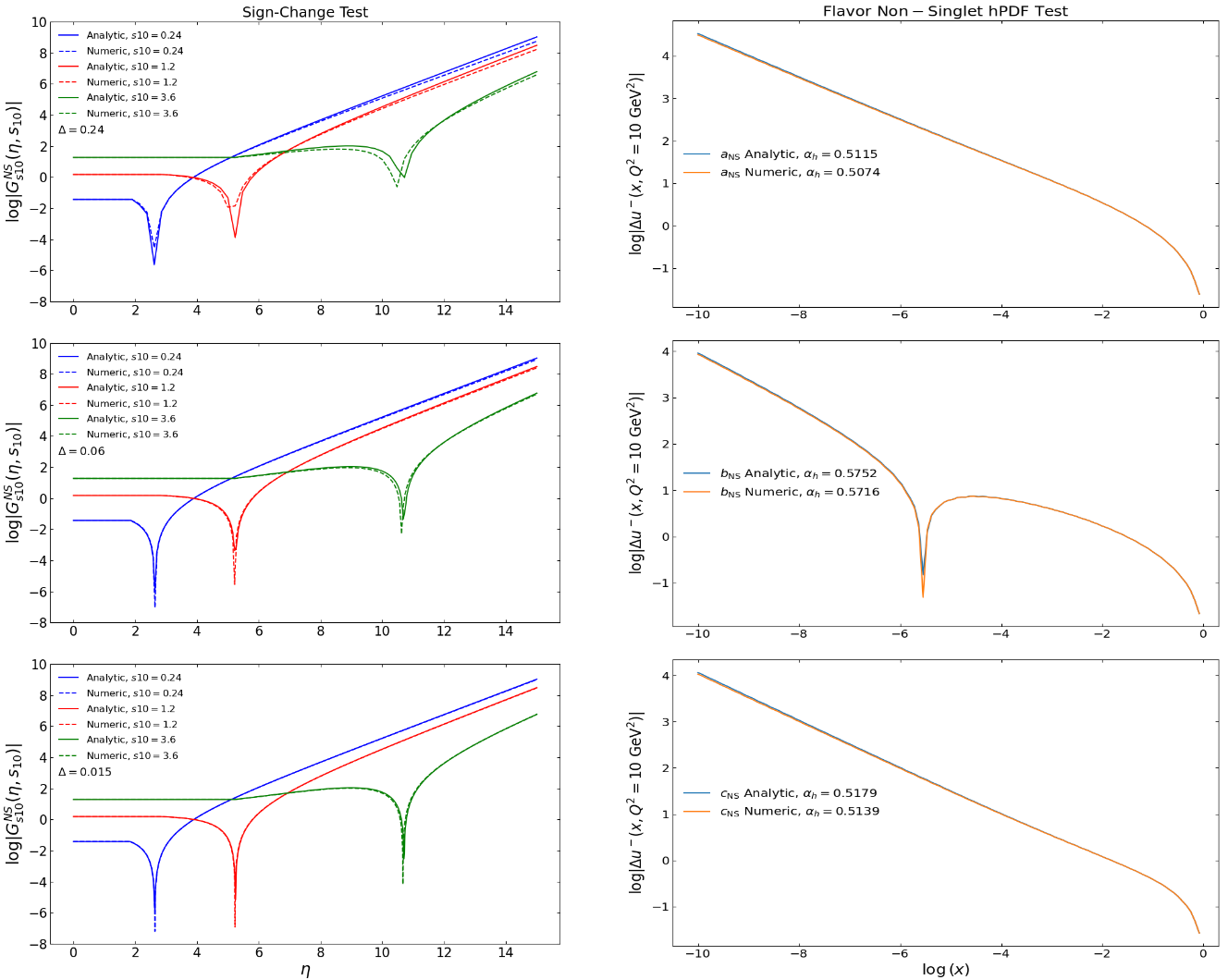}
		\caption{ (Left) A plot of $\ln G_u^{\mathrm{NS}}(s_{10},\eta)$ from the initial condition $b_u^{\mathrm{NS}}=1$ as a function of $\eta$. Each color represents a different fixed value of $s_{10}$. The location of the sign-change in the amplitude, indicated by the cusp, appears to vary with $s_{10}$. Smaller step sizes lead to convergence of the sign-change between the analytic and numeric solutions, and $\Delta\eta=\Delta s_{10} = \Delta < 0.06$ retains small-$x$ agreement. (Right) Three plots of $\ln \Delta u^-(x,Q^2)$, one plot for each basis functions ($a_u^{\mathrm{NS}}$, $b_u^{\mathrm{NS}}$, and $c_u^{\mathrm{NS}}$) as a function of $\ln{(x)}$ for a fixed $Q^2 = 10~\mathrm{GeV}^2$; both the numerical (orange) and analytic (blue) solutions are plotted. Each plot also gives as a measure of the intercept $\alpha_h$ computed from the numerical and analytical solutions.
			\label{Signchange_test}
		}
	\end{centering}
\end{figure}
\begin{figure}
	\begin{centering}
		\includegraphics[width=\textwidth]
		{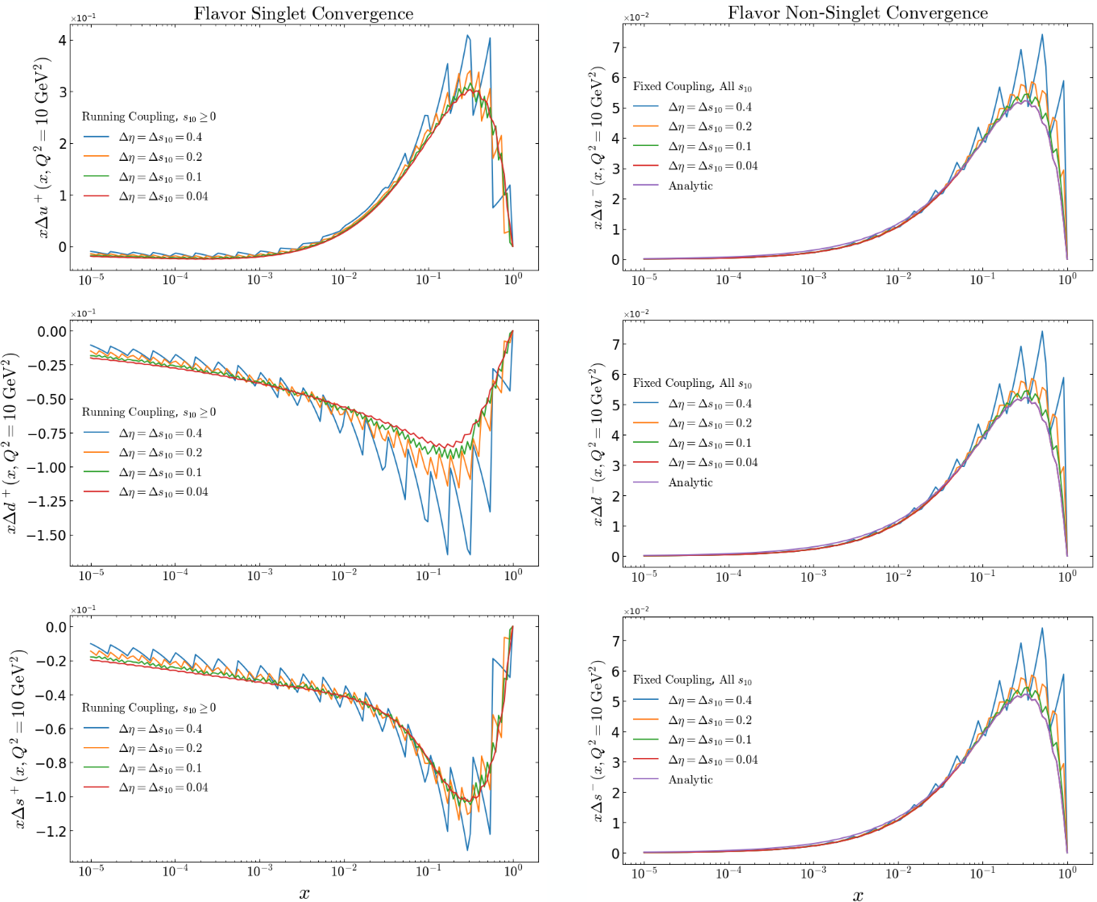}
		\caption{(Left) A numerical computation of $x\Delta q^+(x)$ for a test state of initial conditions. The graph shows the same numerical solution for various choices of step size, $\Delta = \Delta\eta = \Delta s_{10}$. As the step-size $\Delta$ decreases, the numerical solution converges to a single result. (Right) A numerical computation of $x\,\Delta q^-(x)$ that shows the convergence to a single output as $\Delta$ decreases.  For both $x\Delta q^+(x)$ and $x\Delta q^-(x)$, the single output is described by the analytic solution \eqref{e:NS_analytic}.
			\label{Convergences}
		}
	\end{centering}
\end{figure}

I start the cross-check by comparing three specific features of the numerical analytic versions of the polarized dipole amplitudes: the general shape of the polarized dipole amplitude (for fixed $s_{10}$), the location of a sign-change in the $s_{10}$ contributions due to a sign-change that occurs from negative growth after a positive initial condition, and the asymptotic behavior as $x\to 0$; note that the dipole amplitudes' asymptotics should be equal to each other and equal to the constructed hPDF.

Fig.~\ref{NS_Crosscheck} depicts the numerical solution of the polarized dipole amplitudes for a small step-size as a function of $\eta$ and for fixed $s_{10}$, and compares it to the analytic solution counterpart. The left panel of Fig.~\ref{NS_Crosscheck} shows that the general shapes of the two solutions are in agreement so long as the step-size is reasonably small at $\Delta = 0.03$. The analytic solution appears to grow slightly faster than the numerical solution. Taking the logarithm of the dipole amplitudes allows us to observe the small-$x$ asymptotics, and the slope of the lines at large values of $\eta$ gives a measure of the intercept $\alpha_h$, where I find that the numerical and analytic solutions agree within $1.4\%$ for $\Delta = 0.03$. In the center-right panel of Fig.~\ref{NS_Crosscheck} we see the $b_u^{\mathrm{NS}} = 1$ contribution to the polarized dipole amplitude and find the aforementioned zero-crossing represented as a cusp in log-space; this cusp is very sensitive to the large-$x$ (low-$\eta$) behavior of the polarized dipole amplitudes, and the qualitative observation that the two cusps overlap with each other is a sign that the numerical and analytic solution are also in agreement at large $x\sim x_0$. By plotting the dipole amplitudes for larger values of fixed $s_{10}$ we can determine the necessary step-size for retaining the cusp agreement at small $x$; this test is shown in the left panel of Fig.~\ref{Signchange_test}, where I conclude that step-size of $\Delta < 0.06$ will retain agreement with the numerical solution at the level of the polarized dipole amplitudes. This means my use of $\Delta \leq 0.25$ in the phenomenological analyses above is satisfactory.

After examining the two solutions at the level of the polarized dipole amplitudes, I elect to construct the $C$-odd flavor nonsinglet hPDF $\Delta q^-$ (or specifically $\Delta u^-$ for brevity) and plot it in the right panel of Fig.~\ref{Signchange_test}. By taking the logarithm of the hPDFs and studying the small-$x$ behavior, I find that the slopes, and by extend the intercept $\alpha_h$, are in agreement with each other (numerical and analytical) and to their constituent polarized dipole amplitudes; I confirm that the intercepts $\alpha_h$ agree within $1\%$ between the analytic and numerical hPDFs, and are consistent with the intercepts computed at the dipole level. With this, the analytic cross-check between the numerical and analytic solutions of the flavor nonsinglet evolution equations is complete.

The last worthwhile subject of analysis is the numerical convergence of the discretized evolution equations Eq.~\eqref{disc_evo}. As with any numerical solution, the accuracy of its result is largely dependent on the resolution, i.e. the step size. For the polarized dipole amplitudes there are two discrete arguments, $s_{10}$ and $\eta$, resulting a 2D grid ($G[i,j]$) to compute for each polarized dipole amplitude; The numerical solution \eqref{disc_evo} is only valid in the case the entire grid is evenly spaced such that $\Delta s_{10} = \Delta\eta \equiv \Delta$. For any numerical solution to hold merit, the computed solution should converge to a single, consistent output as the step size decreases towards zero. Shown in Fig.~\ref{Convergences} are the calculations of the $\Delta q^{\pm}(x,Q^2)$ (scaled by $x$) as a function of $x$ for fixed $Q^2 = 10~\mathrm{GeV}^2$, with the $C$-even flavor singlet hPDFs plotted on the left and the $C$-odd flavor nonsinglet hPDFs plotted on the right; there are three rows of plots, one for each of the three quark flavors. The resultant curves all come from a single test-state, which is to say they were computed from the same initial condition parameters taken from one good-fitting replica ($\chi^2_{\mathrm{red}}\approx 1$. In each plot are numerous curves that each depict a solution computed from a different step size; as expected from a stable numerical solution, the lines all converge as the step size decreases.  For extra confirmation, I plot the analytical solution for the flavor nonsinglet hPDF and find that as the step size decreases, the curves converge towards the analytic solution. Qualitatively, a step size of $\Delta = 0.04$ is sufficiently converged, meanwhile the global analysis was performed with a step size of $\Delta \approx 0.02$.
	\newpage
	\setlength{\arrayrulewidth}{0.4pt}
\setlength{\arraycolsep}{2pt}
\setlength{\tabcolsep}{1.5pt}
\setlength{\doublerulesep}{2pt}
\renewcommand{\arraystretch}{1.2}
\section{$\boldsymbol{\mathrm{Phenomenology~II\!:~ DIS,~ SIDIS,~ and~} pp~ \mathrm{Data}}$}\label{pheno_2}
Being the next iteration of small-$x$ helicity phenomenology, the background and subject of this work is largely similar to that of \textbf{Chapter}~\ref{pheno_1} and covers the work done in Ref.~\cite{Adamiak:2025dpw}. Having said that, there have been some phenomenological updates between the publication of Ref.~\cite{Adamiak:2023yhz} (the article covered in \textbf{Chapter}~\ref{pheno_1}) and the phenomenology covered here (\cite{Adamiak:2025dpw}) that deserve a mention. This includes more recent DGLAP-based phenomenology \cite{Borsa:2023tqr, Borsa:2024mss, Hunt-Smith:2024khs, Cruz-Martinez:2025ahf}, as well as more recent interest in small-$x$ spin-dependent evolution \cite{Santiago:2024iem, Duan:2024qev, Agrawal:2024xkb, Borden:2024bxa, Kovchegov:2023yzd, Ermolaev:2024aqq, Manley:2024pcl} and sub-eikonal evolution \cite{Altinoluk:2024dba}. More related specifically to the KPS-CTT evolution equations and phenomenology regarding them, I note that the KPS-CTT agree with BER up to two decimal places \cite{Cougoulic:2022gbk,Adamiak:2023okq} and disagree beyond \cite{Borden:2023ugd}, and recent developments have found that this is due to missing quark-to-gluon and gluon-to-quark transition operators in the large-$N_c\&N_f$ limit \cite{Borden:2024bxa}. Including these new operators results in agreement with the three known orders of the spin-dependent DGLAP anomalous dimensions at large-$N_c\&N_f$ \cite{Altarelli:1977zs, Dokshitzer:1977sg, Zijlstra:1993sh, Mertig:1995ny, Moch:1999eb, vanNeerven:2000uj, Vermaseren:2005qc, Moch:2014sna, Blumlein:2021ryt, Blumlein:2021lmf, Davies:2022ofz, Blumlein:2022gpp}, as well as a new contribution to the KPS-CTT evolution equations; the latter point is nuanced and will be discussed more soon, but suffice it to say now that these updates are not included in this analysis.

\subsection{$\boldsymbol{\mathrm{Subject~of~this~work~and~updates~to~formalism}}$}

The subject of this work is a direct sequel to the conclusions of \textbf{Chapter}~\ref{pheno_1}; the global analysis of DIS and SIDIS data using the KPS-CTT evolution equations found that, while it was possible to describe world polarized data very well, the uncertainties at small $x$ grew larger than one would expect from the inherent predictive power of small-$x$ helicity evolution equations. Close examination of the replicas and parameter distributions lead to the conclusion that asymptotic bimodality of solutions was responsible for a diverging uncertainty, and that basis functions constructed from the gluon-dominated polarized dipole amplitudes $\widetilde{G}$ and $G_2$ had the largest influence at small $x$, yet were not properly constrained by DIS and SIDIS data. The large uncertainty bands for $\Delta g(x,Q^2)$\footnote{This chapter makes a change in nomenclature from $\Delta G(x,Q^2)$ to represent the gluon hPDF in \textbf{Chapter}~\ref{pheno_1} to $\Delta g(x,Q^2)$ to be more consistent with existing spin-dependent literature, including the article this chapter is intended to cover \cite{Adamiak:2025dpw}. I instead use $\Delta G(Q^2)$ to represent the integrated spin contributions.} imply that $G_2$ could stand to be better constrained, and the insensitivity of DIS and SIDIS data to $\widetilde{G}$ inherently explains its lack of constraint. One method to better constrain these contributions is to more rigorously model the initial conditions as was done in Ref.~\cite{Dumitru:2024pcv} and phenomenologically tested in Ref.~\cite{Adamiak:2025mdy}. In this analysis, the method of improving the constraints over the $\widetilde{G}^{(0)}$ and $G_2^{(0)}$ initial conditions is to incorporate a new observable that is directly sensitive to these polarized dipole amplitudes. 

In this work I will build upon the analysis of DIS and SIDIS data covered in \textbf{Chapter}~~\ref{pheno_1} by also including polarized proton-proton ($pp$) scattering data to study the impact on the small-$x$ predictions of the hPDFs and $g_1$ structure function. Proton-proton collisions allow for leading order (LO) gluon interactions, and as such one would expect that the double longitudinal spin asymmetries $A_{LL}$ in these reactions would better constrain the gluon-dominated polarized dipole scattering amplitudes $\widetilde{G}$ and $G_2$. I focus specifically on single-inclusive jet production and the spin asymmetry $A_{LL}^{\mathrm{jet}}$ for mid-rapidity jets; the numerator I use (c.f. Eq.~\eqref{ALL_ratio}) is the small-$x$ polarized $pp \to g\,X$ cross-section derived in Ref.~\cite{kovchegov:2024aus}. This small-$x$ polarized cross-section was developed in the pure-glue limit, and because of this I would be expected to employ the large-$N_c$ KPS-CTT evolution equations (as opposed to the large-$N_c\&N_f$ evolution). However, a crucial part of this analysis is the need to resolve individual quark flavor hPDFs which are not a feature provided by the large-$N_c$ evolution equations. To best approximate a large-$N_c$ analogue in the presence of external quarks, I make use of what I dub the ``large-$N_c^{+q}$" evolution equations which takes the large-$N_c\&N_f$ evolution equations \eqref{eq_LargeNcNf} from Ref.~\cite{Cougoulic:2022gbk} and sets $N_f=0$. In this way, quark loops are eliminated, and what remains is flavor-blind evolution with flavor-dependent initial conditions. As with the previous analysis I still employ the running of the coupling (which results in a slight reduction of the intercept $\alpha_h$ and makes the spin integrals \eqref{spin_integrals} integrable \cite{Adamiak:2023yhz}. Teased above was a mention of a new contribution to the KPS-CTT evolution equations from Ref.~\cite{Borden:2024bxa}, but I note now that the new polarized dipole amplitude only couples to the others with a prefactor of $N_f$, thus nullifying its contribution to this analysis.

In addition to the change of evolution equations and inclusion of a new type of data, there has also been an important change in the renormalization scheme. Touched on at the end of \textbf{Chapter}~~\ref{hPDFs_stf_sec1}, the analysis covered in \textbf{Chapter}~\ref{pheno_1} makes use of the ``polarized DIS" scheme. In this analysis I bring in a change that makes my definitions of the hPDFs and observables more in line with that of $\overline{\mathrm{MS}}$ scheme; in accordance with the updated definitions of Ref.~\cite{Borden:2024bxa}, I redefine the $C$-even and $C$-odd quarks hPDFs at DLA as
\begin{equation}\label{deltaqpm_msbar}
	\begin{bmatrix}
		\Delta q^+(x,Q^2) \\
		\Delta q^-(x,Q^2)
	\end{bmatrix}
	= -\frac{N_c}{2\pi^3}\int\limits_{\Lambda^2/s}^1\,\frac{d z}{z}\int\limits_{\mathrm{max}\{1/Q^2, 1/zs\}}^{\mathrm{min}\{1/zQ^2,1/\Lambda^2\}}\,\frac{d x_{10}^2}{x_{10}^2}\,
	\begin{bmatrix}
		Q_q(x_{10}^2,zs) + 2G_2(x_{10}^2,zs)\\
		-G_q^{\mathrm{NS}}(x_{10}^2,zs)
	\end{bmatrix}\,,
\end{equation}
where the notable difference between Eqs.~\eqref{Deltaqm} and \eqref{deltaqpm_msbar} can be found in the different lower limit of the $x_{10}^2$ integral. This expression allows for complete agreement between the KPS-CTT evolution and the $\overline{\mathrm{MS}}$ polarized small-$x$ splitting functions at large-$N_c\&N_f$ to the three known loops (up to a residual scheme transformation) \cite{Borden:2024bxa}. It is important to note that the evolution equations themselves have not been changed (except, of course, for the modification $N_f=0$ to the large-$N_c\&N_f$ evolution equations), nor has the definition of the $g_1$ structure function been altered as it is still expressed by Eq.~\eqref{g_1}. Because the hPDF $\Delta q^+$ has a different definition but $g_1$ does not, I must make clear that there is no longer a direct correspondence to the LO collinear factorization expression from polarized DIS,
\begin{equation}\label{g1_relation_2}
	g_1(x,Q^2) \ne \frac{1}{2}\sum_qZ_q^2\,\Delta q^+(x,Q^2)\,.
\end{equation}
It may also be beneficial to rewrite the SIDIS structure function in terms of the less-subjective polarized dipole amplitudes:
\begin{align}\label{g1h_2}
	&g_1^h (x,z_h,Q^2) = -\sum_q\frac{N_c\,Z_q^2}{16\pi^3}\,D_1^{h/q}(z_h,Q^2) \\
	&\hspace{1.5cm}\times\int\limits_{\Lambda^2/s}^1\,\frac{d z}{z}\int\limits_{1/zs}^{\mathrm{min}\{1/zQ^2,1/\Lambda^2\}}\,\frac{d x_{10}^2}{x_{10}^2}\,\bigl[Q_q(x_{10}^2,zs) + 2G_2(x_{10}^2,zs) \pm G_q^{\mathrm{NS}}(x_{10}^2,zs)\bigr], \notag
\end{align}
where the flavor nonsinglet polarized dipole amplitude can come in with a relative plus or minus depending on whether the tagged hadron comes from a quark or antiquark.

With the new quark hPDF definitions provided, I take this time to write out the new large-$N_c^{+q}$ evolution equations. In truth, they are largely the same as Eq.~\eqref{eq_LargeNcNf}, but there are a number of terms I must remove due to the choice to set $N_f=0$; keep in mind that in writing out Eq.~\eqref{eq_LargeNcNf}, some terms are pseudo-$N_f$ terms, which is to say that I also must remove the $\sum_qQ_q$ terms since they are just a flavor-dependent version of $N_f\,Q$ in the simplified case that $Q_u=Q_d=Q_s\equiv Q$. The new evolution equations then read \cite{Adamiak:2025dpw}: 
\setlength{\baselineskip}{\singlespace}
{\allowdisplaybreaks
	\begin{subequations}\label{eq_LargeNcq}
		\begin{align}
			& Q_q (x^2_{10},zs) = Q_q^{(0)}(x^2_{10},zs) + \frac{N_c}{2\pi} \int\limits^{z}_{1/x^2_{10}s} \frac{d z'}{z'}   \int\limits_{1/z's}^{x^2_{10}}  \frac{d x^2_{21}}{x_{21}^2}  \ \alpha_s \!\!\left( \frac{1}{x_{21}^2} \right) \\
			&\hspace*{3cm} \times \, \bigg[ 2 \, {\widetilde G}(x^2_{21},z's)+ 2 \, {\widetilde \Gamma}(x^2_{10},x^2_{21},z's)+ Q_q (x^2_{21},z's) \notag \\
			&\hspace*{5cm}-  \overline{\Gamma}_q (x^2_{10},x^2_{21},z's) + 2 \, \Gamma_2(x^2_{10},x^2_{21},z's) + 2 \, G_2(x^2_{21},z's)   \bigg] \notag  \\
			&\hspace*{2cm}+ \frac{N_c}{4\pi} \int\limits_{\Lambda^2/s}^{z} \frac{d z'}{z'}   \int\limits_{1/z's}^{\min \left[ x^2_{10}z/z', 1/\Lambda^2 \right]}  \frac{d x^2_{21}}{x_{21}^2} \ \alpha_s\!\! \left( \frac{1}{x_{21}^2} \right) \, \bigg[Q_q (x^2_{21},z's) + 2 \, G_2(x^2_{21},z's) \bigg] ,  \notag  \\[0.5cm]
			&\overline{\Gamma}_q (x^2_{10},x^2_{21},z's) = Q^{(0)}_q (x^2_{10},z's) + \frac{N_c}{2\pi} \int\limits^{z'}_{1/x^2_{10}s} \frac{d z''}{z''}   \int\limits_{1/z''s}^{\min[x^2_{10}, x^2_{21}z'/z'']}  \frac{d x^2_{32}}{x_{32}^2}   \ \alpha_s\!\! \left( \frac{1}{x_{32}^2} \right)   \\
			&\hspace*{3cm} \times \; \, \bigg[ 2\, {\widetilde G} (x^2_{32},z''s)+ \; 2\, {\widetilde \Gamma} (x^2_{10},x^2_{32},z''s) +  Q_q (x^2_{32},z''s) \notag \\
			&\hspace*{5cm}-  \overline{\Gamma}_q (x^2_{10},x^2_{32},z''s) + 2 \, \Gamma_2(x^2_{10},x^2_{32},z''s) + 2 \, G_2(x^2_{32},z''s) \bigg] \notag \\
			&\hspace*{2cm}+ \frac{N_c}{4\pi} \int\limits_{\Lambda^2/s}^{z'} \frac{d z''}{z''}   \int\limits_{1/z''s}^{\min \left[ x^2_{21}z'/z'', 1/\Lambda^2 \right] }  \frac{d x^2_{32}}{x_{32}^2} \ \alpha_s\!\! \left( \frac{1}{x_{32}^2} \right) \, \bigg[Q_q (x^2_{32},z''s) + 2 \, G_2(x^2_{32},z''s) \bigg] , \notag \\[0.5cm]
			& {\widetilde G}(x^2_{10},zs) = {\widetilde G}^{(0)}(x^2_{10},zs) + \frac{N_c}{2\pi}\int\limits^{z}_{1/x^2_{10}s}\frac{d z'}{z'}\int\limits_{1/z's}^{x^2_{10}} \frac{d x^2_{21}}{x^2_{21}} \ \alpha_s\!\! \left( \frac{1}{x_{21}^2} \right)  \\
			&\hspace*{3cm}  \times \; \bigg[3 \, {\widetilde G}(x^2_{21},z's)+ \; {\widetilde \Gamma}(x^2_{10},x^2_{21},z's) +  2\,G_2(x^2_{21},z's) +  2\, \Gamma_2(x^2_{10},x^2_{21},z's)\bigg] \notag \\[0.5cm]
			& {\widetilde \Gamma} (x^2_{10},x^2_{21},z's) = {\widetilde G}^{(0)}(x^2_{10},z's) + \frac{N_c}{2\pi}\int\limits^{z'}_{1/x^2_{10}s}\frac{d z''}{z''}\int\limits_{1/z''s}^{\min[x^2_{10},x^2_{21}z'/z'']} \frac{d x^2_{32}}{x^2_{32}} \ \alpha_s\!\! \left( \frac{1}{x_{32}^2} \right) \,   \\
			&\hspace*{3cm} \times \;\bigg[3 \, {\widetilde G} (x^2_{32},z''s) + {\widetilde \Gamma}(x^2_{10},x^2_{32},z''s) + 2 \, G_2(x^2_{32},z''s)  +  2\,\Gamma_2(x^2_{10},x^2_{32},z''s)\bigg] \notag \\[0.5cm]
			& G_2(x_{10}^2, z s)  =  G_2^{(0)} (x_{10}^2, z s) + \frac{N_c}{\pi} \, \int\limits_{\Lambda^2/s}^z \frac{d z'}{z'} \, \int\limits_{\max \left[ x_{10}^2 , 1/z' s \right]}^{\min \left[ x_{10}^2 z/z', 1/\Lambda^2 \right] } \frac{d x^2_{21}}{x_{21}^2} \ \alpha_s\!\! \left( \frac{1}{x_{21}^2} \right) \\
			&\hspace*{9cm} \times \, \bigg[ {\widetilde G} (x^2_{21} , z' s) + 2 \, G_2 (x_{21}^2, z' s)  \bigg] , \notag \\[0.5cm]
			& \Gamma_2 (x_{10}^2, x_{21}^2, z' s)  =  G_2^{(0)} (x_{10}^2, z' s) + \frac{N_c}{\pi} \!\! \int\limits_{\Lambda^2/s}^{z' x_{21}^2/x_{10}^2} \frac{d z''}{z''} \!\!\!\!\!\!\!\!\! \int\limits_{\max \left[ x_{10}^2 , 1/z''s \right]}^{\min \left[  z'x_{21}^2/z'', 1/\Lambda^2 \right] } \!\! \frac{d x^2_{32}}{x_{32}^2} \ \alpha_s\!\! \left( \frac{1}{x_{32}^2} \right) \\
			&\hspace*{9cm} \times \; \bigg[ {\widetilde G} (x^2_{32} , z'' s) + 2 \, G_2(x_{32}^2, z'' s)  \bigg] \notag \\
			& G_q^{\mathrm{NS}}(x_{10}^2, z )  =  G_q^{\mathrm{NS}\,(0)} (x_{10}^2, z ) + \frac{N_c}{4\pi} \, \int\limits_{\Lambda^2/s}^z \frac{d z'}{z'} \, \int\limits_{1/z's}^{\min \left[ z x_{10}^2/z', 1/\Lambda^2 \right] } \frac{d x^2_{21}}{x_{21}^2} \ \alpha_s\!\! \left( \frac{1}{x_{21}^2} \right) \; G_q^{\mathrm{NS}}(x_{21}^2,z') , 
		\end{align}
	\end{subequations}
}
\setlength{\baselineskip}{\doublespace}
where only the evolution of $\widetilde{G}$ and $\widetilde{\Gamma}$ have been changed. This change makes it so that $\widetilde{G}$ and $\widetilde{\Gamma}$ are now independent of the of the dipole amplitudes $Q_q$ and $\overline{\Gamma}_q$ which means that large-$x$ gluons cannot produce small-$z$ quark loops due to $N_f/N_c$ suppression. The indirect effect of this is that $Q_q^{(0)}$ initial conditions no longer contribute to $G_2$ evolution since $\widetilde{G}$ acted as a ``link" to allow for mixing between the $Q_q$ and $G_2$ dipoles. By extend this also means that $Q_q^{(0)}$ initial conditions cannot generate $\Delta g(x,Q^2)$ basis functions; $\Delta g(x,Q^2)$ basis functions can still be produced from $\widetilde{G}^{(0)}$ and $G_2^{(0)}$ initial conditions, so there is effect on phenomenology (or at least my ability to extract to small-$x$ $\Delta g(x,Q^2)$). At the risk of taxing the reader's patience for reading equations, I quickly write the discretized form of the large-$N_c^{+q}$ evolution equations, including the small-$x$, dipole size, and physicality assumptions discussed in \textbf{Chapter}~\ref{numerics_etc}:
{\allowdisplaybreaks
	\begin{subequations}\label{disc_evo_largeNcq}
		\begin{align}
			& Q_q[i,j] = Q_q[i,j-1] + Q_q^{(0)}[i,j] - Q_q^{(0)}[i,j-1]  \\
			&\qquad\ +\Delta^2\sum\limits_{i' = i}^{j-2-y_0}\alpha_s[i']\Big[\frac{3}{2}\,Q_q[i',j-1] + 2\,\widetilde{G}[i',j-1] + 2\,\widetilde{\Gamma}[i,i',j-1] \notag \\
			&\qquad\qquad\qquad\;-\overline{\Gamma}_q[i,i',j-1] + 3\,G_2[i',j-1] + 2\,\Gamma_2[i,i',j-1]\Big] \notag \\
			&\qquad + \frac{1}{2}\Delta^2\sum\limits_{j' = j-1-i}^{j-2}\alpha_s[i+j'-j+1]\Big[Q_q[i+j'-j+1,j'] + 2\,G_2[i+j'-j+1,j']\Big] \notag \\[0.5cm]
			& \overline{\Gamma}_q[i,k,j] = \overline{\Gamma}_q[i,k-1,j-1] + Q_q^{(0)}[i,j] - Q_q^{(0)}[i,j-1]  \\
			&\qquad\ +\Delta^2\sum\limits_{i' = i}^{j-2-y_0}\alpha_s[i']\Big[\frac{3}{2}Q_q[i',j-1] + 2\,\widetilde{G}[i',j-1] + 2\,\widetilde{\Gamma}[i,i',j-1] \notag \\
			&\qquad\qquad\qquad\;-\overline{\Gamma}_q[i,i',j-1] + 3\,G_2[i',j-1] + 2\Gamma_2[i,i',j-1]\Big] \notag \\[0.5cm]
			&\widetilde{G}[i,j] = \widetilde{G}[i,j-1] + \widetilde{G}^{(0)}[i,j] - \widetilde{G}^{(0)}[i,j-1] \\
			&\qquad +\Delta^2\sum\limits_{i'=i}^{j-2-y_0}\alpha_s[i']\Big[2\,\widetilde{G}[i',j-1] + \widetilde{\Gamma}[i,i',j-1] +2\,G_2[i',j-1] + 2\,\Gamma_2[i,i',j-1] \Big] \notag \\[0.5cm]
			&\widetilde{\Gamma}[i,k,j] = \widetilde{\Gamma}[i,k-1,j-1] + \widetilde{G}^{(0)}[i,j] - \widetilde{G}^{(0)}[i,j-1] \\
			&\qquad +\Delta^2\sum\limits_{i'=i}^{j-2-y_0}\alpha_s[i']\Big[2\,\widetilde{G}[i',j-1] + \widetilde{\Gamma}[i,i',j-1] +2\,G_2[i',j-1] + 2\,\Gamma_2[i,i',j-1] \Big] \notag \\[0.5cm] 
			&G_2[i,j] = G_2[i,j-1] + G_2^{(0)}[i,j] - G_2^{(0)}[i,j-1]  \\
			&\qquad + 2\Delta^2\sum\limits_{j' = j-1-i}^{j-2}\alpha_s[i+j'-j+1]\Big[\widetilde{G}[i+j'-j+1,j'] +2\,G_2[i+j'-j+1,j']\Big] \notag \\[0.5cm]
			&\Gamma_2[i,k,j] = \Gamma_2[i,k-1,j-1] + G_2^{(0)}[i,j] - G_2^{(0)}[i,j-1] \\[0.5cm]
			&G_q^{\mathrm{NS}}[i,j] = G_q^{\mathrm{NS}}[i,j-1] + G_q^{\mathrm{NS}\,(0)}[i,j] - G_q^{\mathrm{NS}\,(0)}[i,j-1] \\
			&\qquad + \frac{1}{2}\Delta^2\Bigg[\sum\limits_{i'=i}^{j-2-y_0}\alpha_s[i']G_q^{\mathrm{NS}}[i',j-1] + \sum\limits_{j' = j-1-i}^{j-2}\alpha_s[i+j'-j+1]\,G_q^{\mathrm{NS}}[i+j'-j+1,j']\Bigg], \notag 
		\end{align}
	\end{subequations}
}
where, again, only the evolution $\widetilde{G}$ and $\widetilde{\Gamma}$ have been modified after setting $N_f = 0$. In further preparation for the following analysis, I mention that the large-$N_c^{+q}$ evolution equations retain an important behavior observed in Ref.~\cite{Adamiak:2023yhz}, namely that basis hPDFs constructed from $G_2^{(0)}$ and $\widetilde{G}^{(0)}$ initial conditions have the two largest contributions at small $x$.  I will show below that these are the two polarized dipole amplitudes for which the $pp$ jet production cross-section is directly sensitive, meaning that not only should the polarized $pp$ data provide much better constrains on the parameters $\widetilde{a},\, \widetilde{b},\, \widetilde{c},\, a_2,\, b_2,\, c_2$, but also that these constrains are going to be especially impactful on the small-$x$ hPDF extractions discussed below.

\subsection{$\boldsymbol{\mathrm{Polarized~ Proton\text{-}Proton~ Scattering~ at~ Small~}x}$}
\begin{figure}[h!]
	\begin{centering}
		\includegraphics[width=\textwidth]{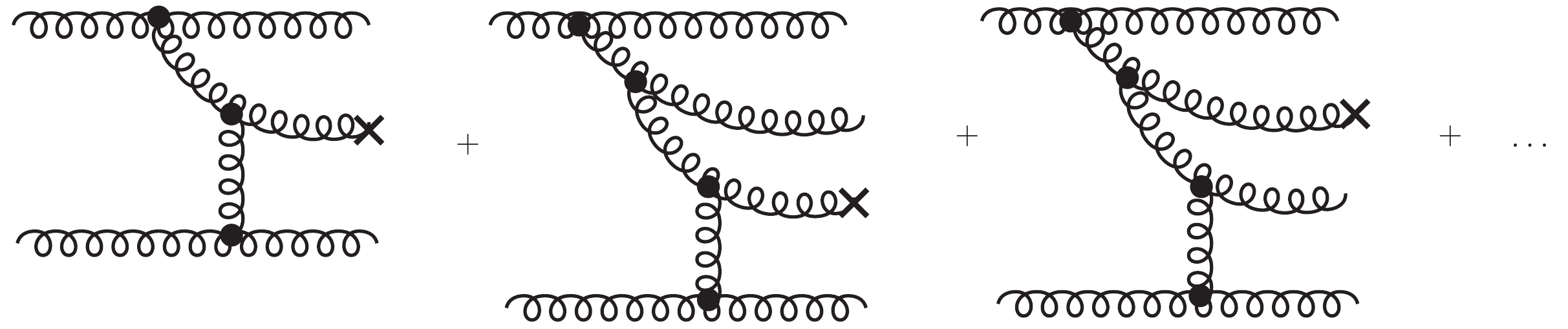}
		\caption{Schematic diagrams for the process $gg\rightarrow gX$. The black dots indicate the relevant sub-eikonal gluon vertices. 
			\label{gg_to_gluon}
		}
	\end{centering}
\end{figure}

\begin{figure}[h!]
	\begin{centering}
		\includegraphics[width=0.8\textwidth]{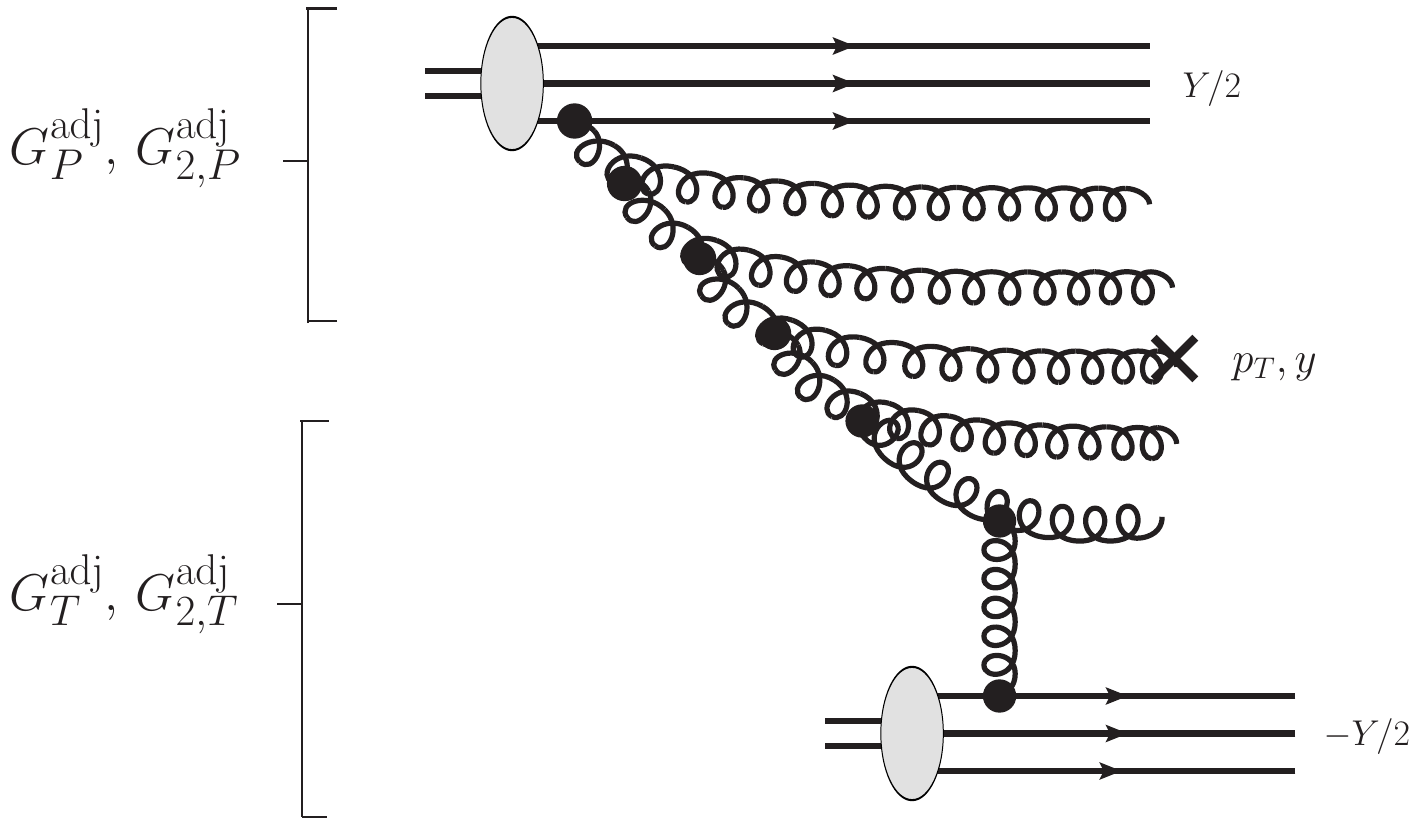}
		\caption{Schematic diagram for the process $pp\rightarrow g X$. The projectile and target protons have rapidity $Y/2$ and $-Y/2$, respectively, and the gluon jet is produced at rapidity $y$ in the mid-rapidity region with transverse momentum $p_T$.
			\label{pp_to_gluon}
		}
	\end{centering}
\end{figure}
At this point there has only been one proton-proton particle production cross-section developed in the small-$x$ formalism, and that is for the $pp\to g\,X$ process in the gluon-only limit \cite{kovchegov:2024aus}:
\begin{align}\label{KLprod}
	&\frac{d\Delta\sigma^{pp\to gX}}{d^2\hat{p}_T d \hat{y}}
	= \frac{C_F}{\pi^4\alpha_s}\frac{1}{s\,\hat{p}_T^2}\int d^2x_{\perp}\,
	e^{-i\hat{\bm{p}}_\perp\cdot\,\bm{x}_\perp}\, 
	\\
	&\hspace{2.5cm}\times (G_{P}^{\mathrm{adj}}\;G_{2,P}^{\mathrm{adj}})\,
	\begin{pmatrix}
		\frac{1}{4}\lvec{\bm{\nabla}}_{\perp}\cdot\rvec{\bm{\nabla}}_{\perp} & \lvec{\bm{\nabla}}_{\perp}^2+\lvec{\bm{\nabla}}_{\perp}\cdot\rvec{\bm{\nabla}}_{\perp} \\
		\rvec{\bm{\nabla}}_{\perp}^2+\lvec{\bm{\nabla}}_{\perp}\cdot\rvec{\bm{\nabla}}_{\perp} & 0
	\end{pmatrix}
	\begin{pmatrix}
		G_{T}^{\mathrm{adj}} \\
		G_{2,T}^{\mathrm{adj}}
	\end{pmatrix}
	, \notag
\end{align}
where the arguments of the dipoles are $G_p^{\mathrm{adj}}\Big(x_{\perp}^2,\sqrt{2}\,p_2^-\,\hat{p}_T\,e^{-\hat{y}}\Big)$ and $G_T^{\mathrm{adj}}\Big(x_{\perp}^2,\sqrt{2}\,p_1^+\,\hat{p}_T\,e^{\hat{y}}\Big)$. In the gluon-only limit protons will interact via gluon exchanges, and the final state produced gluon will be one radiated through this process. Shown in Fig.~\ref{gg_to_gluon} are what typical $gg \to g\,X$ interactions look like. In this figure and in Eq.~\eqref{KLprod} $\sqrt{s}$ is still the CM energy of the collision while $\hat{p}_T = |\hat{\bm{p}}_\bot|$ is the magnitude of the 2-vector transverse momentum of the produced gluon with rapidity $\hat{y}=\tfrac{1}{2}\ln(\hat{p}^-/\hat{p}^+)$\footnote{In addition to changing the nomenclature from $\Delta G(x,Q^2)$ to $\Delta g(x,Q^2)$ for the gluon hPDF, I also make the shift to denote vectors by using a bold typeface: $\underline{x}_{ij}\to \bm{x}_{ij}$}; in this scattering process the momentum fractions of the (massless) partons coming from either proton are defined as $x = \tfrac{\hat{p}_T}{\sqrt{s}}e^{\pm y}$, where the $\pm$ depends on which proton the parton in question came from. Before the terminology and choice of variables/symbols gets too far, I want to point out an important distinction between how the transverse dipole size is defined in the DIS and SIDIS cross-sections and it is defined in this $pp$ cross-section: in DIS and SIDIS the transverse dipole size was denoted as $x_{10}$ because there was only one dipole involved (constructed from Wilson lines at positions $0$ and $1$), however in $pp$ scattering there are two dipoles involved (one interacting with the target and one interacting with the projectile\footnote{The language of ``projectile" and ``target" are common in small-$x$ literature but is less common when discussion hadronic collisions in collinear factorization.}). Because the vector $\bm{x}_{ij}$ does not have foundational meaning in $pp$ collisions (there are no specific positions $\bm{x}_i$ or $\bm{x}_j$), and since the transverse dipole size is Fourier-conjugate to the produced gluon's transverse momentum, I denote the transverse dipole size in Eq.~\eqref{KLprod} as $x_\bot = |\bm{x}_\bot|\sim \mathcal{O}(1/\hat{p}_T)$. One must also keep in mind the fact that the dipoles are only physical when their transverse size remains above the smallest length scale, $x_{10}^2 \geq 1/ zs$ where $z$ is the longitudinal momentum fraction of the softest Wilson line in the dipole. In the case that there are two dipoles (and thus two length scales), I must ensure that $1/x_\bot^2 < \min\{z^{(T)}s, z^{(P)}s\}$, where $z^{(P,T)}$ are the longitudinal momentum fractions with respect to the projectile moving along the $-z\text{-}$axis or the target moving along the $+z\text{-}$axis. Similar to the parton momentum fractions, I define these longitudinal momentum fractions as
\begin{align}\label{pp_z}
	z^{(P)} \equiv \frac{p_T}{\sqrt{s}} e^{y}\,
	, \qquad
	z^{(T)} \equiv \frac{p_T}{\sqrt{s}} e^{- y}
	\: .
\end{align}
The two protons will be colliding with large momentum components $p_1^+$ and $p_2^-$ which correspond to the rapidities $-Y/2$ and $Y/2$; in the center-of-mass frame $p_1^+ = p_2^- = \sqrt{s/2}$. I show in Fig.~\ref{pp_to_gluon} what the overarching $pp\to g\,X$ interaction looks like and specify where the polarized dipole amplitudes fit in to the gluon production process. The polarized dipole amplitudes $G_{P(T)}^{\mathrm{adj}}$ and $G_{2,\,P(T)}$ encode the gluon cascade in the projectile $P$ (target $T$) proton, where one of the gluons is then tagged in the final state.

In the DLA accuracy at which I work, I can conveniently take the tagged final state gluon to be the detected jet within the framework of the narrow jet approximation (NJA) \cite{Kaufmann:2015hma}\footnote{The NJA means as it sounds; the produced jet is narrow, meaning that the hadron fragments are largely collimated and the radius of the jet is small. This applies to both $k_T$-type jets and cone-type jets, although they use different algorithms to calculate the jet radius \cite{blazey2000run,Ellis_1993}}. In the NJA, the single-inclusive jet production cross-section can be written analogously to the usual collinear factorization formula for single-inclusive hadron productions, with the typical hadronic fragmentation functions being replaced by jet functions $\mathcal{J}(z_J,\tfrac{Rp_T}{\mu_F},\mu_R)$, with jet radius R, final state factorization and renormalization scales $\mu_F$ and $\mu_R$, and transverse momentum fractions $z_J$ \cite{Kaufmann:2015hma}. At LO in $\alpha_s$ the jet functions is simply a delta function $\mathcal{J}\approx \delta(1-z_J)$, but there are also higher order terms (in $\alpha_s$); indeed some of these terms have logarithmic enhancements, but these logarithms are those that lead to DGLAP evolution of the jet function under $\ln(R)$ resummation. Logarithms of these kind may lead to better accuracy in future work, but they are definitively not the small-$x$ logarithms resummed in KPS-CTT evolution and the jet radii of the experimental data are not so small to make $\ln(R)$ resummation significant. These logarithms are outside the DLA power counting and I can simply take the leading delta function in place of the jet function.

The small-$x$ polarized $pp\to g\,X$ cross-section for small-$R$ jets then reads
\begin{align}\label{num_xperp}
	&\frac{d\Delta\sigma^{pp\to \mathrm{jet} X}}{d^2p_T dy} 
	= \frac{C_F}{\pi^4}\frac{1}{s\,p_T^2}
	\int d^2x_\bot\,
	e^{-i\bm{p}_\perp\cdot\, \bm{x}_\perp}\,
	\frac{1}{\alpha_s(1/x_{\perp}^2)}
	\\
	&\hspace{2cm}\times \big( G_{P}^{\mathrm{adj}}\;G_{2,P}^{\mathrm{adj}} \big)\,
	\begin{pmatrix}
		\frac{1}{4}\lvec{\bm{\nabla}}_{\perp}\cdot\rvec{\bm{\nabla}}_\perp & \lvec{\bm{\nabla}}_{\perp}^2+\lvec{\bm{\nabla}}_{\perp}\cdot\rvec{\bm{\nabla}}_{\perp} \\
		\rvec{\bm{\nabla}}_{\perp}^2+\lvec{\bm{\nabla}}_{\perp}\cdot\rvec{\bm{\nabla}}_{\perp} & 0
	\end{pmatrix}
	\begin{pmatrix}
		G_{T}^{\mathrm{adj}} \\
		G_{2,T}^{\mathrm{adj}}
	\end{pmatrix}
	\,, \notag
\end{align}
where now I have set $\hat{p}_T\to p_T$ and $\hat{y}\to y$, with $p_T$ and $y$ the transverse momentum and rapidity of the produced jet. The arguments of the dipoles are $G_P^{\mathrm{adj}}(x_\bot^2,\sqrt{s}\,p_T\,e^{-y})$ and $G_T^{\mathrm{adj}}(x_\bot^2,\sqrt{s}\,p_T\,e^{y})$. See that the running of the coupling is incorporated into Eq.~\eqref{num_xperp} through the transverse dipole size in a manner similar to its use in the evolution equations. In the large-$N_c$ limit (and as well in the variant employed here) I can make use of the relations $G^{\mathrm{adj}} = 4\widetilde{G}$ and $G_2^{\mathrm{adj}} = 2G_2$ \cite{Kovchegov:2018znm, Cougoulic:2022gbk} that connect the polarized dipole amplitudes in the adjoint and fundamental representations. I can substitute the adjoint dipoles in Eq.~\eqref{num_xperp} and perform the angular integral over $\bm{x}_\bot$\footnote{I have made use of the fact that the impact-parameter integrated polarized dipole amplitudes are azimuthally symmetry for the target proton.} and obtain a (semi-)final expression for the differential cross-section
\begin{align}\label{jet_numerator}
	&\frac{d\Delta\sigma^{pp\to \mathrm{jet} X}}{d^2p_T d y} 
	= \frac{8\,C_F}{\pi^3}\frac{1}{s\,p_T^2}\int\limits_0^{\infty} d x_{\perp} \frac{1}{\alpha_s(1/x_\perp^2)}\, x_{\perp} J_0(p_T\, x_{\perp}) 
	\\
	&\hspace{2cm} \times  \Bigl[2\,G_{2,P}\bm{\nabla}_{\perp}^2\widetilde{G}_T + 2\,(\bm{\nabla}_{\perp}^2\widetilde{G}_P)G_{2,T} + \frac{\partial}{\partial x_{\perp}}\widetilde{G}_P\frac{\partial}{\partial x_{\perp}}\widetilde{G}_T \notag \\
	&\hspace{4cm}+ 2\,\frac{\partial}{\partial x_{\perp}}G_{2,P}\frac{\partial}{\partial x_{\perp}}\widetilde{G}_T + 2\,\frac{\partial}{\partial x_{\perp}}\widetilde{G}_P\frac{\partial}{\partial x_{\perp}}G_{2,T}\Bigr]\,, \notag
\end{align}
where I have suppressed the arguments of the polarized dipole amplitudes for brevity. A notable observation of Eq.~\eqref{jet_numerator} is the fact that it has quadratic dependence on the polarized dipole amplitudes, whereas the evolution equations, hPDFs, and structure functions are all linear in the polarized dipole amplitudes. This is in fact a necessary cross-check that I recovered the quadratic nature because it is an essential feature of the $pp$ cross-section for gluon production at mid-rapidity; in DIS and SIDIS interactions gluons can only be produced from the (singular) nucleon target, whereas the inclusion of two nucleon targets allow for gluons to be produced from both (quadratic).

From here I make the logical next step to discretize the polarized cross-section, but first, I have to make a change to my logarithmic variables and enforce the crucial small-$x$ and dipole size constraints. Substituting Eq.~\eqref{pp_z} into Eq.~\eqref{log_vars} allows me to define the similar $\eta^{(P,T)}$ where $z$ has been replaced with $z^{(P,T)}$, and likewise I replace $x_{ij}$ with $x_\bot$ and end up with $s_{10}$; the ``10" subscript still does not have direct correspondence with any positions in the jet cross-section, but I keep the convention just for simplicity since the dipoles share the same $s_{10}$-$\eta$ grid regardless of how those variables are calculated. For bookkeeping purposes I will write out below the result of applying the change of variables to the derivatives that appear in Eq.~\eqref{jet_numerator}:
\begin{subequations}
	\begin{align}
		\frac{\partial}{\partial x_{\perp}} 
		&= -\Lambda \sqrt{\frac{2N_c}{\pi}} \exp\!\left(\sqrt{\frac{\pi}{2N_c}} s_{10}\right) \frac{\partial}{\partial s_{10}}\,,
		\\
		\bm{\nabla}_{\perp}^2 
		&= \frac{1}{x_{\perp}}\frac{\partial}{\partial x_{\perp}} + \left( \frac{\partial}{\partial x_{\perp}} \right)^2
		= \Lambda^2\frac{2N_c}{\pi} 
		\exp\!\left(\sqrt{\frac{2\pi}{N_c}} s_{10}\right) 
		\left( \frac{\partial}{\partial s_{10}} \right)^2 \,,
	\end{align}
\end{subequations}

Strictly speaking, all of these changes result in the following expression
\begin{align}\label{pJet_GA_Bessel}
	&\frac{d\Delta\sigma^{pp\to \mathrm{jet} X}}{d^2 p_T\,d y} 
	= \frac{16\,C_F}{\pi^3}\frac{1}{p_T^2\,s}\sqrt{\frac{N_c}{2\pi}}\,\int\limits_{0}^{\mathrm{min}\{\eta^{(P)},\eta^{(T)}\}}
	d s_{10}\,
	\frac{1}{\alpha_s(s_{10})}\, J_0\biggl(\frac{p_T}{\Lambda}
	\exp\Big(-\sqrt{\frac{\pi}{2 N_c}} s_{10}\Big)\biggr)
	\\
	&\hspace{2.5cm}\times \Bigl\{2\,G_2(s_{10},\eta^{(T)})\widetilde{G}^{(\prime\prime)}(s_{10},\eta^{(P)}) + 2\,\widetilde{G}^{(\prime\prime)}(s_{10},\eta^{(T)})G_2(s_{10},\eta^{(P)}) \notag \\ &\hspace{3.5cm}+\widetilde{G}^{(\prime)}(s_{10},\eta^{(T)})\widetilde{G}^{(\prime)}(s_{10},\eta^{(P)})
	+2\,G_2^{(\prime)}(s_{10},\eta^{(T)})\widetilde{G}^{(\prime)}(s_{10},\eta^{(P)}) \notag \\ 
	&\hspace{3.5cm}+ 2\,\widetilde{G}^{(\prime)}(s_{10},\eta^{(T)})G_2^{(\prime)}(s_{10},\eta^{(P)})\Bigr\}
	\bigg|_{\eta^{(P,T)}=\sqrt{\frac{N_c}{2\pi}}\left(\ln\frac{p_T\sqrt{s}}{\Lambda^2}\pm y\right)}\,, \notag
\end{align}
where $G^{('),('')}$ refer to the first and second derivatives of the dipole amplitudes with respect to $s_{10}$, however this is not its true final form just yet. I would like to point out that the sharp upper and lower limits are a result of the above assumptions and approximations; the cutoff $s_{10} < \min\{\eta^{(P,T)}\}$ is a feature of the DLA, and the cutoff $s_{10} >0$ is due to the restriction that $x_\bot < 1/\Lambda$. These sharp cutoffs on the integral cause oscillation artifacts from the $J_0$ Bessel function that are not intended features of the small-$x$ evolution in the DLA. To avoid these artifacts I impose a lower limit on its dynamic argument's integral $s_{10} > \sqrt{\frac{2N_c}{\pi}}\ln(p_T/\Lambda)$, and set $J_0 = 1$. I ultimately achieve a polarized cross-section that is ready for global analysis (at least before discretizing it):
\begin{align}\label{pJet_GA}
	&\frac{d\Delta\sigma^{pp\to \mathrm{jet} X}}{d^2 p_T\, d y}
	= \frac{16\,C_F}{\pi^3}\frac{1}{p_T^2\,s}\sqrt{\frac{N_c}{2\pi}}\,\int\limits_{\sqrt{\frac{2N_c}{\pi}}\ln(p_T/\Lambda)}^{\mathrm{min}\{\eta^{(P)},\eta^{(T)}\}}d s_{10}\,\frac{1}{\alpha_s(s_{10})} \\
	&\hspace{2.5cm}\times \Bigl\{2\,G_2(s_{10},\eta^{(T)})\widetilde{G}^{(\prime\prime)}(s_{10},\eta^{(P)}) + 2\,\widetilde{G}^{(\prime\prime)}(s_{10},\eta^{(T)})G_2(s_{10},\eta^{(P)})\notag \\
	&\hspace{3.5cm}+\widetilde{G}^{(\prime)}(s_{10},\eta^{(T)})\widetilde{G}^{(\prime)}(s_{10},\eta^{(P)}) +2\,G_2^{(\prime)}(s_{10},\eta^{(T)})\widetilde{G}^{(\prime)}(s_{10},\eta^{(P)})\notag \\
	&\hspace{3.5cm}+ 2\,\widetilde{G}^{(\prime)}(s_{10},\eta^{(T)})G_2^{(\prime)}(s_{10},\eta^{(P)})\Bigr\} \bigg|_{\eta^{(P,T)}=\sqrt{\frac{N_c}{2\pi}}\left(\ln\frac{p_T\sqrt{s}}{\Lambda^2}\pm y\right)}\,. \notag
\end{align}
The discretization of this cross-section is trivial since I simply approximate the one integral by a Riemann sum just as with the evolution equations, structure functions, and hPDFs. The only non-trivial part is the novel use of numerical derivatives that have presented themselves in the $pp$ cross-section. I use the $\mathcal{O}(\Delta^3)$ backward finite difference approximation for the first and second derivatives:
\begin{subequations}
	\begin{align}
		f'(x)&\approx \frac{1}{\Delta} \Big( \tfrac{11}{6}f(x) - 3f(x-\Delta) + \tfrac{3}{2}f(x-2\Delta) - \tfrac{1}{3}f(x-3\Delta)\Big) \,,
		\\
		f''(x)&\approx \frac{1}{\Delta^2} \Big( \tfrac{35}{12}f(x) - \tfrac{26}{3}f(x-\Delta) + \tfrac{19}{2}f(x-2\Delta) - \tfrac{14}{3}f(x-3\Delta) + \tfrac{11}{12}f(x-4\Delta) \Big)\,.
	\end{align}
\end{subequations}
The use of numerical derivatives is unavoidable, but in my testing I find that numerical derivatives' uncertainties are more sensitive to the step size $\Delta$ than numerical integrals. The analysis performed in \textbf{Chapter}~\ref{pheno_1} achieved a numerical uncertainty well below $1\%$, however in this analysis I must balance the same computational resources (i.e. the same step size) but with the numerical integrals and derivatives and achieve a numerical uncertainty of approximately $2.5\%$; higher-order finite difference approximations showed negligible improvements beyond this.

\subsection{$\boldsymbol{\mathrm{Results~I\!:~ Data~vs~Theory}}$}

Before diving into the data, I must first make a note regarding some updates that will retroactively affect the large-$N_c\&N_f$ analysis of DIS and SIDIS \cite{Adamiak:2023yhz} discussed in \textbf{Chapter}~\ref{pheno_1} and subsequently affect this analysis as well. I established in \textbf{Chapter}~\ref{Observables} that the longitudinal spin asymmetries use the small-$x$ polarized cross-sections for the numerators but that they also require input from unpolarized structure functions and fragmentation functions for their denominators. In \textbf{Chapter}~\ref{pheno_1} I used the unpolarized functions from the DGLAP-based JAM analysis of Ref.~\cite{Cocuzza:2022jye}. In the time between that analysis was conducted and this one took place it was found that once source of data used in extracting those functions accidentally labeled the systematic uncertainty as a renormalization uncertainty \cite{Anderson:2024evk}. The data set has since been corrected and the analysis \cite{Anderson:2024evk} extracted updated fragmentation functions and imposed new kinematic cutoffs (for the affected SIDIS dataset), both of which I use in this analysis \cite{Adamiak:2025dpw}. I also took these updates and repeated the large-$N_c\&N_f$ analysis of DIS and SIDIS data and updated my various extractions/predictions, so any comparisons to the ``previous" analysis (or ``the analysis from \textbf{Chapter}~\ref{pheno_1}") that follow are understood to be comparing up-to-date analyses as far as fragmentation functions and SIDIS data cuts are concerned.

We already discussed the spin asymmetries in \textbf{Chapter}~\ref{chapter_3}, while in \textbf{Chapter}~\ref{pheno_1} we discussed the cuts on DIS and SIDIS data. As a brief review, and to attend to the updates discussed in the paragraph above, the cuts are as follows: the small-$x$ polarized DIS and SIDIS data occupy the ranges $5\times 10^{-3} < x < 0.1$ and $1.69~\mathrm{GeV}^2 < Q^2 < 10.4~\mathrm{GeV}^2$, and in (for SIDIS data) $W^2_{\mathrm{SIDIS}} > 10~\mathrm{GeV}^2$ and $z_h < 0.8$. I also set $\Lambda = 1~\mathrm{GeV}$ which roughly characterizes the size of the proton; a change to $\Lambda$ is possible if desired and would simply require a redefinition of the $c$ parameters in Eq.~\eqref{IC_params}. For polarized $pp$ data I still enforce a small-$x$ cutoff through $x_{T,P} = \tfrac{p_T}{\sqrt{s}}e^{\pm y} < 0.1$. However, I must be careful to select data that are completely compatible with this small-$x$ cutoff; experimental data inherently presents itself over discrete bins, which in this case are bins of rapidity and transverse momentum. I take the conservative approach and only allow data points in which both $x_T$ and $x_P$ are small across the entire $y$ and $p_T$ range of the given bin (as opposed to, say, the central value of the bins). I still employ the cutoff of $x < x_0 \equiv 0.1$, and tested the applicability of the small-$x$ polarized theory to describe data at large values of $x_0$ and found that the $\chi^2_{\mathrm{red}}$ systematically degraded as the cutoff increased. For cutoffs of $x_0 = 0.125$ and $x_0 = 0.15$ the $\chi^2_{\mathrm{red}}$ was roughly 2 times and 3 times larger than that of the $x_0 = 0.1$ fit that I will discuss below. 

Furthermore, I must take care that any other approximations are tended to with appropriate cuts on $A_{LL}^{\mathrm{jet}}$ data. The NJA approximation is only applicable for jet radii $R\leq 0.7$ \cite{Jager:2004jh}, finding agreement to within $5\text{-}10\%$ with the Monte Carlo simulation from Ref.~\cite{deFlorian:1998qp}; this result encouraged me to disregard one dataset of $R=1$ jets. Additionally, the DGLAP-based JAM analysis from which the unpolarized $pp$ jet production cross-section is taken is only valid for jets with $p_T \geq 8~\mathrm{GeV}$, so this restriction is imposed on $A_{LL}^{\mathrm{jet}}$ as well. While on the subject I also make the note that, while the polarized cross-section has been taken in the gluon-only limit (and supported by the choice to use large-$N_c^{+q}$ evolution), I do not make any such gluon-only modifications the unpolarized jet production cross-section; since the purposed for completing this analysis is specifically to test the impact of $pp$ data on the small-$x$ hPDF extractions, it was decided that artificially removing the equally important quark contributions from the unpolarized cross-section would not aid in this effort. DIS, SIDIS, and $pp$ unpolarized cross-sections are therefore treated consistently and use their unaltered JAM-DGLAP calculations \cite{Cocuzza:2022jye, Zhou:2022wzm}. In the end, there are 122 polarized DIS data points, 104 polarized SIDIS data points, and 14 polarized $pp$ data points, making a combined total of $N_{\mathrm{pts}} = 240$.

I use the same set of DIS and SIDIS data as in \textbf{Chapter}~\ref{pheno_1_data}: measurements from COMPASS \cite{Alekseev:2010hc, Adolph:2015saz, Adolph:2016myg}, EMC \cite{Ashman:1989ig}, HERMES \cite{Ackerstaff:1997ws, Airapetian:2007mh}, SLAC \cite{Anthony:1996mw, Abe:1997cx, Abe:1998wq, Anthony:1999rm, Anthony:2000fn}, and SMC \cite{Adeva:1998vv, Adeva:1999pa} for polarized DIS data, and measurements from COMPASS \cite{COMPASS:2009kiy, COMPASS:2010hwr}, HERMES \cite{HERMES:1999uyx, HERMES:2004zsh}, and SMC \cite{SpinMuon:1998eqa} for polarized SIDIS data. The $A_{LL}^{\mathrm{jet}}$ data come from four different STAR collaboration measurements from Refs.~\cite{STAR:2014wox, STAR:2019yqm, STAR:2021mfd, STAR:2021mqa}. While the DIS and SIDIS data are all the same as the previous analysis, I expect the fit to individual data sets to be slightly different since the evolution equations are different and extra data are being fitted simultaneously. In Tables \ref{t:Chi2_DIS_2}, \ref{t:Chi2_SIDIS_2}, and \ref{t:Chi2_pJet} I give a breakdown of each dataset and their respective $\chi^2_{\mathrm{red}}$; these measurements are a result of 398 replicas that passed a ``filter" of $\chi^2_{\mathrm{red}} \leq 10$ per reaction (DIS, SIDIS, and $pp$) and were fitted to data twice to reach convergence. The previous analysis used a less restrictive replica filter, passing them through so long as the overall $\chi^2_{\mathrm{red}} \leq 10$, but since there are significantly more DIS and SIDIS data than $pp$ data I needed a more restrictive filter this time in order to make sure $pp$ data was being weighted properly. Fitting to data twice means that I did an initial fit to data and then took the final replicas of that fit as a new initial ``guess" for a subsequent fit; this process ensures that the results converge. Testing showed a $1.04\%$ improvement to the $\chi^2_{\mathrm{red}}$ going from a first fit to a second fit, but then showed a negligible $0.18\%$ improvement for a subsequent third fit, signifying that the results had indeed converged; the results that follow are those that came from the second-iteration fit. 
\begin{table}[h!]
	\caption{Summary of the fit to polarized DIS data, separated into $A_1$  (left) and $A_{\parallel}$ (right), along with the  $\chi^2_{\mathrm{red}}$ for each data set.}
	\label{t:Chi2_DIS_2}
	\vspace{0.4cm}
	\begin{tabular}{l|c|c|c} 
		\hline
		{$\boldsymbol{\mathrm{Dataset}~(A_1)}$} & 
		~{$\boldsymbol{\mathrm{Target}}$}~ &
		~$\boldsymbol{N_\mathrm{pts}}$~ & 
		~$\boldsymbol{\chi^2_{\mathrm{red}}}$~  
		\\ \hline     
		COMPASS   \cite{Alekseev:2010hc} &
		$p$ &
		$5$ &
		$0.77$  \\ \hline
		COMPASS   \cite{Adolph:2015saz} &
		$p$ &
		$17$ &
		$0.93$   \\ \hline
		COMPASS  \cite{Adolph:2016myg} &
		$d$ &
		$5$ &
		$0.34$  \\ \hline
		EMC  \cite{Ashman:1989ig} &
		$p$ &
		$5$ &
		$0.23$   \\ \hline
		HERMES   \cite{Ackerstaff:1997ws} &
		$n$ &
		$2$ &
		$1.11$ \\ \hline
		SLAC (E142) \cite{Anthony:1996mw} &
		${}^3 \mathrm{He}$ &
		$1$       & 
		$1.47$    \\ \hline
		SMC  \cite{Adeva:1998vv, Adeva:1999pa} &
		$p$ &
		$6$ &
		$1.26$  \\ 
		&
		$p$ &
		$6$ &
		$0.43$  \\
		&
		$d$ &
		$6$ &
		$0.65$  \\
		&
		$d$ &
		$6$ &
		$2.13$  	\\ \hline\hline
		{$\boldsymbol{\mathrm{Total}}$} &    & 59 & 0.90 \\ \hline
	\end{tabular}
	\qquad\;\;
	\begin{tabular}{l|c|c|c} 
		\hline
		{$\boldsymbol{\mathrm{Dataset}~(A_{\parallel})}$}& 
		~{$\boldsymbol{\mathrm{Target}}$}~ &
		~$\boldsymbol{N_\mathrm{pts}}$~ & 
		~$\boldsymbol{\chi^2_{\mathrm{red}}}$~  
		\\ \hline
		HERMES    \cite{Airapetian:2007mh} &
		$p$ &
		$4$ &
		$1.47$  \\ 
		&
		$d$ &
		$4$ &
		$1.00$  \\ \hline
		SLAC (E143) \cite{Abe:1998wq} &
		$p$ &
		$9$ &
		$0.55$    \\
		&
		$d$ &
		$9$ &
		$1.01$  	\\ \hline
		SLAC (E154) \cite{Abe:1997cx} &
		${}^3 \mathrm{He}$ &
		$5$       &
		$0.69$     \\ \hline
		SLAC (E155) \cite{Anthony:1999rm} &
		$p$ &
		$16$ &
		$1.07$ 	\\ 
		&
		$d$ &
		$16$ &
		$1.57$      \\ \hline\hline
		{$\boldsymbol{\mathrm{Total}}$} &   & 63 & 1.10 \\ \hline
	\end{tabular}
	\label{t:Chi2_DIS_Apa_2}
\end{table}

\begin{table}[t!]
	\begin{center}
		\caption{Summary of the polarized  SIDIS data on $A_1^h$ included in the fit, along with the $\chi^2/N_{\rm pts}$ for each data set.}
		\label{t:Chi2_SIDIS_2}
		\vspace{0.4cm}
		\begin{tabular}{l|c|c|c|c} 
			\hline
			{$\boldsymbol{\mathrm{Dataset}~(A_1^h)}$}& 
			~{$\boldsymbol{\mathrm{Target}}$}~ &
			~{$\boldsymbol{\mathrm{Tagged~Hadron}}$}~ &
			~$\boldsymbol{N_\mathrm{pts}}$~ & 
			~$\boldsymbol{\chi^2_{\mathrm{red}}}$~  
			\\ \hline
			COMPASS \cite{COMPASS:2009kiy}  
			&
			$d$ &
			$\pi^+$ &
			$5$   &
			$0.63$ 	\\ 
			&
			$d$  &
			$\pi^-$ &
			$5$   &
			$0.83$ \\
			&
			$d$   &
			$h^+$ &
			$5$ &
			$1.01$ \\ 
			&
			$d$ &
			$h^-$  &
			$5$  &
			$1.02$ 		\\ 
			&
			$d$     &
			$K^+$ &
			$5$  &
			$1.60$ 		\\
			&
			$d$ &
			$K^-$ &
			$5$ &
			$0.71$ 	\\ \hline
			COMPASS \cite{COMPASS:2010hwr} &
			$p$ &
			$\pi^+$ &
			$5$ &
			$1.94$ 		\\ 
			&
			$p$   &
			$\pi^-$ &
			$5$ &
			$1.18$ \\ 
			&
			$p$  &
			$K^+$ &
			$5$   &
			$0.46$ 	\\
			&
			$p$ &
			$K^-$ &
			$5$   &
			$0.23$ 		\\ \hline	
			HERMES \cite{HERMES:1999uyx} &
			${}^3 \mathrm{He}$  &
			$h^+$  &
			$2$ &
			$0.55$ 	\\ 
			&
			${}^3 \mathrm{He}$ &
			$h^-$  &
			$2$   &
			$0.29$ 	\\ \hline
			HERMES  \cite{HERMES:2004zsh} &
			$p$ &
			$\pi^+$ &
			$2$ &
			$2.75$ \\ 
			&
			$p$ &
			$\pi^-$ &
			$2$  &
			$0.00$    \\ 
			&
			$p$ &
			$h^+$ &
			$2$ &
			$1.25$  \\
			&
			$p$ &
			$h^-$ &
			$2$  &
			$0.19$  \\ 
			&
			$d$  &
			$\pi^+$ &
			$2$ &
			$0.58$ 	 \\
			&
			$d$ &
			$\pi^-$ &
			$2$ &
			$1.23$ 	 \\
			&
			$d$ &
			$h^+$  &
			$2$  &
			$3.03$   \\ 
			&
			$d$  &
			$h^-$ &
			$2$  &
			$1.24$ 	 \\ 
			&
			$d$ &
			$K^+$ &
			$2$  &
			$0.82$ 	 \\ 
			&
			$d$  &
			$K^-$ &
			$2$  &
			$0.25$ 	\\ 
			&
			$d$  &
			$K^+ + K^-$ &
			$2$  &
			$0.36$ 	\\ \hline
			SMC  \cite{SpinMuon:1998eqa} &
			$p$ &
			$h^+$ &
			$7$ &
			$1.22$  \\ 
			&
			$p$ &
			$h^-$ &
			$7$ &
			$1.41$  		\\ 
			&
			$d$ &
			$h^+$ &
			$7$ &
			$0.84$  \\ 
			&
			$d$ &
			$h^-$ &
			$7$ &
			$1.52$   \\ \hline\hline
			{$\boldsymbol{\mathrm{Total}}$} &    &    & 104 & 1.04 \\ \hline
		\end{tabular}
	\end{center}
\end{table}

\begin{figure}[h!] 
	\begin{centering}
		\includegraphics[width=0.8\textwidth]{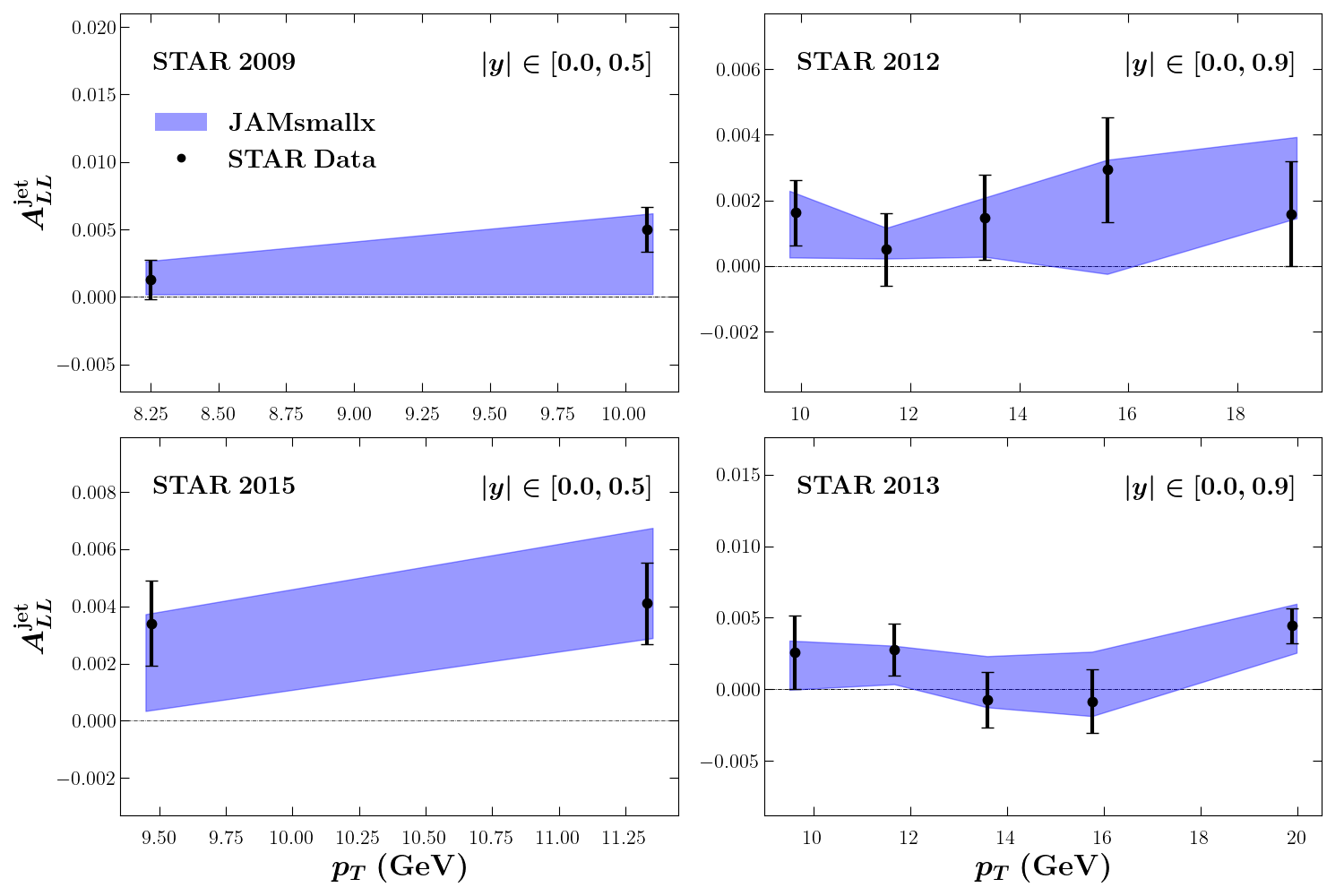}
		\caption{Experimental jet asymmetry $A_{LL}^{\mathrm{jet}}$ data from STAR~\cite{STAR:2014wox, STAR:2019yqm, STAR:2021mfd, STAR:2021mqa} (black circles) as a function of $p_T$ in the kinematic range $p_T \geq 8$~GeV, $-1 < y < +1$, and $x_{T,P} < 0.1$, compared with the theoretical asymmetries (blue bands, 68\% confidence intervals).
			\label{pjet_data_vs_thy_2smx}
		}
	\end{centering}
\end{figure}

\begin{table}[t!]
	\begin{center}
		\caption{Summary of polarized $pp$ data for $A_{LL}^{\mathrm{jet}}$ along with their respective  $\chi^2_{\mathrm{red}}$ values.}
		\label{t:Chi2_pJet}
		\vspace{0.4cm}
		\begin{tabular}{l|c|c} 
			\hline
			{$\boldsymbol{\mathrm{Dataset}~(A_{\mathrm{LL}}^{\mathrm{jet}}})$}& 
			~$\boldsymbol{N_\mathrm{pts}}$~ & 
			~$\boldsymbol{\chi^2_{\mathrm{red}}}$~  
			\\ \hline
			STAR    \cite{STAR:2014wox} &
			$2$ &
			$0.60$  \\ \hline
			STAR \cite{STAR:2019yqm} &
			$5$ &
			$0.30$    \\ \hline
			STAR \cite{STAR:2021mfd} &
			$2$       &
			$0.55$     \\ \hline
			STAR \cite{STAR:2021mqa} &
			$5$ &
			$0.24$ 	\\ \hline\hline
			{$\boldsymbol{\mathrm{Total}}$} & 14 & 0.36 \\ \hline
		\end{tabular}
	\end{center}
\end{table}
In summary, the fit to DIS and SIDIS data is the same as before, $\chi^2_{\mathrm{red}} = 1.02$ (so the comparison plots are not shown here), while the fit to $pp$ data specifically sees $\chi^2_{\mathrm{red}}=0.36$, whose rather low value can be attributed to the small number of data points. The data vs theory plots for $pp$ data are given in Fig.~\ref{pjet_data_vs_thy_2smx}. The overall fit, as measured from the midpoint of all replicas, yields an impressive $\chi^2_{\mathrm{red}} = 0.98$, which means I have successfully integrated $pp$ data into the analysis without degrading the fit to DIS and SIDIS data. 

\subsection{$\boldsymbol{\mathrm{Results~I\!:~ Impact~of}~pp~\mathrm{data}}$}

\begin{figure}[t]
	\begin{centering}
		\includegraphics[width=\textwidth]{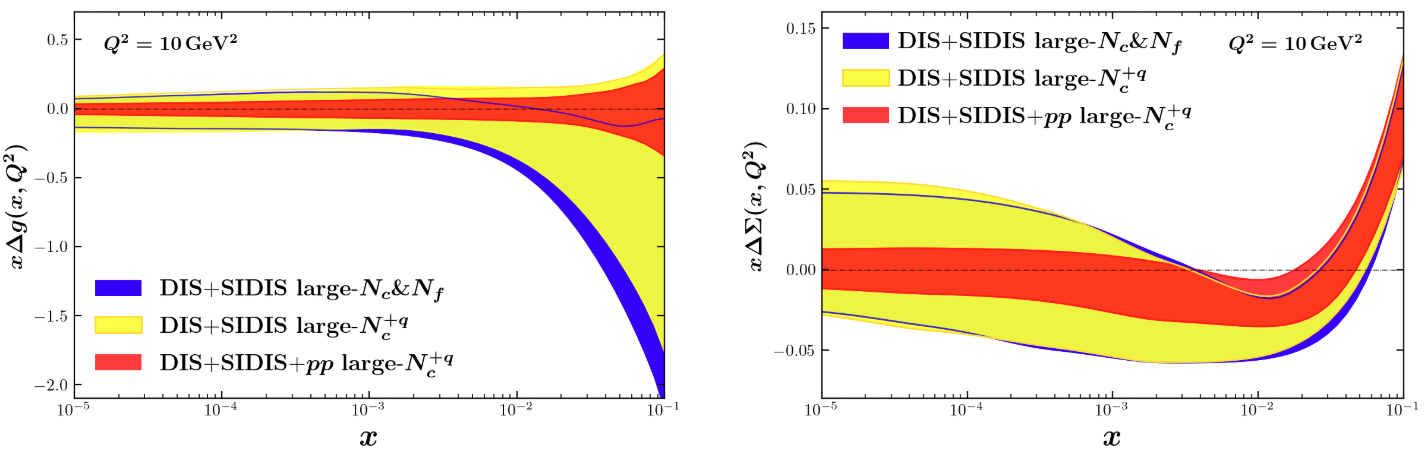}
		\caption{(Left) The gluon hPDF $x\Delta g(x,Q^2)$ and (Right) the flavor singlet quark hPDF combination $x\Delta\Sigma(x,Q^2)$, both as a function of $x$ at $Q^2 = 10$~GeV$^2$. The bands represent 68\% confidence intervals from three analyses:~DIS+SIDIS data with large-$N_c\&N_f$ evolution (blue), DIS+SIDIS data with large-$N_c^{+q}$ evolution (yellow), and DIS+SIDIS+$pp$ data with large-$N_c^{+q}$ evolution equations (red). 
			\label{DeltaG_DeltaSigma_pp}
			1    }
	\end{centering}
\end{figure}

\begin{figure}[h!]
	\begin{centering}
		\includegraphics[width=\textwidth]{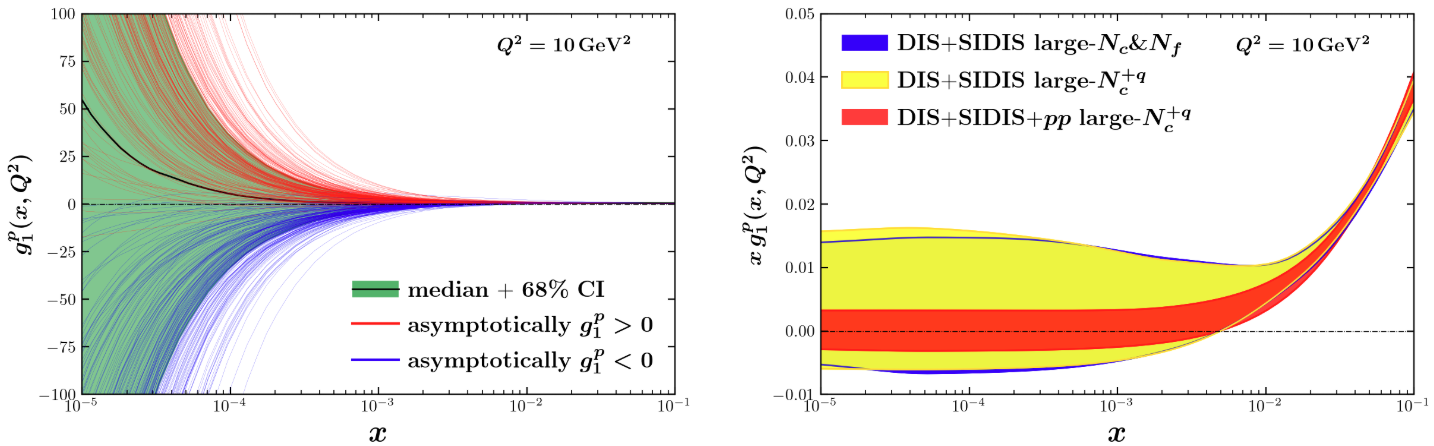}
		\caption{(Left) $g_1^p(x, Q^2)$ structure function versus $x$ at $Q^2=10\,{\rm GeV^2}$ for the DIS+SIDIS+$pp$ analysis with the large-$N_c^{+q}$ evolution equations. Also shown are the replicas that grow asymptotically positive (red) and negative (blue) as $x\to 0$ (measured at $x=10^{-7}$). The black curve is the median value of all replicas while the green band shows their 68\% confidence interval.  
			(Right) $x \,g_1^p(x,Q^2)$ versus $x$ at $Q^2 = 10~\mathrm{GeV}^2$ for the DIS+SIDIS data with large-$N_c\&N_f$ evolution (blue), DIS+SIDIS data with large-$N_c^{+q}$ evolution (yellow), and DIS+SIDIS+$pp$ data with large-$N_c^{+q}$ evolution  (red). The bands also reflect the 68\% confidence intervals for their respective analyses.
			\label{g1_band_comparison_pp}
		}
	\end{centering}
\end{figure}

The results of my analysis for the small-$x$ extractions of the hPDFs $\Delta \Sigma(x,Q^2)$ and $\Delta g(x,Q^2)$ and the $g_1^p(x,Q^2)$ structure functions as functions of $x$ are shown in Figs.~\ref{DeltaG_DeltaSigma_pp} and \ref{g1_band_comparison_pp}. The uncertainty bands, calculated as the $68\%$ confidence intervals, for the helicity distributions are scaled by $x$ so we can see the behaviors at both the larger-$x$ boundary $x\lesssim0.1$ and the smaller-$x$ region as $x \to 0$. These plots compare three important analyses to control for the change in evolution equations and isolate the impact resulting from the inclusion of $pp$ data, all computed for $Q^2 = 10~\mathrm{GeV}^2$; shown in blue is a representation of the previous analysis' fit to DIS and SIDIS data using the large-$N_c\&N_f$ evolution equations, while shown in yellow is an intermediate analysis where I fit to the same DIS and SIDIS data using the new large-$N_c^{+q}$ evolution equations, and lastly in red is the current analysis of DIS, SIDIS, and $pp$ data using the large-$N_c^{+q}$ evolution equations. The left panel of Fig.~\ref{g1_band_comparison_pp} shows the new distribution of $g_1^p$ replicas from the fit to DIS, SIDIS, and $pp$ data, while the right panel shows the three analysis uncertainty bands for the $x$-scaled structure functions to better observe the small-$x$ impact and qualitatively confirm that the uncertainties agree in the data-range of $5\times10^{-3} < x< 0.1$ (which they do).

As an important cross-check, I confirm that the change in evolution equations does not degrade the goodness-of-fit; the DIS+SIDIS analysis using large-$N_c\&N_f$ evolution has $\chi^2_{\mathrm{red}} = 1.017$, which sees a statistically insignificant reduction to $\chi^2_{\mathrm{red}}$ for a DIS+SIDIS analysis using large-$N_c^{+q}$. Since both analyses are capable of describing data well, I have to determine how the change in evolution equations influenced the $C$-even quark hPDFs, the gluon hPDF, and the $g_1^p$ structure function. 

Notably, I find that the change in evolution equations does not affect the observation made in \textbf{Chapter}~\ref{pheno_1} that hPDF basis functions constructed from $\widetilde{G}^{(0)}$ and $G_2^{(0)}$ initial conditions make the largest contributions at small $x$. The small-$x$ dominance of $\widetilde{G}$ and $G_2$ result in two important phenomenological features: ($\bm{1}$) the small-$x$ asymptotics of $g_1^p$, $\Delta\Sigma$, and $\Delta g$ are correlated, and ($\bm{2}$) the small-$x$ uncertainties (including the asymptotic bimodality) are controlled by the data's relative constraining power on $\widetilde{G}$ and $G_2$ polarized dipole amplitudes \cite{Adamiak:2023yhz}. Both of these features remain true upon changing the evolution equations, but their influence on the fits are subtly different. The overall replica split of the asymptotic solutions for $g_1^p$ has a $69\%$ preference for positive solutions when analyzing DIS and SIDIS data with the large-$N_c^{+q}$ evolution equations, which is comparable to the $70\%$ preference for positive solutions when using the large-$N_c\&N_f$ evolution equations \cite{Adamiak:2023yhz}. This split is very similar, but what had a more substantial change is the asymptotic adoption of those replicas; as seen from Figs.~\ref{DeltaG_DeltaSigma_pp} and the right panel of Fig.~\ref{g1_band_comparison_pp}, the small-$x$ uncertainties from the DIS+SIDIS analysis using large-$N_c^{+q}$ evolution are systematically larger than those from the analysis using large-$N_c\&N_f$ evolution. Since the split in replicas is largely unchanged, and the uncertainties agree in the region of $x$ where data is being fitted, $5\times10^{-3} < x< 0.1$, this implies that the large-$N_c^{+q}$ replicas adopt their asymptotics slightly faster once the data runs out and $x < 5\times 10^{-3}$. Additionally, the gluon hPDF (Fig.~\ref{DeltaG_DeltaSigma_pp} left) gets a positive shift near the boundary $x\lesssim0.1$, no longer making it negative-definite. Observe also that the yellow bands in Figs.~\ref{DeltaG_DeltaSigma_pp} and \ref{g1_band_comparison_pp} have a positive mean value as $x\to0$ for $g_1^p$, which is met with a positive mean value as $x\to0$ for $\Delta\Sigma$ and a negative mean value as $x\to0$ for $\Delta g$, agreeing with the (anti-)correlation \cite{Adamiak:2023yhz} of Eq.~\eqref{e:hpdf_correlation}.

To summarize, the change in evolution equations result in slightly larger uncertainty bands across the board, a positive shift in the gluon hPDF $\Delta g(x,Q^2)$ near $x = 0.1$ (while still retaining its overall preference for negative solutions), and the small-$x$ features of hPDF anticorrelation and $\widetilde{G}$ and $G_2$ dominance are unaffected. Any further comparisons below are then strictly caused by the integration of polarized $pp$ data.

Taking a closer look at the $g_1^p(x,Q^2)$ structure function in the left panel of Fig.~\ref{g1_band_comparison_pp} we see that while it looks similar to Fig.~\ref{Plot_g1} from \textbf{Chapter}~\ref{pheno_1} it has some important differences. See immediately that the inclusion of $pp$ data (and its sensitivity to $\widetilde{G}$ and $G_2$) did not, unfortunately, break the asymptotic bimodality as there is still a split of asymptotically positive and negative solutions. The split of those solutions has changed, with a more even split and only slight $54\%$ preference for positive solutions; this even split results in an uncertainty band centered near zero. Crucially, I note that the magnitude of the uncertainty bands has also changed; comparing the $x\,g_1^p(x,Q^2)$ solutions in the right panel of Fig.~\ref{g1_band_comparison_pp} (also noticeable in Fig.~\ref{DeltaG_DeltaSigma_pp}) we see that there has been a dramatic reduction in the uncertainty bands as a result of the constraining power of $pp$ data on the $\widetilde{G}^{(0)}$ and $G_2^{(0)}$ initial conditions. Quantitatively, I find that when comparing apples-to-apples large-$N_c^{+q}$ analysis, the uncertainty band when fitting to DIS, SIDIS, and $pp$ data is reduced by $73\%$ compared to the fit to only DIS and SIDIS data. This $73\%$ reduction was measured at $x=10^{-7}$\footnote{I only compute extractions down to $x=10^{-7}$ because this is the approximate prediction for $x_{\mathrm{sat}}$, the value of $x$ at which saturation effects set in at a given $Q^2$. This value was determined by using the Golec-Biernat Wusthoff parametrization for the proton saturation scale \cite{Golec-Biernat:1998zce} at $Q^2=10~\mathrm{GeV}^2$: I required that $Q_{s,p}^2 (x_{\rm sat}) = Q^2=10~\mathrm{GeV}^2$ and solved for $x_{\rm sat}$, obtaining $x_{\rm sat} \approx 10^{-7}$.}, but I also measured the reduction to be $69\%$ at $x = 10^{-4}$, the lowest $x$ value that the EIC data is expected to produced. These reductions in uncertainty are a direct result of better constraints on the $\widetilde{G}^{(0)}$ and $G_2^{(0)}$ parameters from $pp$ data, evidenced by the standard deviation of the parameters $\widetilde{a}$, $\widetilde{b}$, and $\widetilde{c}$ being reduced on average by $52\%$, and the standard deviation of the parameters $a_2$, $b_2$, and $c_2$ being reduced on average by $23\%$ compared to their standard deviations in the large-$N_c^{+q}$ analysis of only DIS and SIDIS data.

This same $73\%$ reduction of the small-$x$ uncertainties (and also the general centering around zero) are also found in the hPDFs $\Delta\Sigma(x,Q^2)$ and $\Delta g(x,Q^2)$, as seen from Fig.~\ref{DeltaG_DeltaSigma_pp}. Focusing on the left panel, the gluon hPDF has not only been better constrained at small $x$, but there has been a similar reduction in uncertainty across all values of $x < 0.1$. This reduction in uncertainty also results in positivity being satisfied ($\Delta g(x,Q^2) < |G(x,Q^2)|$. The largest change occurs near the boundary $0.01 \lesssim x < 0.1$, where a large number of negative solutions have been removed (compared to the DIS+SIDIS analysis), suggesting that polarized $pp$ discourages large $|\Delta g|$. The helicity distributions as functions of $x$ are only a part of the picture, so perhaps it is time to compute their truncated moments.

\begin{figure}[t!]
	\begin{centering}
		\includegraphics[width=0.8\textwidth]{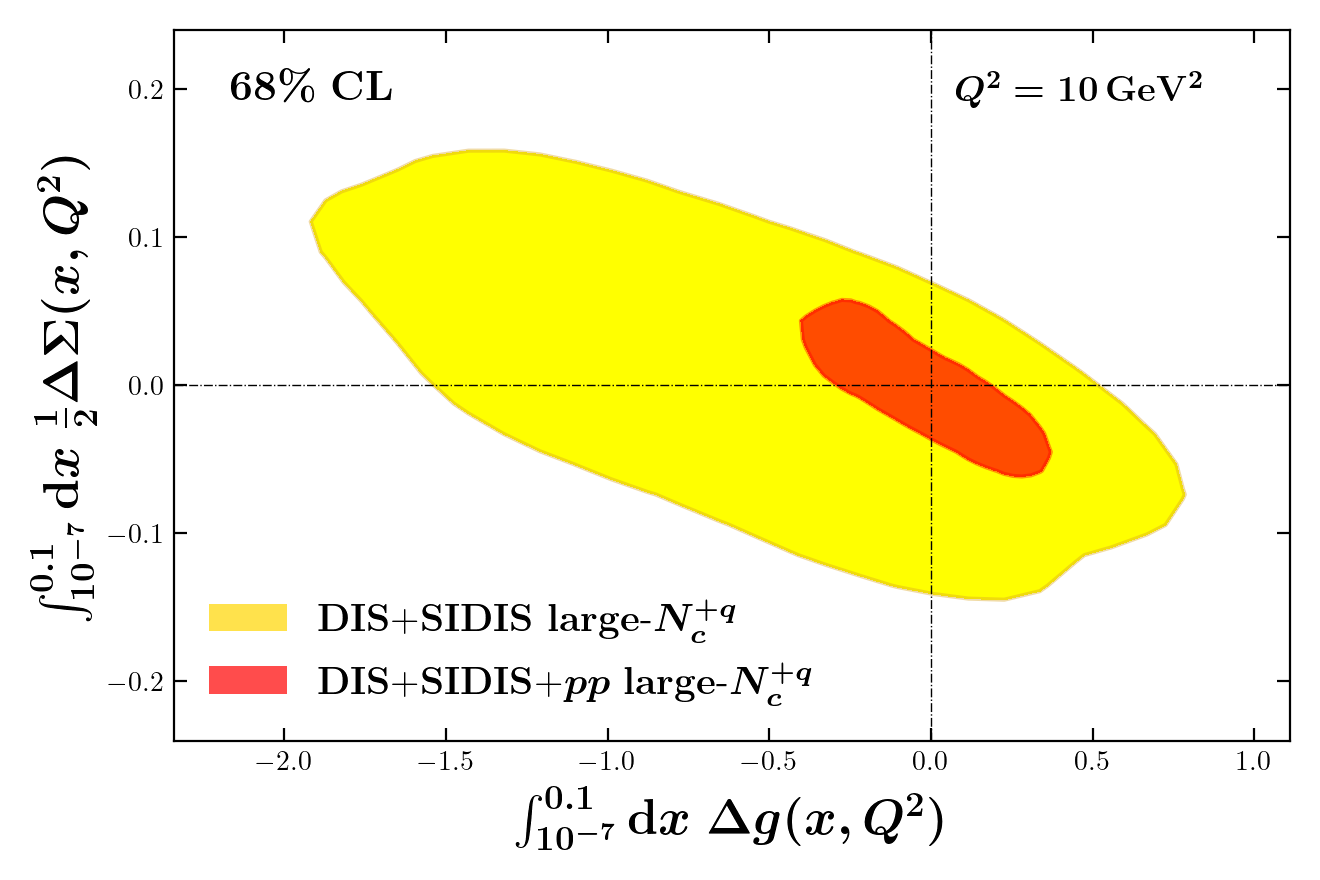}
		\caption{Truncated integrals of $\frac{1}{2}\Delta\Sigma(x,Q^2)$ and $\Delta g(x, Q^2)$ over the range $x\in[10^{-7}, 0.1]$ for the DIS+SIDIS large-$N_c^{+q}$ (yellow) and DIS+SIDIS+$pp$ large-$N_c^{+q}$ (red) analyses, at $Q^2=10~\mathrm{GeV}^2$. The contours contain 68\% of the replicas.
			\label{xmin_integral_comparison_pp}
		}
	\end{centering}
\end{figure}

In Fig.~\ref{xmin_integral_comparison_pp} I show the truncated moments of the parton spin contribution $\tfrac{1}{2}\Delta\Sigma(Q^2)$ and $\Delta G(Q^2)$, where I note that the change in arguments (the missing $x$ dependence) signifies an integral over $x$. For this analysis all truncated integrals are computed at $Q^2 = 10~\mathrm{GeV}^2$. These truncated integrals cover the small-$x$ region of interest $10^{-7} < x < 0.1$, with the quark hPDF integral $\tfrac{1}{2}\Delta\Sigma(Q^2)$ plotted on the vertical axis and the gluon hPDF integral $\Delta G(Q^2)$ is plotted on the horizontal axis; with the DIS+SIDIS analysis plotted in yellow and DIS+SIDIS+$pp$ analysis plotted in red. After including $pp$ data, the resultant removal of a large number of negative helicity replicas (especially for $\Delta g(x,Q^2)$ near $x\lesssim0.1$) results in a significant positive shift in the integrated hPDFs, and that the reduction in uncertainty at the hPDF level likewise affected the truncated moments. In agreement with the anticorrelation \eqref{e:hpdf_correlation}, the near-even split of asymptotic solutions, and the zero-centered hPDFs in Fig.~\ref{DeltaG_DeltaSigma_pp}, the truncated moments are now also centered at zero. I quantify the results below for the truncated moment of the gluon hPDF and truncated moment for the total parton helicity (the sum of the vertical and horizontal axes in Fig.~\ref{xmin_integral_comparison_pp}), and include the comparison with the large-$N_c^{+q}$ analysis of DIS and SIDIS data:
\begin{equation} \label{e:spinDIS}
	\int\limits_{10^{-7}}^{0.1}\!d x \, \Delta g (x, Q^2)\Big|_{\mathrm{(SI)DIS}} \!\approx -0.52 \pm 0.96 \,,  \int\limits_{10^{-7}}^{0.1}\!d x \, \Bigl(\frac{1}{2}\Delta\Sigma + \Delta g \Bigr)(x, Q^2)\Big|_{\mathrm{(SI)DIS}} \!\approx -0.52 \pm 0.90 \,,
\end{equation}
\begin{equation} \label{e:spinpp}
	\int\limits_{10^{-7}}^{0.1}\!d x \,  \Delta g (x, Q^2)\Big|_{+\,pp} \!\approx -0.04 \pm 0.26\,,  \int\limits_{10^{-7}}^{0.1}\!d x \, \Bigl(\frac{1}{2}\Delta\Sigma + \Delta g \Bigr)(x, Q^2)\Big|_{+\,pp} \!\approx -0.04 \pm 0.23\,.
\end{equation}

\begin{figure}[t!]
	\begin{centering}
		\includegraphics[width=\textwidth]{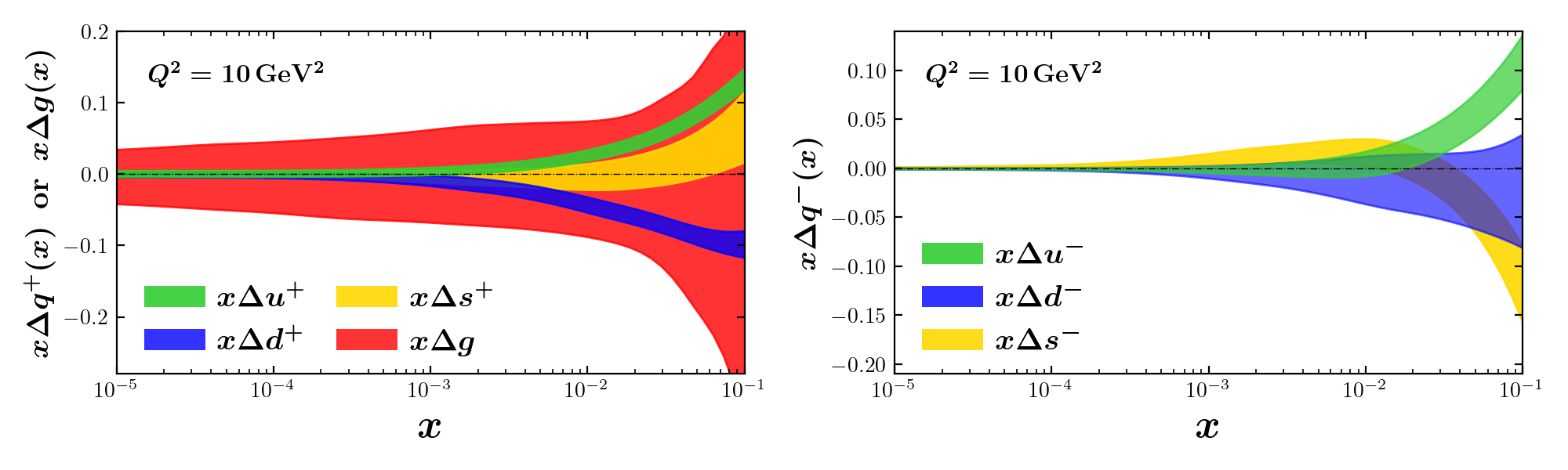}
		\caption{(Left) $C$-even hPDFs $x\Delta u^+$, $x\Delta d^+$, and $x\Delta s^+$ and the gluon hPDF $x\Delta g$, (green, blue, yellow, and red, respectively). (Right) $C$-odd hPDFs $x\Delta u^-$, $x\Delta d^-$, and $x\Delta s^-$ (green, blue, and yellow, respectively). All distributions are at a fixed $Q^2=10$~GeV$^2$. \label{hPDFs_2smx}
		}
	\end{centering}
\end{figure}

For a complete analysis I also provide the ($x$-scaled) $C$-even and $C$-odd quark hPDFs, along with $\Delta g$, as functions of $x$ in Fig.~\ref{hPDFs_2smx}. A direct comparison between Fig.~\ref{hPDFs_2smx} and Fig.~\ref{hPDF_bands} from \textbf{Chapter}~\ref{pheno_1} is inadvisable since multiple modifications are occurring at once, including the change in the scheme (``polarized DIS" vs $\overline{\mathrm{MS}}$), a change in evolution equations, and the inclusion of polarized $pp$ data. I investigated each of these modifications and make the following notes. $\Delta u^{\pm}$ and $\Delta d^{\pm}$ see a slight reduction in their uncertainty and a small decrease in their large-$x$ magnitudes, attributed to both the change in evolution equations and the inclusion of $pp$ data. The changes to $\Delta s^{\pm}$ are more severe and I determine that the change in evolution equations had the largest hand in reducing the uncertainty for $\Delta s^-$, but that the shift in $\Delta s^+$ to be more consistent with zero (as well as the shift of $\Delta s^-$ to be more positive as $x \to 0.1$) is a result of the $pp$ data constraints. Note that while the flavor nonsinglet evolution equation was not changed by setting $N_f = 0$, the flavor nonsinglet polarized dipole amplitudes' evolution is still affected since SIDIS data is sensitive to both the flavor singlet and flavor nonsinglet polarized dipole amplitudes as one can see from Eq.~\eqref{g1h_2}. 

\subsection{$\boldsymbol{\mathrm{Matching~onto~DGLAP\text{-}based}~\Delta g~\mathrm{and~EIC~impact~study}}$}

In line with what was conducted in \textbf{Chapter}~\ref{pheno1_more_testing}, I elect to retry the matching onto DGLAP-based JAM analyses and to study the impact of EIC data when pseudodata is generated from the current analysis.

\begin{figure}[b!]
	\begin{centering}
		\includegraphics[width=\textwidth]{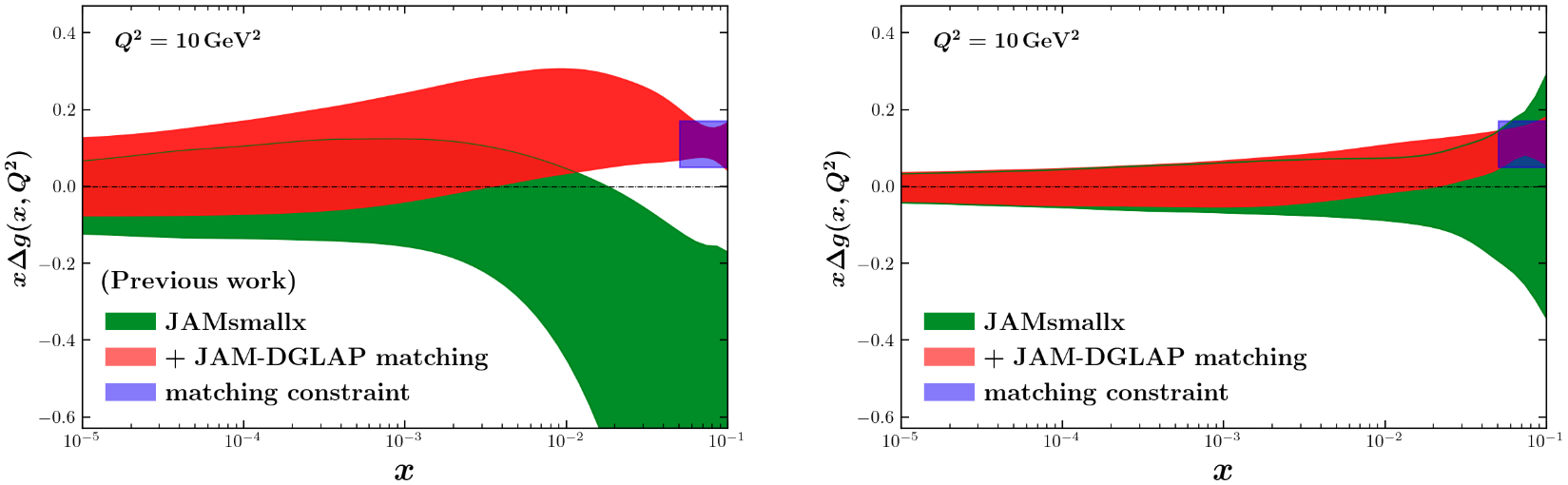}
		\caption{$x\,\Delta g(x, Q^2)$ versus~$x$ with and without matching onto the JAM-DGLAP fits of Refs.~\cite{Zhou:2022wzm, Cocuzza:2022jye, Hunt-Smith:2024khs} (``SU(3)$+$positivity scenario", ``JAM", and ``$+$LQCD", respectively). The baseline JAMsmallx fit (green) is compared with the fit with matching (red), and the blue box is a union of all three JAM-DGLAP $\Delta g(x, Q^2)$ extractions \cite{Zhou:2022wzm,Cocuzza:2022jye, Hunt-Smith:2024khs} in the range $10^{-1.3} < x < 0.1$. The left and right panels provide the results of JAM-DGLAP matching on the large-$N_c\&N_f$ analysis of DIS and SIDIS data and the large-$N_c^{+q}$ analysis of DIS, SIDIS, and $pp$ data, respectively.
			\label{DGLAP_DeltaG}
		}
	\end{centering}
\end{figure}

As discussed before, an entirely new, more rigorous formalism will be required to conduct a true small-to-large-$x$ extraction of hPDFs, but in the meantime I can perform preliminary tests by using the large-$x$ JAM-DGLAP extractions as a constraint for the small-$x$ helicity evolution initial conditions. The matching procedure is the same as before, where I approximate the JAM-DGLAP gluon hPDF extractions near the small-$x$ cutoff $x \lesssim 0.1$, and apply a $\chi^2$ punishment for any JAMsmallx replicas that probe outside this region. The box-matching process here utilizes the same box parameters $10^{-1.3}<x<0.1$ and $0.05<\Delta g(x,Q^2=10~\mathrm{GeV}^2) < 0.1$, which is fortunately a union of the JAM-DGLAP extractions from Ref.~\cite{Zhou:2022wzm}, \cite{Cocuzza:2022jye}, and \cite{Hunt-Smith:2024khs}, in their `SU(3)$+$positivity scenario", ``JAM",  and ``$+$LQCD" fits, respectively, Ref.~\cite{Hunt-Smith:2024khs} being the most recent DGLAP-based analysis. In Fig.~\ref{DGLAP_DeltaG} be show a side-by-side comparison of the ``+DGLAP matching" results from the JAMsmallx analyses of DIS+SIDIS using large-$N_c\&N_f$) (left) and DIS+SIDIS+$pp$ using large-$N_c^{+q}$ (right); both plots have the JAMsmallx baseline fits (without matching) in green, and the result of the box-matching in red, with the JAM-DGLAP $\Delta g$ box in blue. The previous work suggested that the small-$x$ helicity analysis was nearly incompatible with the DGLAP-based JAM extraction, however I find that the inclusion of $pp$ data results in an overlap of my JAMsmallx extractions with the JAM-DGLAP extractions; the blue box in the previous work effectively forced a new solution for small-$x$ replicas (left panel), whereas the blue box in the current analysis simply restricts the spread of JAMsmallx replicas (right panel). This comparison provides an optimistic outlook for matching small-$x$ extractions and large-$x$ extractions of hPDFs. It is also worth mentioning that the matching process in the previous analysis resulted in dramatic positive shift across all values of $x$ for the gluon hPDF, whereas the matching process in the current analysis results in a nearly unaltered, zero-centered uncertainty band as $x\to0$ (compared to Fig.~\ref{DeltaG_DeltaSigma_pp}), allowing for some $\Delta g$ replicas to still grow asymptotically negative. 
\begin{figure}[t] 
	\begin{centering}
		\includegraphics[width=\textwidth]{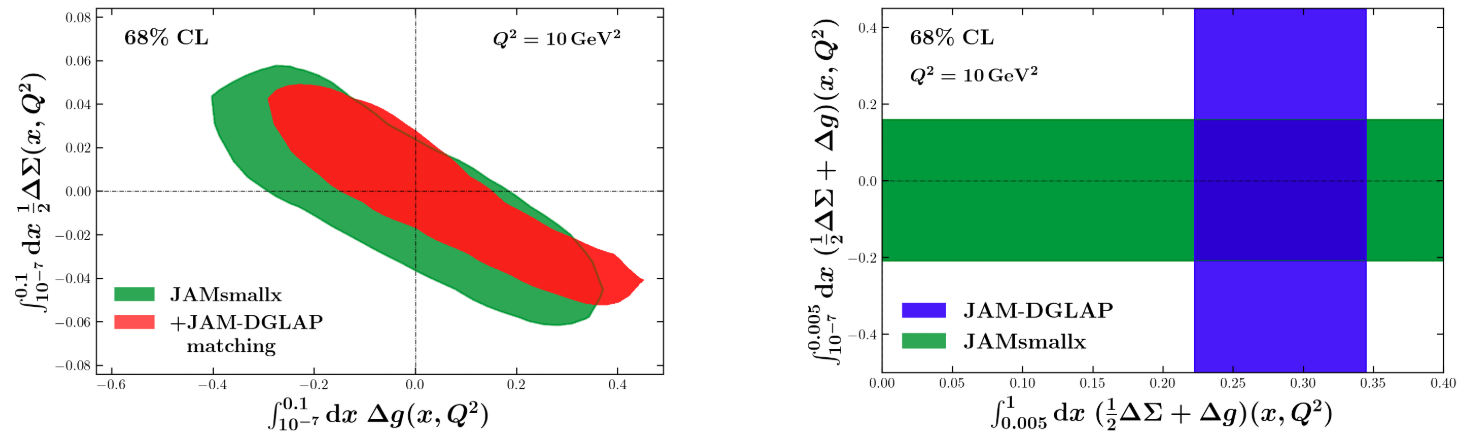}
		\caption{(Left) A comparison of the $10^{-7}< x < 0.1$ truncated integrals of JAMsmallx analyses with and without matching (green and red, respectively) onto DGLAP-based JAM extractions. (Right) An exclusion plot of two truncated integrals:~the vertical axis shows the truncated integral of total parton helicity over the prediction range $10^{-7}< x < 0.005$, while the horizontal axis shows the truncated integral of total parton helicity over the data-driven range $0.005< x< 1$. The JAMsmallx analysis (green) does not fit data for $x > 0.1$, while the JAM-DGLAP analysis (blue) has uncontrolled extrapolation for $x < 0.005$. The overlapping region is the best prediction for total parton helicity that simultaneously satisfies both analyses. 
			\label{DGLAP_trunc}
		}
	\end{centering}
\end{figure}
It was already established that $pp$ data shifted $\Delta g(x,Q^2)$ more positive as $x\to 0.1$, and it is clear that this trend continues with the JAM-DGLAP matching, removing even more negative solution replicas at the boundary $x\lesssim0.1$. It is then natural to compute the new truncated moments over the small-$x$ range $10^{-7}<x<0.1$ (and at $Q^2 = 10\mathrm{GeV}^2$). I plot such truncated integrals of $\Delta\Sigma(Q^2)$ and $\Delta G(Q^2)$ in the left panel of Fig.~\ref{DGLAP_trunc}; in green is the analysis without matching onto DGLAP extractions while in red is the truncated integral for the analysis with matching. Unsurprisingly, the truncated moments gluon helicity and total parton helicity with matching are now trending more positive, in agreement with the right panel of Fig.~\ref{DGLAP_DeltaG}. The extracted truncated moments are
\begin{equation}
	\int\limits_{10^{-7}}^{0.1}\!d x \, \Delta g (x, Q^2)\Big|_{{\rm matching}} \approx 0.08 \pm 0.24 \,, \;\; \int\limits_{10^{-7}}^{0.1}\!d x \,  \Bigl(\frac{1}{2}\Delta\Sigma + \Delta g \Bigr) (x, Q^2)\Big|_{{\rm matching}} \approx 0.08 \pm 0.21\,,
\end{equation}
which still carry the caveat that a full matching procedure would require a composite small-$x$+large-$x$ formalism to simultaneously satisfy both DGLAP large-$x$ and KPS-CTT small-$x$ analyses.

The compatibility seen in the right panel of Fig.~\ref{DGLAP_DeltaG} encouraged me to make a more powerful statement than was warranted in the previous analysis. Plotted in the right panel of Fig.~\ref{DGLAP_trunc} are two truncated moments of the total parton helicity, one from the current JAMsmallx analysis of DIS, SIDIS, and $pp$ data (green) and one from a recent DGLAP-based JAM analysis ~\cite{Cocuzza:2022jye} (blue). The JAMsmallx truncated integrals cover the prediction range $10^{-7}<x<0.005$ and are plotted on the vertical axis, while the JAM-DGLAP truncated integrals cover the data-driven range $0.005 \leq x \leq 1$ and are plotted on the horizontal axis. The $68\%$ confidence intervals for these two analyses are only trustworthy for their respective regions; JAMsmallx analyses do not include data for $x > 0.1$, and JAM-DGLAP analyses have unconstrained uncertainties for $x < 0.005$. The overlapping region indicates values of the truncated moment for total parton helicity that satisfy both JAMsmallx and JAM-DGLAP analyses. The lower-left corner and upper-right corner of the overlapping rectangle therefore act as the lower and upper boundaries of the $68\%$ confidence interval of a ``full-$x$" (technically not full, but the range $10^{-7}<x<1$) total parton helicity contribution to the proton spin. Summing the vertical and horizontal values at the two corners of interest gives a measure of total parton spin contribution at $Q^2 = 10~\mathrm{GeV}^2$ 
\begin{equation}\label{extended_truncation}
	\int\limits_{10^{-7}}^{1}\!d x \,  \Bigl(\frac{1}{2}\Delta\Sigma + \Delta g \Bigr) (x, Q^2)\ \in\ [0.02, 0.51].
\end{equation}
The extended range of $10^{-7}<x<1$ still does not cover the entirety of $x\in[0,1]$, however, for extremely small values of $x$ it is expected that saturation effects will have a large impact. Parton saturation will be a dominant feature for $x < 10^{-7}$ (at $Q^2=10~\mathrm{GeV}^2$), and the expectation\footnote{Reggeon evolution in the unpolarized sector shows that parton PDFs are greatly diminished in the saturation region \cite{Itakura:2003jp}. I presume that this effect will occur for the hPDFs and OAM distributions as well.} is that this will effectively make the parton helicity and orbital angular momentum contributions (OAM) negligibly small in that region. If this is to be the case, then the truncated integral in Eq.~\eqref{extended_truncation} is a reasonable estimate for the total parton helicity contribution to the proton spin sum rule \eqref{spin_sum}. This would imply that the parton OAM contributions to the proton spin are restricted to $L_{q+\bar{q}}(Q^2)+L_G(Q^2) \in [-0.01, 0.48]$ within a $68\%$ confidence interval.

\begin{figure}[t] 
	\begin{centering}
		\includegraphics[width=\textwidth]{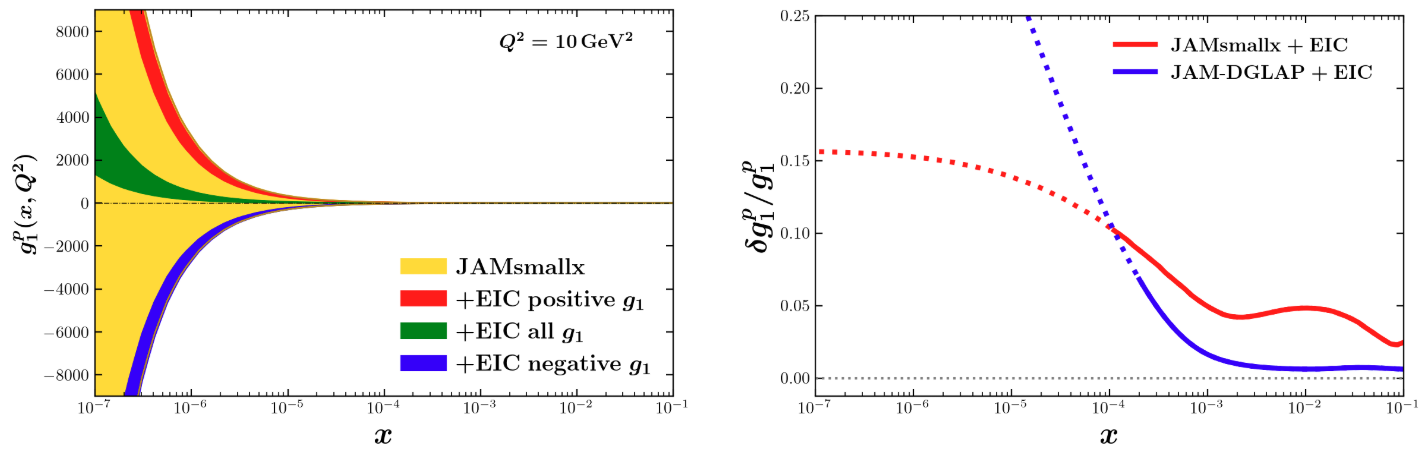}
		\caption{(Left) $g_1^p(x, Q^2)$ at $Q^2=10~\mathrm{GeV}^2$ for the current JAMsmallx global analysis of DIS, SIDIS, and $pp$ experimental data (yellow), along with the three EIC pseudodata fits for the scenarios described in the text (red, green, blue).
			(Right) Relative uncertainties for the EIC impact study using small-$x$ helicity evolution (red) compared to an EIC impact study using DGLAP evolution \cite{Zhou:2021llj} (blue). All results are for the asymptotically positive $g_1^p$ scenario to avoid zero-crossings relative uncertainty denominator.  The lines become dotted in the region of $x$ below which the pseudodata was generated:~the JAMsmallx analysis generated pseudodata down to $x = 10^{-4}$, while the JAM-DGLAP analysis generated pseudodata down to $x = 2\times10^{-4}$.
			\label{EIC_impact_2}
		}
	\end{centering}
\end{figure}

Lastly, for this analysis I perform a new EIC impact study. The EIC will not produce polarized proton-proton data; however, since the extractions of the $g_1^p$ structure function have changed as a result of the new constraints from $pp$ data, the newly generated EIC pseudodata will be slightly different. Following the methodology in \textbf{Chapter}~\ref{pheno1_more_testing}, I generate EIC pseudodata in the kinematic range $10^{-4} <x <0.1$ and $1.69~\mathrm{GeV}^2 <Q^2 <50~\mathrm{GeV}^2$, based on my theoretical asymmetries; pseudodata is again generated from the midpoint $g_1^p$ predictions, the average of asymptotically positive solutions, and the average of asymptotically negative solutions. Shown in the left panel of Fig.~\ref{EIC_impact_2} is the current JAMsmallx baseline fit to DIS, SIDIS, and $pp$ data (yellow), along with the fits that also include the midpoint (green), positive solution (red), and negative solution (blue) EIC pseudodata. The right panel of Fig.~\ref{EIC_impact_2} shows the relative uncertainties of my positive $g_1^p$ JAMsmallx+EIC extraction in red against a JAM-DGLAP+EIC extraction \cite{Zhou:2021llj} in blue; solid lines represent the region in $x$ where data is being fitted (including pseudodata) while the dotted lines are the extrapolation/prediction region. I find that even on top of the $\sim70\%$ uncertainty reduction resulting from $pp$ data constraints, that the EIC pseudodata data down to $x = 10^{-4}$ results in a further $93\%$ reduction in the $g_1^p$ uncertainty band (for the midpoint scenario) at $x = 10^{-7}$. This highlights the importance of $pp$ data even with the very precise DIS and SIDIS measurements that the EIC will provide. Additionally, the smaller-$x$ $10^{-4}\lesssim x \lesssim 5\times10^{-3}$ data provided by the EIC will be able to break the persisting asymptotic bimodality, just as was concluded in \textbf{Chapter}~\ref{pheno1_more_testing}. In more agreement with previous observations, it is clear from the right panel of Fig.~\ref{EIC_impact_2} that the predictive power of the KPS-CTT small-$x$ evolution equations results in smaller, more stable uncertainties well beyond below the data's reach in $x$ --- something that is not observed from the DGLAP-based JAM analysis~\cite{Zhou:2021llj}. 

\subsection{$\boldsymbol{\mathrm{Conclusions\!:~Phenomenology~II}}$}

This chapter summarized the first study of a simultaneous analysis of polarized DIS, SIDIS, and proton-proton scattering data using the KPS-CTT small-$x$ helicity evolution equations \cite{Adamiak:2025dpw}. I employed the pure-gluon small-$x$ $pp\to g\,X$ particle production cross-section in polarized $pp$ collisions \cite{kovchegov:2024aus} to approximate single-inclusive jet production to analyze the double-longitudinal spin asymmetry $A_{LL}^{\mathrm{jet}}$. The pure-gluon nature of the particle production cross-section motivated the use of a large-$N_c$ limit variation which I dubbed the large-$N_c^{+q}$ limit, in an effort to best approximate a quark-less evolution with flavor-dependent initial conditions. With the above methodology, a simultaneous fit to DIS, SIDIS, and $pp$ data was achievable, resulting in an overall goodness-of-fit measurement of $\chi^2_{\mathrm{red}}\approx 0.98$ for $N_{\mathrm{pts}}=240$ data points at $x < x_0 \equiv 0.1$. This new data was directly sensitive to the small-$x$ dominant $\widetilde{G}$ and $G_2$ polarized dipole amplitudes and the new constrains on the dipoles resulted in a $73\%$ reduction in small-$x$ uncertainty for hPDFs and the $g_1^p$ structure function (Figs.~\ref{DeltaG_DeltaSigma_pp} and \ref{g1_band_comparison_pp}), a more even $54\%/46\%$ split between asymptotically positive and negative solutions (Fig.~\ref{g1_band_comparison_pp}), and a positive shift in the gluon helicity PDF $\Delta g(x,Q^2)$ for $x\lesssim 0.1$.

The new split of positive and negative solutions results in hPDF extractions that are centered around zero as $x\to 0$, and by extent the truncated moments of total parton helicity are now consistent with zero (see Fig.~\ref{xmin_integral_comparison_pp}) with still large uncertainties of $\pm 0.25$. This is a significant change in my small-$x$ predictions compared to the negative-definite spin contribution from \textbf{Chapter}~\ref{pheno_1} which required a large positive OAM contribution from small-$x$ partons. An EIC impact study still paints a positive picture for the future of small-$x$ helicity evolution equations, as I concluded that the extra data at $10^{-4} < x < 5\times 10^{-3}$ will not only further reduce the uncertainties by $\sim90\%$, but will also break the bimodality of solutions. 

I made the first preliminary statement about a ``full-x" parton spin contribution after a comparison analysis with the DGLAP-based JAM extractions \cite{Cocuzza:2022jye}. Under the intuition that saturation effects for $x < 10^{-7}$ (at $Q^2  =10~\mathrm{GeV}^2$) will largely diminish parton helicity and OAM contributions, the boundaries of integrated total parton helicity that satisfy both my JAMsmallx truncated moments over the prediction range $10^{-7}<x<0.005$ and the JAM-DGLAP truncated moments over the data-drive range $0.005 < x <1$ can effectively estimate the full-$x$ spin contributions. From these boundaries, I predict that the total parton helicity contribution $\tfrac{1}{2}\Delta\Sigma(Q^2) + \Delta G(Q^2)$ is between $0.02$ and $0.51$ --- see Fig.~\ref{DGLAP_trunc}. To satisfy the Jaffe-Manohar spin sum rule \eqref{spin_sum}, this inherently restricts the total parton OAM contributions to between $-0.01$ and $0.48$ with $68\%$ confidence. 

The future of this analysis lies with the theoretical formulation of the remaining quark contributions to the small-$x$ parton production cross-section in polarized proton-proton collisions. The remaining channels may be sensitive to the quark polarized dipole amplitudes, potentially giving even better constraints on the hPDFs than seen here. With the quark contributions determined, I would then be able to reinstitute the more accurate large-$N_c\&N_f$ evolution equations, including the recent update from Ref.~\cite{Borden:2024bxa}. This in turn would lay the foundations necessary to expand the polarized proton-proton observables to potentially include single-inclusive hadron, isolated direct photon, and weak gauge boson production (See Ref.~\cite{RHICSPIN:2023zxx} for a summary on $A_{LL}^{\mathrm{jet}}$ data from the Relativistic Heavy Ion Collider). Further in the future I may also attempt to include new logarithmic resummations or dijet production \cite{STAR:2021mqa}. Higher order $\ln(R)$ resummation can improve the accuracy of the jet function that I approximated as a delta function early in this analysis, bringing in $z_J$ dependence and potentially improving the fit to $A_{LL}^{\mathrm{jet}}$ \cite{Kaufmann:2014nda, Dasgupta:2014yra, Kang:2016mcy}. I conclude that there are several promising pathways to continue developing the
modern theory and phenomenology of spin at small $x$.

	\newpage
	\section{$\boldsymbol{\mathrm{Conclusions}}$}

The proton spin puzzle, or proton spin `crisis' if you really want to sell your grant proposal, is an important mystery of nuclear physics that has yet to be solved, but the research outline in this dissertation has taken us a step closer to solving the puzzle. As outline from the very beginning in Eqs.~\eqref{spin_sum} and \eqref{spin_integrals}, the angular momentum of the proton will only be fully understood so long as the spin contributions and orbital angular momentum contributions of its constituent quarks and gluons are fully detailed; focusing on the spin contributions, even those will not be understood in their entirety until the small-$x$ behavior of the helicity PDFs $\Delta\Sigma(x,Q^2)$ and $\Delta g(x,Q^2)$ can be predicted with accuracy and precision. \textbf{Chapters}~\ref{intro} and \ref{chapter_2} laid the foundation for how I can achieve such a goal, using parton evolution equations to predict a final state evolved object from a nonperturbative initial conditions, and what the specifics of small-$x$ helicity evolution equations looks like with the KPS-CTT evolution equations \eqref{eq_LargeNcNf}, \eqref{eq_LargeNcq}.

The KPS-CTT evolution equations can be used to describe how polarized dipole scattering amplitudes behave as $x$ goes to zero, provided I can use nonperturbative means or data-fitting to determine their appropriate initial conditions. These evolution equations describe the energy/rapidity/$x$-dependence of polarized dipoles scattering off of a nucleon target; this dependence emerges from the resummation of doubly-logarithmic diagrams, i.e., those that come with the resummation parameters $\alpha_s\ln^2(1/x)\sim1$. Numerical and analytic solutions have shown that, in the large-$N_c$ or large-$N_c\&N_f$ limit, the KSP-CTT evolution equations predict a potentially large and important contribution from small-$x$ partons since they predict power law growth of $\Delta\Sigma(x,Q^2)\sim\Delta g(x,Q^2) \sim (\tfrac{1}{x})^{\alpha_h}$ in the asymptotic limit that $x\to 0$. This intercept $\alpha_h$ is dependent on the limit taken, but the analytical solution in the large-$N_c$ has determined an intercept of $\alpha_h = 3.66074\sqrt{\tfrac{\alpha_sN_c}{2\pi}}$, and is only small enough to be integrable in the case the running of the strong coupling $\alpha_s$ is implemented; this solution does support the accuracy of the KPS-CTT evolution equations since it is found to be within $3\%$ accuracy of the BER infrared evolution equations approach. 

Having understood the powerful potential of the KPS-CTT evolution equations, this dissertation continues the phenomenological research based on such a theoretical framework. I make the effort to constrain the initial conditions of the evolution equations and subsequently make novel predictions about the quark and gluon spins/helicities at extremely small $x=Q^2/(E_{CM}^2+Q^2) \approx Q^2/s$, beyond that which can ever be probed from experiment. \textbf{Chapters}~\ref{chapter_2} and \ref{chapter_3} detailed the phenomenology methodology; a quark target toy model determined that the polarized dipole scattering amplitude's initial conditions share the same general form of being a linear combination of logarithmic variables. I then chose to generalize the initial conditions using a linear combination \textit{ansatz}, such that the initial conditions of the entire set of evolution equations can be parameterized by a 24-parameter initial conditions state. Using the Jefferson Angular Momentum, or JAM, Bayesian Monte-Carlo framework, I can randomly sample the 24 initial condition parameters, construct theoretical asymmetries using the evolution equations and definitions \eqref{all_hPDFs}, and use a chi-squared minimization algorithm to iterate the initial conditions until an optimized combination of parameters has been found that minimizes the chi-squared computed from a comparison of theoretical asymmetries and experimental asymmetries from small-$x$ polarized DIS, SIDIS, or proton-proton scattering data. From the optimized parameters, one can then re-solve the evolution equations and compute the evolved polarized dipoles, and thus the hPDFs, at arbitrarily small $x$ (arbitrary as far as Eqs.~\eqref{eq_LargeNcNf} and \eqref{eq_LargeNcq} are concerned, but I am limited by the finite amount of computational resources).

\textbf{Chapter}~\ref{pheno_1} covered the first phenomenological study to which I contributed, and it was the first analysis of polarized DIS and SIDIS using the revised and updated large-$N_c\&N_f$ KPS-CTT evolution equations (an improvement over the KPS evolution equations). The addition of SIDIS data, for which one needs flavor nonsinglet evolution equations to analyze, allowed for small-$x$ extractions of the $C$-even quark hPDFs $\Delta q^{+} = \Delta q+\Delta\bar{q}$, $C$-odd quark hPDFs $\Delta q^- = \Delta q - \Delta \bar{q}$, and the gluon hPDF $\Delta G(x,Q^2)$ (or $\Delta g(x,Q^2)$, which is more common in polarized literature). The use of specifically the large-$N_c\&N_f$ evolution equations (as opposed to large-$N_c$) allowed for flavor-dependent initial conditions and evolution due to it re-instituting internal quark loops that were otherwise suppressed at large-$N_c$; this is therefore a more realistic limit, allowing for better phenomenology. This work found that an analysis of polarized DIS and SIDIS data using the large-$N_c\&N_f$ KPS-CTT evolution equations was possible, resulting in the overall fit to 226 data points with a goodness quantified by $\chi^2_{\mathrm{red}} = 1.03$. From the 500 sets of optimized parameters I made predictions of the small-$x$ hPDFs and found that the quark hPDFs $\Delta\Sigma(x,Q^2)$ prefer to grow positive as $x\to 0$, while the gluon hPDF grows negative as $x\to0$, resulting in the overall small-$x$ integrated parton helicity to be $-0.64\pm0.60$, implying that small-$x$ OAM contributions may be an important contribution to the spin-sum rule. These conclusions, however, carried the caveat that the small-$x$ uncertainties were very large due to the insensitivity of polarized DIS and SIDIS data to the $\widetilde{G}$ and $G_2$ polarized dipole amplitudes, the dipoles that also happen to be the most dominant contributors in the small-$x$ limit. This insensitivity led to a bimodality of solutions, largely controlled by the $\widetilde{G}^{(0)}$ parameters, where DIS and SIDIS data could not predict the asymptotic sign of the hPDFs. To address these issues, a subsequent iteration of small-$x$ helicity evolution phenomenology was performed that included polarized proton-proton scattering data.

The latest small-$x$ helicity phenomenology, covered in \textbf{Chapter}~\ref{pheno_2}, provides the first ever study of polarized proton-proton scattering data using small-$x$ helicity evolution equations. Because the scattering of polarized protons allows for gluon-gluon interactions at lowest order, these double-longitudinal spin asymmetries at small $x$ will be sensitive to the yet-constrained $\widetilde{G}$ and $G_2$ polarized dipole scattering amplitudes. Within the JAM framework the only polarized proton-proton scattering data available were single-inclusive jet production cross-sections in the form of the double-longitudinal spin asymmetry $A_{LL}^{\mathrm{jet}}$. Within the narrow jet approximation framework, it was found that a single-inclusive hadron production cross-section (in polarized $pp$ collisions) could be modified into a jet production cross-section by replacing the hadronic fragmentation functions with jet functions; conveniently, higher-order terms in the jet functions do not contribute two logarithms of energy, and are thus outside the DLA power counting, allowing me to approximate a parton production cross-section as the jet production cross-section. The only small-$x$ parton production cross-section in polarized $pp$ collisions is that which has been derived in the pure-gluon limit in Ref.~\cite{kovchegov:2024aus}. I further approximate the jet production cross-section with a gluon-only gluon production cross-section and use that to fit to the spin asymmetry. As hypothesized, such a cross-section is indeed sensitive to the gluon dipoles $\widetilde{G}$ and $G_2$, resulting in better constraints on those parameters and much improved small-$x$ uncertainty. As a whole, the hPDF and $g_1^p$ structure function uncertainty bands see a $73\%$ reduction due to the extra constraints from polarized $pp$ data -- trustworthy results considering the incorporation of $pp$ data did not degrade the goodness-of-fit to DIS and SIDIS data, netting an overall average of $\chi^2_{\mathrm{red}} = 0.98$. The sensitivity to $G_2$ resulted in a large removal of negative $\Delta g(x,Q^2)$ solutions as $x\to0.1$, which had did not translate to any small-$x$ behavioral changes, but did result in better compatibility with DGLAP-based large-$x$ JAM analyses and a positive shift in the truncated moment of total parton helicity. This latest analysis now predicts that small-$x$ quark and gluon spin contributions are consistent with zero, $0.08\pm 0.21$, but with the caveat that persisting asymptotic bimodality still makes the small-$x$ uncertainty large. The newfound compatibility with large-$x$ DGLAP-based extractions encouraged a comparison between the JAMsmallx and JAM-DGLAP analyses in their respective best-constrained regions, from which I determined that the truncated parton helicity over an extended range of $10^{-7} < x < 1$ is between $0.02$ and $0.51$ within $68\%$ confidence; under the presumption that parton saturation will largely diminish parton helicity and OAM contributions for $10^{-7}$, this results provides a restriction on parton OAM to contribute between $-0.01$ and $0.48$ to the proton spin to satisfy the Jaffe-Manohar spin sum rule.

Lastly, I make some notes about the future potential of small-$x$ helicity theory and phenomenology. EIC impact studies in both \textbf{Chapters}~\ref{pheno_1} and \ref{pheno_2} paint a positive picture; EIC polarized data provided at the improved kinematic range $10^{-4} < x < 0.1$ and $1.69~\mathrm{GeV}^2 < Q^2 < 50~\mathrm{GeV}^2$ should be precise enough and at small-enough $x$ to dramatically improve the small-$x$ predictions from KPS-CTT, as well as break the asymptotic bimodality that persists. Apart from waiting for the future EIC to be constructed, there are some more readily available avenues to pursue, which include improving the accuracy of current small-$x$ polarized $pp$ parton production cross-sections by reintroducing quark contributions (and by extend a phenomenology that reinstates the large-$N_c\&N_f$ evolution equations), improving the accuracy of the jet function by expanding the DLA power counting to consider $\ln(R)$ resummation (which follows the DGLAP evolution equations), incorporating more polarized $pp$ observables such as single-inclusive hadron, isolated direct photon, and weak gauge boson production. In the short term, I have plans to conduct a principal component analysis of the existing small-$x$ helicity phenomenology research, with the goal of determining hidden correlations of parameters, the relative constraint (or lack thereof) of individual parameters, and other similar properties of the initial condition parameter space to better identify the best next course of action for small-$x$ helicity evolution phenomenology.

	\newpage
	\section{$\boldsymbol{\mathrm{Appendix}}$}

\subsection{$\boldsymbol{\mathrm{Contribution~to~Cold~Nuclear~Matter~Research}}$}

My brief adventure into the field of Cold Nuclear Matter research focused on generalizing the existing decomposition of the Sivers function of a nucleus in the quasi-classical approximation with a rigid rotator model for the nucleus \cite{Kovchegov:2013cva} for the case of nucleons with arbitrary polarization in a vein similar to that which was done for the Boer-Mulders TMD in Ref.~\cite{Kovchegov:2015zha}. This process is best studied in the SIDIS process on a large nucleus.

This work begins with a closer look at the transverse momentum dependent (TMD) quark correlation function in a hadronic state $|h(P,S)\rangle$ with momentum $P$ and spin $S$ defined by
\begin{equation}\label{quark_correlator_1}
	\phi_{\alpha\beta}(x,\underline{k};P,S)\equiv \frac{g_{+-}}{(2\pi)^3}\int d^{2-}r\,e^{i\,k\cdot r}\langle h(P,S)|\,\bar\psi_{\beta}(0)\,\mathcal{U}[0,r]\,\psi_{\alpha}(r)\,|h(P,S)\rangle_{r^+=0},
\end{equation}
where $\alpha,\;\beta$ are Dirac indices, $x$ and $\underline{k}$ are the Bjorken-$x$ and transverse momentum of the parton in question, the separation vector between the quark fields is $r^{\mu} = (0^+,r^-,\underline{r})$, and $\mathcal{U}[0,r]$ is the future/past pointing gauge link described above Eq.~\eqref{gluon_TMD_Wilson}. I work in the same light-front coordinates as discussed in \textbf{Chapter}~\ref{intro}, and $g^{+-} = 1$ and $g^{+-}=2$ are two common choices of light-front metric. The correlator \eqref{quark_correlator_1} can be decomposed into various TMD PDFs (typically themselves referred to as TMD) as \cite{Boer:1997nt,Mulders:1995dh,Meissner:2007rx}
\begin{align}\label{decomp_1}
	&\phi(x,\underline{k};P,S) = \Big(f_1 - \frac{\uline{k}\times\uline{S}}{m}\,f_{1T}^{\bot}\Big)\Big[\tfrac{1}{2}g_{+-}\gamma^-\Big] + \Big(S_{L}g_1+\frac{\uline{k}\cdot\uline{S}}{m}\,g_{1T}\Big)\,\Big[\tfrac{1}{2}g_{+-}\gamma^5\gamma^-\Big] \\
	&\qquad +\Big(S_\bot^i\,h_{1T}+\frac{k_\bot^i}{m}S_Lh_{1L}^\bot + \frac{k_\bot^i}{m}\frac{(\uline{k}\cdot\uline{S})}{m}h_{1T}^\bot\Big)\,\Big[\frac{1}{2}g_{+-}\gamma^5\gamma_{\bot i}\gamma^-\Big] + \Big(\frac{k_\bot^i}{m}h_1^\bot\Big)\,\Big[\frac{i}{2}g_{+-}\gamma_{\bot i}\gamma^-\Big] \notag
\end{align}
where ($\uline{S},S_L$) denote the transverse and longitudinal spin components of the spin vector, $m$ is the hadron mass, and $\uline{k}\times\uline{S}\equiv k_xS_y-k_yS_x$ with ($x,y$) the coordinates in the plane transverse to the beam. The various functions $f$, $g$, and $h$ are the TMD's, with arguments $x$ and $\uline{k}$. I will work with the quark correlator for now, with the intention of isolating the Sivers TMD $f_{1T}^{\bot}$ by tracing the decomposition with $\gamma^+$, related to the production of unpolarized quarks. 

The quasi-classical factorization formula for semi-inclusive DIS on a heavy nucleus is \cite{Kovchegov:2013cva} 
\begin{align}
	&\Phi_{\alpha\beta}^A(x,\uline{k};P,S) = A\frac{g_{+-}}{(2\pi)^5}\sum_{\sigma}\int d^{2+}p\,db^-\,d^2x\,d^2y,\,d^2k'\,e^{-i(\uline{k}-\uline{k}')\cdot(\uline{x}-\uline{y})} \\
	&\hspace{6cm}\times W_{\sigma}(p,b;P,S)\,\phi_{\alpha\beta}^N(\hat{x},\uline{k}'-\hat{x}\uline{p};p\sigma)\,D_{xy}[\infty^-,b^-] \notag
\end{align}
where $\hat{x}\equiv x\tfrac{P^+}{p^+}$ is the quark momentum fraction with respect to the active nucleon, $d^{2+}p\equiv\ d^2p_\bot\,dp^+$, and $b^{\mu}\equiv (0^+,b^-,\tfrac{1}{2}(\uline{x}+\uline{y}))$ is the position of the struck nucleon. $D_{xy}$ is the unpolarized dipole scattering amplitude. The Wigner distribution of the nucleons $|N(p,\sigma)\rangle$ inside the nucleus $|A(P,S)\rangle$ can be written for nucleons with an arbitrary spin state $|S\rangle$ in terms of the light-cone helicity basis $|\pm\rangle$. Likewise, the nucleus itself can be defined for an arbitrary spin state. The Wigner distribution for polarized nucleons in a polarized nucleus is written compactly as
\begin{equation}
	W(p,b, S^N;P,S^A)\equiv \frac{1}{2(2\pi)^3}\int \frac{d^{2+}(\delta p)}{\sqrt{p^+p'^{+}}}e^{-i\delta p\cdot b}\,\langle A(P,S^A)|\,N(p',S^N)\rangle\,\langle N(p,S^N)|\,A(P,S^A)\rangle,
\end{equation}
for nucleon and nucleus spins $S^N$ and $S^A$. The arbitrary spin-state of a nucleon or nucleus can be decomposed as \cite{Diehl:2005jf}
\begin{equation}\label{arb_spin}
	|S\rangle \equiv \cos\frac{\theta}{2}|+\rangle + \sin\frac{\theta}{2}e^{i\phi}|-\rangle,
\end{equation}
while in the nucleon or nucleus' rest frame (RF) the eigen-axis of spin projections are given by the three-vector
\begin{equation}
	\vec{S}\stackrel{R.F}{\equiv} (S_\bot^1,S_\bot^2,\lambda)
\end{equation}
with $\lambda^2+S_T^2 = 1$. The Lorentz-invariant spin vector $S^{\mu}$ can be defined by taking the rest frame picture and boosting it into the frame where the nucleon/nucleus has momentum $p^{N/A}$, such that $p^{(N/A)}_{\mu}S^{(N/A)\,\mu}=0$ and 
\begin{equation}\label{smu}
	S^{\mu}\stackrel{R.F}{\equiv} (0,S_\bot^1,S_\bot^2,\lambda)
\end{equation}
with transverse spin components $\uline{S} = (S_\bot^1,S_\bot^2)$ and longitudinal spin component $\lambda$. By combining Eqs.~\eqref{arb_spin} and \eqref{smu} I can define the outer product
\begin{equation}\label{outer_prod}
	|S\rangle\langle S| = \frac{1}{2}\Big[|+\rangle\,|-\rangle\Big]\Big\{\Big[\bm{1}\Big]-S_{\mu}(p)\,\Big[\hat{\sigma}^{\mu}(p)\Big]\Big\}\,\begin{bmatrix}
		\langle +| \\
		\langle -|
	\end{bmatrix}
\end{equation}
where $\sigma^{\mu}$ is the Lorentz-invariant set of Pauli matrices, $\sigma^{\mu}=(0,\sigma_\bot^1,\sigma_\bot^2,\sigma_\bot^3)$ in the rest frame. With Eq.~\eqref{outer_prod} I can decompose the winger distribution (and even the quark correlator) into its unpolarized and polarized contributions. Defining the nucleus' spin with the vector $S^{\mu}$ and polarization state $\Lambda$, along with the nucleon's spin with the vector $S^{\nu}$ and polarization state $\lambda$, I can invert the spin-vectors as
\begin{equation}
	\vec{S}^{\mu} \Rightarrow \Big[\sigma^{\mu}(P)\Big]_{\Lambda\Lambda'},\qquad \vec{S}^{\nu} \Rightarrow \Big[\sigma^{\nu}(p)\Big]_{\lambda\lambda'}
\end{equation}
and define the following traces
\begin{subequations}
	\begin{align}
		F_{unp} &= \frac{1}{2}\sum_{\lambda\lambda'}\Big[F_{\lambda\lambda'}\,\bm{1}_{\lambda'\lambda}\Big] \\
		F_{pol} & = \frac{1}{2}\sum_{\lambda'\lambda'}\Big[F_{\lambda\lambda'}\,\sigma_{\lambda'\lambda}\Big]
	\end{align}
\end{subequations}
so I can ultimately write the following spin-decompositions
\begin{align}
	\Phi^A_{\Gamma}(x,\uline{k};P,S^{\mu}) &\equiv \Big[\Phi^A_{\Gamma}(x,\uline{k};P)\Big]_{\Lambda\Lambda'} = \Phi^A_{\Gamma,unp}\Big[\bm{1}\Big]_{\Lambda\Lambda'}-\Phi^A_{\Gamma,\,pol,\,\mu}\Big[\sigma^{\mu}\Big]_{\Lambda\Lambda'} \label{quark_A_dec}\\ 
	W(p,b,S^{\nu};P,S^{\mu}) &\equiv \Big[W(p,b,S^{\nu};P)\Big]_{\Lambda\Lambda'} = W(S^{\nu})_{unp}\Big[\bm{1}\Big]_{\Lambda\Lambda'} - W(S^{\nu})_{pol,\,\mu}\Big[\sigma^{\mu}\Big]_{\Lambda\Lambda'} \\
	W(p,b,S^{\nu};P) &\equiv W_{\lambda\lambda'}(p,b;P) = W^{unp}\Big[\bm{1}\Big]_{\lambda\lambda'} - W^{pol,\,\nu}\Big[\sigma^{\nu}\Big]_{\lambda\lambda'},
\end{align}
Where the subscript $\Gamma$ is a placeholder for the spin-projection used to identify the produced quark's polarization state. I am, admittedly, being very cavalier with my notation above, but one must consider that the Wigner distribution carries both the nucleus and the nucleon carry spin information. To keep them separate I denote the spin contribution and spin-index of the nucleus to be held in the subscript while the nucleons' are held in the superscript; this means that, for the sake of this appendix, multiplication notation of the likes of $W^{\nu}\sigma^{\nu}$ must be accepted. This means that the Wigner distribution for a polarized nucleon in a polarized nucleus has a double decomposition:
\begin{align}\label{W_decomp}
	W(p,b,S^{\nu};P,S^{\mu}) \equiv \Big[W_{\lambda\lambda'}(p,b;P)\Big]_{\Lambda\Lambda'}&\Rightarrow W_{unp}^{unp}\Big[\bm{1}\Big]_{\lambda\lambda'}\Big[\bm{1}\Big]_{\Lambda\Lambda'} - W_{unp}^{pol,\,\nu}\Big[\sigma_{\nu}\Big]_{\lambda\lambda'}\Big[\bm{1}\Big]_{\Lambda\Lambda'}\\
	& \qquad - W_{pol,\,\mu}^{unp}\Big[\bm{1}\Big]_{\lambda\lambda'}\Big[\sigma^{\mu}\Big]_{\Lambda\Lambda'} + W_{pol,\,\mu}^{pol,\,\nu}\Big[\sigma_{\nu}\Big]_{\lambda\lambda'}\Big[\sigma^{\mu}\Big]_{\Lambda\Lambda'}. \notag
\end{align}
From Eq.~\eqref{quark_correlator_1} we see that the correlator of the nucleus $\Phi^A$ is dependent on the quark correlator of the nucleon $\phi^N$. Both being quark correlators, one should expect that the nucleon correlator follows the same decomposition as Eq.~\eqref{quark_A_dec}, though I should clarify that it carries the polarization state of the nucleon, not the nucleus. Putting this all together I can write out the quark correlator of a polarized nucleus as
\begin{align}
	\Big[\Phi^A_{\Gamma}(x,\uline{k};P)\Big]_{\Lambda\Lambda'} &= A\frac{2\,g_{+-}}{(2\pi)^5}\int d^2p\,dp^+\,d^2b_\bot\,db^-\,d^2\uline{k}'\,d^2\uline{r}\,e^{-i(\uline{k}-\uline{k}'-\hat{x}\uline{p})\cdot\uline{r}} \\
	&\hspace{2cm} \times \Big[W_{\lambda\lambda'}(p,b;P)\Big]_{\Lambda\Lambda'}\,\phi^N_{\Gamma\,\lambda'\lambda}(\hat{x},\uline{k}';p) S_{(r_T,b_T)}^{[\infty^-,b^-]} ,\notag 
\end{align}
where $\uline{r}\equiv \uline{x}-\uline{y}$. I trace over the nucleon polarization states $\lambda,\,\lambda'$ and write the final spin-decomposition
\begin{align}\label{pol_nuc}
	&\Big[\Phi^A_{\Gamma}(x,\uline{k};P)\Big]_{\Lambda\Lambda'} = A\frac{2\,g_{+-}}{(2\pi)^5}\int d^2p\,dp^+\,d^2b_\bot\,db^-\,d^2\uline{k}'\,d^2\uline{r}\,e^{-i(\uline{k}-\uline{k}'-\hat{x}\uline{p})\cdot\uline{r}}\;\;S_{(r_T,b_T)}^{[\infty^-,b^-]}\\
	&\hspace{0.8cm} \times \Bigg[ \Big(W_{unp}^{unp}\,\phi^{N\,unp}_{\Gamma} - W_{unp}^{pol,\nu}\,\phi^{N\,pol,\,\nu}_{\Gamma}\Big)\Big[\bm{1}\Big]_{\Lambda\Lambda'} - (W_{pol,\,\mu}^{unp}\,\phi^{N\,unp}_{\Gamma} - W_{pol,\,\mu}^{pol,\,\nu}\,\phi^{N\,pol,\,\nu}_{\Gamma}\Big)\Big[\sigma^{\mu}\Big]_{\Lambda\Lambda'}\Bigg]. \notag
\end{align}
At this point I can identify two specific terms: the terms multiplying by the identity matrix are the terms corresponding to the Wigner distribution of an unpolarized nucleus, while the terms scaled by $\sigma^{\mu}$ are those relating to the polarized nucleus. The unpolarized nucleus case has already been completed in Ref.~\cite{Kovchegov:2015zha}, with the correlator given in Eq. (39) of that paper and the unpolarized quark TMD $f_1^A$ and Boer-Mulders TMD $h_1^{\bot A}$ given by Eqs. (61) and (74), respectively. The novel work done here is a solution for a TMD from a polarized nucleus, specifically the Sivers TMD. I follow the outline laid out in \cite{Kovchegov:2015zha}.

Assuming that the polarized nucleus can be modeled as a rigid rotator, then we can understand the vectors at play here: there is the momentum of the struck nucleon in the nucleus $p$, the position of that nucleus $b$, the spin-vectors $S^A$ and $S^N$. With those laid out, I can identify the multiple symmetries that apply to this system and constrain the form of the Wigner distributions. The Wigner distribution is built purely from the nucleon wave functions, and because of this it possess unbroken discrete symmetries of parity and time reversal. The Wigner distribution for a polarized nucleus but unpolarized nucleus in its rest frame depends only on three vectors, $W_{pol,\,\mu}^{unp}(\vec{p}, \vec{b}, \vec{S}^A)$; $\mathcal{P}$-symmetry means that it should be invariant under $\vec{p}\to-\vec{p}$ and $\vec{b}\to-\vec{b}$, while $\mathcal{T}$-symmetry means that it should be invariant under $\vec{p}\to-\vec{p}$ and $\vec{S}^A\to-\vec{S}^A$. I must impose the rotational invariance on the polarized part of the Wigner distribution:
\begin{align}
	-S^{\mu}W_{pol,\,\mu}^{unp} &\stackrel{R.F.}{=} \vec{S}^A\cdot \vec{W}_{pol}^{unp}(\vec{p},\vec{b}) \Rightarrow \\ \notag
	&=\big(\vec{S}^A\cdot\vec{b}\big)\,W_1(\vec{p},\vec{b}) + \big(\vec{S}^A\cdot\vec{p}\big)\,W_2(\vec{p},\vec{b}) + \big(\vec{S}^A\cdot(\vec{b}\times\vec{p})\big)\,W_{OAM}^{unp,pol}, \notag
\end{align}
however I must take into account the aforementioned symmetries. The prefactors to the Wigner distributions $W_1$ and $W_2$ are both $\mathcal{P}$-odd, so the only remaining term is $W_{OAM}^{unp,pol}$, in which OAM refers to the orbital angular momentum prefactor, and the super-script ``$unp,pol$" refers to the nucleon, nucleus polarizations. There is more that can be done as far as restricting the Wigner distributions; the Wigner distributions should have manifest rotational invariance as well, meaning that I can restrict $W_{OAM}^{unp,pol}(\vec{p},\vec{b})$ (and in fact all Wigner distributions) to $W_{OAM}^{unp,pol}[\vec{p}^2,\vec{b}^2, (\vec{p}\cdot\vec{b})]$. One can notice, however, that $(\vec{p}\cdot\vec{b})$ is $\mathcal{T}$-odd, so I can restrict the arguments even further such that all Wigner distributions can only depend on the arguments 
\begin{equation}
	W(\vec{p},\vec{b}) = W[\vec{p}^2,\vec{b}^2,(\vec{p}\cdot\vec{b})^2].
\end{equation}
These arguments are not boost-invariant quantities, but I can recover their boost invariant analogs by defining the longitudinal momentum fraction of the nucleon in the nucleus as $\alpha = p^+/P^+$ and noting that, in the nuclear rest frame, $\vec{p} = [\uline{p},M_A(\alpha-1/A)$ and $\vec{b} = [\uline{b},-g_{+-}P^+b^-/M_A]$. Then I can separate the transverse, longitudinal, and cross-term variables. In the end, all Wigner distributions can only depend on 
\begin{align}\label{Wigner_arg}
	&W(\alpha,\uline{p};b^-,\uline{b}) \Rightarrow W\big[p_T^2,b_T^2;(\uline{p}\cdot\uline{b})_T^2;\big(\alpha - \frac{1}{A}\big)^2,(P^+b^-)^2;(\uline{p}\cdot\uline{b})\big(\alpha - \frac{1}{A}\big)(P^+b^-)\big]
\end{align}
In truth, the integral as written in Eq.~\eqref{pol_nuc} will not treat each of these arguments equally; there is no sensitivity to the direction $\uline{b}$ after the $d^2b_\bot$ integral, and the linear $(\vec{p}\cdot\vec{b})$ argument would not survive this integral, meaning that the arguments $(\vec{p}\cdot\vec{b})_T^2$ and $(\uline{p}\cdot\uline{b})\big(\alpha - \frac{1}{A}\big)(P^+b^-)$ do not contribute. I keep these arguments, though, because both of these statements are only true in the case that I make use of the eikonal symmetric dipole scattering amplitude $S_{(r_T,b_T)}^{[\infty^-,b^-]}$. This scattering amplitude has the definition
\begin{equation}\label{eikonal_s}
	S_{xy}[\infty^-,b^-]  = \mathrm{exp}\Big[-\frac{1}{2}|x-y|_T^2\,Q_s^2(|\tfrac{x+y}{2}|_T)\Big(\frac{R^-(|\tfrac{x+y}{2}|_T)-b^-}{2R^-(|\frac{x+y}{2}|_T)}\Big)\,\ln\frac{1}{|x-y|_T\Lambda}\Big]
\end{equation}
which is a function of the saturation scale $Q_s^2(b_T)$ and the radius of the nucleon $R^-(b_T)$. It has been found that sub-eikonal contributions such as medium-induced drift can affect this dipole amplitude \cite{Antiporda:2021hpk, Bahder:2024jpa}, and will be sensitive to the direction of the transverse position of the nucleon $\uline{b}$. In this case I would indeed need the extra Wigner distribution arguments, so I elect to keep them here, even if the arguments are suppressed in future equations to save space.

I can rewrite the rotational invariance in terms of the boost-invariant quantities and thus identify the spin-dependent structure that contributes to the nuclear TMDs related to the Wigner distribution for an unpolarized nucleon in a polarized nucleus
\begin{equation}
	-S^{\mu}W_{pol,\,\mu}^{unp}(\alpha,\uline{p};b^-,\uline{b}) = \big[\Lambda(\uline{b}\times\uline{p}) - \frac{1}{M_A}(g_{+-}P^+b^-)(\uline{p}\times\uline{S}^A) + \big(\alpha - \frac{1}{A}\big)M_A(\uline{S}^A\times\uline{b})]W_{OAM}^{unp,pol},
\end{equation}
where I have suppressed the arguments of the Wigner distribution for brevity, and $\Lambda$ and $\uline{S}^A$ are the longitudinal and transverse components of the nucleus' spin vector.

I can now need to determine the spin-structure of the Wigner distribution for a polarized nucleon in a polarized nucleus. This Wigner distribution is dependent on all four vectors, $\vec{p}$,$\vec{b}$, $\vec{S}^N$, and $\vec{S}^A$. I impose rational invariance on this Wigner distribution in a similar way as above, taking care to only keep terms spin-structures that are $\mathcal{P}$-even and $\mathcal{T}$-even and remembering that the nucleus carries the index $\mu$ while the nucleon carries the index $\nu$:
\begin{align}\label{wigner_decomp}
	S^{\mu}S_{\nu}W_{pol,\,\mu}^{pol,\,\nu} &\stackrel{R.F}{=} \Big[\vec{S}^A\cdot\vec{S}^N\Big]W_{SS} + \Big[\big(\vec{S}^A\cdot(\vec{b}\times\vec{p})\big)\big((\vec{b}\times\vec{p})\cdot\vec{S}^N\big)\Big]W_{SL} \\
	&\qquad +M_A^2\Big[\big(\vec{S}^A\cdot\vec{b}\big)\big(\vec{b}\cdot \vec{S}^N\big)\Big]W_{SbS} + \frac{1}{M_A^2}\Big[\big(\vec{S}^A\cdot\vec{p}\big)\big(\vec{p}\cdot\vec{S}^N\big)\Big]W_{SpS}. \notag
\end{align}
Transforming these terms into their boost-invariant quantities is not for the faint of heart, but after meticulously caring for each term and performing some vector algebra, I combine like terms and separate the Wigner distributions into their transverse and longitudinal contributions. These Wigner distributions are
\begin{subequations}
	\begin{align}
		W_{SS}^L&\equiv W_{SS} + g_{+-}^2(P^+b^-)^2\,W_{SbS} + \big(\alpha - \frac{1}{A}\big)^2\,W_{SpS} + (b_T^2p_T^2-(\uline{b}\cdot\uline{p}^2)\,W_{SL} \\[0.3cm]
		W_{SS}^T &\equiv W_{SS} + \Big(\frac{g_{+-}^2}{M_A^2}(P^+b^-)^2p_T^2 + 2g_{+-}(P^+b^-)\big(\alpha - \frac{1}{A}\big)(\uline{b}\cdot\uline{p}) + M_A^2\big(\alpha - \frac{1}{A}\big)^2b_T^2\Big)\,W_{SL} \\[0.5cm]
		W_{SbS}^L &\equiv \Big(\frac{g_{+-}}{M_A^2}(P^+b^-)p_T^2 + (\alpha-\frac{1}{A}\big)(\uline{b}\cdot\uline{p})\Big)\,W_{SL} - g_{+-}(P^+b^-)\,W_{SbS} \\[0.3cm]
		W_{SbS}^T &\equiv W_{SbS} - \big(\alpha - \frac{1}{A}\big)^2\,W_{SL} \\[0.5cm]
		W_{SpS}^L & \equiv \big(\alpha-\frac{1}{A}\big)\,W_{SpS} - \Big(g_{+-}(P^+b^-)(\uline{b}\cdot\uline{p}) + M_A^2\big(\alpha-\frac{1}{A}\big)b_T^2\Big)\,W_{SL} \\[0.3cm]
		W_{SpS}^T &\equiv W_{SpS} - g_{+-}^2(P^+b^-)^2\,W_{SL},
	\end{align}
\end{subequations}
along with $W_{SL}$ itself, such that I can fully decompose the structure that contributes to the quark TMD of a polarized nucleus:
\begin{align}
	&-S^{\mu}\,W_{pol,\,\mu}^{unp} + S^{\mu}S_{\nu}W_{pol,\,\mu}^{pol,\,\nu} \stackrel{R.F.}{=} \\
	&\hspace{2.5cm}\Big[\Lambda\,(\uline{b}\times\uline{p}) - \frac{1}{M_A}g_{+-}(P^+b^-)(\uline{p}\times\uline{S}^A) + M_A\,\big(\alpha-\frac{1}{A}\big)\,(\uline{S}^A \times \uline{b})\Big]\,W_{OAM}^{unp,pol}  \notag \\
	&\hspace{3cm}- g_{+-}(P^+b^-)\big(\alpha-\frac{1}{A}\big)\Big[(\uline{S}^A\cdot\uline{p})(\uline{b}\cdot\uline{S}^N) + (\uline{S}^A\cdot\uline{b})(\uline{p}\cdot\uline{S}^N)\Big]\,W_{SL} \notag \\
	&\hspace{3cm} + M_A^2(\uline{S}^A\cdot\uline{b})(\uline{b}\cdot\uline{S}^N)\,W_{SbS}^T + \frac{1}{M_A^2}(\uline{S}^A\cdot\uline{p})(\uline{p}\cdot\uline{S}^N)\,W_{SpS}^T \notag\\
	&\hspace{3cm}  + M_A\Lambda(\uline{S}^N\cdot\uline{b})\,W_{SbS}^L + M_A\lambda(\uline{S}^A\cdot\uline{b})\,W_{SbS}^L \notag \\
	&\hspace{3cm} + \frac{1}{M_A}\Lambda(\uline{S}^N\cdot\uline{p})\,W_{SpS}^L + \frac{1}{M_A}\lambda(\uline{S}^A\cdot\uline{p})\,W_{SpS}^L \notag \\
	&\hspace{3cm} + (\uline{S}^A\cdot\uline{S}^N)\,W_{SS}^T + \Lambda\lambda\,W_{SS}^L \notag .
\end{align}
I also write this in the notation of Eq.~\eqref{W_decomp}; for the case of an unpolarized nucleon I get
\begin{subequations}\label{w_unp_decomp}
	\begin{align}
		W_{pol, 0}^{unp}(p,b) &\stackrel{R.F}{=} 0 \\[0.3cm]
		W_{pol,\,3}^{unp}(p,b) &\stackrel{R.F}{=}\epsilon^{mn}_T\,b_\bot^mp_\bot^n\,W_{OAM}^{unp,pol} \\[0.3cm]
		W_{pol,\bot_k}^{unp}(p,b) &\stackrel{R.F}{=} -\Big(\frac{g_{+-}}{M_A}(P^+b^-)p_\bot^m + M_A\big(\alpha-\frac{1}{A}\big)b_\bot^m\Big)\,\epsilon_{T,\,k}^m\,W_{OAM}^{unp,pol}
	\end{align}
\end{subequations}
while for the case of a polarized nucleon I get
\begin{subequations}\label{w_pol_decomp}
	\begin{align}
		W_{pol,\,\mu}^{pol,0}(p,b) &\stackrel{R.F}{=} 0 \\[0.3cm]
		W_{pol,\,3}^{pol,\,3}(p,b) &\stackrel{R.F}{=} W_{SS}^L \\[0.3cm]
		W_{pol,\,\bot k}^{pol,\,3}(p,b) &\stackrel{R.F}{=} M_A\,b_{\bot k}\,W_{SbS}^L + \frac{1}{M_A}\,p_{\bot k}\,W_{SpS}^L \\[0.3cm]
		W_{pol,\,3}^{pol,\,\bot j}(p,b) &\stackrel{R.F}{=} M_A\,b_\bot^j\,W_{SbS}^L + \frac{1}{M_A}\,p_\bot^j\,W_{SpS}^L \\[0.3cm]
		W_{pol,\,\bot k}^{pol,\,\bot j}(p,b) &\stackrel{R.F}{=} \delta_k^j\,W_{SS}^T + M_A^2\,b_\bot^j\,b_{\bot k}\,W_{SbS}^T + \frac{1}{M_A^2}\,p_\bot^j\,p_{\bot k}\,W_{SpS}^T \\ 
		&\qquad\qquad - g_{+-}(P^+b^-)\big(\alpha-\frac{1}{A}\big)\,b_\bot^j\,p_{\bot k}\,W_{SL} \notag
	\end{align}
\end{subequations}

Now that I are getting into more details, I should take a step back because we have not yet discussed the spin-decomposition of quark correlator, $(\phi^N,\Phi^A)_{\Gamma}^{unp}$ and $(\phi^N,\Phi^A)_{\Gamma}^{pol,\,\nu}$. This is known, and I present its decomposition as taken from the rest-frame \cite{Kovchegov:2015zha}:
\begin{subequations} \label{quark_cor_dec}
	\begin{align}
		\phi_\Gamma^{unp} &= f_1\,\Big[\frac{1}{4}g_{+-}\mathrm{Tr}[\gamma^-\Gamma]\Big] + \Big(\frac{k_\bot^i}{m}\,h_a^\bot\Big)\Big[\frac{i}{4}g_{+-}\mathrm{Tr}[\gamma_{\bot i}\gamma^-\Gamma]\Big] \\[0.3cm]
		\phi_{\Gamma}^{pol,\,0} &\stackrel{R.F}{=} 0 \\[0.3cm]
		\phi_{\Gamma}^{pol,\,3} &\stackrel{R.F}{=} g_1 \Big[\frac{1}{4}g_{+-}\mathrm{Tr}[\gamma^5\gamma^-\Gamma]\Big] + \Big(\frac{k_\bot^i}{m}\,h_{1L}^{\bot}\Big)\Big[\frac{1}{4}g_{+-}\mathrm{Tr}[\gamma^5\gamma_{\bot i}\gamma^-\Gamma]\Big] \\[0.5cm]
		\phi_{\Gamma}^{pol,\, \bot j} &\stackrel{R.F}{=}\big(-\frac{k_\bot^i}{m}\epsilon_T^{ij}f_{1T}^\bot\big)\Big[\frac{1}{4}g_{+-}\mathrm{Tr}[\gamma^-\Gamma]\Big] + \big(\frac{k_\bot^j}{m}g_{1T}\big)\Big[\frac{1}{4}g_{+-}\mathrm{Tr}[\gamma^5\gamma^-\Gamma]\Big]  \\
		&\hspace{1cm} + \big(\delta^{ij}h_{1T} + \frac{k_\bot^i k_\bot^j}{m^2}h_{1T}^\bot\big)\Big[\frac{1}{4}g_{+-}\mathrm{Tr}[\gamma^5\gamma_{\bot i}\gamma^-\Gamma]\Big] \notag
	\end{align}
\end{subequations}
where I note that while this equation was written using the symbol $\phi$, it applies to both the nucleon and nucleus quark correlators. We now have all the information to tackle the polarized part of the quark correlator for a polarized nucleus:
\begin{align}\label{nuc_cor_pol}
	&\Phi^A_{\Gamma\,pol,\,\mu}(x,\uline{k};P) = A\frac{2\,g_{+-}}{(2\pi)^5}\int d^2p\,dp^+\,d^2b_\bot\,db^-\,d^2\uline{k}'\,d^2\uline{r}\,e^{-i(\uline{k}-\uline{k}'-\hat{x}\uline{p})\cdot\uline{r}}\;\\
	&\hspace{5cm} \times  \Big(W_{pol,\,\mu}^{unp}\,\phi^{N\,unp}_{\Gamma} - W_{pol,\,\mu}^{pol,\,\nu}\,\phi^{N\,pol,\,\nu}_{\Gamma}\Big)\;S_{(r_T,b_T)}^{[\infty^-,b^-]}. \notag
\end{align}

Since I am interested in the Sivers TMD $f_{1T}^{\bot A}$, Eqs.~\eqref{quark_cor_dec} say that I must trace over $\Gamma = \gamma^+$ and only take interest for $\mu = \bot_k$. While this results in a direct relationship $\Phi_{\gamma^+\,pol,\,\bot k}^A \propto f_{1T}^{\bot A}$, there are multiple nucleon TMDs that will contribute since there is still a product over the $\nu$ index; the trace over $\gamma^+$ results in non-zero contributions from $f_1^N$ and $f_{1T}^{\bot N}$. Of the traces found in Eq.~\eqref{quark_cor_dec}, $\Gamma = \gamma^+$ only provides a non-zero contribution for $\mathrm{Tr}[\gamma^-\Gamma]$, so for $g_{+-} = 1$,
\begin{align}\
	\phi_{\gamma^+}^{N\,unp}(\hat{x},\uline{k}';p) &\stackrel{R.F.}{=} f_1^N\Big[\frac{1}{4}g_{+-}\mathrm{Tr}[\gamma^-\gamma^+]\Big] = f_1^N \\
	\phi_{\gamma^+}^{N\,pol,\,0}(\hat{x},\uline{k}';p)  &\stackrel{R.F.}{=}0 \\
	\phi_{\gamma^+}^{N\,pol,\,3}(\hat{x},\uline{k}';p)  &\stackrel{R.F.}{=} 0 \\
	\phi_{\gamma^+}^{N\,pol,\,\bot j}(\hat{x},\uline{k}';p)  &\stackrel{R.F.}{=} -\frac{{k'}_\bot^N}{m}\epsilon_T^{nj}\,f_{1T}^{\bot N}.
\end{align}

I then rewrite Eq.~\eqref{nuc_cor_pol} as
\begin{align}
	&-\frac{k_\bot^\ell}{M_A}\epsilon_{T,k}^\ell\,f_{1T}^{\bot A} = -A\frac{2\,g_{+-}}{(2\pi)^5}\int d^2p\,dp^+\,d^2b_\bot\,db^-\,d^2\uline{k}'\,d^2\uline{r}\,e^{-i(\uline{k}-\uline{k}'-\hat{x}\uline{p})\cdot\uline{r}}\;\;S_{(r_T,b_T)}^{[\infty^-,b^-]} \\
	&\hspace{2cm} \times  \Bigg\{ \Big(\frac{g_{+-}}{M_A}(P^+b^-)p_\bot^m + M_A\big(\alpha-\frac{1}{A}\big)b_\bot^m\Big)\,\epsilon_{T,\,k}^m\,W_{OAM}^{unp,pol}\times f_1^N , \notag \\
	&\hspace{3cm} + \Big[\delta_k^j\,W_{SS}^T + M_A^2\,b_\bot^j\,b_{\bot k}\,W_{SbS}^T + \frac{1}{M_A^2}\,p_\bot^j\,p_{\bot k}\,W_{SpS}^T \notag \\
	&\hspace{4cm}- g_{+-}(P^+b^-)\big(\alpha-\frac{1}{A}\big)\,b_\bot^j\,p_{\bot k}\,W_{SL}\Big] \times  \frac{{k'}_\bot^n}{m}\epsilon_T^{nj}\,f_{1T}^{\bot N}\Bigg\} \notag .
\end{align}
From here, all that is left to do is contract the free index $k$ on both sides with $-k_{\bot}^m\epsilon_{T,k}^m$ and solve for the nuclear Sivers TMD to obtain
\begin{align}
	&f_{1T}^{\bot A} = A\frac{2\,g_{+-}\,M_A}{(2\pi)^5\,k_T^2}\int d^2p\,dp^+\,d^2b_\bot\,db^-\,d^2\uline{k}'\,d^2\uline{r}\,e^{-i(\uline{k}-\uline{k}'-\hat{x}\uline{p})\cdot\uline{r}}\;\;S_{(r_T,b_T)}^{[\infty^-,b^-]} \\
	&\hspace{.5cm} \times  \Bigg\{ \Big(\frac{g_{+-}}{M_A}(P^+b^-)(\uline{k}\cdot\uline{p}) + M_A\big(\alpha-\frac{1}{A}\big)(\uline{k}\cdot\uline{b})\Big)\,W_{OAM}^{unp,pol}\times f_1^N , \notag \\
	&\hspace{1.5cm} + \Big[(\uline{k}\cdot\uline{k}')\,W_{SS}^T - M_A^2\,(\uline{b}\times\uline{k}')(\uline{k}\times\uline{b})\,W_{SbS}^T - \frac{1}{M_A^2}(\uline{p}\times\uline{k}')(\uline{k}\times\uline{p})\,W_{SpS}^T \notag \\
	&\hspace{2.5cm}+ g_{+-}(P^+b^-)\big(\alpha-\frac{1}{A}\big)\,\Big((\uline{b}\times\uline{k}')(\uline{k}\times\uline{p})+(\uline{p}\times\uline{k}')(\uline{k}\times\uline{b})\Big)\,W_{SL}\Big] \times  \frac{1}{m}\,f_{1T}^{\bot N}\Bigg\} \notag .
\end{align}
I can do some more vector algebra to better identify the vector projections between the nucleus momentum $\uline{k}$, quark momentum $\uline{k}'$, and nucleon position/impact parameter $\uline{b}$. It may also be useful to reinstitute the original Wigner functions from Eq.~\eqref{wigner_decomp} since I can better correlate the vector projections with the spin-decomposition. Doing so allows me to write the final form for the nuclear Sivers function written in terms of nucleon TMDs with arbitrary spin-states:
\begin{align}\label{final_sivers}
	&f_{1T}^{\bot A} = A\frac{2\,g_{+-}}{(2\pi)^5}\frac{M_A}{k_T^2}\int d^2p\,dp^+\,d^2b_\bot\,db^-\,d^2\uline{k}'\,d^2\uline{r}\,e^{-i(\uline{k}-\uline{k}'-\hat{x}\uline{p})\cdot\uline{r}}\;\;S_{(r_T,b_T)}^{[\infty^-,b^-]} \\
	&\hspace{1cm} \times  \Bigg\{ \bigg(\frac{g_{+-}}{M_A}(P^+b^-)(\uline{k}\cdot\uline{p}) + M_A\big(\alpha-\frac{1}{A}\big)(\uline{k}\cdot\uline{b})\bigg)\,W_{OAM}^{unp,pol}\times f_1^N , \notag \\
	&\hspace{1.5cm} + \bigg((\uline{k}\cdot\uline{k}')\,W_{SS} + M_A^2\,\Big[b_T^2(\uline{k}\cdot\uline{k}')-(\uline{k}\cdot\uline{b})(\uline{k}'\cdot\uline{b})\Big]\,W_{SbS} \notag \\
	&\hspace{2cm}+ \frac{1}{M_A^2}\,\Big[p_T^2(\uline{k}\cdot\uline{k}') - (\uline{k}\cdot\uline{p})(\uline{k}'\cdot\uline{p})\Big]\,W_{SpS} \notag \\
	&\hspace{2cm}+ \Big[\frac{g_{+-}^2}{M_A^2}(P^+b^-)^2(\uline{k}\cdot\uline{p})(\uline{p}\cdot\uline{k}') + M_A^2\big(\alpha-\frac{1}{A}\big)^2(\uline{k}\cdot\uline{b})(\uline{b}\cdot\uline{k}') \notag \\
	&\hspace{3cm} + g_{+-}(P^+b^-)\big(\alpha-\frac{1}{A}\big)\Big((\uline{k}\cdot\uline{p})(\uline{b}\cdot\uline{k}') + (\uline{k}\cdot\uline{b})(\uline{k}'\cdot\uline{p})\Big)\Big]\,W_{SL}\bigg) \times  \frac{1}{m}\,f_{1T}^{\bot N}\Bigg\} \notag .
\end{align}
Where I note that each of the Wigner distributions can only be written in terms of the arguments given by Eq.~\eqref{Wigner_arg}, and the arguments of the TMDs are $f_{1T}^{\bot A}(x,k_T)$, $f_1^N(\hat{x},k_T')$, and $f_{1T}^{\bot N}(\hat{x},k'_T)$.

The purpose of this work is to facilitate further exploration of medium-induced drift effects in Cold Nuclear Matter. At present, Eq.~\eqref{final_sivers} can be simplified under the impression that I only take the eikonal limit of the dipole scattering amplitude; because it is completely independent of the direction of $\uline{b}$ and only dependent on the magnitude $b_T$,any terms that are linear in $\uline{b}$ survive the integral over $d^2b_\bot$ due to $\uline{b}\to-\uline{b}$ antisymmetry. If I remove those terms then I get the simpler expression
\begin{align}\label{sivers_eik}
	&f_{1T}^{\bot A}\bigg|_{\mathrm{eik}} = A\frac{2\,g_{+-}}{(2\pi)^5}\frac{M_A}{k_T^2}\int d^2p\,dp^+\,d^2b_\bot\,db^-\,d^2\uline{k}'\,d^2\uline{r}\,e^{-i(\uline{k}-\uline{k}'-\hat{x}\uline{p})\cdot\uline{r}}\;\;S_{\mathrm{eik}}^{[\infty^-,b^-]} \\
	&\hspace{1cm} \times  \Bigg\{ \frac{g_{+-}}{M_A}(P^+b^-)(\uline{k}\cdot\uline{p})\,W_{OAM}^{unp,pol}\times f_1^N , \notag \\
	&\hspace{1.5cm} + \frac{1}{m_N}\Big[(\uline{k}\cdot\uline{k}')\widetilde{W}_{SS} - \frac{1}{M_A^2}(\uline{k}\cdot\uline{p})(\uline{p}\cdot\uline{k}')\widetilde{W}_{SpS}\Big] \times  \,f_{1T}^{\bot N}\Bigg\} \notag .
\end{align}
where I have take the liberty of defining the new Wigner distributions
\begin{subequations}
	\begin{align}
		\widetilde{W}_{SS} &\equiv W_{SS} + \frac{1}{2}b_T^2M_A^2\,W_{SbS} + \frac{1}{M_A^2}p_T^2\,W_{SpS} + \frac{1}{2}b_T^2M_A^2\,\big(\alpha-\frac{1}{A}\big)^2\,W_{SL} \\[0.5cm]
		\widetilde{W}_{SpS} &\equiv W_{SpS} - g_{+-}^2(P^+b^-)^2\,W_{SL}
	\end{align}
\end{subequations}
which have more restricted arguments 
\begin{equation}
	W(p,b) \Rightarrow W\Big[p_T^2,b_T^2;\big(\alpha-\frac{1}{A}\big)^2, (P^+b^-)^2\Big].
\end{equation}

In the future I will be able to take Eq.~\eqref{final_sivers} and apply the sub-eikonal correction to the dipole scattering amplitude and investigate how that impacts the spin-structure and quantify the change in the Sivers contribution due to the sub-eikonal effect. The best inferred form the sub-eikonal dipole scattering amplitude is given by
\begin{equation}
	S_{\uline{b},\uline{r}}[\infty^-, b^-]\Bigg|_{\mathrm{sub\text{-}eik}} =\Bigg[\bm{1} + \frac{1}{E\,r_T^2}\frac{\uline{r}\cdot\uline{u}(b^-)}{1-u_{\parallel}(\uline{b})}\Bigg] e^{-\frac{1}{2}r_T^2\,Q_s^2(b_T)\Big(\frac{R^-(b_T)-b^-}{2R^-(b_T)}\Big)\,\ln\frac{1}{r_T\Lambda}}
\end{equation}
where $\uline{u}$ and $u_{\parallel}$ is the momentum of the nucleon which I can approximate as a point moving inside the rigid rotator of the nucleus; this momentum is what constitutes the ``medium-induced" part of the sub-eikonal drift correction, and it will depend more precisely on the nucleon position than just its transverse magnitude. A study of this effect is left for future work.
	\newpage

	\addcontentsline{toc}{section}{REFERENCES}
	\setlength{\baselineskip}{\singlespace}

	%
\end{document}